%% file: tesis-lanl.tex
\documentclass[oneside,epsfig,12pt]{book}
\usepackage[centertags]{amsmath}
\usepackage{amscd}
\usepackage{amsfonts}
\usepackage{amssymb}
\usepackage{amsthm}
\usepackage{graphicx}
\usepackage[spanish]{babel}
\usepackage{cite}
\usepackage{epsfig}

\frontmatter
\begin{document}
\thispagestyle{empty}
\begin{center}
\vskip 3cm
{\bf\Large PhD Thesis}\\[.25cm]
{\bf\large Departamento de F\'{\i}sica\\[.09cm]
Facultad de Ciencias Exactas\\[.1cm]
Universidad Nacional de La Plata}\\[4cm]
{\bf\LARGE Propagation of High Energy}\\[.2cm]
{\bf\LARGE Galactic Cosmic Rays}\\[.3cm] 
{\Large (in Spanish)}\\[5 cm]
{\bf\Large Juli\'an Candia\\[.3cm]
Advisor: Esteban Roulet\\[.1cm]
Co-Advisor: Luis N. Epele\\[.1cm]
\vskip 2 cm {\large La Plata, 2004}}
\end{center}
\newpage
\include{abstract}
\thispagestyle{empty}
\begin{center}
\vskip 3cm
{\bf\Large Tesis Doctoral}\\[.25cm]
{\bf\large Departamento de F\'{\i}sica\\[.09cm]
Facultad de Ciencias Exactas\\[.1cm]
Universidad Nacional de La Plata}\\[4cm]
{\bf\LARGE Propagaci\'on de rayos c\'osmicos gal\'acticos\\[.2cm] 
de muy alta energ\'{\i}a}\\[5 cm]
{\bf\Large Juli\'an Candia\\[.3cm]
Director: Esteban Roulet\\[.1cm]
Co-Director: Luis N. Epele\\[.1cm]
\vskip 2 cm {\large La Plata, 2004}}
\end{center}
\newpage
\newpage
{\it A mi abuela Chela y a mis viejos.

Y en homenaje a la tenacidad de mis ancestros.} 
\newpage
\newpage
\chapter{Prefacio}
A\'un cuando el fen\'omeno de los rayos c\'osmicos viene siendo estudiado en la f\'{\i}sica desde hace casi un siglo,  
sigue siendo en la actualidad un campo de intensa investigaci\'on te\'orica y experimental.
Los rayos c\'osmicos de baja energ\'{\i}a, que pueden ser detectados mediante globos, cohetes y sat\'elites, son 
esencialmente el \'unico tipo de materia de origen c\'osmico del que puede tenerse evidencia directa, mientras que, 
en el otro extremo del espectro, los rayos c\'osmicos ultra-energ\'eticos pueden alcanzar energ\'{\i}as de decenas de 
Joules y constituyen, en efecto, las part\'{\i}culas m\'as 
energ\'eticas conocidas en el Universo.  

M\'as all\'a del inter\'es intr\'{\i}nseco asociado a este fen\'omeno, la f\'{\i}sica de rayos c\'osmicos est\'a 
en \'{\i}ntima relaci\'on con otros campos de investigaci\'on (la astronom\'{\i}a, la f\'{\i}sica de part\'{\i}culas, 
la astrof\'{\i}sica, la cosmolog\'{\i}a, la f\'{\i}sica nuclear, la f\'{\i}sica de plasmas, etc.), de modo que el
progreso en esta disciplina tambi\'en contribuye significativamente al desarrollo de otras \'areas de la f\'{\i}sica
y la astronom\'{\i}a.

Pese a la notable antig\"uedad de este problema, su explicaci\'on no est\'a a\'un bien establecida. En verdad, sabemos que 
el fen\'omeno de los rayos c\'osmicos depende del rango de energ\'{\i}as que se considere, 
y requiere desarrollar, en cada caso,
la explicaci\'on apropiada que d\'e cuenta de su naturaleza, de los mecanismos que los originan, y de la manera 
en que se propagan desde las fuentes hasta nuestra ubicaci\'on como observadores. 
El Cap\'{\i}tulo 1 presenta una breve introducci\'on general a la f\'{\i}sica de rayos c\'osmicos; all\'{\i} se brinda 
una revisi\'on hist\'orica de la disciplina, y luego se presentan los aspectos esenciales acerca de la aceleraci\'on,
la propagaci\'on y la detecci\'on de los rayos c\'osmicos.   

Los Cap\'{\i}tulos siguientes contienen las contribuciones originales de esta Tesis Doctoral,
en la que se exploran diversos aspectos relacionados con la propagaci\'on de rayos c\'osmicos gal\'acticos de muy alta
energ\'{\i}a (es decir, en el rango $E\simeq 10^{15}-10^{18}$~eV, que comprende la rodilla y la segunda rodilla del espectro)
utilizando diferentes herramientas te\'oricas, tanto anal\'{\i}ticas como num\'ericas. 
 
En el Cap\'{\i}tulo 2 se estudia la propagaci\'on de rayos c\'osmicos en los campos magn\'eticos turbulentos 
de la galaxia; all\'{\i} se muestra que la rodilla y la segunda rodilla del espectro pueden 
ser explicadas en este escenario como un {\it crossover} entre dos reg\'{\i}menes difusivos 
diferentes, a saber, la difusi\'on transversal (que domina a bajas energ\'{\i}as) y la 
antisim\'etrica (o {\it drift}) que prevalece a energ\'{\i}as altas. 
Mientras que la supresi\'on de los protones produce la primera rodilla, la de la componente pesada 
(fundamentalmente, n\'ucleos de hierro) da lugar a la segunda. 
La apropiada inclusi\'on de los efectos de drift en las ecuaciones de transporte 
difusivo no s\'olo explica en forma natural la rodilla y la segunda rodilla del espectro, 
sino tambi\'en el comportamiento observado para la composici\'on y las anisotrop\'{\i}as
en el rango de energ\'{\i}as que se extiende desde la rodilla hasta el tobillo del espectro. 
Por otro lado, se lleva a cabo un estudio del comportamiento del tensor de difusi\'on en condiciones
de alta turbulencia. Una aplicaci\'on particular de este estudio es la descripci\'on detallada de la difusi\'on de rayos c\'osmicos 
en la galaxia para energ\'{\i}as en la regi\'on de la rodilla, aunque los resultados satisfacen relaciones 
de escala que permiten ser aplicados en otros problemas de inter\'es astrof\'{\i}sico.
Los resultados que se presentan en este Cap\'{\i}tulo fueron publicados en los siguientes art\'{\i}culos:
\begin{itemize}
\item J. Candia, E. Roulet and L.N. Epele, {\it Turbulent diffusion and drift in galactic 
magnetic fields and the explanation of the knee in the cosmic ray spectrum}, 
Journal of High Energy Physics JHEP {\bf{12}} (2002) 033.
\item J. Candia, S. Mollerach and E. Roulet, {\it Cosmic ray drift, the second knee and galactic 
anisotropies}, Journal of High Energy Physics JHEP {\bf{12}} (2002) 032.
\item J. Candia, S. Mollerach and E. Roulet, {\it Cosmic ray spectrum and anisotropies from the knee to the second knee}, 
Journal of Cosmology and Astroparticle Physics JCAP {\bf{05}} (2003) 003.
\item J. Candia and E. Roulet, {\it Diffusion and drift of cosmic rays in highly turbulent magnetic fields}, 
Journal of Cosmology and Astroparticle Physics JCAP {\bf{10}} (2004) 007.
\end{itemize}

El Cap\'{\i}tulo 3 investiga el impacto en la predicci\'on de los flujos de neutrinos de altas energ\'{\i}as 
que resulta de considerar escenarios en los que la rodilla del espectro de rayos c\'osmicos 
es un efecto dependiente de la rigidez magn\'etica, como corresponde, por ejemplo, al escenario de la difusi\'on 
turbulenta y drift. Se encuentra que el fondo de neutrinos atmosf\'ericos 
se reduce significativamente, facilitando la b\'usqueda de fuentes astrof\'{\i}sicas.
As\'{\i}, se pone en evidencia la necesidad de resolver el origen de la rodilla, con el fin de 
interpretar correctamente las observaciones en los nuevos telescopios de neutrinos.
Este fen\'omeno se estudia en el canal usual de trazas lept\'onicas (asociado a la
detecci\'on de neutrinos mu\'on), y luego en 
el canal complementario de las cascadas hadr\'onica y electromagn\'etica (asociado a la detecci\'on 
de neutrinos electr\'on), que se propone como 
un nuevo m\'etodo para aislar el flujo de neutrinos atmosf\'ericos prompt. 
Estas investigaciones dieron lugar a las siguientes publicaciones:   
\begin{itemize}
\item J. Candia and E. Roulet, {\it Rigidity dependent knee and cosmic ray induced high energy neutrino fluxes}, 
Journal of Cosmology and Astroparticle Physics JCAP {\bf{09}} (2003) 005.
\item J.F. Beacom and J. Candia, {\it Shower power: isolating the prompt atmospheric neutrino flux using 
electron neutrinos}, Journal of Cosmology and Astroparticle Physics (en prensa).
\end{itemize}

El Cap\'{\i}tulo 4 explora en detalle el modelo en el que la rodilla del espectro 
se debe a efectos de fotodesintegraci\'on nuclear
por la interacci\'on con fotones \'opticos y UV blandos en la regi\'on de las fuentes. Este escenario ofrece una explicaci\'on 
alternativa a la rodilla del espectro, y predice, contrariamente a lo que resulta de los escenarios dependientes de la
rigidez, una composici\'on que se torna m\'as liviana por encima de la rodilla. Este trabajo fue publicado en el art\'{\i}culo
siguiente:    
\begin{itemize}
\item J. Candia, L.N. Epele and E. Roulet, {\it Cosmic ray photodisintegration and the knee 
of the spectrum}, Astroparticle Physics {\bf{17}} (2002) 23. 
\end{itemize}

El Cap\'{\i}tulo 5 contiene las Conclusiones de esta Tesis. Finalmente, 
se presentan los Ap\'endices y la Bibliograf\'{\i}a. 

Esta Tesis Doctoral se desarroll\'o con el apoyo de una beca de postgrado interna otorgada por el 
Consejo Nacional de Investigaciones Cient\'{\i}ficas y T\'ecnicas (CONICET) en el per\'{\i}odo 04/2000\ --\ 03/2004, y 
un subsidio otorgado por la Fundaci\'on Antorchas en el per\'{\i}odo 05/2003\ --\ 04/2004. Este trabajo fue concluido
en Fermilab (Batavia IL, USA) en el marco del Programa para doctorandos latinoamericanos, en el per\'{\i}odo
03/2004\ --\ 08/2004. 

\newpage
\chapter{Agradecimientos}
\begin{itemize}
\item A mi Director, Esteban Roulet, con la admiraci\'on, el respeto y la gratitud de un disc\'{\i}pulo a su maestro.
\item A mi Co-Director, Luis Epele, y a mis colaboradores, Silvia Mollerach y John Beacom.
\item A los grupos de trabajo que integr\'e, o con quienes estuve en contacto durante mi doctorado: 
el Grupo de Fenomenolog\'{\i}a de Altas Energ\'{\i}as de Luis Epele y Tere Dova, 
el Grupo de F\'{\i}sica Computacional de Ezequiel Albano,
el Grupo de Part\'{\i}culas del Centro At\'omico Bariloche, y los Grupos de F\'{\i}sica Te\'orica y Astrof\'{\i}sica del
Fermilab.  
\item A Carlos Na\'on, a Virginia Man\'{\i}as, y a los dem\'as amigos del barrio de la Oficina 25.
\item A mi familia y a mis amigos, por el apoyo, el afecto y la compa\~n\'{\i}a de siempre.  
\end{itemize}
\newpage
\tableofcontents
\mainmatter
\include{capitulo1}
\include{capitulo2}

\include{capitulo3}
\include{capitulo4}
%\backmatter
\include{conclusiones}

\include{apendices}

\end{document}

%% file: abstract.tex
\begin{center}
\vskip 7cm
{\bf\LARGE Abstract}\\[1.5cm]
\end{center}
We explore several aspects related to the propagation of high energy cosmic rays (CRs) of galactic
origin (i.e. in the range $E\simeq 10^{15}-10^{18}$~eV). In particular, we study in detail 
the diffusion/drift scenario, a rigidity-dependent model in which the knee of the spectrum is explained 
as due to the escape of the light CR component. We show that this scenario also explains the second knee 
(which is produced by the leakage of heavy nuclei), as well as the most recent observations of composition and 
anisotropies (amplitude and first harmonic in right ascension) in the range from the knee to the
ankle. Moreover, the behavior of the diffusion tensor for charged particles propagating in highly turbulent
magnetic fields is determined numerically.    

We also investigate the high energy neutrino fluxes induced by CRs, and we show that a 
rigidity-dependent scenario for the CR knee has the effect of reducing the background of
atmospheric neutrinos, hence encouraging the search for high energy astrophysical neutrino sources.  
After exploring this effect in the detection channel of lepton tracks, we investigate the 
less studied ``shower'' channel. As we show here, the shower detection channel is an effective means 
for isolating the prompt atmospheric neutrino component, and it is an important complement to the 
usually-considered track channel.   

Finally, we consider in detail the nuclear photodisintegration model as an alternative explanation
for the knee in the CR spectrum. In this scenario, CR nuclei emitted by a source are disintegrated 
during their propagation through a surrounding optical/UV photon field. Hence, the composition is predicted to
turn lighter across the knee, in constrast to the results obtained in rigidity-dependent scenarios.  
 

%% file: capitulo1.tex
\chapter{La f\'{\i}sica de rayos c\'osmicos}
\begin{center}
\begin{minipage}{5.6in}
\textsl{
Aqu\'{\i} se ofrece un breve resumen de los aspectos esenciales de la f\'{\i}sica de rayos c\'osmicos,
con el objetivo de presentar un contexto a las contribuciones originales de esta Tesis.
Se incluye una lista de publicaciones de referencia
que, desde diferentes enfoques, describen los progresos en este campo de investigaci\'on.}
\end{minipage}
\end{center}

\section{Introducci\'on}
Existe una extensa bibliograf\'{\i}a que cubre el desarrollo hist\'orico de la f\'{\i}sica de rayos c\'osmicos, que expone con 
gran detalle diversos aspectos de esta disciplina, a saber, el origen, la propagaci\'on y la detecci\'on de los rayos 
c\'osmicos en los diferentes rangos de energ\'{\i}a relevantes, y que resume las principales contribuciones que se han 
desarrollado, desde la teor\'{\i}a o el experimento, para comprender este fen\'omeno. 

Los vol\'umenes de Longair \cite{lon92} presentan la astrof\'{\i}sica de altas energ\'{\i}as a un nivel de pregrado, 
poniendo especial \'enfasis en el estudio de los rayos c\'osmicos; Gaisser \cite{gai90} desarrolla la f\'{\i}sica de rayos 
c\'osmicos desde su conexi\'on con la f\'{\i}sica de part\'{\i}culas, y dedica una considerable atenci\'on 
a diversos aspectos de la detecci\'on (desarrollo de lluvias atmosf\'ericas, parametrizaci\'on de los observables, detectores,
t\'ecnicas de simulaci\'on, etc.); Berezinskii {\it et al.} \cite{ber90} presentan un estudio abarcativo 
de toda la astrof\'{\i}sica de rayos c\'osmicos, que resume el estado de la disciplina hasta 1990;   
mientras que Sokolsky \cite{sok89} se ocupa, en particular, de los rayos c\'osmicos ultra-energ\'eticos. 
Un libro m\'as reciente, de Schlickeiser \cite{sch02}, trata con gran detalle los aspectos asociados a la propagaci\'on.

Con respecto a los trabajos de revisi\'on, aqu\'{\i} citaremos a Blandford y Eichler \cite{bla87} en mecanismos de aceleraci\'on, 
a Cesarsky \cite{ces80,ces87} en propagaci\'on, y a Wiebel-Sooth y Biermann \cite{wie98} en la compilaci\'on de datos experimentales. 
Trabajos de revisi\'on m\'as breves y recientes
que cubren diversos aspectos de la f\'{\i}sica de rayos c\'osmicos son, por ejemplo, los de Boratav y Watson \cite{bor00}, 
Kampert \cite{kam01a}, Biermann y Sigl \cite{bie01}, Roulet \cite{rou03} y H\"orandel \cite{hoe04}; 
existe adem\'as un gran n\'umero de publicaciones de revisi\'on 
referidas espec\'{\i}ficamente a los rayos c\'osmicos ultra-energ\'eticos (por ejemplo, \cite{yos98,nag00,oli00,wat00,bha00,cro04}), 
un t\'opico de gran inter\'es actual, pero que escapa al tema espec\'{\i}fico de esta Tesis.   
Finalmente, deben mencionarse los {\it Proceedings of the ICRC (International Cosmic Ray Conference)}, 
las reuniones internacionales bienales que convocan a una parte significativa de la comunidad cient\'{\i}fica del \'area. 
Las reuniones m\'as recientes a la fecha fueron las de Utah (1999), Hamburg (2001) y Tsukuba (2003).

\section{Una breve revisi\'on hist\'orica}
El estudio de los rayos c\'osmicos se origin\'o a comienzos del siglo XX, vinculado a las investigaciones sobre radioactividad 
que se llevaban a cabo con gran \'exito en esa \'epoca. 
Un instrumento muy utilizado en esos experimentos era el electroscopio, 
que permit\'{\i}a medir la ionizaci\'on del gas contenido en el aparato (y, en consecuencia, la cantidad
de radiaci\'on recibida) a partir de la velocidad con que las l\'aminas cargadas del instrumento se aproximaban entre s\'{\i}. 
Una peculiar observaci\'on fue que las l\'aminas del electroscopio resultaban descargarse a\'un en ausencia de fuentes 
radioactivas; con el fin de explicarla, se realizaron diversos experimentos que, en 
principio, buscaban atribuir el fen\'omeno a la radioactividad natural de la superficie terrestre, a la del recipiente, 
o bien a posibles emanaciones radioactivas en el propio gas. 

En 1910, Wulf encontr\'o que la radiaci\'on descend\'{\i}a un $40\%$ 
al subir a la torre Eiffel, de 330~m de altura. Este experimento suger\'{\i}a, de todos modos, que la radiaci\'on 
observada no era emitida por la superficie de la tierra; 
usando el coeficiente de absorci\'on de los rayos $\gamma$ en el aire, ya conocido 
en ese momento, se esperaba que la radiaci\'on descendiera a la mitad en s\'olo 80~m, de forma que deb\'{\i}a ser 
totalmente despreciable en el punto m\'as alto de la torre.
La evidencia definitiva sobre el origen c\'osmico de la radiaci\'on provino de los 
vuelos en globo realizados por Hess a partir de 1912. En particular, en su vuelo del 7 de agosto de 1912, alcanz\'o una 
altitud de 5~km y pudo observar que la tasa de ionizaci\'on resultaba varias veces mayor que la medida a nivel del 
mar \footnote{La conclusi\'on de las observaciones de Hess di\'o lugar al descubrimiento definitivo de los rayos c\'osmicos: 
``The results of the present observations seem to be most readily explained by the assumption that a radiation of very high 
penetrating power enters our atmosphere from above, and still produces in the lower layers a part of the ionisation observed 
in closed vessels.'' \cite{hes12}.}; \'estos resultados luego fueron confirmados por Kolh\"orster en vuelos que realiz\'o 
entre 1913 y 1914, alcanzando altitudes de hasta 9 km.  
Durante los a\~nos siguientes, se estudi\'o este fen\'omeno asumiendo que se trataba de alg\'un tipo de
radiaci\'on $\gamma$ con un poder de penetraci\'on mayor que el observado en la radioactividad natural, 
lo que incidentalmente sugiri\'o a Millikan, en 1925, llamarla con el nombre que perdura hasta nuestros d\'{\i}as.
En 1927 fue descubierto el efecto geomagn\'etico (es decir, que la ionizaci\'on producida por los rayos c\'osmicos 
depende de la latitud), lo que di\'o lugar a establecer definitivamente hacia 1936 que los rayos c\'osmicos primarios 
eran part\'{\i}culas cargadas; m\'as tarde se demostr\'o que los rayos c\'osmicos eran fundamentalmente protones, 
aunque tambi\'en conten\'{\i}an n\'ucleos de elementos m\'as pesados, y una proporci\'on peque\~na (inferior al $1\%$) de 
electrones relativistas.    

El desarrollo de nuevos instrumentos de medici\'on, como por ejemplo la c\'amara de niebla de Skobeltsyn, o el 
detector Geiger-M\"uller, ambos inventados en 1929, hizo posible profundizar las investigaciones sobre la naturaleza de
los rayos c\'osmicos, y promovi\'o a la vez una gran evoluci\'on en la f\'{\i}sica de part\'{\i}culas. En efecto, 
durante las dos d\'ecadas siguientes, los rayos c\'osmicos representaron un acelerador de part\'{\i}culas natural, 
es decir, una fuente gratuita e inagotable de part\'{\i}culas de alta energ\'{\i}a con la atm\'osfera como blanco; 
s\'olo era necesario construir el detector para estudiar los productos resultantes de las colisiones. Siguiendo esa 
l\'{\i}nea de investigaci\'on, en ese per\'{\i}odo fueron descubiertos los positrones, muones, piones, kaones e hyperones; 
m\'as tarde, 
a partir de los a\~nos `50, la tecnolog\'{\i}a de los aceleradores lleg\'o a producir haces de part\'{\i}culas de 
energ\'{\i}as comparables a la contenida en los rayos c\'osmicos, con la ventaja adicional de poder manipular y controlar 
los eventos producidos, y entonces la f\'{\i}sica de part\'{\i}culas se traslad\'o definitivamente 
a los grandes laboratorios.          

\begin{figure}[t]
\centerline{{\epsfxsize=3.5in\epsfxsize=4.2in\epsffile{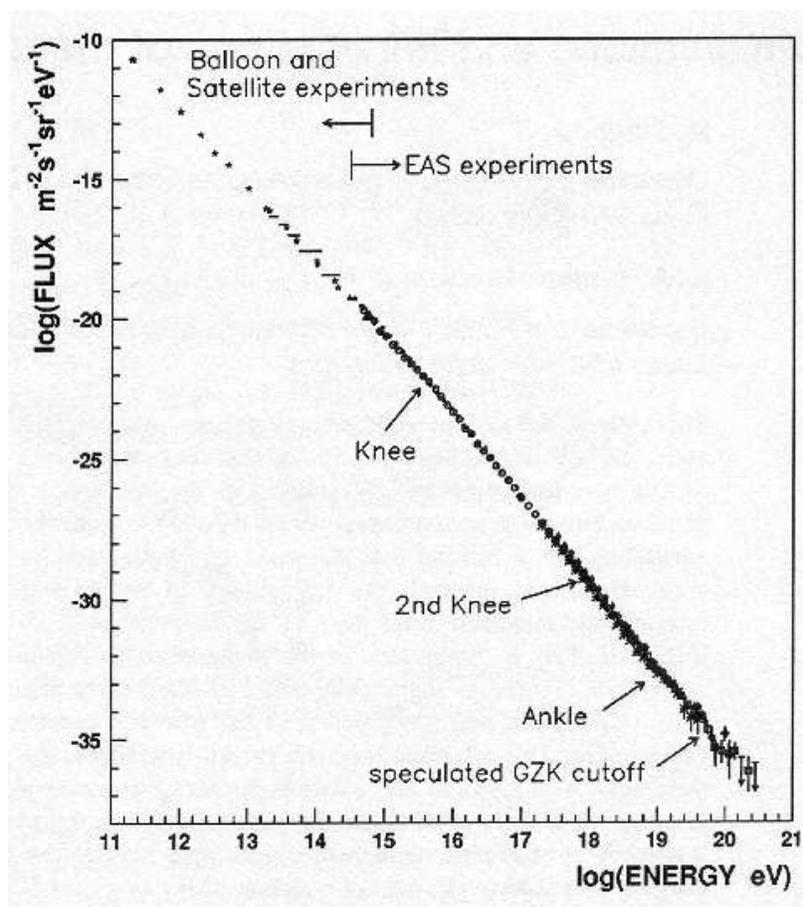}}}
\caption{El espectro de rayos c\'osmicos en funci\'on de la energ\'{\i}a. El \'{\i}ndice espectral es aproximadamente 
constante; sin embargo, exhibe cambios, peque\~nos pero bien definidos, que dan lugar a la rodilla, la segunda rodilla y 
el tobillo del espectro, se\~nalados en la figura (de \cite{nag00}).}  
\label{spectrum}
\end{figure}

Entre las particularidades m\'as notables del fen\'omeno de los rayos c\'osmicos, sin duda la m\'as llamativa es el espectro
(ver la fig.\ref{spectrum}), que se presenta como una ley de potencias que se extiende por m\'as de 11 d\'ecadas en la 
energ\'{\i}a, desde $10^9$ hasta m\'as all\'a de $10^{20}$~eV, con un \'{\i}ndice espectral esencialmente constante.  
El espectro ${\rm d}N/{\rm d}E\propto E^{-\alpha}$ tiene un \'{\i}ndice espectral 
$\alpha\simeq 2.7$ a bajas energ\'{\i}as, luego un cambio hacia $\alpha\simeq 3$ que ocurre en $E_r\simeq 3\times 10^{15}$~eV,
un fen\'omeno conocido como la rodilla del espectro, un segundo cambio con $\alpha\simeq 3.3$, que aparece para  
$E_{sr}\simeq 4\times 10^{17}$~eV, la segunda rodilla, y finalmente un endurecimiento del espectro hacia $\alpha\simeq 2.7$
en $E_t\simeq 5\times 10^{18}$~eV, el tobillo. 

Para energ\'{\i}as por debajo de $10^{14}$~eV, los rayos c\'osmicos primarios pueden
ser observados en forma directa mediante detectores a bordo de globos, cohetes y sat\'elites. Sin embargo, a energ\'{\i}as 
mayores \'esto no es posible en general, debido a que el flujo cae r\'apidamente con la energ\'{\i}a; por ejemplo, el flujo 
integral en la regi\'on de la rodilla es menor a 1 part\'{\i}cula por m$^2$ por a\~no. Para su detecci\'on, en cambio, 
debe recurrirse a la reconstrucci\'on de las lluvias atmosf\'ericas ({\it extensive air showers, EAS}), es decir, observando
la cascada de part\'{\i}culas secundarias producidas por la incidencia de los rayos c\'osmicos sobre la atm\'osfera de la 
tierra. Las lluvias atmosf\'ericas comenzaron a ser exploradas en los a\~nos `30, a partir de los trabajos de Rossi
\footnote{Rossi parece haber sido el primero en sospechar la existencia de las lluvias atmosf\'ericas, en base a sus 
observaciones en Eritrea. En 1934, escribi\'o: ``It would seem ... from time to time there arrive upon the equipment 
very extensive groups of particles which produce coincidences between counters even rather distant from each other''
\cite{ros34}.} y de Auger y sus colaboradores \cite{aug38,aug39}, midiendo la coincidencia temporal en arreglos 
de contadores Geiger-M\"uller. Las t\'ecnicas experimentales fueron progresando con el tiempo, permitiendo acceder a
la reconstrucci\'on de eventos de mayores energ\'{\i}as; el grupo del MIT \cite{bas53} incorpor\'o los centelladores 
como detectores en un arreglo que, operando entre 1954 y 1957, permiti\'o obtener el espectro en el rango desde 
$3\times 10^{15}$ hasta $10^{18}$~eV \cite{cla61}. 
Este arreglo, llamado Harvard Agassiz Station, sirvi\'o adem\'as como prototipo para el arreglo de gran altitud en 
Chacaltaya (Bolivia) y el de Volcano Ranch en New Mexico (Estados Unidos). 
En este \'ultimo, Linsley detect\'o el primer evento por encima de $10^{20}$~eV \cite{lin63}; la detecci\'on de lluvias 
atmosf\'ericas mediante arreglos de superficie extendidos fue, m\'as tarde, exitosamente utilizada en otros sitios: 
Haverah Park (Gran Breta\~na), Narribri (Australia), Yakutsk (Rusia) y AGASA (Jap\'on).   

Tambi\'en fueron explorados m\'etodos alternativos de detecci\'on; en particular, Suga \cite{sug62} y Chudakov \cite{chu62}
discutieron en 1962 la posibilidad de utilizar la atm\'osfera terrestre como un centellador gigante, una idea que 
fue luego investigada en el terreno experimental por Greisen en 1965 \cite{gre65}, y por Tanahashi y colaboradores 
en 1968 \cite{har70}. En 1976, se detect\'o en Volcano Ranch la emisi\'on fluorescente de las lluvias en coincidencia
con el arreglo de detectores de superficie \cite{ber77}, dando lugar al desarrollo de experimentos basados s\'olo en esa
t\'ecnica, a saber, los exitosos Fly's Eye y HiRes situados en Utah (Estados Unidos). 

Una nueva generaci\'on de experimentos est\'a
por comenzar con el Observatorio Auger en Malarg\"ue (Argentina) \cite{auger}, 
que consiste en un arreglo h\'{\i}brido formado por 
detectores de superficie y de fluorescencia; la operaci\'on combinada de las dos t\'ecnicas es potencialmente muy poderosa,
y permitir\'{\i}a eliminar las discrepancias sistem\'aticas que se observan al comparar las observaciones de experimentos que 
utilizan s\'olo un m\'etodo de detecci\'on, como es el caso, por ejemplo, de AGASA y HiRes. Por otra parte, la superficie
cubierta por los detectores ser\'a de 3000~km$^2$, considerablemente mayor que los 100~km$^2$ cubiertos por
AGASA. La publicaci\'on de los primeros resultados experimentales se espera para mediados de 2005, y corresponder\'a a una
estad\'{\i}stica comparable a la acumulada por AGASA desde su apertura en 1990 hasta la fecha. 

Otros experimentos han sido proyectados y se encuentran en diversas fases de planeamiento. Como parte del Proyecto
Auger, se planea construir otro Observatorio en Utah (Estados Unidos), con el prop\'osito de cubrir toda la esfera
celeste. Otro proyecto, basado en la t\'ecnica de la fluorescencia, es el Telescope Array Project \cite{tap}, 
que permitir\'{\i}a la
reconstrucci\'on estereosc\'opica de todos los eventos ultra-energ\'eticos desde $10^{18}~$eV, con un excelente potencial 
para discriminar la especie del rayo c\'osmico primario. Finalmente, las propuestas m\'as ambiciosas consideran la 
observaci\'on de la luz fluorescente desde sat\'elites en el espacio, dando lugar al proyecto Airwatch/OWL \cite{air-owl}.
La observaci\'on desde el espacio ofrecer\'{\i}a, por un lado, la ventaja de cubrir un \'area mucho mayor que desde la tierra, 
mientras que, adem\'as, la absorci\'on de los fotones emitidos ser\'{\i}a mucho menor, incrementando la sensibilidad y 
reduciendo los errores sistem\'aticos. La realizaci\'on de cualquiera de estos grandes proyectos quedar\'a, 
muy probablemente, supeditada a los resultados que se obtengan en los pr\'oximos a\~nos en el Observatorio Auger Sur.   
    
\section{La aceleraci\'on de rayos c\'osmicos}

Explicar el origen de los rayos c\'osmicos requiere, por un lado, especificar un mecanismo de aceleraci\'on que produzca 
protones y n\'ucleos con espectros de ley de potencias, $dN/dE\propto E^{-\beta}$, con los \'{\i}ndices espectrales adecuados 
a las observaciones. Como veremos m\'as adelante, en el Cap\'{\i}tulo 2, la difusi\'on normal domina el transporte de 
rayos c\'osmicos a bajas energ\'{\i}as (debajo de la rodilla), y de ello se espera que $\beta=\alpha-1/3$, donde $\alpha$ es
el \'{\i}ndice del espectro observado. Siendo que, t\'{\i}picamente, se observa $\alpha\simeq 2.7$ para los protones y
las especies nucleares m\'as abundantes, 
deber\'{\i}a esperarse $\beta\simeq 2.4$ para el \'{\i}ndice de producci\'on en las fuentes
\footnote{Una observaci\'on notable es que el mecanismo de aceleraci\'on de part\'{\i}culas hasta energ\'{\i}as 
ultrarrelativistas parece ser similar para una variedad de diferentes fuentes astrof\'{\i}sicas; por ejemplo, 
los espectros t\'{\i}picos de fuentes de radio corresponden a espectros de electrones con \'{\i}ndice 
$\beta\approx 2.6\pm 0.4$, y los espectros continuos de cu\'asares en longitudes de onda \'opticas y X corresponden a 
$\beta\approx 3$.}. 
Por otra parte, se necesita un mecanismo que permita acelerar rayos c\'osmicos hasta energ\'{\i}as extremadamente altas; 
a\'un dejando de lado los problemas espec\'{\i}ficos de los rayos c\'osmicos ultra-energ\'eticos, 
se requiere un mecanismo que acelere part\'{\i}culas, al menos, hasta la regi\'on del 
tobillo del espectro. Y finalmente, asumiendo que la efectividad del mecanismo sea del orden del $\sim\%$, 
se necesita, adem\'as, que la  
luminosidad total de las fuentes sea mayor, en alrededor de 2 \'ordenes de magnitud, que la correspondiente a los rayos c\'osmicos.

Aqu\'{\i} veremos que un mecanismo de aceleraci\'on apropiado es el que resulta de procesos difusivos en frentes de onda 
de choque; en particular, la aceleraci\'on en remanentes de supernova ({\it supernova remnants, SNRs}) satisface los 
requerimientos que hemos mencionado, y constituye un candidato preferencial en la explicaci\'on del origen de los 
rayos c\'osmicos. La propuesta original de este mecanismo fue desarrollada por Fermi en 1949 \cite{fer49}, y en su
formulaci\'on actual se conoce con el nombre de ``mecanismo de Fermi de segundo orden''. 
La idea b\'asica parte de considerar la transferencia de energ\'{\i}a 
cin\'etica adquirida por part\'{\i}culas cargadas que colisionan con objetos masivos en movimiento; llamando 
$\xi=\Delta E/E$ a la energ\'{\i}a relativa ganada por una part\'{\i}cula en cada colisi\'on, 
al cabo de $n$ colisiones la energ\'{\i}a de la part\'{\i}cula es 
\begin{equation}
E=E_0(1+\xi)^n\ ,          
\end{equation}
donde $E_0$ es la energ\'{\i}a inicial. Si $T_c$ es el tiempo caracter\'{\i}stico entre colisiones sucesivas, y $T_{esc}$ el
tiempo caracter\'{\i}stico de escape de la regi\'on de aceleraci\'on, la probabilidad de escape luego de una colisi\'on es
$P_{esc}=T_c/T_{esc}$; entonces, la densidad de part\'{\i}culas aceleradas hasta una energ\'{\i}a mayor o igual a $E$ resulta
\begin{equation}
N(\geq E)\propto\sum_{m=n}^\infty(1-P_{esc})^m={{\left(1-P_{esc}\right)^n}\over{P_{esc}}}\ ,          
\end{equation}
donde $n=\log(E/E_0)/\log(1+\xi)$. Entonces, la densidad queda dada por  
\begin{equation}
N(E)\propto E^{-\left(P_{esc}/\xi+1\right)}\ ;           
\end{equation} 
vemos que este proceso conduce a un espectro de ley de potencias hasta una energ\'{\i}a m\'axima $E_{max}$ dada por 
\begin{equation}
E_{max}\simeq E_0(1+\xi)^{min(T_A,T_{esc})/T_c}\ ,           
\end{equation} 
donde $T_A$ es la vida del acelerador. 

En la concepci\'on original de Fermi, la realizaci\'on astrof\'{\i}sica de este mecanismo era la reflexi\'on de las 
part\'{\i}culas cargadas en ``espejos magn\'eticos'' asociados con irregularidades en el campo magn\'etico de la galaxia 
(es decir, nubes magnetohidrodin\'amicas no relativistas propag\'andose en el medio interestelar). 
Como la nube tiene una masa mucho mayor que la part\'{\i}cula, su velocidad no cambia con la colisi\'on; en el sistema 
inercial asociado a la nube, el choque es el\'astico y la part\'{\i}cula se refleja conservando su energ\'{\i}a. 
Transformando al sistema del observador, en el que la velocidad de la nube es $V$, la de la part\'{\i}cula es $v\approx c$,
y $\theta$ es el \'angulo formado entre la velocidad de la nube y la velocidad de salida de la part\'{\i}cula, resulta
\begin{equation}
{{\Delta E}\over{E}}={{2Vv\cos\theta}\over{c^2}}+2\left({{V}\over{c}}\right)^2\ .           
\end{equation}    
La ganancia media de energ\'{\i}a se obtiene ahora promediando sobre el \'angulo $\theta$; debe tenerse en cuenta que, de la 
cinem\'atica de la colisi\'on, la probabilidad de un encuentro con un dado \'angulo $\theta$ viene dada por 
$P(\theta)\propto 1+(V/c)\cos\theta$ \cite{lon92}, de modo que los choques frontales (que corresponden a $0\leq\theta<\pi/2$)  
son m\'as probables que los choques desde atr\'as (que corresponden a $\pi/2<\theta\leq\pi$). El resultado es
\begin{equation}
\langle{{\Delta E}\over{E}}\rangle={{8}\over{3}}\left({{V}\over{c}}\right)^2\ ,
\end{equation} 
de modo que la ganancia media por colisi\'on es de segundo orden en $V/c$. Siendo que las velocidades 
estoc\'asticas de las nubes interestelares en la galaxia son t\'{\i}picamente $V/c\leq 10^{-4}$, este resultado dar\'{\i}a 
una ganancia de energ\'{\i}a insuficiente; por otra parte, asumiendo $T_c\sim 1/c\rho\sigma$ (es decir, considerando las
nubes como centros dispersores de densidad $\rho$ y secci\'on eficaz $\sigma$), el \'{\i}ndice espectral $\beta$ resulta 
depender de propiedades locales, en lugar de ser universal.  

\begin{figure}[t]
\centerline{{\epsfxsize=3.9in\epsfxsize=4.4in\epsffile{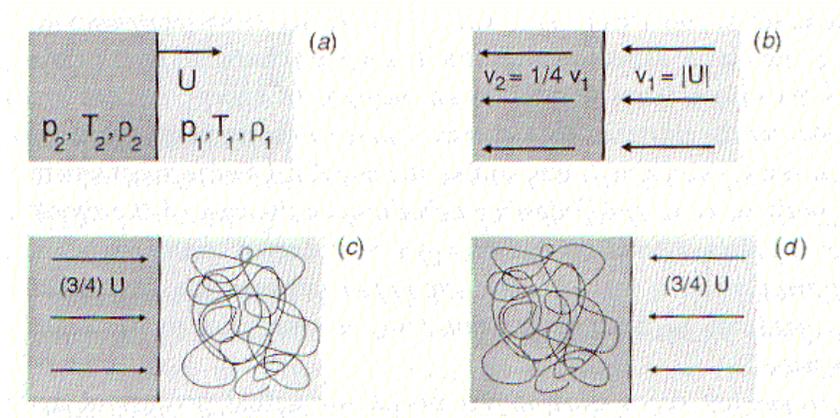}}}
\caption{La aceleraci\'on de part\'{\i}culas en la expansi\'on de frentes de onda de choque (de \cite{lon92}). 
(a) La propagaci\'on del 
frente de ondas entre el upstream (a la derecha) y el downstream (a la izquierda). (b) El flujo de gas 
interestelar a trav\'es del frente de ondas, en el sistema en el que el frente est\'a en reposo. 
(c) El flujo de gas en el sistema del upstream, con el frente de onda incidiendo sobre una distribuci\'on de 
part\'{\i}culas isotr\'opica. (d) El flujo de gas en el sistema del downstream, con el frente de onda incidiendo sobre 
una distribuci\'on de part\'{\i}culas isotr\'opica.}
\label{shock}
\end{figure} 
Hacia fines de los a\~nos `70, se encontr\'o \cite{axf77,kry77,bel78,bla78} que
estos problemas pod\'{\i}an corregirse considerando la colisi\'on con frentes de onda de choque relativistas, dando lugar a un 
mecanismo de aceleraci\'on universal con una ganancia media por colisi\'on de primer orden en $V/c$. 
Como sabemos, una perturbaci\'on en un gas produce ondas que se propagan a la velocidad del sonido y transmiten
informaci\'on sobre la fuente; si \'esta se mueve a una velocidad mayor a la del sonido, la perturbaci\'on del 
medio se propaga en la forma de una onda de choque (con la velocidad de la fuente), que representa una discontinuidad
entre el estado termodin\'amico del medio perturbado ({\it downstream}) y el del medio no perturbado ({\it upstream}).    
Denotaremos la presi\'on, temperatura y densidad del upstream con $p_1,T_1$ y $\rho_1$, respectivamente; a las variables
termodin\'amicas del downstream las llamaremos $p_2,T_2$ y $\rho_2$. La fig.\ref{shock}(a) muestra el avance del 
frente de onda de choque con velocidad $U=Mc_s$, donde $c_s$ es la velocidad del sonido en el gas y $M\gg 1$ el 
n\'umero de Mach. 
En el sistema de referencia del frente de onda, el gas interestelar incide desde el upstream con velocidad $v_1=|U|$
y se aleja del frente en el downstream con velocidad $v_2$. 
En ese sistema de referencia, se pueden relacionar las propiedades termodin\'amicas del downstream relativas al upstream, 
planteando  la conservaci\'on de la masa (es decir, la ecuaci\'on de 
continuidad), de la energ\'{\i}a (la ecuaci\'on de Bernoulli) y del impulso, para el flujo de gas  
desde el upstream hacia el downstream. Si consideramos un gas ideal, siendo $\gamma\equiv c_p/c_v$ el cociente de 
calores especif\'{\i}cos, resulta $\rho_2/\rho_1=v_1/v_2=(\gamma+1)/(\gamma-1)$; en particular, para un gas completamente
ionizado, $\gamma={{5}\over{3}}$ y resulta $v_2=v_1/4$ (ver fig.\ref{shock}(b)).
       
Consideremos ahora una distribuci\'on de part\'{\i}culas de alta energ\'{\i}a delante del frente de ondas. 
En el sistema en el que el gas est\'a en reposo, la distribuci\'on de part\'{\i}culas es isotr\'opica, debido al scattering. 
Entonces, visto desde el upstream (ver la fig.\ref{shock}(c)), el gas en el downstream se aproxima con velocidad 
$V={{3}\over{4}}U$ y la distribuci\'on de velocidades de las part\'{\i}culas que se encuentran con el frente de ondas es 
isotr\'opica. Si consideramos ahora el proceso opuesto, en el que las part\'{\i}culas pasan a trav\'es del frente desde
el downstream hacia el upstream, y lo observamos desde el sistema del downstream (ver la fig.\ref{shock}(d)), se 
encuentra lo mismo: la distribuci\'on de part\'{\i}culas es isotr\'opica y el upstream se aproxima con velocidad 
$V={{3}\over{4}}U$. Es decir, una part\'{\i}cula que pasa desde el downstream hacia el upstream experimenta un 
incremento $\Delta E$ en su energ\'{\i}a, que resulta igual al del proceso inverso. En el mecanismo original de Fermi, 
el cambio medio de energ\'{\i}a en una colisi\'on era de orden $V/c$, pero deb\'{\i}an promediarse colisiones frontales (que
produc\'{\i}an ganancias de energ\'{\i}a) con colisiones desde atr\'as (que produc\'{\i}an p\'erdidas); del promedio, 
resultaba una ganancia neta de orden $(V/c)^2$. En cambio, aqu\'{\i} el mecanismo asegura una completa simetr\'{\i}a
en el pasaje de part\'{\i}culas de un lado a otro del frente de ondas, ganando energ\'{\i}a en cada colisi\'on.     
Al ir y volver, la energ\'{\i}a ganada resulta \cite{lon92}
\begin{equation}
\langle{{\Delta E}\over{E}}\rangle={{4}\over{3}}{{V}\over{c}}\ ,
\end{equation} 
que es de primer orden en $V/c$; por otra parte, el \'{\i}ndice espectral resulta 
$\beta=2+4/M^2$. Este resultado es muy satisfactorio, porque reproduce los valores esperados y, adem\'as, permite 
explicar la universalidad observada en este fen\'omeno; la presencia de frentes de onda supers\'onicos es 
plausible en cualquier fuente astrof\'{\i}sica de part\'{\i}culas de alta energ\'{\i}a, tales como remanentes de supernova, 
n\'ucleos gal\'acticos activos y fuentes de radio extendidas.    

Como hemos comentado m\'as arriba, un escenario ampliamente difundido y aceptado en la actualidad es el que atribuye la 
aceleraci\'on de los rayos c\'osmicos (exceptuando a los ultra-energ\'eticos) a los remanentes de supernova; dependiendo de
si los frentes de onda se desarrollan en el medio interestelar, o bien en el viento estelar de una estrella 
predecesora o en el de una compa\~nera \cite{bie95}, los protones podr\'{\i}an ser acelerados hasta una energ\'{\i}a 
m\'axima en el rango entre $E_c=10^{15}$ y $10^{17}$~eV, mientras que, para n\'ucleos de carga $Ze$, esta energ\'{\i}a 
m\'axima vendr\'{\i}a dada por $E_{cZ}\simeq ZE_c$ \footnote{Otros mecanismos permitir\'{\i}an 
acelerar rayos c\'osmicos hasta energ\'{\i}as a\'un mayores, como es el caso, por ejemplo, de la aceleraci\'on 
electrost\'atica en p\'ulsares j\'ovenes. Sin embargo, estos mecanismos de aceleraci\'on alternativos tienen por lo general 
dificultades para predecir espectros de ley de potencias, y usualmente son propuestos s\'olo para explicar el origen de los 
rayos c\'osmicos ultra-energ\'eticos.}. Debe indicarse, sin embargo, que a\'un no ha sido posible obtener evidencia directa
de la aceleraci\'on de protones en remanentes de supernova \cite{but02}. En el marco del paradigma de las supernovas,
el tobillo en $E_t\simeq 5\times 10^{18}$~eV aparece naturalmente como un {\it crossover} entre una regi\'on dominada por los
rayos c\'osmicos gal\'acticos (para $E<E_t$) y otra dominada por los rayos c\'osmicos extragal\'acticos (para $E>E_t$). 

\begin{figure}[t]
\centerline{{\epsfxsize=3.1in\epsfxsize=3.5in\epsffile{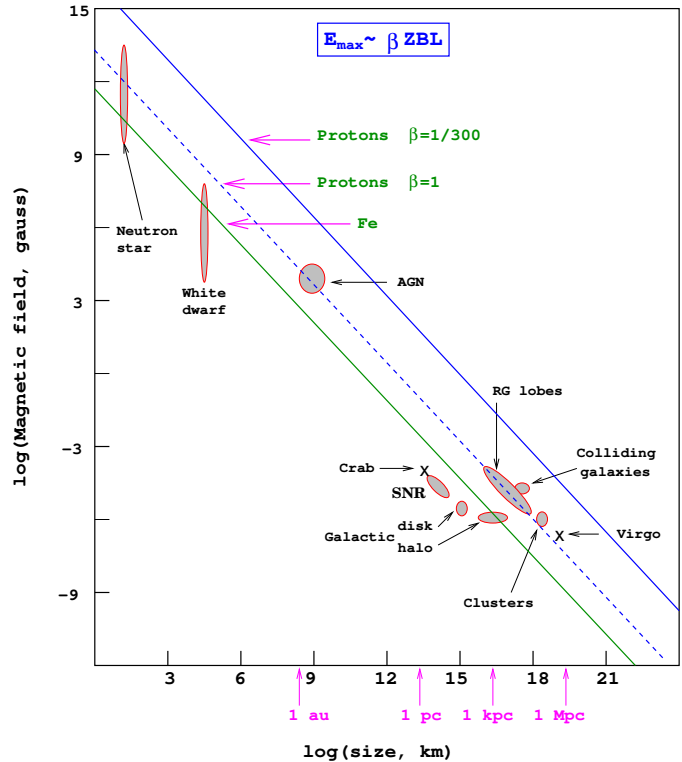}}}
\caption{El plot de Hillas: el campo magn\'etico y el tama\~no caracter\'{\i}sticos de 
diversos objetos astrof\'{\i}sicos, comparados con la energ\'{\i}a m\'axima que puede adquirir una part\'{\i}cula 
en los sitios de aceleraci\'on (de \cite{bha00}).}
\label{hillas}
\end{figure}
La energ\'{\i}a m\'axima alcanzable en un proceso difusivo de aceleraci\'on en frentes de onda de choque est\'a dada por
$E_{max}=kZeUBL$ \cite{hil84,dru94}, donde $k$ est\'a asociado a la eficiencia del mecanismo de aceleraci\'on ($k\leq 1$), 
$B$ es el campo magn\'etico en la regi\'on de aceleraci\'on, y $L$ el 
tama\~no del acelerador. En el caso de aceleraci\'on \'optima, que corresponde a tomar $k=1$ y $U=c$, se obtiene   
\footnote{Las unidades astron\'omicas
de longitud usadas habitualmente son el parsec o segundo-parallax (pc), el a\~no luz (ly) y la unidad astron\'omica (AU),
relacionadas por 1~pc = 3.26 ly = 2.06$\times 10^5$~AU = 3.09$\times 10^{16}$~m .}
\begin{equation}
E_{max}=0.9\times 10^{18}\ Z\left({{B}\over{{\rm \mu G}}}\right)\left({{L}\over{\rm kpc}}\right)\ {\rm eV}.
\label{emax}
\end{equation}
En particular, este l\'{\i}mite es muy ilustrativo cuando se lo grafica en un diagrama $B$ vs $L$ comparado 
con los valores t\'{\i}picos de diferentes objetos astrof\'{\i}sicos, tal como se muestra en la fig.\ref{hillas}. 
Las rectas corresponden a $E_{max}=10^{20}$~eV para protones y n\'ucleos de Fe, y diferentes valores de $\beta=U/c$; 
s\'olo las fuentes ubicadas en la regi\'on superior satisfacen la condici\'on impuesta por la ec.~(\ref{emax}).
Este gr\'afico, conocido como ``plot de Hillas'', pone en evidencia las dificultades que aparecen para encontrar 
objetos astrof\'{\i}sicos que provean un mecanismo apropiado para acelerar rayos c\'osmicos hasta las m\'as altas 
energ\'{\i}as observadas. 

\section{La propagaci\'on de rayos c\'osmicos}
Como veremos, la abundancia qu\'{\i}mica de los rayos c\'osmicos de bajas energ\'{\i}as, que puede obtenerse mediante 
t\'ecnicas de detecci\'on directa, provee importante informaci\'on acerca de los potenciales sitios de aceleraci\'on y, 
fundamentalmente, sobre diferentes aspectos referidos a la propagaci\'on de los rayos c\'osmicos en la galaxia. 

La espectroscop\'{\i}a astron\'omica busca determinar las abundancias qu\'{\i}micas de los elementos en diferentes objetos 
celestes; en particular, las l\'{\i}neas de absorci\'on debido a iones, \'atomos y mol\'eculas en la fot\'osfera solar 
permiten medir las abundancias en el sol. De la teor\'{\i}a de la evoluci\'on estelar, sabemos que la abundancia c\'osmica 
t\'{\i}pica se adquiere m\'as bien r\'apidamente, de modo que la abundancia solar es representativa de la de toda la 
galaxia \footnote{El sol tiene una edad de $4.6\times 10^9$~a\~nos y la galaxia, al menos, el doble; las estrellas m\'as 
viejas presentan un d\'eficit en elementos pesados, pero puede asumirse que la abundancia c\'osmica no se modific\'o 
notablemente desde la \'epoca de la formaci\'on del sol.}. Otro medio para determinar las abundancias es el
estudio de los meteoritos; si bien se encuentran diferencias muy significativas en las abundancias en diferentes 
tipos de meteorito, que probablemente reflejan sus diversas procedencias, existe una clase de meteorito particular,
los condr\'{\i}ticos, que tienen abundancias muy similares a los de la fot\'osfera solar. \'Esto resulta
sumamente valioso para extraer informaci\'on sobre las abundancias de algunos de los elementos m\'as raros de la tabla 
peri\'odica, que no pueden ser bien determinadas mediante t\'ecnicas espectrosc\'opicas. 

\begin{figure}[t]
\centerline{{\epsfxsize=3.8in\epsfxsize=4.2in\epsffile{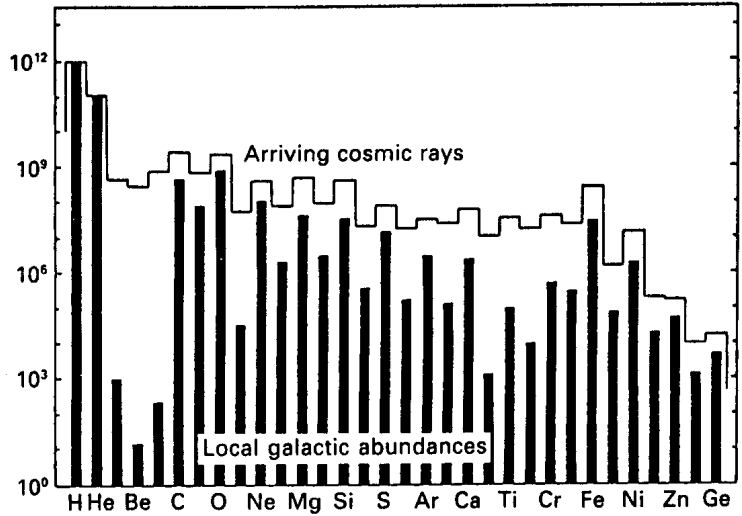}}}
\caption{Comparaci\'on entre la abundancia gal\'actica local y la observada en los rayos 
c\'osmicos (de \cite{lun84}).}
\label{abundances}
\end{figure} 

La fig.\ref{abundances} muestra una comparaci\'on entre la abundancia gal\'actica local y la observada en los rayos 
c\'osmicos; la impresi\'on general es que las abundancias son muy similares, de modo que los rayos c\'osmicos deben
ser acelerados a partir de un material de composici\'on qu\'{\i}mica muy similar a la observada en el sistema solar. 
Sin embargo, se observa en los rayos c\'osmicos un marcado exceso de Li, Be y B, y tambi\'en en el grupo de elementos 
pesados justo debajo de Fe. Otras sobreabundancias muy significativas, que no se muestran en la figura, son las 
de los is\'otopos $^2$He y $^3$He; por ejemplo, las abundancias de $^2$He/$^1$He y de $^3$He/$^4$He en los rayos 
c\'osmicos son unas 5000 veces mayores que las abundancias c\'osmicas correspondientes.     
Como veremos, estas anomal\'{\i}as pueden explicarse como debidas al proceso de fragmentaci\'on nuclear
({\it spallation}) de las especies m\'as pesadas; adem\'as, \'esto permite determinar la regi\'on de la galaxia en la 
que los rayos c\'osmicos est\'an confinados. 

La ecuaci\'on de transporte para n\'ucleos de alta energ\'{\i}a propag\'andose a trav\'es del gas interestelar es
\begin{equation}
{{\partial N_i}\over{\partial t}}=
\nabla(\hat{D}\nabla N_i-VN_i)+{{\partial}\over{\partial E}}(b(E)N_i)+Q_i-
{{N_i}\over{\tau_i^d}}+{{N_i}\over{\tau_i^{sp}}}+\sum_{j>i}{P_{ji}\over{\tau_j^{sp}}}N_j\ .
\label{transp} 
\end{equation}
Aqu\'{\i}, $N_i=N_i(E,{\bf x},t)$ es la densidad de n\'ucleos de la especie $i-$\'esima 
(donde el sub\'{\i}ndice ordena las especies de acuerdo con su masa, en forma creciente), $\hat{D}$ el tensor de difusi\'on,
$V$ es una velocidad de convecci\'on (que puede representar, por ejemplo, un viento gal\'actico),  
$\partial(b(E)N_i)/\partial E$ toma en cuenta ganancias y p\'erdidas continuas de energ\'{\i}a, donde $b(E)=-dE/dt$
(es decir, $b(E)<0$ corresponde a una aceleraci\'on sistem\'atica promedio, como la debida, por ejemplo, a la aceleraci\'on 
estoc\'astica de Fermi, mientras que $b(E)>0$ representa p\'erdidas continuas de
energ\'{\i}a debidas a la emisi\'on de radiaci\'on por bremsstrahlung, a la emisi\'on sincrotr\'onica, a las 
p\'erdidas por ionizaci\'on, etc.), $Q_i$ representa la tasa de inyecci\'on de part\'{\i}culas en las fuentes por unidad 
de volumen, y $\tau_i^d$ es la vida media de decaimiento radiactivo para los n\'ucleos inestables. 
Los dos \'ultimos t\'erminos describen el efecto de ganancias y p\'erdidas debidas a la fragmentaci\'on nuclear, 
siendo $\tau_i^{sp}$ la vida media de la especie $i-$\'esima asociada a este proceso; a $N_i$ contribuye
la fragmentaci\'on de todas las especies m\'as pesadas, como indica la suma sobre $j>i$ en el \'ultimo t\'ermino, 
que acopla las ecuaciones correspondientes a diferentes componentes nucleares.  
Un caso particular de gran inter\'es es el correspondiente al estado estacionario, en el que la dependencia expl\'{\i}cita 
con el tiempo desaparece.  

Basados en la ecuaci\'on general de transporte dada por la ec.~(\ref{transp}), adoptaremos un modelo muy simplificado 
(conocido en la literatura como modelo {\it slab} \cite{lon92}) con el fin de ilustrar la manera con que el proceso de 
spallation explica las sobreabundancias observadas. Despreciaremos la difusi\'on, la convecci\'on, las ganancias y p\'erdidas
continuas de energ\'{\i}a, la inyecci\'on de nuevas part\'{\i}culas, y el decaimiento radioactivo; s\'olo consideraremos la
propagaci\'on rectil\'{\i}nea de los n\'ucleos, agrupados en el grupo liviano $L$ (formado por Li, Be y B) y el grupo 
medio M (constitu\'{\i}do fundamentalmente por C, N y O). Siguiendo la trayectoria de un grupo de n\'ucleos que 
se propaga con velocidad $v={\rm d}x/{\rm d}t$, veremos c\'omo es la dependencia espacial de la densidad $N$ entre la fuente y el 
observador, suponiendo que la producci\'on continua en las fuentes da lugar a un estado estacionario.    
Es conveniente adoptar como variable a la columna de densidad $\xi\equiv\bar\rho x=\bar\rho vt$, 
donde $\bar\rho$ es la densidad promedio en la regi\'on de propagaci\'on.  
Entonces, las ecuaciones para la densidad de los grupos $L$ y $M$ se reducen a  
\begin{equation}
{{{\rm d}N_M(\xi)}\over{{\rm d}\xi}}=-{{N_M(\xi)}\over{\xi_M}}
\label{grupoM}
\end{equation}
y
\begin{equation}
{{{\rm d}N_L(\xi)}\over{{\rm d}\xi}}=-{{N_L(\xi)}\over{\xi_L}}+{{P_{ML}}\over{\xi_M}}N_M(\xi)\ ;
\label{grupoL}
\end{equation}
asumiremos, adem\'as, que $N_L(0)=0$, es decir, que la densidad del grupo liviano producido en las fuentes es despreciable. 
El camino libre medio de spallation viene dado por $\xi_{L,M}=m_p/\sigma_{L,M}$; reemplazando las secciones eficaces de 
spallation, $\sigma_L\simeq 200$~mb y $\sigma_M\simeq 280$~mb, resultan $\xi_L\simeq 84$~kg m$^{-2}$ y 
$\xi_M\simeq 60$~kg m$^{-2}$, respectivamente. Para la probabilidad de fragmentaci\'on promedio, puede tomarse el 
valor $P_{ML}\simeq 0.28$. Integrando las ecs.~(\ref{grupoM}) y (\ref{grupoL}), se obtiene
\begin{equation}  
{{N_L(\xi)}\over{N_M(\xi)}}={{P_{ML}\xi_L}\over{(\xi_L-\xi_M)}}\left[\exp\left({{\xi}\over{\xi_M}}-
{{\xi}\over{\xi_L}}\right)-1\right]\ .
\end{equation}  
De los datos experimentales, sabemos que $N_L(\xi)/N_M(\xi)=0.25$; reemplazando este valor en la ecuaci\'on precedente,
resulta $\xi= 48$~kg m$^{-2}$. 
El mismo tipo de c\'alculo puede hacerse para la producci\'on de $^3$He por spallation de $^4$He, y se obtiene 
$\xi= 50$~kg m$^{-2}$, es decir, un resultado muy similar al obtenido para los n\'ucleos del grupo L. 

Cuando se busca explicar la sobreabundacia del grupo pesado (Mn, Cr, V) por spallation de los n\'ucleos de Fe, aparece 
una dificultad; la secci\'on eficaz del Fe es mucho mayor que la de elementos m\'as livianos, y resulta en un camino libre 
medio mucho menor, $\xi_{Fe}= 22$~kg m$^{-2}$, de modo que la abundancia del Fe quedar\'{\i}a excesivamente suprimida 
luego de atravesar una columna de densidad $\xi\approx 50$~kg m$^{-2}$. La soluci\'on se obtiene de considerar que 
los n\'ucleos no recorren una \'unica columna de densidad, como en el modelo slab, sino asumiendo una distribuci\'on de 
longitudes de recorrido, $P(\xi)$. 

La ecuaci\'on de difusi\'on y escape es una versi\'on simplificada de la ecuaci\'on de
transporte (\ref{transp}), en la que se considera que los rayos c\'osmicos escapan de la galaxia con un tiempo 
caracter\'{\i}stico $\tau_e$; es decir, 
\begin{equation}
{{\partial N}\over{\partial t}}=D\nabla^2N-{{N}\over{\tau_e}}\ .
\label{diffloss}
\end{equation}   
%\begin{figure}[t]
%\centerline{{\epsfxsize=4.in\epsfxsize=4.7in\epsffile{abund2.eps}}}
%\caption{Abundancia isot\'opica inferida en las fuentes (columna derecha) en comparaci\'on con la observada en la
%vecindad de la tierra (columna izquierda); en esta \'ultima, se distingue adem\'as la fracci\'on que fue producida 
%en las fuentes (zona blanca) de la parte que proviene de la fragmentaci\'on nuclear de especies m\'as pesadas durante la 
%propagaci\'on (zona negra) (de \cite{sha91}).}
%\label{spall}
%\end{figure} 
En el modelo conocido como {\it leaky box}, se asume una densidad homog\'enea $N$ en toda la regi\'on de propagaci\'on; 
luego, el t\'ermino difusivo en la ec.~(\ref{diffloss}) se anula, y resulta una distribuci\'on exponencial de columnas de 
densidad $P(\xi)\propto\exp(-\xi/\xi_e)$. Alternativamente,     
asumiendo la difusi\'on como el t\'ermino preponderante, es decir, tomando $\tau_e\to\infty$, se obtiene una distribuci\'on
gaussiana. En cualquier caso, resulta que la ecuaci\'on de transporte conduce a distribuciones de caminos de densidad que
resuelven el problema que aparec\'{\i}a con la abundancia del Fe, y que permiten, adem\'as, reproducir adecuadamente todas las 
abundancias observadas. 
%La fig.\ref{spall} muestra, para cada especie nuclear,
%la abundancia inferida en las fuentes (columna derecha) en comparaci\'on con la observada en la
%vecindad de la tierra (columna izquierda); en esta \'ultima se distingue, adem\'as, la fracci\'on que fue producida 
%en las fuentes (zona blanca) de la parte que proviene de la fragmentaci\'on nuclear de especies m\'as pesadas 
%durante la propagaci\'on (zona negra).        
De deducir la abundancia en las fuentes y compararla con los valores t\'{\i}picos para el sistema solar, se encuentra
que los elementos cuyo primer potencial de ionizaci\'on es mayor a 10~eV tienen una abundancia suprimida por un factor 
$\sim 5$; \'esto sugiere la existencia de alg\'un mecanismo de supresi\'on para la producci\'on de elementos con 
potenciales de primera ionizaci\'on altos. 
Por otro lado, en experimentos que resuelven la carga y la masa de las 
part\'{\i}culas se observan otras anomal\'{\i}as, como por ejemplo la sobreabundancia isot\'opica en $^{22}$Ne/$^{20}$Ne, 
$^{25,26}$Mg/$^{24}$Mg y $^{29,30}$Si/$^{28}$Si, sugiriendo que al menos una parte de los rayos c\'osmicos se origina
en medios ricos en neutrones. 

Hasta aqu\'{\i}, hemos visto que la abundancia observada para las especies estables se explica a partir de la fragmentaci\'on 
nuclear; esta explicaci\'on permite, adem\'as, determinar la abundancia en las fuentes y la columna de densidad recorrida 
por los rayos c\'osmicos en su propagaci\'on. Sin embargo, para determinar el volumen de confinamiento y el tiempo de
residencia (o de escape), es necesario recurrir a la abundancia relativa de los is\'otopos radioactivos.  

En el disco gal\'actico, la densidad del medio interestelar es 
$\rho\simeq m_pn_d$, donde $m_p$ es la masa del prot\'on y $n_d\simeq 1{\rm cm}^{-3}$; 
entonces, si los rayos c\'osmicos est\'an confinados a la regi\'on del disco, 
una columna de densidad $\xi\approx 50$~kg m$^{-2}$ corresponde a $\tau_e\approx 3\times 10^6$~a\~nos. 
Notemos que este tiempo es mucho mayor a $t=c\ L_G\approx 3\times (10^3-10^4)$~a\~nos, 
el tiempo que tardar\'{\i}a una 
part\'{\i}cula relativista en atravesar la galaxia, tomando $L_G\approx 1$--10~kpc como valores indicativos de la 
dimensi\'on de la galaxia. \'Esto implica que los rayos c\'osmicos de bajas energ\'{\i}as est\'an efectivamente
confinados por los campos magn\'eticos gal\'acticos. 

Una determinaci\'on m\'as precisa del tiempo de escape puede obtenerse
a partir de la abundancia del is\'otopo radioactivo del berilio, $^{10}$Be, cuya vida media es 
$\tau_d\approx 3.9\times 10^6$~a\~nos, es decir, similar a la primera estimaci\'on de $\tau_e$ que hemos hecho.
El berilio es producido por la fragmentaci\'on de carbono y ox\'{\i}geno; el $10\%$ del total producido es la especie 
radioactiva $^{10}$Be, que luego decae a $^{10}$B por decaimiento $\beta$, mientras que el resto corresponde a las
especies estables $^7$Be y $^9$Be. En el modelo estacionario leaky box, las densidades vienen dadas por 
\begin{equation}
-{{N}\over{\tau_i^e}}-{{N}\over{\tau_i^d}}-{{N}\over{\tau_i^{sp}}}+C_i=0\ ,
\end{equation}
donde, para simplificar la notaci\'on, definimos $C_i$ como la tasa de producci\'on de la especie $i-$\'esima: 
\begin{equation}
C_i=\sum_{j>i}{{P_{ij}}\over{\tau_j^{sp}}}N_j\ .  
\end{equation} 
Asumiendo que los is\'otopos de berilio son producidos por spallation de los elementos del grupo M (es decir, C, N y O), 
las densidades resultan  
\begin{equation}
N_i={{C_i}\over{1/\tau_i^e+1/\tau_i^d+1/\tau_i^{sp}}}\ ;
\label{isot}
\end{equation} 
para los is\'otopos estables, esta expresi\'on requiere tomar el l\'{\i}mite $\tau_i^d\to\infty$.   
El valor observado de la abundancia del is\'otopo inestable $^{10}$Be relativa a la abundancia total del Be producido 
es 0.028 \cite{sim83}; reemplazando este valor en la soluci\'on dada por la ec.~(\ref{isot}), 
resulta que el tiempo de escape es t\'{\i}picamente $\tau_e\approx 10^7$~a\~nos, correspondiendo a una densidad 
media $\bar n\approx 0.2$~cm$^{-3}$.  \'Esto implica que las part\'{\i}culas pasan mucho tiempo en el halo que 
rodea la galaxia, cuya densidad (en el caso de un halo extendido de $\sim 10$~kpc
de altura) es t\'{\i}picamente $n\approx 10^{-2}$~cm$^{-3}$. 
El tiempo de escape decrece con la energ\'{\i}a de las part\'{\i}culas, dado que el confinamiento debido a los 
campos magn\'eticos gal\'acticos se torna menos efectivo a energ\'{\i}as mayores.   
De la ec.~(\ref{diffloss}) estacionaria, resulta que $\tau_e\sim L^2/D$, donde $L$ es una dimensi\'on caracter\'{\i}stica 
de la regi\'on de propagaci\'on; como veremos en el pr\'oximo Cap\'{\i}tulo, la difusi\'on a energ\'{\i}as por debajo de la 
rodilla tiene una dependencia con la energ\'{\i}a $D\propto E^{1/3}$, de donde resulta que $\tau_e\propto E^{-1/3}$. 

\section{La detecci\'on de rayos c\'osmicos}

La detecci\'on de los rayos c\'osmicos de alta energ\'{\i}a ($E\geq 10^{14}$~eV) es indirecta, y requiere reconstruir la 
cascada de part\'{\i}culas secundarias que produce en la atm\'osfera la incidencia de la part\'{\i}cula primaria.  

%\begin{figure}[t]
%\centerline{{\epsfxsize=3.5in\epsfxsize=4in\epsffile{nucl.eps}}}
%\caption{Esquema de la colisi\'on de un prot\'on de alta energ\'{\i}a con un n\'ucleo (de \cite{lon92}).}
%\label{nucl}
%\end{figure} 
Consideremos, en primer lugar, la incidencia de un prot\'on de alta energ\'{\i}a sobre un n\'ucleo atmosf\'erico de masa 
$Am_p$. Como el radio nuclear, dado por $R\approx 1.2\ A^{1/3}$~fm (1~fm = 10$^{-15}$~m), es mucho mayor que el prot\'on 
relativista (por ejemplo, la longitud de onda de de Broglie es $\sim 0.01$~fm para un prot\'on de 10~GeV), el prot\'on 
realiza un {\it scattering} m\'ultiple con los nucleones individuales en el interior del n\'ucleo. En cada interacci\'on, 
se producen principalmente piones y kaones, emitidos fundamentalmente en la direcci\'on de incidencia; 
a la vez, estas part\'{\i}culas 
secundarias pueden tambi\'en iniciar otras colisiones dentro del n\'ucleo. Cada nucle\'on que colisiona con el prot\'on 
incidente o con las part\'{\i}culas secundarias es, por lo general, removido del n\'ucleo; si el n\'ucleo resultante es 
de una especie inestable, se fragmentar\'a en n\'ucleos estables m\'as livianos. De cada fragmento excitado, adem\'as, pueden 
emitirse neutrones. Una colisi\'on entre un prot\'on de energ\'{\i}a $E>1$~GeV y un n\'ucleo 
atmosf\'erico produce t\'{\i}picamente $2(E/{\rm GeV})^{1/4}$ part\'{\i}culas relativistas cargadas. 
%El proceso completo se ilustra esquem\'aticamente en la fig.\ref{nucl}.

%\begin{figure}[t]
%\centerline{{\epsfxsize=3.5in\epsfxsize=4in\epsffile{slant.eps}}}
%\caption{Definici\'on de las variables usadas en el estudio de lluvias atmosf\'ericas (de \cite{gai90}).}
%\label{slantfig}
%\end{figure} 
En el estudio de lluvias atmosf\'ericas, suele utilizarse como variable al {\it slant depth} o {\it grammage},
que representa la coordenada longitudinal $X$ que mide la columna de densidad atravesada desde la parte superior de la 
atm\'osfera. 
%(ver la fig.\ref{slantfig}). 
La profundidad atmosf\'erica vertical $X_v$ viene dada por 
\begin{equation}
X_v=\int_h^\infty\rho(h')dh'=X\ \cos\theta\ ,
\label{slant}
\end{equation}
donde $\rho(h')$ es el perfil de densidad de la atm\'osfera, y $\theta$ es 
el \'angulo de inclinaci\'on de la lluvia, medido desde la vertical.
El camino libre medio de un prot\'on de alta energ\'{\i}a en la atm\'osfera es 
$X_{p-atm}\approx 80$~g cm$^{-2}$, 
valor mucho menor que la columna de densidad total de la atm\'osfera en la superficie al nivel del mar, 
$X_{sup}\approx 1000$~g cm$^{-2}$.

Si la part\'{\i}cula primaria es un n\'ucleo, en lugar de un prot\'on, la colisi\'on con los n\'ucleos atmosf\'ericos 
puede generar fragmentos relativistas que desarrollan luego cascadas por separado; en efecto, para $E>10^{17}$~eV se
observan EAS que tienen m\'as de un eje. Por otro lado, la interacci\'on del primario ocurre m\'as arriba en la atm\'osfera, 
debido a que la secci\'on eficaz aumenta con la masa del n\'ucleo. Una regla sencilla, conocida como ``principio de 
superposici\'on'', considera cada n\'ucleo relativista de masa $Am_p$ como un grupo de $A$ nucleones independientes, 
ya que la energ\'{\i}a de ligadura entre los nucleones puede despreciarse; en ese caso, la secci\'on eficaz de interacci\'on 
con los n\'ucleos atmosf\'ericos depende linealmente de la masa.        

%\begin{figure}[t]
%\centerline{{\epsfxsize=3.1in\epsfxsize=3.5in\epsffile{eas.eps}}}
%\caption{Esquema del desarrollo de una lluvia atmosf\'erica (de \cite{lon92}).}
%\label{lluvia}
%\end{figure} 
Los nucleones secundarios, piones cargados y kaones contin\'uan propag\'andose en la atm\'osfera, generando nuevas colisiones 
nucleares hasta llegar a la energ\'{\i}a por part\'{\i}cula umbral para la producci\'on de piones m\'ultiples, alrededor 
de 1~GeV. A este proceso se lo llama cascada nucle\'onica o hadr\'onica y, en \'el, la energ\'{\i}a inicial del primario 
incidente se degrada en un gran n\'umero de hadrones de baja energ\'{\i}a.      
Sin embargo, los piones y kaones tambi\'en pueden decaer antes de interactuar. 
En particular, los piones neutros tienen una vida
media muy corta ($\tau=1.78\times 10^{-16}$~s) y decaen r\'apidamente via $\pi^0\to2\gamma$. Los fotones resultantes producen
pares electr\'on-positr\'on, y \'estos producen, a la vez, fotones via {\it bremsstrahlung}; el camino libre medio de ambos
procesos en la atm\'osfera es similar, $\lambda_{EM}\approx 280$~m. Esta cascada de fotones, electrones y positrones que 
se reconvierten sucesivamente entre s\'{\i} recibe el nombre de lluvia electromagn\'etica; t\'{\i}picamente, exhibe un 
crecimiento inicial exponencial, desarrolla un n\'umero m\'aximo de part\'{\i}culas proporcional a la energ\'{\i}a de la 
part\'{\i}cula que inicia la cascada y, m\'as all\'a del m\'aximo, se aten\'ua r\'apidamente. Cuando la energ\'{\i}a por 
part\'{\i}cula cae por debajo de la energ\'{\i}a cr\'{\i}tica $E_c$ (que es aqu\'ella para la cual las p\'erdidas por 
bremsstrahlung igualan a las p\'erdidas por ionizaci\'on), los electrones y positrones pierden energ\'{\i}a en la 
producci\'on de iones y electrones libres; para electrones en aire, $E_c=83$~MeV. Con respecto a los fotones, por otra parte, 
la secci\'on eficaz de producci\'on de pares tambi\'en decrece y se hace del orden de las correspondientes al scattering 
Compton y absorci\'on fotoel\'ectrica.           
Los piones cargados tambi\'en pueden decaer, via las reacciones $\pi^+\to\mu^++\nu_\mu$ y $\pi^-\to\mu^-+\bar\nu_\mu$,
cuya vida media es $\tau=2.55\times 10^{-8}$~s. Los muones, a su vez, pueden decaer via $\mu^+\to e^++\nu_e+\bar\nu_\mu$
y $\mu^-\to e^-+\bar\nu_e+\nu_\mu$ (con $\tau=2.2\times 10^{-6}$~s) e iniciar otras lluvias electromagn\'eticas. 
%La fig.\ref{lluvia} muestra un esquema de las componentes hadr\'onica, electromagn\'etica y mu\'onica en el desarrollo de 
%una lluvia atmosf\'erica.  

Como caracter\'{\i}sticas generales, diremos que una lluvia atmosf\'erica se desarrolla, alcanza un m\'aximo de $N_M$
part\'{\i}culas a una profundidad $X_M$ y luego se aten\'ua, debido fundamentalmente a p\'erdidas por ionizaci\'on. 
Cuanto mayor es $E_0$, la energ\'{\i}a de la part\'{\i}cula primaria, mayores resultan $N_M$ y $X_M$; una estimaci\'on  
(precisa dentro del $\sim 25\%$) es $E_0\simeq 1.4\ N_M$~GeV. El frente de una lluvia de part\'{\i}culas tiene un espesor 
t\'{\i}picamente no mayor a algunos metros, y una extensi\'on que depende de $E_0$; por ejemplo, las lluvias verticales 
producidas por rayos c\'osmicos ultra-energ\'eticos tienen su m\'aximo desarrollo a nivel del mar, involucran $\sim 10^{11}$
part\'{\i}culas y su extensi\'on lateral es de varios km. Hasta el m\'aximo, la mayor parte de la energ\'{\i}a de la lluvia 
est\'a contenida en los electrones (y positrones); luego, esta componente se aten\'ua r\'apidamente, y la energ\'{\i}a pasa a 
estar contenida fundamentalmente en los muones.  

La tarea de extraer la energ\'{\i}a y masa de la part\'{\i}cula primaria a partir de los par\'ametros observados en la lluvia 
atmosf\'erica (el n\'umero total de electrones, muones y hadrones, la forma de sus distribuciones laterales de densidad,
la reconstrucci\'on de la altura del m\'aximo de la lluvia, etc.) requiere adoptar modelos que describan el desarrollo
de las cascadas; con ese fin, se han desarrollado varios programas de simulaci\'on Monte Carlo como, por ejemplo, 
CORSIKA y AIRES. Los modelos QCD fenomenol\'ogicos que se incorporan a los programas de simulaci\'on (por ejemplo, 
QJSJET, VENUS y SYBILL) quedan constre\~nidos
por el gran cuerpo de datos experimentales disponibles a partir de los experimentos de colisiones de iones pesados (en el
CERN y Brookhaven) y de colisiones prot\'on-antiprot\'on (en el CERN SPS y el Tevatron del Fermilab). Sin embargo, 
las interacciones que resultan de rayos c\'osmicos con energ\'{\i}as por encima de la rodilla involucran energ\'{\i}as
de centro de masa mayores a las disponibles en el Tevatron; en consecuencia, los modelos de interacci\'on hadr\'onica que 
se incorporan a los c\'odigos de simulaci\'on deben extrapolar los resultados experimentales conocidos por varios \'ordenes de
magnitud, y \'esto introduce grandes incertezas en los resultados. En este contexto, los experimentos de rayos c\'osmicos 
pueden ofrecer la posibilidad de testear las simulaciones de lluvias atmosf\'ericas y, en consecuencia, los modelos de
interacci\'on hadr\'onica en regiones no accesibles a los laboratorios.

%\begin{figure}[th!]
%\centerline{{\epsfxsize=4.2in\epsfxsize=4.7in\epsffile{kascade.eps}}}
%\caption{Esquema del experimento KASCADE. A la izquierda, el arreglo de detectores de electrones, fotones
%y muones, que cubre un \'area de $200\times 200$~m$^2$. A la derecha, el t\'unel de trazas de muones (arriba)
%y el detector central (abajo) (de \cite{kam01a}).}
%\label{kascade}
%\end{figure}  
Como hemos comentado m\'as arriba, en la f\'{\i}sica de rayos c\'osmicos se ha desarrollado, a lo largo de casi un siglo
de investigaciones, un numeroso y variado conjunto de experimentos; a modo de ejemplo, aqu\'{\i} describiremos brevemente
el experimento KASCADE ({\it Karlsruhe Shower Core and Array Detector}), 
en operaci\'on desde 1996 en Karlsruhe, Alemania.   
%La fig.\ref{kascade} muestra un esquema de las instalaciones del experimento. 
Un arreglo de centelladores de $200\times 200$~m$^2$ mide las componentes
electromagn\'etica y mu\'onica. Un sistema central de detectores, de 320~m$^2$,
combina un gran calor\'{\i}metro hadr\'onico con varios detectores de muones, mientras que, 
adicionalmente, las trazas de muones de alta energ\'{\i}a son medidas por un sistema 
subterr\'aneo de 128~m$^2$. Este experimento combinado permite medir 
simult\'aneamente las cascadas electromagn\'etica, mu\'onica y hadr\'onica;
adem\'as de ser, probablemente, el experimento que provee las medidas m\'as
precisas y confiables en el rango $E\sim 10^{14}-10^{17}$~eV, la 
informaci\'on ``redundante'' que es capaz de proveer presenta la posibilidad de 
poner a prueba los modelos fenomenol\'ogicos de interacci\'on hadr\'onica en procesos de muy alta energ\'{\i}a. 
Como parte de las perspectivas futuras del experimento, se proyecta su extensi\'on para permitir la 
detecci\'on de lluvias atmosf\'ericas para energ\'{\i}as del primario de hasta $10^{18}$~eV, en un 
proyecto conocido como KASCADE-Grande \cite{kam03,hau04}. 

\newpage

%% file: capitulo2.tex
\chapter{La difusi\'on de rayos c\'osmicos en la galaxia}

\begin{center}
\begin{minipage}{5.6in}
\textsl{
En este Cap\'{\i}tulo, estudiamos en detalle la propagaci\'on de rayos c\'osmicos 
en los campos magn\'eticos turbulentos de la galaxia, teniendo en cuenta su 
car\'acter esencialmente difusivo. En este escenario,
la apropiada inclusi\'on de los efectos de drift macrosc\'opico provee una explicaci\'on
natural de las observaciones del espectro, la composici\'on y las anisotrop\'{\i}as
(amplitud y fase en ascenci\'on recta) de los rayos c\'osmicos en el rango $E\simeq 10^{15}-10^{18}$~eV.
En efecto, la rodilla y la segunda rodilla del espectro resultan del crossover entre el r\'egimen dominado
por la difusi\'on transversal (a bajas energ\'{\i}as) y el dominado por los drifts (a energ\'{\i}as altas); 
veremos que la rodilla aparece como consecuencia de la supresi\'on de los protones, mientras que la segunda rodilla 
resulta de la supresi\'on de los n\'ucleos de hierro. Por otra parte, la tendencia hacia componentes m\'as pesadas 
en la composici\'on, y la dependencia con la energ\'{\i}a de las anisotrop\'{\i}as, reproducen adecuadamente
las observaciones m\'as recientes.
Finalmente, se presenta un estudio del tensor de difusi\'on en condiciones
de alta turbulencia, relevante en el contexto del transporte de rayos c\'osmicos gal\'acticos de muy alta energ\'{\i}a, 
aunque tambi\'en \'util en otras aplicaciones de inter\'es astrof\'{\i}sico.}
\end{minipage}
\end{center}

\section{El escenario de la difusi\'on turbulenta y drift}

\subsection{Las componentes regular y turbulenta \\ del campo magn\'etico gal\'actico}
\begin{figure}
\centerline{{\epsfxsize=3.5in \epsffile{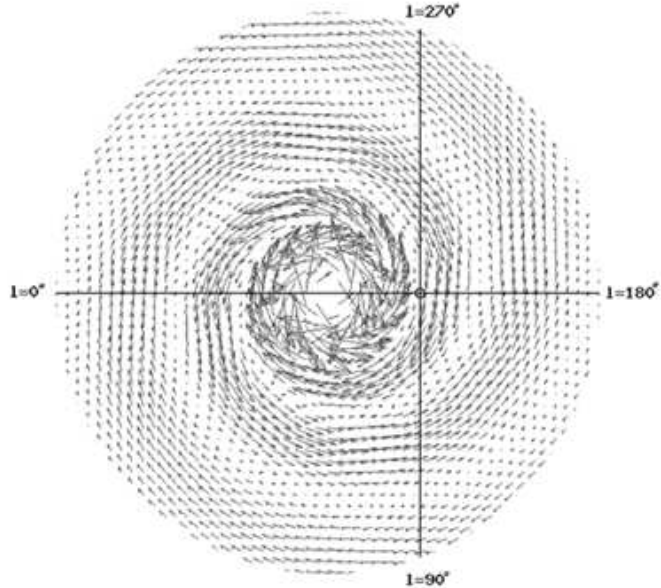}}}
\caption{Esquema del campo regular del disco de la V\'{\i}a L\'actea. El sol, ubicado sobre el plano gal\'actico, 
a una distancia de 8.5~kpc del centro, aparece indicado en coordenadas gal\'acticas (de \cite{one01}).}
\label{field}
\end{figure}

Como otras galaxias espirales, la V\'{\i}a L\'actea tiene un campo magn\'etico regular
de gran escala ${\bf{B_0(x)}}$ en el que las l\'{\i}neas de campo tienden a seguir la
estructura de los brazos espirales. Este campo se observa haciendo uso del 
efecto Faraday, es decir, midiendo la rotaci\'on en el plano de polarizaci\'on de 
la radiaci\'on producida en p\'ulsares y fuentes de radio extragal\'acticas 
\cite{ruz88,han01}. Diferentes estructuras del campo magn\'etico regular de 
galaxias espirales son descriptas por una variedad de modelos. Una primera 
clasificaci\'on distingue entre campos bisim\'etricos (BSS) y axisim\'etricos (ASS),
definidos de modo que los campos BSS (ASS) son antisim\'etricos (sim\'etricos) con
respecto a una rotaci\'on de $\pi$ radianes en torno de un eje normal al plano 
gal\'actico que pasa por el centro. Por ejemplo, en una galaxia espiral con dos 
brazos, el campo invierte su signo entre ellos en un modelo BSS, mientras que
mantiene su orientaci\'on en un modelo ASS. En el caso particular de la V\'{\i}a 
L\'actea, el campo magn\'etico regular en el disco gal\'actico es predominantemente 
toroidal, alineado con la estructura espiral y con inversiones de orientaci\'on 
entre brazos contiguos, de modo que parece corresponder a un modelo BSS.
La fig.\ref{field} muestra esquem\'aticamente una estructura BSS
que podr\'{\i}a corresponder al campo regular del disco de la V\'{\i}a L\'actea, e indica
tambi\'en la posici\'on del sol en coordenadas gal\'acticas (ubicado sobre el
plano gal\'actico, a una distancia de 8.5~kpc del centro).   
Adicionalmente, los modelos de campo magn\'etico pueden clasificarse como 
sim\'etricos (S) o antisim\'etricos (A) con respecto al plano gal\'actico; en particular, 
el campo magn\'etico regular en el disco gal\'actico de la V\'{\i}a L\'actea es usualmente 
considerado como sim\'etrico con respecto al plano gal\'actico, de modo que corresponder\'{\i}a
a un modelo BSS-S. 

La detecci\'on de una significativa emisi\'on de ondas de radio no t\'ermicas a 
grandes latitudes gal\'acticas indicar\'{\i}a la existencia de una corona con 
campos magn\'eticos formando un halo, que podr\'{\i}a extenderse 
posiblemente hasta una distancia de 10 kpc desde el plano gal\'actico.
Aunque en este halo magn\'etico la orientaci\'on del campo regular no es conocida,
parece ser dominante la componente toroidal. La estructura del campo magn\'etico 
regular del halo podr\'{\i}a ser similar a la del campo en el disco gal\'actico (por
ejemplo, si su origen est\'a relacionado con un viento gal\'actico que de alg\'un 
modo extiende hacia el halo las propiedades del campo del disco), pero tambi\'en
podr\'{\i}a ser diferente (por ejemplo, considerando que es producida en forma independiente
mediante un mecanismo de d\'{\i}namo \footnote{El mecanismo de d\'{\i}namo, 
sugerido originalmente por Larmer en 1919 \cite{lar19}, 
explica el origen de estructuras de campo magn\'etico a gran escala como resultado de una realimentaci\'on
constructiva en el acoplamiento entre el campo magn\'etico y el campo de velocidades en un flu\'{\i}do 
magnetohidrodin\'amico turbulento. Bajo ciertas geometr\'{\i}as particulares, las corrientes inducen campos 
magn\'eticos que, a su vez, realimentan y amplifican las corrientes iniciales, hasta alcanzar un estado de 
saturaci\'on debido a mecanismos disipativos.}). 

Considerando las principales caracter\'{\i}sticas del campo magn\'etico regular que hemos
mencionado anteriormente, aqu\'{\i} asumiremos que la regi\'on de propagaci\'on de 
los rayos c\'osmicos en la galaxia es un cilindro de radio $R=20$~kpc y altura $2H$, 
con un halo de tama\~no $H=10$~kpc. El \'angulo que forman los brazos espirales respecto
de la direcci\'on azimutal es peque\~no (menor a $10^\circ$), de modo que
es una buena aproximaci\'on suponer que el campo magn\'etico regular est\'a orientado seg\'un
la direcci\'on azimutal. 
En consecuencia, resulta natural adoptar coordenadas cil\'{\i}ndricas
$(r,\phi,z)$, tomando ${\bf{B_0}}=B_0\hat{\phi}$ y asumiendo por simplicidad que el
sistema posee simetr\'{\i}a azimutal. Puede notarse, adem\'as, que un campo 
azimutal independiente de la coordenada $\phi$ tiene divergencia nula, de manera que la consistencia
con las ecuaciones de Maxwell est\'a asegurada (equivalentemente, desde un punto de
vista m\'as f\'{\i}sico, las l\'{\i}neas de campo son circunferencias cerradas, 
de modo que no hay fuentes ni sumideros de l\'{\i}neas de campo).   

Para reproducir las inversiones de orientaci\'on del campo propias del modelo BSS-S, 
adoptaremos para la componente regular del disco gal\'actico la expresi\'on  
\begin{equation}
B_0^{disco}(r,z)=B_d\sqrt{{{1+(r_{obs}/r_c)^2}}\over{1+(r/r_c)^2}} 
\sin\left({{\pi r}\over{4~{\rm kpc}}}\right){\rm
th}^2\left(\frac{r}{1\ {\rm kpc}}\right){{1}\over{\cosh(z/z_d)}}\ ,
\label{b0d}
\end{equation}
donde $r_c=4$~kpc representa un radio caracter\'{\i}stico que suaviza el comportamiento
radial global $1/r$ en la regi\'on del centro gal\'actico, $z_d$ es una altura caracter\'{\i}stica
del disco gal\'actico, y $B_d$ es una constante que fija el valor del campo en nuestro 
punto de observaci\'on ($r_{obs}=8.5$~kpc, $z_{obs}=0$), que es negativo (ya que el campo magn\'etico
gal\'actico local est\'a dirigido aproximadamente hacia la longitud $l\simeq 90^\circ$, 
es decir, en la direcci\'on de $-\hat{\phi}$). El factor th$^2(r)$ ha sido introducido s\'olo para
asegurar que no haya singularidades en el eje $r=0$, dado que, de lo contrario, aparecer\'{\i}a
una corriente de deriva ({\it{drift}}) singular espuria a lo largo de ese eje 
(ver la ec.~(\ref{uzeq}) m\'as adelante). 
Adem\'as, puede notarse que el factor sinusoidal introduce inversiones de orientaci\'on 
que ocurren para $r=4,$ 8, 12 y 16 kpc.   

La componente regular correspondiente al halo puede describirse mediante una estructura
de campo con inversiones radiales y sim\'etrica con respecto al plano gal\'actico 
(que llamaremos $R-S$), o bien por un modelo sin inversiones radiales (que llamaremos $NR$);
en este \'ultimo caso, cuatro diferentes configuraciones resultan de considerar la orientaci\'on
relativa del campo en el halo con respecto a la del disco, y de la simetr\'{\i}a con respecto al
plano gal\'actico, que denotaremos con $NR-S\pm$ y $NR-A\pm$ (donde el signo $\pm$ es el del 
producto $B_0^{halo}\cdot B_0^{disco}$ evaluado en $r=r_{obs}$ y $z>0$). Debe notarse que, al
considerar un halo con inversiones radiales, estamos suponiendo que el halo tiene su origen
en un fen\'omeno asociado al disco gal\'actico (por ejemplo mediante un viento gal\'actico, como
hemos mencionado anteriormente), de modo que la estructura del campo magn\'etico deber\'{\i}a ser
similar. \'Esto justifica que, en este caso, sea suficiente considerar s\'olo la configuraci\'on sim\'etrica
$R-S$, mientras que un halo sin inversiones radiales podr\'{\i}a tener un origen independiente del
campo del disco y requiere el estudio de varias configuraciones diferentes.  
Recientes estudios que utilizan el efecto Faraday podr\'{\i}an indicar que la estructura regular 
del halo magn\'etico es la de un campo azimutal con orientaci\'on invertida arriba y debajo del
plano gal\'actico, que podr\'{\i}a corresponder a una configuraci\'on de un d\'{\i}namo A0 \cite{han03}. 
Estos resultados podr\'{\i}an sugerir que el modelo $NR-A$ es el m\'as apropiado, 
aunque debe tenerse en cuenta que la 
evidencia observacional no est\'a a\'un suficientemente bien establecida, de modo que 
todas las configuraciones que hemos descripto resultan plausibles. 

Si la componente regular del halo tiene inversiones radiales, consideraremos
que estar\'a dada por   
\begin{equation}
B_0^{halo}(r,z)=B_h {{1}\over{\sqrt{1+(r/r_c)^2}}} {{1}\over{\sqrt{1+(z/z_h)^2}}} 
\sin\left({{\pi r}\over {4~{\rm  kpc}}}\right){\rm
th}^2\left(\frac{r}{1\ {\rm kpc}}\right)  \ \ ,
\label{b0hr}
\end{equation}   
mientras que si su estructura no tiene inversiones radiales, asumiremos
\begin{equation}
B_0^{halo}(r,z)=B_h {{1}\over{\sqrt{1+(r/r_c)^2}}} {{1}\over{\sqrt{1+(z/z_h)^2}}} 
{\rm
th}^2\left(\frac{r}{1\ {\rm kpc}}\right)\ {\mathcal{R}} \ \ ,
\label{b0h}
\end{equation}
donde $z_h$ es una altura caracter\'{\i}stica del halo, y $B_h$ es una constante que fija el
valor y la orientaci\'on local del campo del halo. La funci\'on $\mathcal{R}$ es simplemente la unidad 
para el caso del halo sim\'etrico (con respecto al plano gal\'actico), mientras que es tomada como 
${\mathcal{R}}(z)={\rm tanh}(z/z_d)$ en el caso de los modelos antisim\'etricos. 

Las ecs. (\ref{b0d})--(\ref{b0h}) proveen una descripci\'on simple pero realista para la estructura regular a gran
escala de los campos magn\'eticos de la galaxia, y contienen las caracter\'{\i}sticas esenciales (como,
por ejemplo, las inversiones radiales de orientaci\'on) relevantes para la determinaci\'on de los
efectos de drift en la propagaci\'on de los rayos c\'osmicos. 

Superpuesto a las estructuras regulares del campo magn\'etico del disco gal\'actico y el halo extendido, 
existe un campo magn\'etico irregular ${\bf B_r(x)}$ estoc\'astico ({\it random}) 
producido por la turbulencia del plasma interestelar.
La escala m\'axima de las irregularidades es $L_{max}\simeq 100$~pc, y localmente su intensidad es comparable
a la de la componente gal\'actica regular,  $\langle B_r\rangle/\langle|{\bf B_0}|\rangle\sim1$\ --\ 3.
Los datos observacionales \cite{arm81} muestran que el espectro 
de inhomogeneidades en la densidad del gas interestelar
podr\'{\i}a ser el mismo que el correspondiente a la componente irregular 
del campo magn\'etico, y \'esto corresponde a un espectro de Kolmogorov \cite{kol41}. 
En consecuencia, las componentes de Fourier del campo magn\'etico random contienen una densidad de energ\'{\i}a magn\'etica
d$E_r/{\rm d}k\propto k^{-5/3}$ (con $k\geq2\pi/L_{max}$). Asumiremos que la intensidad de la componente
irregular posee una dependencia espacial suave, an\'aloga a la envolvente asumida para los campos regulares,
dada por         
\begin{equation}
B_r(r,z)=B_r^*\sqrt{{1+(r_{obs}/r_c)^2}\over{1+(r/r_c)^2}} 
{{1}\over{\cosh(z/z_r)}}\ ,
\label{brand}
\end{equation}
donde $z_r$ es la escala caracter\'{\i}stica del perfil vertical y $B_r^*$ la amplitud cuadr\'atica media (rms) local. 

En el contexto de este trabajo, 
estamos interesados en describir la propagaci\'on de part\'{\i}culas ultrarrelativistas, cuya
velocidad $v\simeq c$ es mucho mayor que la velocidad de Alfv\'en ($v_A\simeq 5\times 10^4$~m/s) caracter\'{\i}stica de la 
propagaci\'on de las ondas magn\'eticas en el medio interestelar \footnote{En magnetohidrodin\'amica, se considera la
propagaci\'on de tres tipos de movimiento ondulatorio: las ondas de sonido, que son longitudinales y se propagan en
la direcci\'on del campo con velocidad $v_s=\sqrt{kT/m_i}$, donde $T$ es la temperatura de los electrones t\'ermicos y
$m_i$ la masa de los iones en el plasma; las ondas de Alfv\'en, que son transversales y se propagan en la direcci\'on
del campo con velocidad $v_A=B/\sqrt{4\pi\rho}$, donde $\rho$ es la densidad del plasma; y las ondas magnetos\'onicas,
que son longitudinales y se propagan perpendicularmente al campo con velocidad $\sqrt{v_s^2+v_A^2}$. En las condiciones 
del medio interestelar, es habitual asumir que $v_s\ll v_A$ \cite{ber90}, de modo que la velocidad de Alfv\'en es 
caracter\'{\i}stica de la propagaci\'on de las perturbaciones magnetohidrodin\'amicas.}. 
En consecuencia, est\'a bien justificada la suposici\'on de 
considerar perturbaciones magn\'eticas est\'aticas, despreciando los campos el\'ectricos.   

\subsection{La ecuaci\'on de difusi\'on}

Una part\'{\i}cula de carga $Ze$, propag\'andose en un campo magn\'etico regular uniforme ${\bf B_0}$, 
describe una trayectoria helicoidal, caracterizada por el \'angulo de pitch $\theta$ (definido como 
el \'angulo entre la velocidad instant\'anea y la direcci\'on del campo) y el radio de Larmor, dado por
\begin{equation}
r_L\equiv {pc\over Ze|\bf{B_0}|}\simeq{{E/Z}\over{10^{15}~{\rm eV}}}\left({|{\bf{B_0}|}
\over{\mu{\rm G}}}\right)^{-1} {\rm pc}\ ,
\label{rl} 
\end{equation}
donde la \'ultima expresi\'on es v\'alida en el l\'{\i}mite relativista. La componente de la velocidad paralela
a ${\bf B_0}$ es $v_\parallel=c\ {\rm cos}\theta$, mientras que el radio de la h\'elice es 
\footnote{Otra definici\'on del radio de Larmor resulta de reemplazar 
$p\to p_\perp=p\ {\rm sen}\theta$ en la ec.~(\ref{rl}), de modo que $a=r_L$; esta elecci\'on es la usual en los libros de texto 
sobre electromagnetismo (por ej., Jackson \cite{jac99}, p. 586).} $a=r_L\ {\rm sen}\theta$.
La rigidez magn\'etica de la part\'{\i}cula se define como
\begin{equation}
R\equiv {pc\over Ze}\ ;
\end{equation}
para una part\'{\i}cula relativista, $R=E[{\rm eV}]/Z$~Volts. 

En presencia de un campo magn\'etico random ${\bf B_r}$,
las part\'{\i}culas se dispersan en las irregularidades del campo estoc\'astico y cambian el \'angulo de pitch;
en este proceso, conocido como ``dispersi\'on del \'angulo de pitch'' (${\it pitch\ angle\ scattering}$), 
la velocidad, la rigidez y el radio de Larmor (\ref{rl}) permanecen constantes. La dispersi\'on
del \'angulo de pitch ocurre con probabilidad m\'axima en condiciones de resonancia (es decir, cuando la escala de las irregularidades
es del orden de $r_L$), de modo que es un mecanismo de isotropizaci\'on eficiente siempre que $r_L<L_{max}$.  
Por ejemplo, en el campo magn\'etico gal\'actico, caracterizado por una amplitud $|{\bf B_0}|$ de algunos $\mu$G y 
una escala m\'axima de turbulencia $L_{max}\simeq 100$~pc, 
la dispersi\'on del \'angulo de pitch es eficiente para protones hasta energ\'{\i}as por encima de $10^{17}$~eV.  

Por consiguiente, para energ\'{\i}as compatibles con la condici\'on de resonancia, $r_L<L_{max}$, 
el transporte de rayos c\'osmicos a trav\'es 
de los campos magn\'eticos gal\'acticos turbulentos queda bien descripto por medio de una ecuaci\'on de difusi\'on,
en la cual las componentes del tensor de difusi\'on dependen tanto del campo regular como de la turbulencia
estoc\'astica. Despreciando la fragmentaci\'on nuclear, las p\'erdidas de energ\'{\i}a de cualquier tipo 
(por ejemplo, por ionizaci\'on, bremsstrahlung, emisi\'on sincrotr\'onica, etc.), efectos convectivos (que 
podr\'{\i}an deberse, por ejemplo, a un posible viento gal\'actico), procesos de reaceleraci\'on, etc., la ecuaci\'on
estacionaria de transporte (\ref{transp}) resulta
\begin{equation}
\nabla\cdot{\bf J}=Q(\textbf{x})\ ,
\label{diffeq1} 
\end{equation}
donde $Q$ es el t\'ermino que describe a las fuentes, y donde la corriente de rayos c\'osmicos,
dada en t\'erminos de la densidad $N(\textbf{x})$, es
\begin{equation}
J_i=-D_{ij}(\textbf{x})\nabla_jN(\textbf{x})\ .
\label{jcr}
\end{equation}
En general, el tensor de difusi\'on $D_{ij}$ puede escribirse como \footnote{La estructura del tensor de 
difusi\'on, ec.~(\ref{dij}), est\'a explicada en el Ap\'endice A. Aqu\'{\i} asumimos $D_A>0$ para la difusi\'on de part\'{\i}culas
con carga positiva, mientras que las de carga negativa tienen asociado el signo opuesto.}
\begin{equation}
D_{ij}=\left(D_{\parallel}-D_{\perp}\right)b_ib_j+D_{\perp}\delta_{ij}+D_A\epsilon_{ijk}b_k\ ,
\label{dij}
\end{equation}
donde ${\bf{b}}={\bf{B_0}}/{|\bf{B_0}|}$ es un vector unitario a lo largo del campo 
magn\'etico regular, $\delta_{ij}$ es la delta de Kronecker, y $\epsilon_{ijk}$ es el
tensor antisim\'etrico de Levi-Civita. Los t\'erminos sim\'etricos contienen los coeficientes de difusi\'on
que corresponden a las direcciones paralela y perpendicular al campo regular ($D_\parallel$ y $D_\perp$) 
y describen la difusi\'on debida a las fluctuaciones turbulentas de peque\~na escala. 
El t\'ermino antisim\'etrico contiene el coeficiente de difusi\'on de Hall ($D_A$) y es el responsable de la corriente
macrosc\'opica de drift \footnote{El 
drift macrosc\'opico resulta del comportamiento colectivo de una distribuci\'on de part\'{\i}culas, y no debe ser confundido 
con el drift microsc\'opico, asociado a la curvatura y los gradientes en ${\bf B_0}$.}
de las part\'{\i}culas en propagaci\'on. La ecuaci\'on de difusi\'on puede ser reescrita como 
\begin{equation}
-\nabla_iD_{Sij}\nabla_jN+u_i\nabla_iN=Q\ ,
\label{diffeq2}
\end{equation}
donde $D_{Sij}= \left(D_\parallel-D_\perp\right) b_ib_j+D_\perp\delta_{ij}$ contiene la parte sim\'etrica del
tensor de difusi\'on, mientras que $u_i=-\epsilon_{ijk}\nabla_k\left(D_Ab_j\right)$ es la velocidad de drift.
Por otro lado, la corriente de rayos c\'osmicos dada por la ec.~(\ref{jcr}) puede tambi\'en reescribirse como
\begin{equation}
{\bf J}=-D_\perp\nabla_\perp N-D_\parallel\nabla_\parallel N+D_A{\bf b}\times\nabla N\ ,
\label{curr1}
\end{equation}
donde $\nabla_\parallel\equiv {\bf b}({\bf b}\cdot\nabla)$ y $\nabla_\perp\equiv\nabla-\nabla_\parallel$.
En esta expresi\'on se observa expl\'{\i}citamente que la corriente macrosc\'opica de drift (dada por el \'ultimo t\'ermino 
de la ec.~(\ref{curr1})) es ortogonal tanto a la direcci\'on del campo regular como al gradiente de densidad de rayos c\'osmicos. 

Bajo la suposici\'on de que el campo magn\'etico regular est\'a dirigido 
en la direcci\'on azimutal (es decir, $b_r=b_z=0$, $b_{\phi}=\pm1$) y de que el sistema posee
simetr\'{\i}a azimutal (${{\partial}/{\partial\phi}}=0$), resulta $\nabla_\perp N\equiv\nabla N$ y la 
corriente macrosc\'opica de rayos c\'osmicos queda dada por
\begin{equation}
{\bf J}=-D_\perp\nabla N+D_A{\bf b}\times\nabla N\ ,
\label{curr2}
\end{equation}
es decir, desaparece el t\'ermino que involucra a $D_\parallel$ y la corriente resulta perpendicular al 
campo magn\'etico regular. Dicho en otros t\'erminos, las simetr\'{\i}as asumidas excluyen la
difusi\'on en la direcci\'on del campo (o sea, en la direcci\'on que corresponde a los brazos espirales).
Reemplazando (\ref{curr2}) en (\ref{diffeq1}), la ecuaci\'on de difusi\'on en coordenadas cil\'{\i}ndricas resulta   
\begin{equation}
\left[-{{1}\over{r}}{{\partial}\over{\partial r}}\left(rD_{\perp}
{{\partial}\over{\partial r}}\right)-
{{\partial}\over{\partial z}}\left(D_{\perp}{{\partial}\over{\partial z}}\right)+
u_r{{\partial}\over{\partial r}}+u_z{{\partial}\over{\partial
z}}\right] N(r,z)=Q(r,z)\ , 
\label{difeqcil}
\end{equation}  
donde las velocidades de drift son 
\begin{equation}
u_r=-{{\partial(D_Ab_{\phi})}\over{\partial z}}
\label{ureq}
\end{equation}
y
\begin{equation}
u_z={{1}\over{r}}{{\partial(rD_Ab_{\phi})}\over{\partial r}}\ . 
\label{uzeq}
\end{equation}
En estas expresiones, vemos que un cambio de signo en $b_\phi$ (como resulta de una inversi\'on en la orientaci\'on 
de los campos regulares) afecta las velocidades de drift y, por consiguiente, 
la difusi\'on en esa regi\'on.

Sobre la superficie exterior $\Sigma$ correspondiente a los bordes del halo, 
la densidad de rayos c\'osmicos debe anularse, dado que las part\'{\i}culas pueden escapar libremente 
hacia el espacio intergal\'actico. Por lo tanto, las ecs.~(\ref{difeqcil})--(\ref{uzeq}) deben considerarse con 
la condici\'on de borde $N|_{\Sigma}=0$. 

Con respecto a la distribuci\'on espacial de las fuentes, asumiremos que est\'an contenidas en un disco delgado de
altura $2h_s=400$~pc, de modo que el t\'ermino de las fuentes puede escribirse como $Q(r,z)=q(r)\theta(h_s-|z|)$, 
donde $\theta$ es la funci\'on de Heaviside y $q(r)$ es una funci\'on de distribuci\'on radial. 
Como se describe m\'as adelante, en la Secci\'on 2.1.4, la ecuaci\'on
de difusi\'on puede resolverse num\'ericamente mediante la discretizaci\'on de Crank-Nicolson y el m\'etodo de
Liebmann \cite{pre92}.   

\subsection{C\'alculo de los coeficientes de difusi\'on}
La difusi\'on en la direcci\'on del campo magn\'etico regular, debida a la dispersi\'on
del \'angulo de pitch, conduce a un coeficiente de difusi\'on dado por 
\begin{equation}
D_\parallel={{c}\over{3}}\lambda_\parallel\ ,
\label{dpar1}
\end{equation}
donde $\lambda_\parallel$ es el camino libre medio en la direcci\'on paralela al campo \cite{ptu93}. 
En esta expresi\'on, $\lambda_\parallel$ depende de la densidad de energ\'{\i}a asociada a los modos del campo 
magn\'etico random en escalas del orden de $r_L$, es decir,
\begin{equation} 
\lambda_\parallel\propto\left.{{r_L}\over{{\rm d}E_r/{\rm d\ ln}k}}\right|_{k=2\pi/r_L}\ ,
\label{dpar2} 
\end{equation} 
donde ${\rm d}E_r/{\rm d}k$ es el espectro de Fourier de la densidad de energ\'{\i}a contenida en el 
campo magn\'etico random. 

En general, una dada componente del tensor de difusi\'on est\'a asociada a ciertos procesos f\'{\i}sicos particulares 
que pueden no contribuir a las otras componentes. La difusi\'on transversal al campo magn\'etico regular es debida tanto 
a la dispersi\'on del \'angulo de pitch (es decir, el mecanismo que prevalece en el caso de la difusi\'on paralela) como 
a la propia difusi\'on de las l\'{\i}neas de campo, que arrastran con ellas a las part\'{\i}culas en la direcci\'on perpendicular 
a ${\bf B_0}$ (un proceso conocido como {\it field line random walk}).   
Cuando el nivel de turbulencia es peque\~no, la difusi\'on en la direcci\'on transversal es mucho m\'as lenta 
que en la direcci\'on paralela; en cambio, en el l\'{\i}mite de turbulencia muy alta, la anisotrop\'{\i}a debida al campo 
regular tiende a desaparecer, y los movimientos paralelo y perpendicular resultan similares.   

Sin embargo, no existe un tratamiento te\'orico general de este problema que permita, una vez definidos los 
campos, la determinaci\'on de los coeficientes de difusi\'on. Considerando el l\'{\i}mite de baja turbulencia, que permite un 
tratamiento perturbativo del campo magn\'etico random, los coeficientes de difusi\'on resultan    
\begin{equation}
D_{\parallel} = {{v r_L}\over{3}}{\omega\tau_{\parallel}}\ ,
\label{dpar}
\end{equation}
\begin{equation}
D_{\perp} = {{v r_L}\over{3}}{{\omega\tau_{\perp}}
\over{1+\left(\omega \tau_{\perp} \right)^2}}
\label{dperpeq}
\end{equation}
y
\begin{equation}
D_A={{v r_L}\over{3}}{{\left(\omega\tau_A\right)^2}
\over{1+\left(\omega \tau_A \right)^2}}\ ,
\label{daeq}
\end{equation}
donde $v\simeq c$ es la velocidad de las part\'{\i}culas, $\omega\equiv v/r_L$ es la frecuencia angular 
de Larmor, y $\tau_\parallel=\tau_\perp=\tau_A\equiv\tau$ representa un tiempo caracter\'{\i}stico de descorrelaci\'on 
de las trayectorias. Debido a que en este tipo de an\'alisis se asume un \'unico mecanismo de scattering como responsable de todas 
las descorrelaciones, interviene una \'unica escala temporal $\tau$ para los tres coeficientes de difusi\'on. 
\'Esto ocurre, por ejemplo,
cuando se considera la evoluci\'on de una distribuci\'on de part\'{\i}culas casi isotr\'opica bajo un proceso de 
dispersi\'on homog\'eneo, en el que $\tau$ representa el tiempo caracter\'{\i}stico de isotropizaci\'on \cite{ise79}. Una 
estructura similar se obtiene tambi\'en para los coeficientes de difusi\'on asociados a procesos de colisi\'on 
en un plasma en equilibrio t\'ermico \cite{bal94}, o bien, considerando $\tau$ como un tiempo caracter\'{\i}stico de
dispersi\'on y tratando las perturbaciones magn\'eticas como centros dispersores constitu\'{\i}dos 
por esferas r\'{\i}gidas \cite{gle69}, 
en lo que se conoce como el ``resultado de dispersi\'on cl\'asico'' \cite{par65,cha70,for75}.

En condiciones de alta turbulencia, en cambio, el tratamiento perturbativo no puede aplicarse, y debemos recurrir a un 
enfoque anal\'{\i}tico de otra naturaleza \cite{bie97}, o bien a simulaciones num\'ericas \cite{gia99,cas02}. 
Debe notarse que el problema de la difusi\'on de rayos c\'osmicos en la galaxia requiere tratar con
niveles de alta turbulencia debido a que, por un lado, la intensidad del campo magn\'etico turbulento es del
mismo orden que la del campo regular y, por otra parte, a que en las regiones en las que el campo regular tiene
inversiones de orientaci\'on su amplitud es peque\~na, de modo que all\'{\i} prevalece largamente la componente estoc\'astica. 

Utilizando el formalismo de Kubo \cite{kub57,for77,bie97}, 
los coeficientes de difusi\'on $D_{ij}$ pueden obtenerse, alternativamente,
de la descorrelaci\'on entre diferentes componentes de la velocidad de una part\'{\i}cula,  
promediada sobre una distribuci\'on isotr\'opica de muchas part\'{\i}culas y sobre muchas configuraciones de campo 
random diferentes. En efecto, resulta\footnote{La f\'ormula de Kubo aparece discutida en el Ap\'endice B.}
\begin{equation}
D_{ij}=\int_0^\infty{\rm{d}}t\ R_{ij}(t)\ ,
\label{kubo}
\end{equation}
con $R_{ij}(t)=\langle v_{0i}v_j(t)\rangle$ (donde $\langle ...\rangle$ corresponde al promedio sobre distribuciones y
configuraciones de campo, 
y con $v_i(t=0)\equiv v_{0i}$). En \cite{bie97}, las funciones de descorrelaci\'on se obtienen a partir de las 
expresiones de $R_{ij}$ correspondientes a trayectorias helicoidales, agregando factores {\it ad hoc} de modulaci\'on exponencial  
para representar el efecto de la turbulencia. Es decir, asumiendo que el campo regular est\'a dirigido seg\'un ${\bf\hat{z}}$, 
se adoptan las expresiones
\begin{equation}
R_{xx}(t)=R_{yy}(t)={{c^2}\over{3}}\cos\omega t\ e^{-t/\tau_\perp}\ ,
\label{rxx}
\end{equation}
\begin{equation}
R_{xy}(t)=-R_{yx}(t)={{c^2}\over{3}}\sin\omega t\ e^{-t/\tau_A}\ 
\label{rxy}
\end{equation}
y
\begin{equation}
R_{zz}(t)={{c^2}\over{3}}\ e^{-t/\tau_\parallel}\ ,
\label{rzz}
\end{equation}
donde las tres escalas temporales de descorrelaci\'on, $\tau_\parallel, \tau_\perp$ y $\tau_A$, son en principio diferentes. 
Si bien este enfoque pretende ser general, y v\'alido a\'un en condiciones de alta turbulencia \cite{bie97}, debe notarse 
que las expresiones (\ref{rxx})-(\ref{rzz}) est\'an fuertemente inspiradas en el comportamiento que se observa a baja
turbulencia; el rango de validez de este enfoque no es evidente de inmediato, y ser\'a reconsiderado m\'as adelante, 
en la Secci\'on 2.4. Sin embargo, resulta interesante observar que, reemplazando las ecs.~(\ref{rxx})-(\ref{rzz}) para
$R_{ij}$ en la integral (\ref{kubo}), se reobtienen los coeficientes de difusi\'on dados por
las ecs.~(\ref{dpar})-(\ref{daeq}), donde $D_\parallel=D_{zz}, D_\perp=D_{xx}=D_{yy}$ y $D_A=D_{xy}=-D_{yx}$. 
Es decir que este enfoque, que incorpora diferentes escalas de descorrelaci\'on, puede
ser visto como una generalizaci\'on de los resultados perturbativos. 

Para poder aplicar las ecs.~(\ref{dpar})--(\ref{daeq}) al c\'alculo de los coeficientes de difusi\'on en un 
problema concreto, todav\'{\i}a debemos especificar de alg\'un modo los tiempos de descorrelaci\'on.
Una alternativa es recurrir a hip\'otesis adicionales que simplifiquen el problema, vinculando las escalas 
temporales de descorrelaci\'on. Por ejemplo, en \cite{bie97} se asume que el tiempo de descorrelaci\'on
asociado a la difusi\'on perpendicular es el mismo que el correspondiente a la difusi\'on antisim\'etrica. 
Asumiendo, adem\'as, que la contribuci\'on principal a la difusi\'on 
perpendicular es la debida a la caminata aleatoria de las l\'{\i}neas de campo, se llega a la expresi\'on  
\begin{equation}
\omega\tau_\perp=\frac{2}{3}\frac{r_L}{D_{flrw}}\ ,
\label{omtau}
\end{equation}      
que relaciona la escala temporal de descorrelaci\'on $\tau_\perp$ con el coeficiente de difusi\'on $D_{flrw}$ asociado
a la caminata aleatoria de las l\'{\i}neas de campo. Luego, estimando $D_{flrw}$ anal\'{\i}tica o num\'ericamente en
funci\'on del nivel de turbulencia y de la forma del espectro de turbulencia asumido, se pueden calcular
los coeficientes $D_\perp$ y $D_A$. Sin embargo, la ec.~(\ref{omtau}) conduce a discrepancias evidentes con los resultados 
de \cite{gia99}, obtenidos mediante simulaciones extensivas Monte Carlo, sugiriendo que las hip\'otesis adicionales asumidas en 
\cite{bie97} no son completamente apropiadas.  

Recientemente, Casse et al. \cite{cas02} extendieron los resultados num\'ericos de \cite{gia99} y determinaron la 
difusi\'on paralela y transversal de part\'{\i}culas de prueba relativistas propag\'andose en campos estoc\'asticos de 
Kolmogorov de gran intensidad. En este estudio se observa que, por lo general, la extrapolaci\'on de los 
resultados perturbativos es inapropiada para describir la difusi\'on de 
part\'{\i}culas en campos altamente turbulentos. Por ejemplo, aunque la ec.~(\ref{dpar}) para la difusi\'on paralela 
parece concordar bien con los resultados obtenidos en \cite{cas02},  
la ec.~(\ref{dperpeq}) para la difusi\'on transversal (con $\tau_\parallel=\tau_\perp$) 
es claramente inapropiada, como es de esperar del hecho de que, en general, la caminata aleatoria de las l\'{\i}neas de campo 
no puede despreciarse.      

En este contexto, lo que haremos es extraer directamente el coeficiente de difusi\'on transversal de los
resultados num\'ericos de \cite{cas02}. Estos resultados son aplicables en el r\'egimen de inter\'es para
el estudio de la regi\'on de la rodilla, ya que el par\'ametro de energ\'{\i}a relevante, definido como
$\rho\equiv r_L/L_{max}$, toma el valor $\rho\simeq 0.02$ para protones cerca de la rodilla, que est\'a
contenido en el rango $0.001\leq 2\pi\rho\leq 10$ que se explora en \cite{cas02}.
Definiendo el nivel de turbulencia como $\eta\equiv B_r^2/(B_0^2+B_r^2)$, 
se encuentra que una expresi\'on aproximada para $D_\perp$, 
v\'alida en el rango $0.1\leq\eta\leq 1$, es
\begin{equation}
D_\perp\simeq 8.7\times 10^{27}e^{3.24\eta}\left(1-\eta\right)^{1/6}\left(
{{r_L}\over{\rm{pc}}}\right)^{1/3}{{\rm{cm^2}}\over{\rm{s}}}\ .  
\label{dperp}
\end{equation}
Aunque esta expresi\'on se obtiene de simulaciones num\'ericas, la dependencia con la rigidez a trav\'es de 
una potencia $1/3$ aparece tambi\'en en expresiones te\'oricas, deducidas en el l\'{\i}mite de baja turbulencia,
bajo la suposici\'on de un espectro de Kolmogorov para la componente irregular del campo magn\'etico \cite{ptu93}.

Para caracterizar los efectos de drift partiremos de la ec.~(\ref{daeq}), que
permite obtener el valor de $D_A$ en casos extremos, tanto de muy alta como de muy baja turbulencia. En el caso de
turbulencia muy alta ($\omega\tau_A\to 0$) resulta $D_A\to 0$, y los efectos de drift tienden a ser despreciables.
Este l\'{\i}mite es de particular inter\'es en las regiones con la inversi\'on de orientaci\'on del campo regular, en las que
la turbulencia tiende a suprimir los efectos de drift. Por otra parte, en el l\'{\i}mite de turbulencia muy baja 
($\omega\tau_A\gg 1$) la expresi\'on obtenida tiende al resultado conocido de la aproximaci\'on cuasi-lineal, es decir,
$D_A= vr_L/3$ \cite{ptu93,bie97}. Con el fin de obtener una transici\'on 
suave entre los dos comportamientos l\'{\i}mite, consideraremos las ecs.~(\ref{dpar})-(\ref{daeq}) 
con una escala caracter\'{\i}stica temporal com\'un para todos los procesos de 
descorrelaci\'on \footnote{M\'as arriba hemos hecho notar
que $\tau_\parallel\neq\tau_\perp$. Sin embargo, con el fin de obtener una prescripci\'on que regule el comportamiento 
de $D_A$ (conectando el r\'egimen de baja turbulencia con el de muy alta turbulencia), asumiremos, en lo que sigue, 
que todas las escalas temporales de descorrelaci\'on son iguales.}.
En particular, resulta que 
\begin{equation}
{{D_\parallel}\over{D_\perp}}=1+\left(\omega\tau\right)^2\ .
\label{dpdp}
\end{equation} 
Este cociente, calculado en \cite{cas02} en funci\'on de la rigidez y del nivel de turbulencia, es esencialmente 
independiente de la energ\'{\i}a en el rango estudiado; se encuentra que la simple relaci\'on aproximada 
$\omega\tau\simeq 4.5(|{\bf B_0}|/B_r)^{1.5}$, sustitu\'{\i}da en la ec.~(\ref{dpdp}), 
reproduce razonablemente los resultados num\'ericos. Este resultado para $\omega\tau$ puede ahora ser reemplazado
en la ec.~(\ref{daeq}) (cambiando $\tau_A\to\tau$), y conduce a 
\begin{equation}
D_A\simeq 3\times 10^{28}\left(1+0.049\left({{\eta}\over{1-\eta}}\right)^{3/2}
\right)^{-1}\left({{r_L}\over{\rm{pc}}}\right){{\rm{cm^2}}\over{\rm{s}}}\ .  
\label{da}
\end{equation}
De este modo, se obtiene una expresi\'on para el coeficiente de difusi\'on de Hall que exhibe el 
comportamiento apropiado en los l\'{\i}mites de muy alta y de muy baja turbulencia. 
Tal como se muestra en el estudio num\'erico de la Secci\'on 2.4, la prescripci\'on (\ref{da}) es adecuada 
en este contexto, y los resultados no cambian sustancialmente al adoptar una parametrizaci\'on 
m\'as precisa. 

\subsection{Resoluci\'on num\'erica de la ecuaci\'on de difusi\'on} 

Adoptando un modelo de campo magn\'etico gal\'actico determinado, eligiendo valores particulares para el
conjunto de par\'ametros que interviene en cada modelo, y especificando una funci\'on de distribuci\'on 
radial para las fuentes, la difusi\'on de rayos c\'osmicos se obtiene de resolver la ecuaci\'on que resulta de las 
ecs.~(\ref{difeqcil})--(\ref{uzeq}) con la condici\'on de borde $N|_{\Sigma}=0$.   
La ecuaci\'on diferencial queda definida una vez calculados los coeficientes de difusi\'on y sus derivadas,
que son determinados en funci\'on de la rigidez y del nivel de turbulencia mediante las ecs.~(\ref{dperp}) y (\ref{da}). 

Una soluci\'on anal\'{\i}tica de este problema queda naturalmente exclu\'{\i}da, debido a la complejidad de las
ecuaciones que deben resolverse. 
La resoluci\'on num\'erica se llev\'o a cabo usando la discretizaci\'on de Crank-Nicholson y 
aplicando el m\'etodo de Liebmann \cite{pre92}, luego implementados en un programa FORTRAN. 
A continuaci\'on, describiremos el esquema de discretizaci\'on empleado en este trabajo.

Comencemos por suponer que se busca resolver una ecuaci\'on diferencial de la forma
\begin{equation}
{\mathcal{L}}u=0\ ,
\label{ecdif1}
\end{equation}
donde $\mathcal{L}$ es un operador diferencial que act\'ua sobre la funci\'on $u(x)$, definida en el rango
$x_A\leq x\leq x_B$, con condiciones de borde de Dirichlet $u(x_A)=u_A$, $u(x_B)=u_B$. Si incorporamos a $u$ una dependencia 
temporal, $u(x)\to u(x,t)$, y planteamos la ecuaci\'on diferencial dependiente del tiempo, 
\begin{equation}
{{\partial u}\over{\partial t}}={\mathcal{L}}u\ ,
\label{ecdif2}
\end{equation} 
con una condici\'on inicial arbitraria (compatible con las condiciones de borde, que deben satisfacerse para todo $t$),
podemos encontrar la soluci\'on de la ec.~(\ref{ecdif1}) como la soluci\'on de equilibrio a la que relaja (\ref{ecdif2}) 
para $t\to\infty$. Este procedimiento es conocido como m\'etodo de relajaci\'on. 

Implementando la t\'ecnica de diferencias finitas, discretizamos la variable espacial en un lattice regular de $N+1$ sitios,
con $N=(x_B-x_A)/\Delta$, mientras que al tiempo se lo discretiza con un \'{\i}ndice entero $n\geq 0$. En este punto, 
debe adoptarse un esquema de discretizaci\'on adecuado para representar al operador diferencial $\mathcal{L}$; 
la estabilidad del c\'alculo num\'erico y la rapidez con que los resultados relajan a la soluci\'on asint\'otica dependen, 
en general, del esquema elegido. En particular, para ${\mathcal{L}}={\rm d}^2/{\rm d}x^2$, 
un esquema de discretizaci\'on estable 
de la ec.~(\ref{ecdif2}) es el de Crank-Nicholson, dado por
\begin{equation}
u_i^{n+1}-u_i^n={{u_{i+1}^{n+1}-2u_i^{n+1}+u_{i-1}^{n+1}}\over{\Delta^2}}\ .
\label{cranck}
\end{equation}  
Resolver la ec.~(\ref{cranck}) implica la resoluci\'on de un conjunto de $N-1$ ecuaciones lineales simult\'aneas 
que involucran los $u_i^m$ con $0\leq i\leq N$ y $m=n,n+1$. Agrupando los t\'erminos en (\ref{cranck}) apropiadamente,
resulta 
\begin{equation}
u_{i-1}^{n+1}-(\Delta^2+2)u_i^{n+1}+u_{i+1}^{n+1}=-\Delta^2u_i^n\ \ \ \ (1\leq i\leq N-1)\ ,
\label{tridag}
\end{equation}  
un sistema tridiagonal que puede ser resuelto f\'acilmente para cada paso temporal \cite{pre92}.  
Pidiendo que las diferencias entre dos configuraciones temporales sucesivas sean menores a una cota predeterminada,
se contin\'ua el esquema de c\'alculo hasta alcanzar la convergencia. 

Si $\mathcal{L}$ es un operador diferencial que puede escribirse como una suma de operadores, 
${\mathcal{L}}=\sum_i{\mathcal{L}}_i$ (tales que, para cada uno por separado, se conoce un esquema de
discretizaci\'on estable que resuelve la ecuaci\'on diferencial), puede seguirse el m\'etodo de divisi\'on 
de operadores, o divisi\'on temporal, que indica actuar en forma sucesiva con cada operador
${\mathcal{L}}_i$. En particular, consideremos una ecuaci\'on diferencial parcial lineal de segundo orden inhomog\'enea, 
\begin{equation}
{{\partial^2 u}\over{\partial x^2}}+{{\partial^2 u}\over{\partial y^2}}+
\phi{{\partial u}\over{\partial x}}+\psi{{\partial u}\over{\partial y}}+\eta=0\ ,
\label{ecdif3}
\end{equation} 
con condiciones de borde de Dirichlet o de Neumann. Esta ecuaci\'on puede escribirse como 
\begin{equation}
{\mathcal{L}}u+\eta=0\ ,
\label{ecdif4}
\end{equation}
donde ${\mathcal{L}}={\mathcal{L}}_{1x}+{\mathcal{L}}_{1y}+{\mathcal{L}}_{2x}+{\mathcal{L}}_{2y}$,
con ${\mathcal{L}}_{1x}={\rm d}^2/{\rm d}x^2$, ${\mathcal{L}}_{2x}=\phi\times{\rm d}/{\rm d}x$, 
y expresiones an\'alogas para $x\to y$.
Generando el primer paso temporal mediante ${\mathcal{L}}_{1x}$, resulta  
$$
u_{ij}^{n+1}-u_{ij}^n={{u_{(i+1)j}^{n+1}-2u_{ij}^{n+1}+u_{(i-1)j}^{n+1}}\over{\Delta^2}}+
{{u_{i(j+1)}^n-2u_{ij}^n+u_{i(j-1)}^n}\over{\Delta^2}}+
$$
\begin{equation}
\phi_{ij}{{u_{(i+1)j}^n-u_{(i-1)j}^n}\over{2\Delta}}+
\psi_{ij}{{u_{i(j+1)}^n-u_{i(j-1)}^n}\over{2\Delta}}+\eta_{ij}\ ,
\end{equation}   
que conduce a un sistema tridiagonal de ecuaciones.
En los pasos temporales siguientes de la serie, se actualizan del mismo modo los t\'erminos que corresponden a 
${\mathcal{L}}_{1y}$, ${\mathcal{L}}_{2x}$ y ${\mathcal{L}}_{2y}$.  

La ec.~(\ref{ecdif3}) corresponde, en efecto, a la ecuaci\'on de difusi\'on (\ref{difeqcil}), identificando 
$u(x,y)\to N(r,z)$, $\phi\to 1/r+\left(\partial D_\perp/\partial r-u_r\right)/D_\perp$, 
$\psi\to \left(\partial D_\perp/\partial z-u_z\right)/D_\perp$, y $\eta\to -Q$, donde $u_r$ y $u_z$ 
son las velocidades de drift dadas por las ecs.~(\ref{ureq}) y (\ref{uzeq}).
El pasaje de coordenadas cil\'{\i}ndricas a rectangulares presenta la dificultad de que el mapeo no es biyectivo 
en $r=0$; la condici\'on de borde $N|_{\Sigma}=0$ que debe aplicarse sobre la superficie exterior $\Sigma$ 
(correspondiente a los bordes del halo, de radio $R$ y altura $2H$) se traduce en condiciones de borde 
homog\'eneas de Dirichlet en $r=R$ y $z=\pm H$. Seg\'un el esquema de c\'alculo que describimos, faltar\'{\i}a
especificar la condici\'on de borde sobre $r=0$, uno de los bordes del lattice en el plano $r-z$. Una soluci\'on
se obtiene mediante la extensi\'on a la regi\'on no f\'{\i}sica $r<0$; extendiendo $\phi,\psi$
y $\eta$ al rango $-R\leq r\leq R$ como funciones pares de la variable $r$, 
podemos imponer la condici\'on $N=0$ a los bordes $r=\pm R$ y el esquema de c\'alculo 
queda bien definido.   

\section{El espectro y la composici\'on de rayos c\'osmicos \\ en la regi\'on de la rodilla}

En el contexto del modelo de difusi\'on turbulenta que hemos descripto en la Secci\'on anterior, una explicaci\'on
para la rodilla del espectro de rayos c\'osmicos surge de un modo natural como consecuencia del cambio de
r\'egimen entre la difusi\'on transversal, dominante a bajas energ\'{\i}as, y la difusi\'on de Hall, que prevalece a 
energ\'{\i}as altas \footnote{En el Ap\'endice C se describe un modelo simplificado, cuya soluci\'on anal\'{\i}tica
reproduce las caracter\'{\i}sticas esenciales de la rodilla del espectro.}. 
\begin{figure}[t]
\centerline{{\epsfxsize=5.8in \epsfysize=2.9in \epsffile{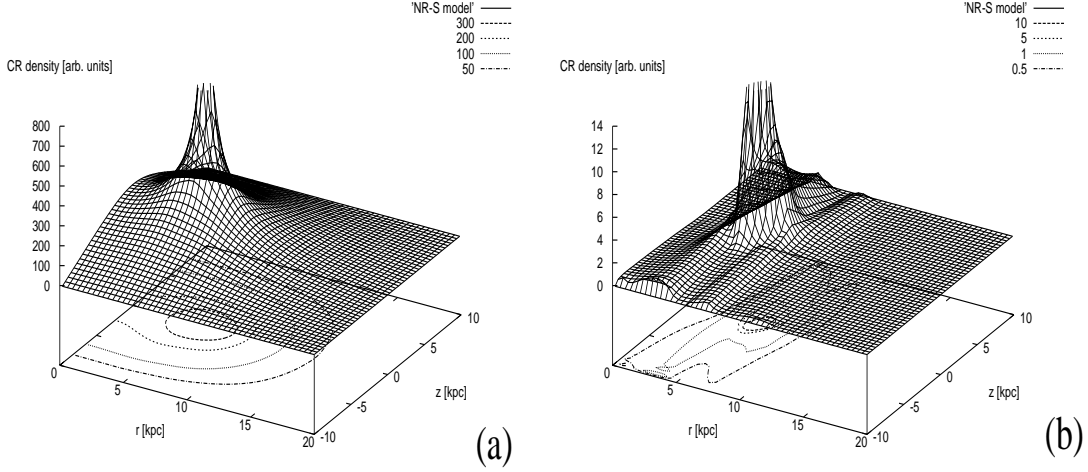}}}
\caption{Distribuci\'on espacial de la densidad de rayos c\'osmicos en la galaxia en el modelo $NR-S+$ 
(halo sin inversiones radiales y sim\'etrico con respecto al plano gal\'actico),
para una fuente en $r_s=6$~kpc: 
(a) a bajas energ\'{\i}as (debajo de la rodilla), el r\'egimen dominado por la difusi\'on transversal;
(b) a energ\'{\i}as altas (arriba de la rodilla), el r\'egimen dominado por los drifts. 
Los picos muy agudos que aparecen alrededor de la ubicaci\'on de la fuente han sido truncados.} 
\label{f1P2}
\end{figure}
Como se observa de las ecs.~(\ref{dperp}) y (\ref{da}), la dependencia de los coeficientes
de difusi\'on con la energ\'{\i}a viene dada por $D_\perp ({\bf x})\simeq D_\perp^0({\bf x})(E/E_0)^{1/3}$ y
$D_A({\bf x})\simeq D_A^0({\bf x})E/E_0$. A bajas energ\'{\i}as ($E\ll Z~E_r$), la difusi\'on transversal es el
mecanismo dominante que gobierna el transporte de rayos c\'osmicos, y conduce a 
${\rm d}N/{\rm d}E\propto(D_\perp^0/D_\perp){\rm d}Q/{\rm d}E\propto E^{-\beta-1/3}$, donde $\beta$ es el
\'{\i}ndice espectral de la fuente, mientras que a energ\'{\i}as altas ($E\gg Z~E_r$) se tiene en cambio que
${\rm d}N/{\rm d}E\propto(D_A^0/D_A){\rm d}Q/{\rm d}E\propto E^{-\beta-1}$. De este modo vemos que, si el
\'{\i}ndice espectral de los flujos producidos en las fuentes es $\beta\simeq 2.4$, 
se obtiene la pendiente espectral correcta debajo de la rodilla (${\rm d}N/{\rm d}E\propto E^{-\alpha}$, 
con $\alpha\simeq 2.7$). Cada componente de rayos c\'osmicos de carga $Z$ comienza a verse afectada por
los drifts a la energ\'{\i}a $E\simeq Z~E_r$, y su espectro se torna progresivamente m\'as empinado, 
con el \'{\i}ndice espectral
cambiando finalmente en $\Delta\alpha\simeq 2/3$ en alrededor de una d\'ecada de energ\'{\i}a. 
Veremos que la envolvente del
espectro total, obtenido de la suma de las diferentes componentes nucleares, reproduce muy adecuadamente el
cambio de un espectro $\propto E^{-2.7}$ debajo de la rodilla, a uno $\propto E^{-3}$ arriba de ella.     
Adicionalmente, como todas las componentes livianas quedan fuertemente suprimidas arriba de energ\'{\i}as 
del orden de $10^{17}$~eV, la componente pesada dominante (correspondiente a n\'ucleos del grupo del hierro) 
cambia gradualmente su espectro hasta que 
el espectro total se torna $\propto E^{-2.7-2/3}$ por encima de $10^{17}$~eV. As\'{\i}, este escenario reproduce 
tambi\'en la segunda rodilla
observada en $E_{sr}\simeq 4\times10^{17}$~eV y el espectro $\propto E^{-3.3}$ que se observa hasta el tobillo,
tal como se describir\'a en detalle m\'as adelante, en la Secci\'on 2.3.  

\begin{figure}[t]
\centerline{{\epsfysize=2.5in \epsffile{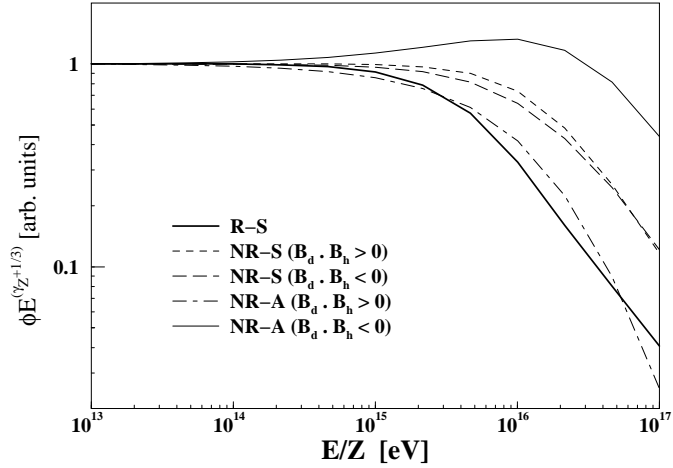}}}
\caption{Espectro diferencial normalizado versus $E/Z$ para los diferentes modelos de campo magn\'etico regular 
investigados. $R-S$ denota un halo con inversiones radiales y sim\'etrico con respecto al plano gal\'actico, 
mientras que $NR-S(A)$ se refiere al halo sin inversiones radiales y sim\'etrico (antisim\'etrico) con respecto a
este plano. Adem\'as, en estos casos la orientaci\'on del halo relativa a la componente regular del disco gal\'actico
se distingue mediante el signo de $B_d\cdot B_h$, como se indica.}  
\label{f2P2}
\end{figure}

Con el objetivo de ilustrar el rol de los drifts en la difusi\'on de rayos c\'osmicos, consideremos en primer
lugar una fuente localizada en $r_s=6$~kpc, tomando para el campo magn\'etico random una intensidad rms
local $B_r^*=2$~$\mu$G y una escala vertical $z_r=3$~kpc. Las figs. \ref{f1P2}(a) y (b) muestran la distribuci\'on espacial 
de la densidad de rayos c\'osmicos en la galaxia, seg\'un resulta para una estructura de campo regular $NR-S+$,
para energ\'{\i}as debajo y arriba de la rodilla, respectivamente. En concordancia con las discusiones precedentes, 
en las figuras se observa que a bajas energ\'{\i}as la difusi\'on de rayos c\'osmicos se debe principalmente a 
fluctuaciones turbulentas de peque\~na escala (gobernadas por $D_\perp$), mientras que a mayores energ\'{\i}as 
el bulk de los rayos c\'osmicos se desplaza de acuerdo a los drifts que resultan del campo magn\'etico regular
de gran escala, dando lugar a evidentes asimetr\'{\i}as de gran escala en la distribuci\'on de densidad.    

La influencia de los efectos de drift en el espectro de rayos c\'osmicos depender\'a,
en general, del modelo de campo gal\'actico, del conjunto particular de par\'ametros especificados para el modelo (es decir,
las amplitudes y escalas verticales asumidas para las componentes regular y estoc\'astica del campo magn\'etico) y de la
distribuci\'on de las fuentes. Debe notarse que un par\'ametro relevante en los coeficientes de difusi\'on es el
radio de Larmor de las part\'{\i}culas; \'esto implica que el impacto de los efectos del drift depender\'an de
la rigidez de los rayos c\'osmicos, es decir, del valor de $E/Z$.     

\begin{figure}[t]
\centerline{{\epsfysize=2.5in \epsffile{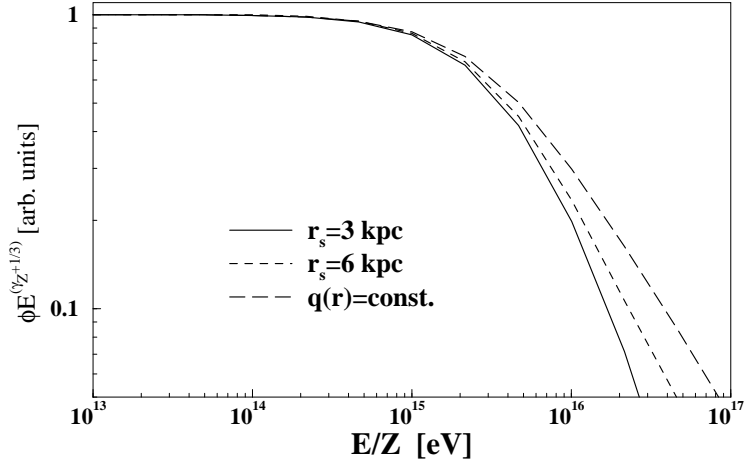}}}
\caption{Espectro diferencial normalizado versus $E/Z$ para el caso $R-S$ y diferentes distribuciones radiales de fuentes:
homog\'eneamente extendida ($q(r)=$cte.) y localizada ($q(r)\propto \delta(r-r_s)$, con $r_s=3$ y 6~kpc).} 
\label{f3P2}
\end{figure}

La fig.\ref{f2P2} muestra el flujo de rayos c\'osmicos normalizado como funci\'on de $E/Z$ para todos los modelos de
campo magn\'etico gal\'actico descriptos anteriormente (ver la Secci\'on 2.1.1), con la fuente localizada en $r_s=6$~kpc,
y con $B_r^*=2$~$\mu$G, $z_r=5$~kpc para el campo magn\'etico irregular. Puede notarse que en \'esta y las pr\'oximas
figuras, el flujo de rayos c\'osmicos aparece multiplicado por $E^{\beta_Z+1/3}$, donde $\beta_Z$ se refiere al
\'{\i}ndice espectral de la componente de carga $Z$ en la fuente. Como la difusi\'on transversal depende de la rigidez 
a trav\'es de una potencia de $1/3$, ec.~(\ref{dperp}), 
resulta ${\rm d}N_Z/{\rm d}E\propto E^{-\gamma_Z+1/3}$ a bajas energ\'{\i}as, debajo de la rodilla.  De la figura se ve que 
la mayor\'{\i}a de los modelos considerados muestran la tendencia apropiada que da cuenta de la rodilla, ya que el
cambio de \'{\i}ndice espectral aparece, por lo general, en el valor apropiado de $E/Z$ (es decir, $E\simeq E_r$ para
protones, y a mayores energ\'{\i}as para n\'ucleos m\'as pesados). Los efectos de drift son particularmente notables
para el caso del halo antisim\'etrico, como podr\'{\i}a anticiparse de la ec. (\ref{ureq}) teniendo en cuenta que, para 
$b_\phi\simeq\pm{\rm signo}(z)$, aparece una velocidad de drift singular $u_r\simeq \mp 2D_A\delta(z)$ sobre
el plano gal\'actico. Siendo que all\'{\i} es donde est\'an localizadas las fuentes, es razonable esperar
que esta contribuci\'on adicional a los drifts afecte los resultados significativamente. Por otra parte, en los
modelos antisim\'etricos el halo contribuye con una velocidad de drift vertical $u_z$ que converge o diverge desde el
plano gal\'actico (dependiendo de la orientaci\'on del campo del halo relativa a la del disco), a diferencia de lo
que ocurre en los modelos sim\'etricos, en los que el drift vertical tiene la misma orientaci\'on a ambos lados
del plano. En modelos antisim\'etricos, se obtiene una rodilla m\'as pronunciada para $B_d\cdot B_h>0$,
mientras que, por el contrario, una protuberancia ({\it bump}) justo antes de la rodilla se observa para $B_d\cdot B_h<0$.  
El pronunciado bump que muestra la fig.\ref{f2P2} para este caso est\'a claramente en desacuerdo con los espectros observados
(comparar con los datos experimentales de la fig.\ref{f6P2}(a)), 
de modo que, en principio, este modelo particular estar\'{\i}a desfavorecido por las observaciones. Sin embargo,
Kempa et al. \cite{kem74} hicieron notar que los datos experimentales podr\'{\i}an sugerir la existencia de un bump 
menos pronunciado justo antes de la rodilla. Desde esta observaci\'on, realizada hace tres d\'ecadas, la existencia del bump
sigue dando lugar a controversias; de cualquier forma, es interesante comentar que caracter\'{\i}sticas de este tipo 
podr\'{\i}an originarse en efectos de drift que aparecen en algunas configuraciones particulares de los campos 
magn\'eticos regulares en la galaxia.

La fig.\ref{f3P2} muestra el caso $R-S$ para una distribuci\'on de fuentes extendida $q(r)=$~cte., 
y para fuentes localizadas en $r_s=3$~kpc y $r_s=6$~kpc, con el mismo campo irregular que se us\'o en la 
fig.\ref{f1P2} (es decir, $B_r^*=2$~$\mu$G y $z_r=3$~kpc).
Como es de esperar, los efectos de drift son m\'as intensos cuanto mayor es la distancia a la que est\'a situada la fuente,
debido a que los drifts pueden remover del plano gal\'actico a los rayos c\'osmicos que se propagan a lo largo de todo
el camino entre la fuente y el observador. El mismo comportamiento cualitativo ha sido tambi\'en observado en los
otros casos estudiados. 

\begin{figure}[t]
\centerline{{\epsfysize=2.5in \epsffile{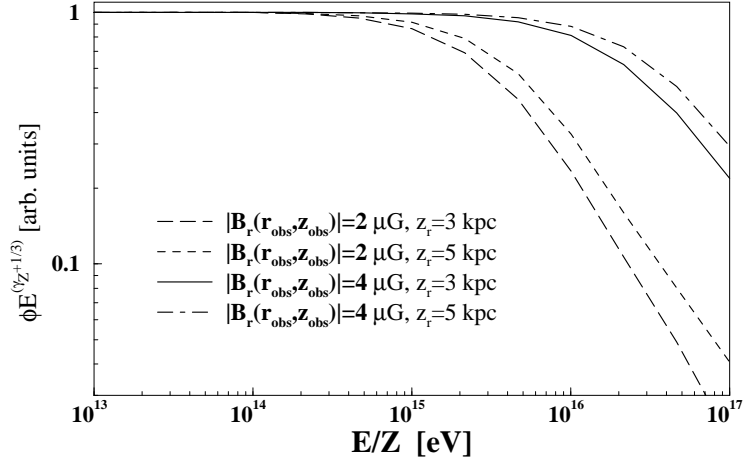}}}
\caption{Espectro diferencial normalizado versus $E/Z$ para el caso $R-S$ y una fuente localizada en $r_s=6$~kpc. 
Los par\'ametros asociados al campo irregular toman los valores $B_r^*=2$ \'o 4~$\mu$G, y $z_r=3$ \'o 5 kpc,
tal como se indica.}
\label{f4P2}
\end{figure}

\begin{table}[t]
\begin{tabular}{|r|r|r||r|r|r||r|r|r||r|r|r|} \hline
\multicolumn{1}{|c|}{Z}&\multicolumn{1}{|c|}{$f_Z$}&\multicolumn{1}{|c||}{$\alpha_Z$}&
\multicolumn{1}{|c|}{Z}&\multicolumn{1}{|c|}{$f_Z$}&\multicolumn{1}{|c||}{$\alpha_Z$}&
\multicolumn{1}{|c|}{Z}&\multicolumn{1}{|c|}{$f_Z$}&\multicolumn{1}{|c||}{$\alpha_Z$}&
\multicolumn{1}{|c|}{Z}&\multicolumn{1}{|c|}{$f_Z$}&\multicolumn{1}{|c|}{$\alpha_Z$}\\ \hline
 1 & 0.3775 & 2.71 & 8 & 0.0679 & 2.68 & 15 & 0.0012 & 2.69 & 22 & 0.0049 & 2.61 \\
 2 & 0.2469 & 2.64 & 9 & 0.0014 & 2.69 & 16 & 0.0099 & 2.55 & 23 & 0.0027 & 2.63 \\
 3 & 0.0090 & 2.54 & 10& 0.0199 & 2.64 & 17 & 0.0013 & 2.68 & 24 & 0.0059 & 2.67 \\
 4 & 0.0020 & 2.75 & 11& 0.0033 & 2.66 & 18 & 0.0036 & 2.64 & 25 & 0.0058 & 2.46 \\
 5 & 0.0039 & 2.95 & 12& 0.0346 & 2.64 & 19 & 0.0023 & 2.65 & 26 & 0.0882 & 2.59 \\
 6 & 0.0458 & 2.66 & 13& 0.0050 & 2.66 & 20 & 0.0064 & 2.70 & 27 & 0.0003 & 2.72 \\
 7 & 0.0102 & 2.72 & 14& 0.0344 & 2.75 & 21 & 0.0013 & 2.64 & 28 & 0.0043 & 2.51 \\ \hline
\end{tabular}
\caption{Abundancias fraccionales (para $E=1$~TeV) e \'{\i}ndices espectrales a bajas energ\'{\i}as (debajo de la rodilla) 
correspondientes a las componentes nucleares de rayos c\'osmicos desde hidr\'ogeno hasta n\'{\i}quel (de \cite{wie98,hoe03}).}
\label{tabla1}
\end{table}

La fig.\ref{f4P2} exhibe el efecto de variar los par\'ametros que describen al campo random. 
Los datos corresponden al caso 
$R-S$ y los par\'ametros asociados al campo irregular toman los valores $B_r^*=2$ \'o 4~$\mu$G, y $z_r=3$ \'o 5 kpc. 
Como se espera, los efectos del drift son m\'as intensos cuanto mayor es la supresi\'on del campo 
random, ya sea asumiendo una amplitud
menor, o bien disminuyendo su escala vertical caracter\'{\i}stica. Un comportamiento an\'alogo se observa en las
otras configuraciones de campo relevantes, debido, en todos los casos, a la mayor supresi\'on de $D_A$ bajo 
condiciones de mayor turbulencia. 

Como estamos tratando con un escenario dependiente de la rigidez, los resultados obtenidos dependen de $E/Z$; para comparar 
nuestros resultados con los datos experimentales, debe calcularse la contribuci\'on de cada componente de carga $Z$ 
convolucionando el espectro producido por las fuentes con los efectos de modulaci\'on que resultan de la propagaci\'on en 
la galaxia. Finalmente, sumando sobre todas las componentes nucleares desde hidr\'ogeno hasta n\'{\i}quel \footnote{
La inclusi\'on de elementos m\'as pesados es generalmente considerada irrelevante en este
contexto, dado que son elementos mucho menos abundantes. De todos modos, en \cite{hoe03} se sugiere que estos n\'ucleos 
`superpesados' (con $Z>28$) pueden ser dominantes entre la rodilla y la segunda rodilla. En la Secci\'on 2.3 consideraremos 
en detalle el espectro de rayos c\'osmicos en esa regi\'on de energ\'{\i}as, y discutiremos el posible rol de la componente
superpesada.} (es decir, para $1\leq Z\leq 28$), 
obtenemos la predicci\'on del modelo para el espectro total de rayos c\'osmicos, que luego puede ser
comparado con los espectros medidos por diferentes experimentos (KASCADE, CASABLANCA, DICE, PROTON y Akeno) 
\cite{swo00,fow01,swo02,kam99,kam01b}. 

A bajas energ\'{\i}as (debajo de la rodilla) el espectro observado localmente para cada componente de carga $Z$ viene dado por
\begin{equation}
{{\rm d}N_Z\over{\rm d}E}=f_Z\phi_0\left(\frac{E}{E_0}\right)^{-\alpha_Z}\ ,
\label{crwiebel}
\end{equation}
donde $\phi_0$ es el flujo total de rayos c\'osmicos correspondiente a la energ\'{\i}a de referencia $E_0$, y
$f_Z$ es la abundancia fraccional de la componente nuclear de carga $Z$ a la misma energ\'{\i}a ($\sum_Zf_Z=1$). 
Aqu\'{\i} adoptaremos $E_0=1$~TeV,
las abundancias e \'{\i}ndices espectrales que resultan de los datos experimentales compilados en \cite{wie98,hoe03}   
(reproducidos en la Tabla \ref{tabla1}), y la normalizaci\'on $\phi_0=3.5\times 10^{-13} {\rm m^{-2}s^{-1}sr^{-1}eV^{-1}}$. 
Asumiremos que los espectros producidos por las fuentes 
tienen \'{\i}ndices espectrales ($\beta_Z$) constantes (es decir, independientes de la energ\'{\i}a), consistentes con
los \'{\i}ndices observados ($\alpha_Z$) a bajas energ\'{\i}as. Adem\'as, en la extrapolaci\'on de los flujos no 
tomamos en cuenta posibles energ\'{\i}as de corte asociadas al mecanismo de aceleraci\'on en las fuentes. 

\begin{figure}[t]
\centerline{{\epsfysize=2.5in \epsffile{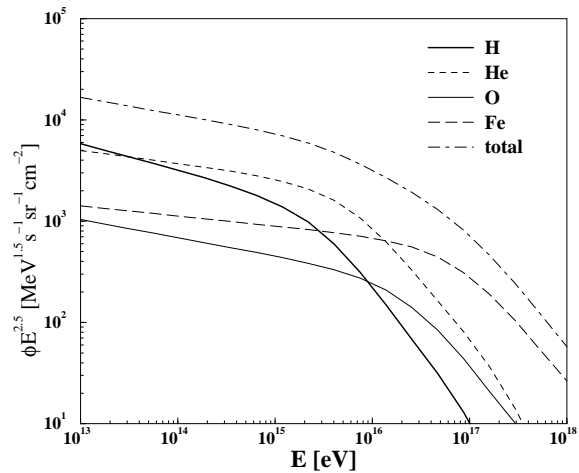}}}
\caption{Principales contribuciones al flujo total de rayos c\'osmicos, provenientes de protones y n\'ucleos
de helio, ox\'{\i}geno y hierro, obtenidas en el modelo $R-S$ para una fuente localizada en $r_s=6$~kpc.}
\label{f5P2}
\end{figure}

\begin{figure}[t]
\centerline{{\epsfxsize=5.8in \epsffile{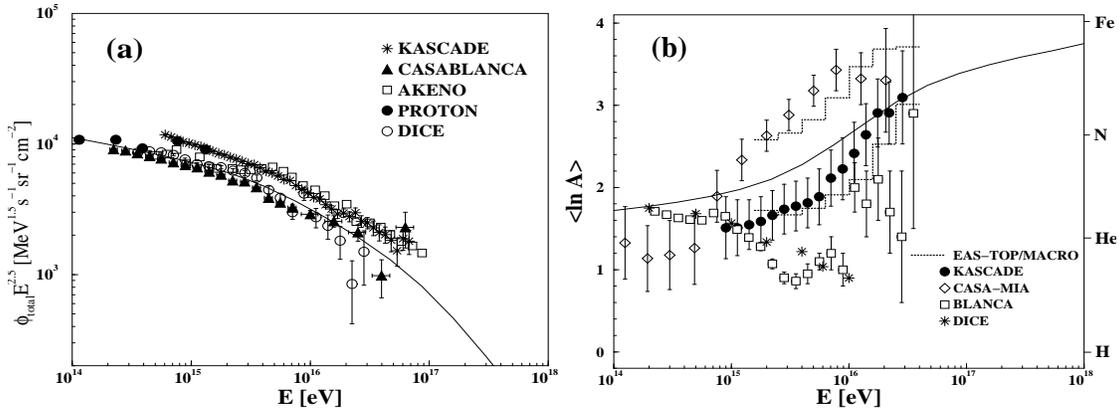}}}
\caption{Comparaci\'on entre las observaciones experimentales y las predicciones te\'oricas obtenidas
en el modelo $R-S$ para una fuente localizada en $r_s=6$~kpc:
(a) espectro total de rayos c\'osmicos;  
(b) composici\'on m\'asica media $\langle\rm{ln}\ A\rangle$.}
\label{f6P2}
\end{figure}

La fig.\ref{f5P2} muestra las principales contribuciones al flujo total de rayos c\'osmicos, 
provenientes de protones y n\'ucleos
de helio, ox\'{\i}geno y hierro, seg\'un se obtiene en el modelo $R-S$ para una fuente localizada en $r_s=6$~kpc. Los 
par\'ametros para el campo random son $B_r^*=2$~$\mu$G y $z_r=3$~kpc. 

La fig.\ref{f6P2}(a) muestra el flujo total 
correspondiente a este caso en comparaci\'on con algunas observaciones; como puede verse claramente, el acuerdo entre las 
predicciones y los datos experimentales es muy satisfactorio. 

La fig.\ref{f6P2}(b) muestra la dependencia de la composici\'on m\'asica media $\langle\rm{ln}\ A\rangle$ 
con la energ\'{\i}a, correspondiente al mismo caso de 
la figura anterior, comparada con las mediciones de diferentes experimentos \cite{swo00,fow01,swo02,kam99,kam01b,agl03}.       
Siendo que el escape de part\'{\i}culas hacia el exterior de la galaxia depende de $E/Z$ y que, por lo tanto,
es m\'as efectivo para las part\'{\i}culas m\'as livianas, una predicci\'on natural de este 
escenario es que la composici\'on se torne m\'as pesada arriba de la rodilla. Esta tendencia hacia componentes m\'as 
pesadas es la que manifiestan los datos de algunos experimentos, como por ejemplo las m\'as recientes observaciones 
de KASCADE \cite{kam01b} y EAS-TOP/MACRO \cite{agl03} \footnote{Es particularmente notable el acuerdo con los \'ultimos resultados
de EAS-TOP/MACRO, dados a conocer luego de la publicaci\'on de nuestro trabajo: ``The present data explain therefore 
the observed knee in the cosmic ray primary spectrum as due to the steepening of the spectrum of a
light component (p, He) at $E_0 \approx 4 \times 10^{15}$ eV, of $\Delta\gamma = 0.7 \pm 0.4$'' \cite{agl03}.}. 

Aqu\'{\i}, sin embargo, debemos hacer notar que los resultados reportados por experimentos que utilizan diferentes t\'ecnicas 
de detecci\'on no son compatibles en todos los casos, evidenciando discrepancias incluso en el comportamiento cualitativo 
de las observaciones. Si bien los datos m\'as recientes de varios experimentos parecen confirmar la tendencia 
hacia el predominio de las componentes m\'as pesadas por encima de la rodilla, deben tenerse en cuenta tambi\'en 
otros resultados recientes \cite{swo00} que sostienen la tendencia opuesta y que obligan a la cautela.   
En la fig.\ref{f6P2}(b) se puede ver que, seg\'un algunos conjuntos de datos, la composici\'on m\'asica media se hace m\'as
liviana en $\sim 10^{15}$~eV, y comienza a crecer hacia componentes m\'as pesadas s\'olo a partir de $\sim 3\times 10^{15}$~eV. 
Una depresi\'on ({\it dip}) de esta naturaleza 
podr\'{\i}a resultar de un bump en el espectro de la componente de protones en ese mismo rango de energ\'{\i}as.      

De las discusiones y figuras precedentes, es claro que puede 
obtenerse una variedad de diferentes espectros modificando el valor de los par\'ametros que intervienen en el modelo, 
o bien considerando otros modelos de campo gal\'actico. Sin embargo, un an\'alisis cuantitativo 
detallado de los diferentes resultados posibles no tiene mayor sentido en el marco de este trabajo. 
Por un lado, los datos experimentales muestran a\'un errores estad\'{\i}sticos y sistem\'aticos considerables, 
mientras que, por otra parte, las predicciones te\'oricas tambi\'en se apoyan en un cierto n\'umero de consideraciones 
simplificadoras. Por ejemplo, la ecuaci\'on de transporte fue considerada como puramente difusiva, despreciando la 
fragmentaci\'on de los n\'ucleos en el medio interestelar, la reaceleraci\'on, o la convecci\'on debido 
a la posible existencia de un viento gal\'actico. Adem\'as, la dependencia espacial de las diferentes componentes del 
campo magn\'etico gal\'actico fue tratada en t\'erminos muy simples; 
aunque los modelos que hemos 
adoptado son plausibles y capturan las caracter\'{\i}sticas esenciales de los campos magn\'eticos en la galaxia, una 
descripci\'on m\'as realista exigir\'{\i}a abandonar las simetr\'{\i}as del sistema, incorporar par\'ametros adicionales, etc.
La tendencia de los resultados obtenidos, sin embargo, muestra un acuerdo notable con las observaciones experimentales. 

\section{El espectro y las anisotrop\'{\i}as de rayos c\'osmicos desde la rodilla hasta la segunda rodilla}

En la Secci\'on precedente, hemos visto que el modelo de difusi\'on turbulenta y drift provee una explicaci\'on
natural para la rodilla del espectro de rayos c\'osmicos, basada en propiedades bien establecidas sobre la 
propagaci\'on de part\'{\i}culas cargadas en campos magn\'eticos regulares y turbulentos. 
El espectro de una componente de carga $Z$ comienza a verse afectado por los drifts a partir de $E\simeq Z~E_r$, 
con el \'{\i}ndice espectral cambiando finalmente en $\Delta\alpha\simeq 2/3$ en alrededor de una d\'ecada de energ\'{\i}a. 
De este modo, la rodilla en el espectro total (es decir, el cambio de \'{\i}ndice $2.7\to 3$ a partir de 
$E_r\simeq 3\times10^{15}$~eV) aparece debido a la supresi\'on de las componentes livianas, fundamentalmente protones
y n\'ucleos de helio. En esta Secci\'on, veremos que este escenario permite explicar tambi\'en la segunda rodilla
del espectro total (el cambio de \'{\i}ndice $3\to 3.3$ a partir de $E_{sr}\simeq 4\times10^{17}$~eV) como resultado
de la supresi\'on de las componentes pesadas (correspondientes al grupo del hierro); incorporando, como es usual, 
una componente de rayos c\'osmicos extragal\'acticos que domina a partir del tobillo ($E_t\simeq 5\times10^{18}$~eV),
mostraremos que este escenario es capaz de reproducir el espectro observado hasta la regi\'on de ultra-alta energ\'{\i}a.  

Con respecto a la composici\'on, en la regi\'on de la segunda rodilla no se dispone de medidas precisas (la magnitud
de los errores en las medidas por encima de $10^{16}$~eV, que pueden verse en la fig.\ref{f6P2}(b), da una buena 
noci\'on acerca de las limitaciones experimentales en este aspecto). Sin embargo, existen algunas observaciones de
inter\'es que permiten, hasta cierto punto, un contraste entre la teor\'{\i}a y el experimento; en particular,
la determinaci\'on de Haverah Park de las abundancias nucleares usando la distribuci\'on lateral de lluvias atmosf\'ericas
\cite{ave03} (que sugiere que, en el rango $2\times 10^{17}$--$10^{18}$~eV, s\'olo un $\sim 30$\% de los rayos c\'osmicos
es liviano), y la observaci\'on de HIRES-MIA de un cambio de composici\'on hacia una mezcla m\'as liviana 
a partir de la segunda rodilla \cite{abu01}. 

En este contexto, las observaciones referidas a la anisotrop\'{\i}a del flujo de rayos c\'osmicos representan una herramienta
adicional de gran importancia. En particular, la colaboraci\'on AGASA \cite{hay99,tes01} report\'o una anisotrop\'{\i}a 
del $\sim 4\%$ para $E\simeq 0.8$--$2\times 10^{18}$~eV, con un exceso observado en una direcci\'on cercana 
al centro gal\'actico y un d\'eficit cercano a la direcci\'on del anticentro gal\'actico. Adicionalmente, ha sido 
observado alg\'un exceso de eventos en la direcci\'on de la regi\'on de Cygnus (a lo largo del brazo espiral de Ori\'on). 
En los datos de SUGAR se ha encontrado tambi\'en un exceso desde una direcci\'on cercana al centro gal\'actico \cite{bel01}, 
y con una direcci\'on separada apenas unos grados del exceso reportado por AGASA, aunque de una extensi\'on espacial menor 
(consistente con una fuente puntual). Debe notarse que estas observaciones ocurren justamente en la regi\'on de 
energ\'{\i}as en la que la componente gal\'actica de rayos c\'osmicos comienza a atenuarse para dar lugar al predominio 
de la componente extragal\'actica. \'Esto sugiere que la explicaci\'on
de estas observaciones podr\'{\i}a estar directamente vinculada con el mismo proceso que es responsable de la atenuaci\'on 
de la componente gal\'actica, y veremos que, en efecto, el escenario de difusi\'on turbulenta y drift las puede reproducir 
adecuadamente.

Como un \'ultimo examen de este escenario, estudiaremos (para diferentes modelos de campo magn\'etico) la dependencia 
con la energ\'{\i}a de las predicciones sobre el flujo, la amplitud de anisotrop\'{\i}a y la fase del primer arm\'onico  
en ascenci\'on recta, comparando los resultados con las observaciones experimentales en todo el rango que se extiende desde la
rodilla hasta el tobillo del espectro. Encontraremos que las predicciones muestran la tendencia adecuada para dar cuenta
de las observaciones, dando robustez al modelo de difusi\'on turbulenta y drift como posible explicaci\'on de los fen\'omenos
observados en los rayos c\'osmicos gal\'acticos de muy alta energ\'{\i}a. 

Siguiendo los pasos de la Secci\'on anterior, consideraremos que la componente gal\'actica de rayos c\'osmicos se obtiene de
resolver la ecuaci\'on de difusi\'on correspondiente, luego de asumir un dado modelo de campo magn\'etico regular, una realizaci\'on
particular de los par\'ametros que describen los campos, y una distribuci\'on radial para las fuentes. Adicionalmente, 
ahora incorporaremos una componente extragal\'actica dada por
\begin{equation}
\left({{\rm{d}N}\over{\rm{d}E}}\right)_{XG}=1.7\times 10^{-33}\ \left({{E}\over{10^{19}{\rm eV}}}\right)^{-2.4}
{\rm m^{-2}s^{-1}sr^{-1}eV^{-1}}\ .  
\label{xgflux}
\end{equation}
La amplitud de esta componente se ajust\'o para reproducir la normalizaci\'on correcta de los flujos observados para 
$E\simeq 10^{19}$~eV. Si consideramos que el flujo extragal\'actico es isotr\'opico en el borde de la galaxia,
entonces, en ausencia de efectos de reaceleraci\'on, tambi\'en ser\'a isotr\'opico en el interior 
(es decir, la densidad de rayos c\'osmicos no puede ser 
incrementada localmente, dentro de la galaxia, s\'olo por procesos difusivos \footnote{\'Esto puede entenderse como
una consecuencia del teorema de Liouville \cite{lem33,cla96}, o bien puede verse directamente de la ecuaci\'on de difusi\'on, 
notando que la soluci\'on en ausencia de fuentes y con condiciones de borde isotr\'opicas es la de una densidad
constante, de modo que los efectos difusivos en la galaxia no pueden producir un incremento local de densidad
en la componente extragal\'actica.}). En consecuencia,    
la dependencia con la energ\'{\i}a de la componente extragal\'actica debe ser similar a la que se infiere para
la producci\'on en las fuentes (es decir, dada por un \'{\i}ndice espectral $\beta\simeq 2.4$). 

\begin{figure}[t]
\centerline{{\epsfxsize=3.5in \epsfysize=2.7in \epsffile{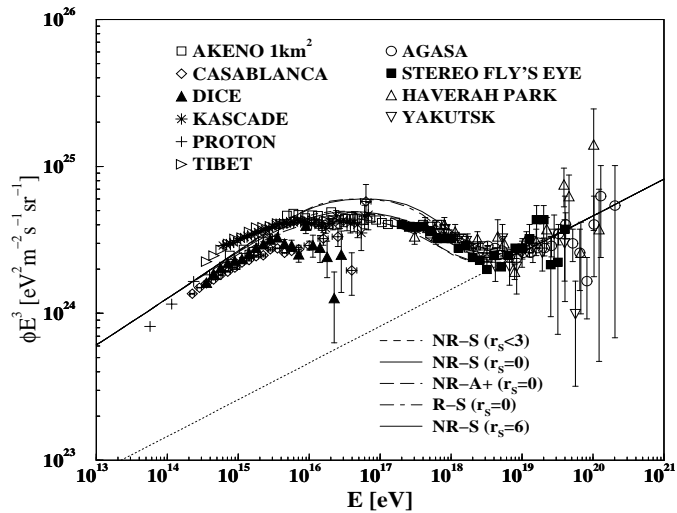}}}
\caption{Espectros de rayos c\'osmicos obtenidos para diferentes modelos de campo magn\'etico y distribuciones de
fuentes, en comparaci\'on con los datos experimentales. La recta punteada representa el flujo 
extragal\'actico, ec.~(\ref{xgflux}).}
\label{f1P3}
\end{figure}

La fig.\ref{f1P3} muestra los espectros que resultan de considerar una componente gal\'actica (calculada, como
en la Secci\'on anterior, para diferentes modelos de campo magn\'etico y diferentes distribuciones de fuentes) y la
extragal\'actica, ec.~(\ref{xgflux}) 
\footnote{Con el fin de preservar la presentaci\'on de las figuras en su contexto original 
(es decir, seg\'un la versi\'on en la que han aparecido publicadas), se opt\'o por sacrificar la uniformidad de presentaci\'on 
a lo largo de esta Tesis. As\'{\i}, por ejemplo, los espectros de la fig.\ref{f6P2}(a) se presentan en t\'erminos
de $\phi E^{2.5}$ [MeV$^{1.5}$cm$^{-2}$s$^{-1}$sr$^{-1}$], y en la fig.\ref{f1P3}, en cambio, como  
$\phi E^3$ [eV$^2$m$^{-2}$s$^{-1}$sr$^{-1}$].}. La elecci\'on de los par\'ametros para cada modelo estuvo 
guiada s\'olo por el requerimiento de reproducir correctamente la primera rodilla, dentro del rango de
valores plausibles. Tal como muestra la figura, el acuerdo con los datos experimentales es, en general, muy bueno.

\begin{figure}[t]
\centerline{{\epsfxsize=3in \epsfysize=2.6in \epsffile{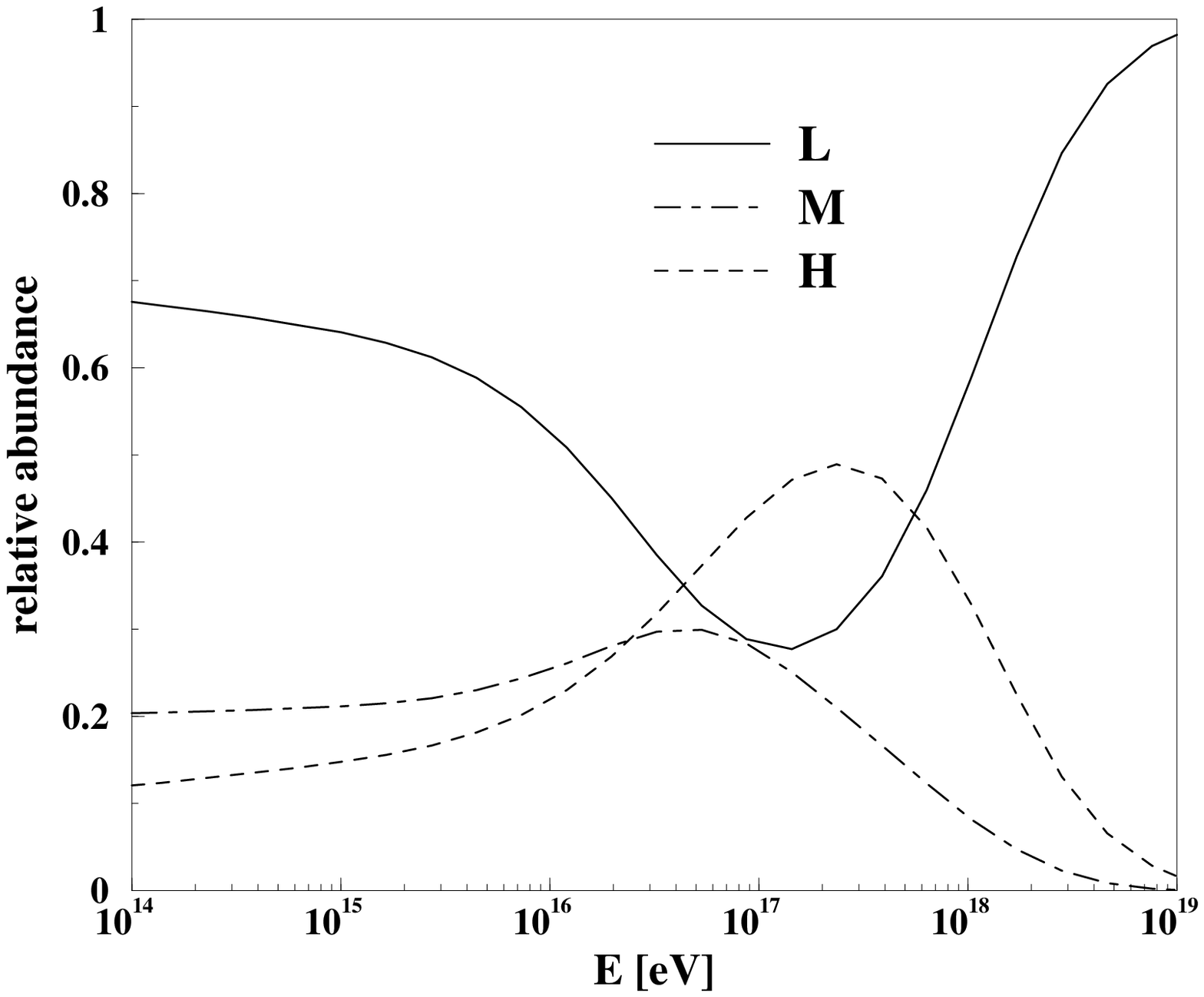}}}
\caption{Abundancias relativas de las diferentes componentes de rayos c\'osmicos, asumiendo que el flujo extragal\'actico 
est\'a constitu\'{\i}do por protones. Los elementos est\'an agrupados en componentes livianas ($1\leq Z\leq 5$), 
intermedias ($6\leq Z\leq 19$) y pesadas ($20\leq Z\leq 28$). Los datos corresponden al mismo modelo de campo magn\'etico 
y a la misma distribuci\'on de fuentes que en la fig.\ref{f3P3}.}
\label{f2P3}
\end{figure}

Las abundancias relativas de las diferentes componentes, 
agrupadas en livianas ($1\leq Z\leq 5$), intermedias ($6\leq Z\leq 19$)
y pesadas ($20\leq Z\leq 28$), se muestran en la fig.\ref{f2P3} como funci\'on de la energ\'{\i}a, asumiendo que el flujo 
extragal\'actico est\'a constitu\'{\i}do por protones. De all\'{\i} resulta que en nuestro modelo la composici\'on de rayos
c\'osmicos para $E\simeq 5\times 10^{17}$~eV consiste en un $\sim 50\%$ de n\'ucleos del grupo del hierro (fundamentalmente
Fe y Mn), en alrededor de un $\sim 20\%$ de n\'ucleos de masa intermedia, mientras que el resto corresponde esencialmente 
a n\'ucleos gal\'acticos de helio y a protones extragal\'acticos. Estos valores son consistentes con la 
determinaci\'on de Haverah Park de las abundancias nucleares \cite{ave03}, que observa s\'olo un $\sim 30\%$ de componentes
livianas en el rango $2\times 10^{17}$--$10^{18}$~eV. Por otra parte, la observaci\'on de HIRES-MIA de un cambio de 
composici\'on hacia una mezcla m\'as liviana a partir de la segunda rodilla \cite{abu01} se explica en este escenario 
como debida a la supresi\'on del grupo de n\'ucleos gal\'acticos pesados, que da lugar al predominio de los protones extragal\'acticos
hacia la regi\'on del tobillo. 

Consideremos ahora las anisotrop\'{\i}as que resultan en este escenario. Como ya hemos comentado, \'estas ser\'an 
producidas s\'olo por la componente gal\'actica, ya que, si suponemos que la componente extragal\'actica es originalmente 
isotr\'opica, conservar\'a su isotrop\'{\i}a luego de que los efectos de propagaci\'on en la galaxia sean tenidos en cuenta.
La componente gal\'actica de rayos c\'osmicos de carga $Z_i$ contribuye a la anisotrop\'{\i}a total una cantidad dada por 
\cite{ber90}
\begin{equation}
\delta_i=\frac{3\ {\bf J_i}}{c\ N_i}\ ,
\label{anis}
\end{equation} 
donde ${\bf J_i}$ es la corriente macrosc\'opica correspondiente a esa componente, dada por una expresi\'on 
an\'aloga a la ec.~(\ref{curr2}), pero involucrando a la densidad $N_i$ que corresponde a los rayos c\'osmicos de
carga $Z_i$. La anisotrop\'{\i}a total viene dada por $\delta=\sum_if_i^*\delta_i$, donde $f_i^*\equiv N_i/N$ son las
abundancias relativas de todas las especies nucleares, y donde $N$ es la densidad de rayos c\'osmicos total, 
obtenida de sumar las contribuciones gal\'actica y extragal\'actica \footnote{N\'otese aqu\'{\i} la diferencia con las 
abundancias gal\'acticas $f_Z$ que se introdujeron en la Secci\'on anterior (ec.~(\ref{crwiebel})), que no tienen en 
cuenta la componente extragal\'actica.}. La presencia de la componente
extragal\'actica isotr\'opica tiende a reducir las fracciones $f_i^*$ de las componentes gal\'acticas, con el efecto de
suprimir el crecimiento de la amplitud de la anisotrop\'{\i}a total hacia la regi\'on del tobillo. 

Como hemos mencionado antes, a bajas energ\'{\i}as la difusi\'on est\'a dominada por $D_\perp$; de
la ec.~(\ref{curr2}) se ve que, en ese caso, la contribuci\'on a la anisotrop\'{\i}a de los rayos c\'osmicos de carga $Z_i$
es $\delta_i\propto D_\perp\nabla N_i/N_i$, de forma que $\delta_i\propto E^{1/3}$. Sin
embargo, a mayores energ\'{\i}as hay un crossover hacia el r\'egimen de dominio del drift, conduciendo al comportamiento
$\delta_i\propto E$. A\'un a mayores energ\'{\i}as, a partir de $\sim 10^{18}$~eV, la densidad de rayos c\'osmicos 
comienza gradualmente a quedar dominada por la anisotrop\'{\i}a intr\'{\i}nseca de la componente extragal\'actica (que, por 
simplicidad, fue considerada despreciable en este trabajo). 

De la ec.~(\ref{curr2}) se ve que la contribuci\'on a la corriente macrosc\'opica que resulta de la difusi\'on transversal 
es en la direcci\'on de $\nabla N$ (y ortogonal a las curvas de nivel de la densidad), mientras que la 
contribuci\'on del drift es normal a $\nabla N$ (y, por lo tanto, paralela a las curvas de nivel).            

En lo que sigue, estudiaremos la anisotrop\'{\i}a asociada a una \'unica componente gal\'actica\footnote{Para simplificar 
la notaci\'on, omitiremos el sub\'{\i}ndice que distingue entre las diferentes contribuciones gal\'acticas.}; 
m\'as adelante, analizaremos la anisotrop\'{\i}a total que resulta de considerar el conjunto de todas las componentes 
gal\'acticas y la extragal\'actica. Como la corriente macrosc\'opica es perpendicular al campo magn\'etico regular,
est\'a contenida en el plano $r-z$; en consecuencia, el vector de anisotrop\'{\i}a tambi\'en est\'a contenido en este plano,
con sus componentes dadas por
\begin{equation}
\delta_r=\frac{3}{c\ N}\left(-D_\perp\frac{\partial N}{\partial r}+
b_\phi D_A\frac{\partial N}{\partial z}\right)
\end{equation} 
y
\begin{equation} 
\delta_z=\frac{3}{c\ N}\left(-D_\perp\frac{\partial N}{\partial z}-
b_\phi D_A\frac{\partial N}{\partial r}\right)\ .
\end{equation}  

\begin{figure}[t!]
\centerline{{\epsfxsize=4.5truein \epsffile{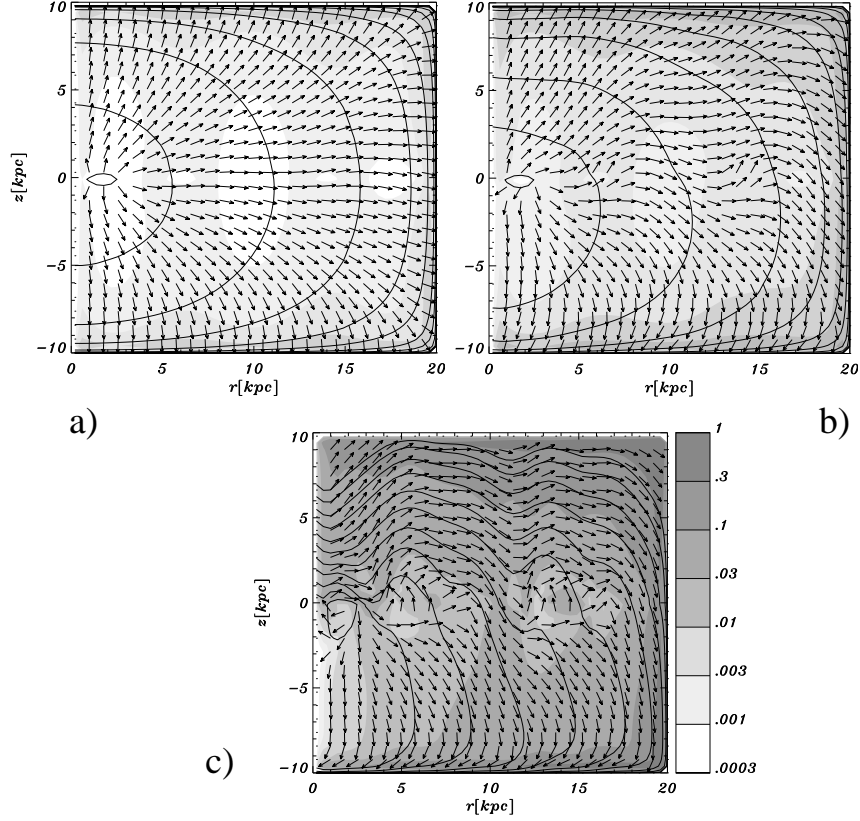}}}
\caption{Curvas de nivel de densidad para rayos c\'osmicos de diferentes energ\'{\i}as,
correspondientes a: a) $E/Z=10^{14}$~eV, b) $E/Z=E_r$, y c) $E/Z=1.5\times 10^{18}$~eV/26. 
Cada dos curvas de nivel, la densidad cambia en un orden de magnitud. Las flechas representan
la direcci\'on de los vectores de anisotrop\'{\i}a, mientras que los sombreados indican la amplitud
de las anisotrop\'{\i}as. El modelo corresponde a una fuente constante contenida en un disco de
radio 3~kpc en el plano gal\'actico, y la estructura de campo magn\'etico descripta en el texto.}  
\label{f3P3}
\end{figure}

\begin{figure}[t!]
\centerline{{\epsfxsize=5.5truein \epsffile{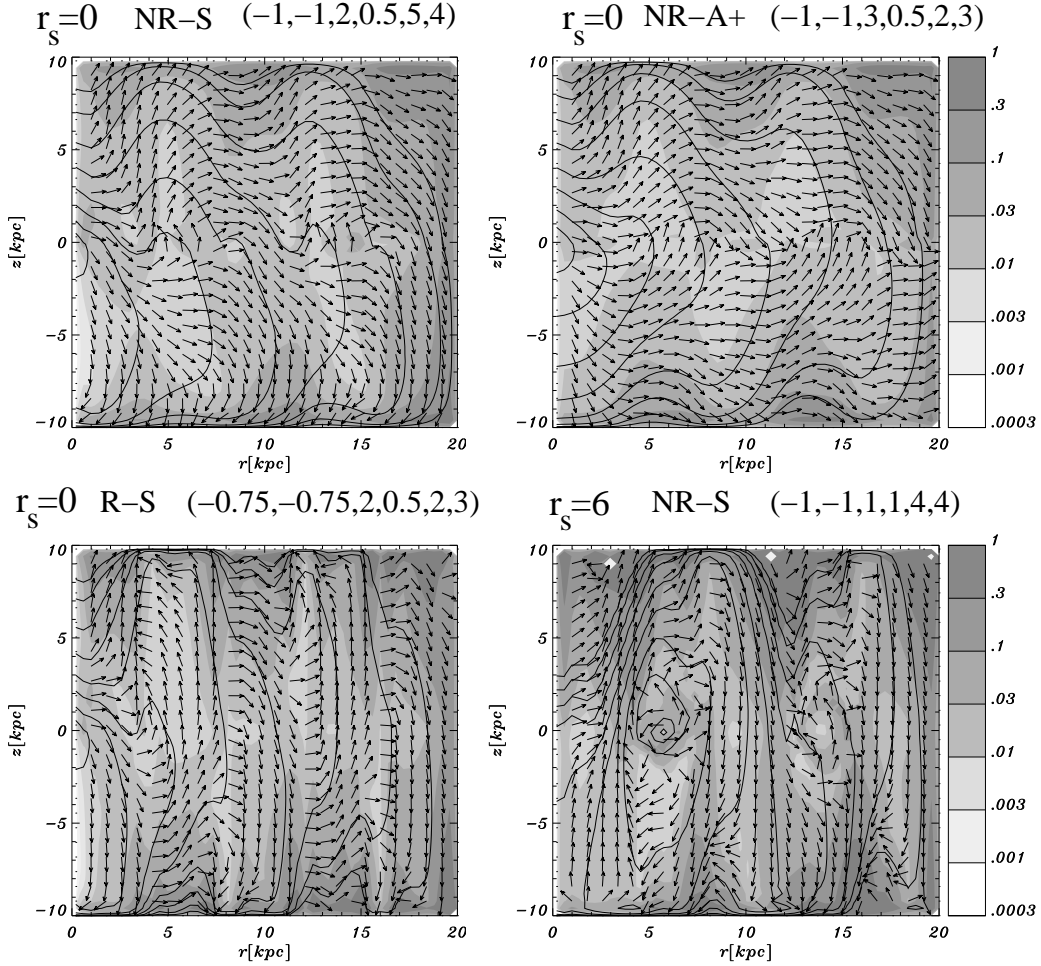}}}
\caption{Densidades, corrientes y anisotrop\'{\i}as para diferentes modelos de campo magn\'etico y
distintas distribuciones de fuentes, en el r\'egimen dominado por los drifts. Entre par\'entesis se indican los valores de 
$(B_0^{*disco},B_0^{*halo},B_r^*,z_d,z_h,z_r)$ en $\mu$G y kpc, respectivamente.}
\label{f4P3}
\end{figure}

Las figs.\ref{f3P3} y \ref{f4P3} muestran, para diferentes modelos de campo magn\'etico y distintas distribuciones
de fuentes, las curvas de nivel asociadas a la densidad de rayos c\'osmicos,
as\'{\i} como tambi\'en la direcci\'on de los vectores de anisotrop\'{\i}a y sus amplitudes (indicadas por medio de los 
sombreados). Los resultados dependen del valor de $E/Z$, es decir que corresponden a diferentes energ\'{\i}as de acuerdo
a la especie nuclear particular que se considere. La importancia relativa de la difusi\'on transversal y el drift para
las distintas regiones de la galaxia puede ser inferida directamente a partir de la direcci\'on entre las curvas de nivel y las
flechas. 

La fig.\ref{f3P3} resulta de considerar, para diferentes energ\'{\i}as, un modelo $NR-S+$ con par\'ametros \footnote{Definiendo
$B_0^{*disco}\equiv B_0^{disco}(r_{obs},z_{obs})$ y $B_0^{*halo}\equiv B_0^{halo}(r_{obs},z_{obs})$,
donde $(r_{obs},z_{obs})=(8.5\ {\rm kpc},0)$.}
$(B_0^{*disco}=-0.75,B_0^{*halo}=-0.75,B_r^*=1.5,z_d=0.5,z_h=5,z_r=5)$ (los campos medidos en $\mu$G y 
las escalas verticales en kpc), y con una fuente constante contenida en un disco de
radio 3~kpc en el plano gal\'actico. 
La fig.\ref{f3P3}(a) corresponde a energ\'{\i}as por debajo de la primera rodilla ($E/Z=10^{14}$~eV), donde
la difusi\'on est\'a casi completamente dominada por $D_\perp$. La fuente corresponde a un disco uniforme de radio $r=3$~kpc
contenido en el plano gal\'actico. Puede verse que las curvas de nivel resultantes son sim\'etricas con respecto al 
plano gal\'actico (comparar con la fig.\ref{f1P2}(a)), mientras que las anisotrop\'{\i}as son de amplitud peque\~na 
($\sim 10^{-4}$) y perpendiculares a las curvas de nivel. La fig.\ref{f3P3}(b) corresponde a $E/Z=E_r$, donde 
los efectos de drift comienzan a ser m\'as notables y la anisotrop\'{\i}a local es del orden de $\sim 10^{-3}$.  
Finalmente, la fig.\ref{f3P3}(c) corresponde a la rigidez $E/Z=1.5\times 10^{18}$~eV/26 (es decir, muestra el 
comportamiento que corresponde, por ejemplo, a n\'ucleos de hierro para $E=1.5\times 10^{18}$~eV). En este r\'egimen, las 
densidades est\'an en gran medida determinadas por los drifts, de forma que reflejan las asimetr\'{\i}as del 
campo magn\'etico regular a gran escala (comparar con la fig.\ref{f1P2}(b)). 
La amplitud de anisotrop\'{\i}a local es, en este caso, del orden de $10^{-2}$.    

En la fig.\ref{f4P3}, se considera el r\'egimen dominado por los drifts (con $E/Z=1.5\times 10^{18}$~eV/26,
la rigidez correspondiente a la fig.\ref{f3P3}(c)), para fuentes localizadas en diferentes posiciones radiales
$r_s$, y para diferentes par\'ametros y estructuras de campo magn\'etico. Si bien todas estas realizaciones 
ajustan los datos del espectro local de rayos c\'osmicos adecuadamente (ver la fig.\ref{f1P3}), 
se observa que las densidades globales, 
las corrientes macrosc\'opicas y las anisotrop\'{\i}as var\'{\i}an significativamente, por lo general, de un 
modelo a otro. En consecuencia, las anisotrop\'{\i}as locales observadas pueden proveer una herramienta muy \'util 
para establecer las propiedades generales del campo magn\'etico.    
 
Resulta importante se\~nalar que, tanto el exceso de eventos hacia el centro gal\'actico como el d\'eficit hacia el anticentro 
gal\'actico observados por AGASA \cite{hay99}, podr\'{\i}an ser reproducidos mediante un flujo difusivo de rayos c\'osmicos
que est\'e localmente dirigido hacia el anticentro gal\'actico, con una amplitud que d\'e lugar a la anisotrop\'{\i}a 
observada del $\sim 4\%$. En el rango de energ\'{\i}as relevante a estas observaciones, 
la difusi\'on de Hall domina sobre la difusi\'on transversal; localmente, $b_\phi=-1$, de modo que
\begin{equation}
\delta_r\simeq -\frac{3\ D_A}{c\ N}\frac{\partial N}{\partial z}\ ,\ \  
\delta_z\simeq \frac{3\ D_A}{c\ N}\frac{\partial N}{\partial r}\ .
\label{aesti}
\end{equation}
Entonces, vemos que la anisotrop\'{\i}a local estar\'a dirigida aproximadamente hacia la direcci\'on radial si, localmente,  
$|\partial N/\partial z|\gg |\partial N/\partial r|$. Por otra parte, volviendo sobre la ec.~(\ref{daeq}), que describe 
al coeficiente de difusi\'on antisim\'etrico, podemos estimar la amplitud de la anisotrop\'{\i}a:  
\begin{equation}
|\delta_r|\simeq \frac{r_L}{h}\frac{(\omega\tau_A)^2}{1+(\omega\tau_A)^2}\simeq 0.04\ \frac{E}{10^{18}{\rm eV}}
\frac{26}{Z}\frac{\rm kpc}{h}\frac{\mu {\rm G}}{|{\bf B_0}|}\frac{(\omega\tau_A)^2}{1+(\omega\tau_A)^2}\ ,
\label{aest}
\end{equation}
donde $h\equiv |\partial\ln N/\partial z|^{-1}$ es la escala vertical caracter\'{\i}stica de la variaci\'on local de la
densidad de rayos c\'osmicos.  
Al considerar la mezcla de todas las especies nucleares, debe evaluarse $\delta (E)=\sum_if_i^*(E)\delta_i(E)$; como
hemos comentado, el drift tiende a remover primero las componentes m\'as livianas, pero a mayores energ\'{\i}as (hacia la 
regi\'on del tobillo) la presencia de la componente de protones extragal\'actica tiende a reducir las 
abundancias relativas de todas las especies nucleares gal\'acticas (ver la fig.\ref{f2P3}). Si bien en la componente
gal\'actica dominan los n\'ucleos del grupo del hierro, debe notarse que la anisotrop\'{\i}a, para una dada energ\'{\i}a, 
es mayor para las componentes m\'as livianas (de la ec.~(\ref{aest}), se ve que $\delta_Z\sim 1/Z$). 
Podemos escribir
\begin{equation}
\delta\simeq\delta_{Fe}\sum_Zf_Z^*{{26}\over{Z}}\ , 
\end{equation}
donde la suma es sobre las componentes gal\'acticas. 
Considerando, como valores indicativos para $1.5\times 10^{18}$~eV, abundancias del orden del 
$\sim 1\%$ para protones (gal\'acticos), $\sim 3\%$ para n\'ucleos de helio, $\sim 5\%$ para n\'ucleos de ox\'{\i}geno 
y $\sim 20\%$ para n\'ucleos de hierro, se obtiene que $\delta\simeq\delta_{Fe}$.
De este modo, la anisotrop\'{\i}a de los n\'ucleos de hierro a esta energ\'{\i}a resulta un buen estimador de la 
anisotrop\'{\i}a total, y ser\'a lo que discutiremos en lo que sigue 
(antes de tratar con el caso detallado en el que sumamos la contribuci\'on de todas las componentes). En particular, 
las figs.\ref{f3P3}(c) y \ref{f4P3} muestran $\delta_{Fe}$  a esta energ\'{\i}a para diferentes realizaciones particulares de 
campo magn\'etico y fuentes que reproducen localmente las observaciones de anisotrop\'{\i}a de AGASA que ya hemos mencionado. 

De la ec.~(\ref{aest}), se ve que las anisotrop\'{\i}as crecen linealmente con la energ\'{\i}a y pueden alcanzar los valores
observados, del orden de $\delta\simeq 0.04$, si la escala vertical caracter\'{\i}stica para las variaciones de la 
densidad de rayos c\'osmicos es del orden del kpc. En los modelos de campos magn\'eticos sim\'etricos aparecen, de un modo
natural, gradientes verticales significativos en la densidad de rayos c\'osmicos sobre el plano gal\'actico, ya que en esos 
modelos los drifts verticales est\'an igualmente orientados en ambos hemisferios. 

Para que el vector de anisotrop\'{\i}a apunte en la direcci\'on radial hacia afuera, el valor local de 
$\partial N/\partial z$ debe ser negativo (ver la ec.~(\ref{aesti})). Para \'esto, las corrientes macrosc\'opicas verticales 
deben estar esencialmente orientadas en la direcci\'on negativa, algo que es m\'as pronunciado, por ejemplo, en modelos
donde un halo sim\'etrico est\'a dirigido en la direcci\'on de $-\hat\phi$. Por otra parte, para tener localmente 
$|\partial N/\partial r|<|\partial N/\partial z|$, es conveniente situar la fuente en una posici\'on no muy cercana a
nuestra ubicaci\'on. El modelo $NR-A+$ (fig.\ref{f4P3}, arriba a la derecha) muestra, adem\'as, la interesante 
caracter\'{\i}stica de que los drifts tienden a converger hacia el plano gal\'actico, resultando en un drift dirigido
hacia afuera para $z\simeq 0$. Por el contrario, un modelo de halo con inversiones radiales en la orientaci\'on del campo
(o bien, un modelo en el que se considera solamente el campo regular del disco gal\'actico) exhibe drifts verticales 
muy pronunciados (fig.\ref{f4P3}, abajo a la izquierda). El factor $(\omega\tau_A)^2/(1+(\omega\tau_A)^2)$ en las 
ecs.~(\ref{daeq}) y (\ref{aest}) refleja la supresi\'on de $D_A$ en presencia de alta turbulencia. Si bien tiende a la 
unidad para $B_r\ll |{\bf B_0}|$, puede hacerse peque\~no si la turbulencia es alta, con el efecto de suprimir las 
anisotrop\'{\i}as.

Otras propuestas, elaboradas con el fin de explicar el exceso de eventos en la direcci\'on del centro gal\'actico, 
asocian la anisotrop\'{\i}a en esta direcci\'on a rayos c\'osmicos que se propagan en forma casi rectil\'{\i}nea desde una 
fuente central \cite{cla00,med01,bed02,bos03}. 
Sin embargo, a\'un en el caso de que esos rayos c\'osmicos fueran protones, la deflexi\'on producida por los
campos magn\'eticos gal\'acticos (regulares y random) exceder\'{\i}a largamente el desplazamiento de $10^\circ$ que se
observa entre la direcci\'on del exceso de eventos y la del centro gal\'actico; para producir una deflexi\'on tan
peque\~na, la fuente deber\'{\i}a estar a una distancia no mayor de $\sim 2$~kpc \cite{med01}. 
La hip\'otesis que atribuye el exceso a neutrones producidos a partir de protones o n\'ucleos acelerados en fuentes en el 
centro gal\'actico (mediante la producci\'on de fotopiones o por procesos de fragmentaci\'on \footnote{En el Cap\'{\i}tulo 4 
se estudiar\'a en detalle este tipo de procesos, en el contexto de una explicaci\'on alternativa para la rodilla del 
espectro de rayos c\'osmicos.}) ha sido explorada como un escenario alternativo \cite{hay99,med01,bos03}. La longitud 
caracter\'{\i}stica de decaimiento de los neutrones es $\sim 10$~kpc~$(E/10^{18}$~eV), 
de modo que, para $E\simeq 10^{18}$~eV, es justamente del orden de la distancia galactoc\'entrica. Adem\'as, esta propuesta 
tambi\'en explicar\'{\i}a porqu\'e no se observa ning\'un exceso significativo a menores energ\'{\i}as. 
Sin embargo, disponer de una fuente de neutrones tan luminosa no es particularmente natural; en cambio, la 
energ\'{\i}a m\'axima de los protones gal\'acticos en el escenario de la difusi\'on turbulenta y drift no necesita exceder 
$\sim 10^{17}$~eV, un requerimiento mucho m\'as plausible. Por otra parte, tampoco resulta evidente, en el escenario 
que propone el decaimiento de neutrones, c\'omo puede explicarse el d\'eficit observado en la direcci\'on del anticentro.  
En lo que respecta al exceso de eventos observados en la direcci\'on de Cygnus, aqu\'{\i} s\'olo haremos notar que, 
en principio, una explicaci\'on similar a la que hemos venido desarrollando en esta Secci\'on puede encontrarse en t\'erminos 
de la difusi\'on de rayos c\'osmicos a lo largo del brazo espiral. Como ya hemos comentado, esto requerir\'{\i}a remover 
la condici\'on de simetr\'{\i}a cil\'{\i}ndrica para permitir la difusi\'on paralela de rayos c\'osmicos en la direcci\'on 
azimutal, gobernada por $D_\parallel$ (ver las ecs.~(\ref{curr1}) y (\ref{curr2})). En cualquier caso, una predicci\'on 
general que resulta en este escenario es que deber\'{\i}a observarse un d\'eficit en la direcci\'on opuesta, algo que resulta
de potencial inter\'es para las futuras observaciones en el hemisferio sur con el observatorio AUGER \cite{auger}. 
 
En la discusi\'on precedente, hemos analizado en detalle la relaci\'on entre el vector de anisotrop\'{\i}a y las 
corrientes macrosc\'opicas de difusi\'on transversal y de drift; adem\'as, estudiamos la dependencia de las anisotrop\'{\i}as 
con diferentes estructuras de campo regular y, finalmente, tomando la anisotrop\'{\i}a del hierro como estimativa de la 
anisotrop\'{\i}a total a energ\'{\i}as del orden de $\sim 10^{18}$~eV, mostramos que el escenario de difusi\'on turbulenta
es capaz de reproducir las observaciones de anisotrop\'{\i}a de AGASA. Desde un enfoque m\'as cuantitativo, en lo 
que sigue consideraremos, para diferentes realizaciones de campo magn\'etico y distintas distribuciones de fuentes, 
la dependencia con la energ\'{\i}a de las medidas de amplitud de anisotrop\'{\i}a y fase del primer arm\'onico en 
ascenci\'on recta, para todo el rango de energ\'{\i}as que comprende las dos rodillas 
(es decir, aproximadamente desde $\sim 10^{15}$~eV hasta $\sim 10^{18}$~eV). 

\begin{figure}[t!]
\centerline{{\epsfxsize=5.5truein \epsffile{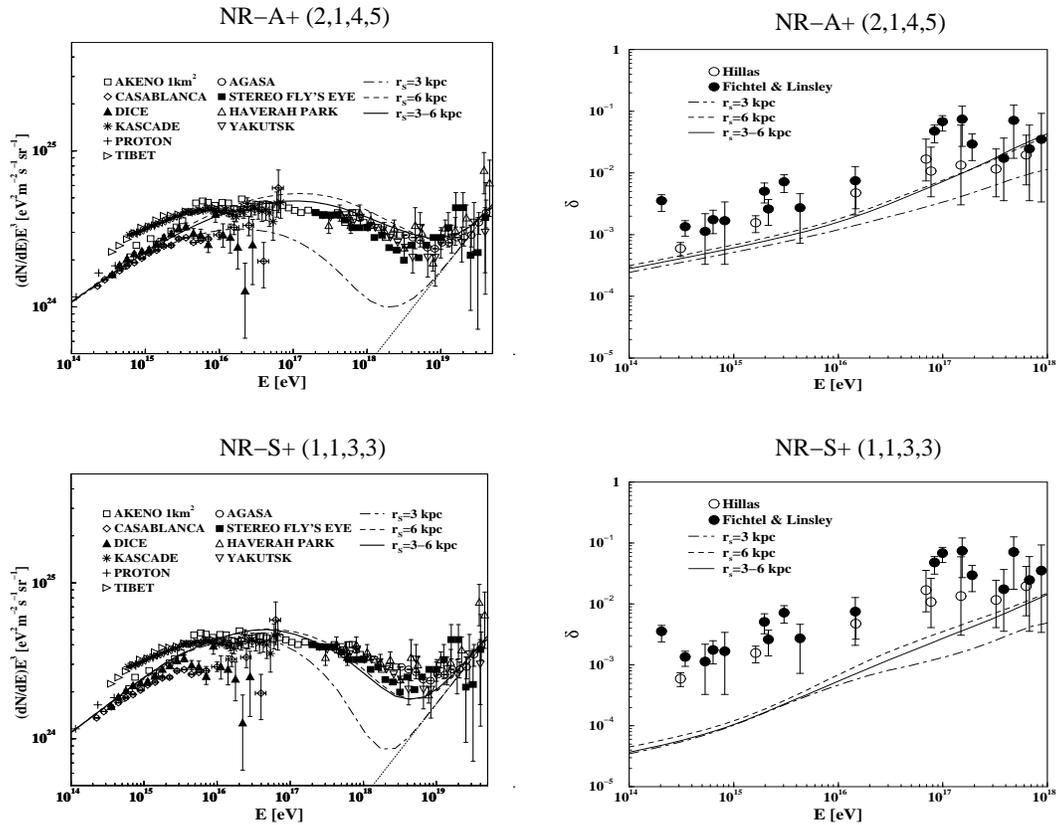}}}
\caption{Espectros y amplitudes de anisotrop\'{\i}a calculados para diferentes modelos de campo magn\'etico gal\'actico, 
con $B_0^{*disco}=B_0^{*halo}=-1$, y los par\'ametros ($B_r^*,z_d,z_h,z_r$) 
en $\mu$G y kpc, respectivamente, indicados sobre cada figura. 
Las rectas punteadas corresponden al flujo extragal\'actico dado por la ec.~(\ref{xgflux}). Tambi\'en se muestran los 
datos experimentales relevantes.}   
\label{f1P4}
\end{figure}  

En la fig.\ref{f1P4}, los espectros y las amplitudes de anisotrop\'{\i}a obtenidas para los modelos de 
campo magn\'etico $NR-A+$ y $NR-S+$ se comparan con las observaciones de diferentes 
experimentos \footnote{Los puntos experimentales de anisotrop\'{\i}a se obtienen de la 
amplitud del primer arm\'onico en ascenci\'on recta
$A$, que se relaciona con la anisotrop\'{\i}a $\delta$ a trav\'es de $A=\delta\cos d\cos\lambda$, donde $d$ es la
declinaci\'on a la cual las observaciones son realizadas, mientras que $\lambda$ es la latitud de la direcci\'on 
donde el flujo es m\'aximo. Los puntos que aparecen en el gr\'afico corresponden a $A/\cos d$ para los diferentes
experimentos, de forma que, en verdad, representan una cota inferior para $\delta$, dado que $\cos\lambda$ no puede
ser recuperado mediante un an\'alisis arm\'onico en ascenci\'on recta.}\cite{nag00,hil84,fic86}. 
Debe notarse que, a partir de aqu\'{\i}, estamos tratando con las anisotrop\'{\i}as totales, calculadas con la 
inclusi\'on de todas las componentes nucleares gal\'acticas 
y la extragal\'actica de protones. En esta figura, se consideran tanto fuentes localizadas (en $r_s=3$ y 6~kpc) como 
extendidas en un anillo de densidad uniforme (para $3\leq r_s\leq 6$~kpc). 
Como ya hemos observado anteriormente, los efectos de drift son
m\'as notables para fuentes m\'as lejanas, ya que las corrientes de drift pueden remover del plano gal\'actico a los rayos 
c\'osmicos que se propagan a lo largo de todo el camino entre la fuente y el observador. Por otra parte, las distribuciones
de fuentes extendidas tienden a producir espectros m\'as planos (es decir, muestran un cambio de \'{\i}ndice espectral menos
abrupto en la regi\'on de la rodilla), debido a que los gradientes de densidad tienden a ser menores. Finalmente, los espectros
de la fig.\ref{f1P4} tambi\'en muestran que, cuando se considera un anillo homog\'eneo extendido de fuentes, 
el flujo observado dominante es el producido cerca del borde m\'as cercano al punto de observaci\'on.            

De la fig.\ref{f1P4} se observa que, asumiendo valores plausibles para los par\'ametros de los modelos de campo, 
el escenario de difusi\'on turbulenta y drift es capaz de reproducir no s\'olo las observaciones del espectro de rayos
c\'osmicos, sino tambi\'en las medidas de anisotrop\'{\i}a. De las ecs.~(\ref{curr2}) y (\ref{anis}) puede verse que   
$|\nabla N|$ juega un papel crucial en el incremento de la amplitud de la anisotrop\'{\i}a, de modo que fuentes cercanas
tienden a producir mayores anisotrop\'{\i}as que otras m\'as lejanas. El efecto de considerar distribuciones de fuentes 
extendidas puede ser comprendido como el resultado de una suma de los resultados producidos por fuentes localizadas, pesada
por la contribuci\'on relativa de cada una a la densidad total de rayos c\'osmicos observada. De esta forma, puede suceder
que s\'olo una peque\~na regi\'on dentro de la distribuci\'on extendida sea responsable de las contribuciones dominantes al 
espectro y la anisotrop\'{\i}a observadas. O bien, puede ocurrir que diversas regiones de la distribuci\'on 
de fuentes contribuyan con corrientes macrosc\'opicas comparables en magnitud pero orientadas seg\'un direcciones muy 
diferentes, dando como resultado que la anisotrop\'{\i}a aparezca suprimida en algunas regiones, e incrementada en otras.   
En cualquier caso, los resultados que aparecen en la fig.\ref{f1P4} para fuentes localizadas deber\'{\i}an ser indicativos
de la tendencia de los que se obtendr\'{\i}an para otras distribuciones de fuentes.    

Aunque algunos resultados para el caso $NR-S+$ exhiben anisotrop\'{\i}as algo bajas, debe notarse que, a\'un en el caso 
de fuentes localizadas, estamos en realidad tratando con distribuciones anulares alrededor del centro gal\'actico, ya que 
nuestro enfoque asume desde el comienzo la simetr\'{\i}a azimutal del sistema. Removiendo esta suposici\'on simplificadora 
y resolviendo la ecuaci\'on de difusi\'on tridimensional para una fuente puntual cercana, podr\'{\i}a obtenerse un incremento 
adicional de la anisotrop\'{\i}a. En efecto, ya ha sido reconocido en la literatura 
\cite{dor84,ber90} que algunas fuentes cercanas discretas (por ejemplo, p\'ulsares y remanentes de supernova tales 
como Vela, Loop III y Geminga) podr\'{\i}an proveer la contribuci\'on dominante a las anisotrop\'{\i}as observadas a bajas 
energ\'{\i}as. Una estimaci\'on precisa de este efecto requiere, sin embargo, indicar la distribuci\'on
espacial, la luminosidad, la edad y la evoluci\'on de las fuentes; en consecuencia, no parece necesario buscar un ajuste
preciso de nuestros resultados a los datos experimentales debajo de la rodilla, aunque es interesante observar que las 
predicciones en este escenario tienden en general a exhibir un ajuste muy aceptable a las observaciones, especialmente en
la regi\'on de altas energ\'{\i}as. 

Al sumar sobre todas las componentes gal\'acticas para obtener el espectro total de rayos c\'osmicos, hemos estado 
considerando la contribuci\'on de todas las especies nucleares desde hidr\'ogeno hasta n\'{\i}quel. De todos modos, 
siguiendo las sugerencias de \cite{hoe03}, tambi\'en exploramos la posible contribuci\'on de los elementos `superpesados' 
hasta el uranio (es decir, con $28<Z\leq 92$). El \'{\i}ndice espectral $\alpha_Z$ de un elemento con $Z\leq 28$
puede medirse a partir del espectro observado a bajas energ\'{\i}as (debajo de $\sim 10^{14}$~eV), y de \'el puede inferirse
el \'{\i}ndice $\beta_Z$ correspondiente a la producci\'on en las fuentes. En cambio,
para los elementos superpesados no hay informaci\'on observacional disponible acerca de los \'{\i}ndices espectrales. 
En \cite{hoe03}, se propuso obtenerlos mediante la extrapolaci\'on de la posible dependencia con $Z$ de los \'{\i}ndices 
espectrales asociados a los elementos m\'as livianos, usando una parametrizaci\'on dada por $-\alpha_Z=A+BZ^C$ para
ajustar los datos. Mientras que esta expresi\'on (en la que intervienen tres par\'ametros a ajustar) favorece una 
extrapolaci\'on no lineal (con $C=1.51\pm 0.13$), tambi\'en el caso de una extrapolaci\'on lineal ($C\equiv 1$) fue 
considerada, encontr\'andose un buen ajuste con los \'{\i}ndices espectrales medidos \cite{hoe03}. Asumiendo la 
extrapolaci\'on lineal de los \'{\i}ndices espectrales, nosotros encontramos que la contribuci\'on de las especies
superpesadas es completamente despreciable en el escenario de la difusi\'on turbulenta y drift. Estos elementos 
superpesados solamente tendr\'{\i}an alg\'un efecto notable con las extremas extrapolaciones no lineales sugeridas en
\cite{hoe03} (seg\'un las cuales el \'{\i}ndice espectral del uranio, por ejemplo, resultar\'{\i}a $\alpha_U\simeq 1.9$), 
pero en nuestro modelo \'esto conducir\'{\i}a a un marcado exceso en el espectro arriba de $\sim 10^{17}$~eV, 
excepto que se considerara un cutoff dependiente de la rigidez en el espectro producido en las fuentes, para energ\'{\i}as cercanas 
a $E\simeq Z\times 10^{17}$~eV. En cambio, en \cite{hoe03} se asume la hip\'otesis {\it ad hoc} de que el \'{\i}ndice 
espectral en la regi\'on de la rodilla cambia en $\Delta\alpha\simeq 2$, y la supresi\'on del espectro resultante 
termina siendo claramente excesiva por encima de la segunda rodilla \footnote{En el Cap\'{\i}tulo 3 se discutir\'a, en
otro contexto, el ajuste al espectro de rayos c\'osmicos en escenarios dependientes de la rigidez. Teniendo en cuenta
s\'olo componentes pesadas hasta $Z=28$, se observa que un cambio de \'{\i}ndice espectral de $\Delta\alpha\simeq 2$ 
conduce a una supresi\'on excesiva en el espectro total a partir de $\sim 10^{17}$~eV (ver la fig.\ref{f1P5}). 
Con la extrapolaci\'on no lineal de los \'{\i}ndices espectrales de elementos superpesados, se evidencia un problema similar 
por encima de la segunda rodilla \cite{hoe03}.}. Por otra parte, ser\'{\i}a natural que el 
espectro producido en las fuentes fuera universal (en t\'erminos de la rigidez), de modo que las diferencias observadas en los 
\'{\i}ndices espectrales a bajas energ\'{\i}as s\'olo reflejasen los efectos del proceso de fragmentaci\'on que 
afecta a la propagaci\'on de los n\'ucleos pesados. Entonces, no deber\'{\i}a esperarse en este caso ning\'un incremento 
de n\'ucleos superpesados por encima de la rodilla.   
   
\begin{figure}[t!]
\centerline{{\epsfxsize=3.15truein \epsffile{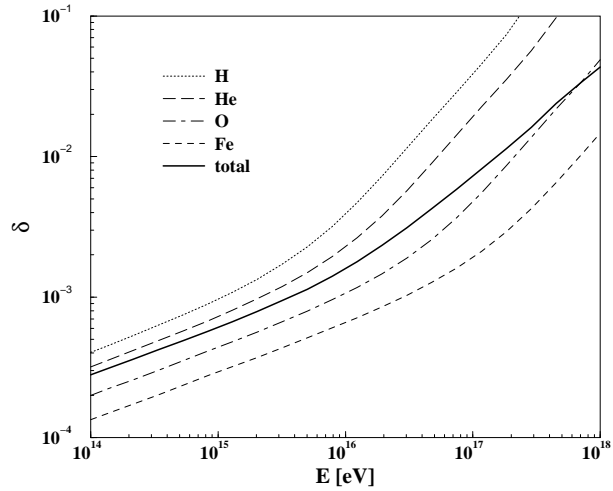}}}
\caption{Amplitudes de anisotrop\'{\i}a correspondientes a las principales componentes de rayos c\'osmicos gal\'acticos: 
protones y n\'ucleos de helio, ox\'{\i}geno y hierro. Para comparar, tambi\'en se muestra la amplitud de anisotrop\'{\i}a 
total, que involucra a las componentes gal\'actica y extragal\'actica. Los datos corresponden a  
un anillo homog\'eneo extendido de fuentes ($3\leq r_s\leq 6$~kpc) para el caso $NR-A+$, con los mismos par\'ametros  
que en la fig.\ref{f1P4}.}
\label{f2P4}
\end{figure}

\begin{figure}[t!]
\centerline{{\epsfxsize=3truein \epsffile{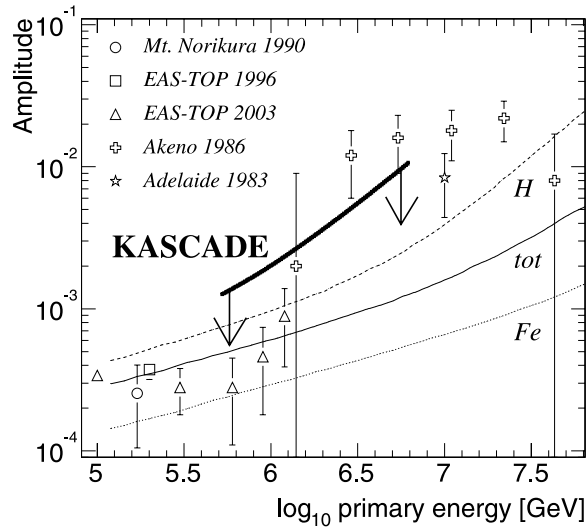}}}
\caption{Amplitudes de anisotrop\'{\i}a en el escenario de la difusi\'on turbulenta y drift, en comparaci\'on 
con los recientes l\'{\i}mites experimentales de KASCADE (de \cite{ant04,kam04,hoe04b}).}
\label{anisKASC}
\end{figure}

Las amplitudes de anisotrop\'{\i}a para diferentes componentes gal\'acticas de rayos c\'osmicos se muestran en la 
fig.\ref{f2P4} como funci\'on de la energ\'{\i}a, para una fuente anular extendida ($r_s=3-6$~kpc), el caso $NR-A+$
y los mismos par\'ametros que se usaron en la fig.\ref{f1P4}. Como ya hemos discutido antes, el r\'egimen dominado 
por los drifts es un fen\'omeno dependiente de la rigidez que incrementa la anisotrop\'{\i}a de todas las componentes 
gal\'acticas. Hemos visto tambi\'en que, aunque los elementos m\'as livianos (particularmente, los protones y los 
n\'ucleos de helio) est\'an fuertemente suprimidos a altas energ\'{\i}as debido al escape de la galaxia, a\'un as\'{\i} 
contribuyen significativamente a la anisotrop\'{\i}a total. Por ejemplo, los n\'ucleos de helio contribuyen, para el
caso de la fig.\ref{f2P4}, alrededor del $30-40\%$ de la anisotrop\'{\i}a total de los rayos c\'osmicos gal\'acticos 
en todo el rango de energ\'{\i}a entre $10^{14}$ y $10^{18}$~eV, aunque la abundancia cae debajo del $10\%$ para 
$10^{18}$~eV. 

La fig.\ref{anisKASC} muestra estos mismos resultados en comparaci\'on con los l\'{\i}mites experimentales 
de KASCADE \cite{ant04,kam04,hoe04b}, obtenidos del reciente an\'alisis de $\sim 10^8$ eventos en el rango de energ\'{\i}a
$(0.7-6)\times 10^{15}$~eV. Si bien la cota de KASCADE es compatible con observaciones anteriores, se 
espera que este experimento (o bien su sucesor, KASCADE-Grande) incremente la estad\'{\i}stica y pueda poner a prueba
las predicciones de este escenario \cite{ant04,kam04,hoe04b}. Por otra parte, ser\'{\i}a de gran inter\'es que se pudieran 
resolver las anisotrop\'{\i}as de las diferentes componentes por separado, ya que eso permitir\'{\i}a 
discriminar entre algunos de los modelos propuestos hasta el momento. KASCADE, en particular, podr\'{\i}a ser capaz de 
realizar este tipo de observaciones en un futuro pr\'oximo \cite{ant04,engel}.    

\begin{figure}[t!]
\centerline{{\epsfxsize=3truein \epsffile{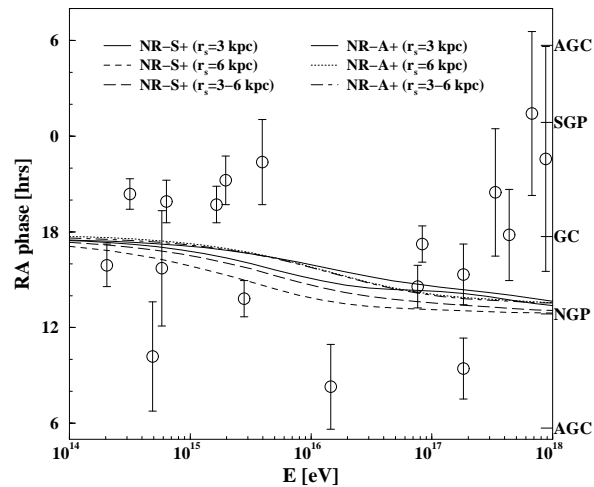}}}
\caption{Fase de anisotrop\'{\i}a del primer arm\'onico en ascenci\'on recta como funci\'on de la energ\'{\i}a, 
correspondiente a los mismos casos de la fig.\ref{f1P4}. Los datos medidos por diferentes experimentos tambi\'en se
muestran para comparar. A la derecha, se indican direcciones de referencia en la galaxia (ACG: Anticentro gal\'actico,
PNG: Polo norte gal\'actico, etc.), tomando en cuenta que $\delta$ est\'a contenida en el plano $r-z$.}    
\label{f3P4}
\end{figure}

En la fig.\ref{f3P4} se grafica la fase de la anisotrop\'{\i}a del primer arm\'onico en ascenci\'on recta en funci\'on de la 
energ\'{\i}a, comparada con las observaciones compiladas en \cite{hil84,fic86}, para los mismos casos que se trataron en la 
fig.\ref{f1P4}. A bajas energ\'{\i}as, la difusi\'on transversal domina y la direcci\'on de m\'axima intensidad corresponde al 
centro gal\'actico, como ya hemos visto antes (ver la fig.\ref{f3P3}(a)). A energ\'{\i}as mayores, la fase de la 
anisotrop\'{\i}a pasa a depender de la geometr\'{\i}a detallada del campo regular adoptado; en los modelos considerados en 
esta figura, las diferencias no son grandes (y tienden a indicar un exceso m\'as bien hacia el polo norte gal\'actico),
pero, tal como hemos visto antes para otros modelos de campo magn\'etico, el m\'aximo de la anisotrop\'{\i}a puede 
estar en la direcci\'on del centro gal\'actico para energ\'{\i}as cercanas a $10^{18}$~eV (ver figs.\ref{f3P3}(c)-\ref{f4P3}).  

\section{Una reconsideraci\'on de los coeficientes de difusi\'on}

Como hemos visto en las secciones precedentes, la apropiada inclusi\'on de los efectos de drift en la propagaci\'on de 
rayos c\'osmicos en los campos magn\'eticos turbulentos de la galaxia es un elemento crucial que permite explicar el
espectro, la composici\'on y las anisotrop\'{\i}as observadas en la regi\'on que comprende las dos rodillas. La difusi\'on turbulenta
y el drift son caracter\'{\i}sticos del transporte de rayos c\'osmicos tambi\'en en otros escenarios de inter\'es, como
por ejemplo en la propagaci\'on de rayos c\'osmicos en los medios interplanetario o intergal\'actico, o en el mecanismo de Fermi de 
aceleraci\'on en diversos objetos astrof\'{\i}sicos.   
Sin embargo, la informaci\'on disponible sobre el comportamiento del tensor de difusi\'on es muy limitada;
recientes investigaciones num\'ericas \cite{gia99,cas02} muestran que la extrapolaci\'on de los resultados anal\'{\i}ticos, 
obtenidos perturbativamente para peque\~nos niveles de turbulencia, no provee una descripci\'on apropiada 
de la difusi\'on en condiciones de alta turbulencia. Complementando esas investigaciones, aqu\'{\i} calcularemos los coeficientes de 
difusi\'on paralela y transversal, parametrizando los resultados mediante relaciones de escala que puedan ser luego utilizadas en 
diferentes aplicaciones. Siguiendo un procedimiento an\'alogo, 
tambi\'en evaluaremos num\'ericamente el coeficiente antisim\'etrico, que es responsable, como ya hemos visto, 
de los efectos de drift, y que no ha sido estudiado previamente en condiciones de alta turbulencia.  
Mientras que en \cite{gia99,cas02} s\'olo ha sido estudiado el caso de fluctuaciones en el campo magn\'etico random dadas por un 
espectro de Kolmogorov, aqu\'{\i} tambi\'en consideraremos otros tipos de turbulencia (a saber, los espectros de turbulencia de 
Kraichnan y de Bykov-Toptygin). Por otra parte, analizaremos la propuesta anal\'{\i}tica de \cite{bie97} 
(en donde se representa el efecto de la turbulencia mediante factores {\it ad hoc} de modulaci\'on exponencial, 
ecs.~(\ref{rxx})-(\ref{rzz})) y discutiremos su rango de validez.      
 
Siguiendo el enfoque de \cite{gia99}, consideraremos part\'{\i}culas cargadas propag\'andose en un campo magn\'etico 
${\bf B}({\bf r})=B_0{\bf\hat{z}}+{\bf B_r}({\bf r})$, donde el primer t\'ermino representa un campo regular uniforme 
dirigido seg\'un la direcci\'on $z$, mientras que el segundo corresponde a la componente random.
Un campo turbulento isotr\'opico y espacialmente homog\'eneo puede ser aproximado num\'ericamente mediante
la suma de un n\'umero grande ($N_m$) de ondas planas con direcciones del vector de ondas, polarizaciones y fases 
elegidas al azar \cite{bat60,gia99}, es decir,   
\begin{equation}
{\bf B_r}({\bf r})=\sum_{n=1}^{N_m}\sum_{\alpha=1}^2A(k_n){\bf\hat{\xi}}_n^\alpha
\cos({\bf k}_n\cdot{\bf r}+\phi_n^\alpha)\ ,
\label{bfourier}
\end{equation}  
donde las dos direcciones de polarizaci\'on ortogonales ${\bf\hat{\xi}}_n^\alpha$ ($\alpha=1,2$) est\'an contenidas en el 
plano perpendicular a la direcci\'on del vector de ondas (o sea, ${\bf\hat{\xi}}_n^\alpha\perp{\bf k}_n$, con el fin de asegurar 
que $\nabla\cdot{\bf B}=0$). La distribuci\'on de n\'umeros de onda se toma de acuerdo a un espaciamiento logar\'{\i}tmico 
constante entre $k_{min}=2\pi/L_{max}$ y $k_{max}=2\pi/L_{min}$, donde $L_{min}$ y $L_{max}$ son las escalas de turbulencia
m\'{\i}nima y m\'axima, respectivamente. La densidad de energ\'{\i}a de la componente turbulenta se toma como 
${\rm d}E_r/{\rm d}k\propto k^{-\gamma}$, donde el \'{\i}ndice espectral $\gamma$ queda determinado por el tipo de mecanismo 
que genera la turbulencia. La amplitud de las ondas planas, en consecuencia, est\'a dada por 
$A^2(k_n)={\mathcal{N}}\langle B_r^2\rangle k_n^{-\gamma}(k_n-k_{n-1})$, donde ${\mathcal{N}}$ es una constante de 
normalizaci\'on tal que $\sum_nA^2(k_n)=\langle B_r^2\rangle$.
Como hemos mencionado m\'as arriba, aqu\'{\i} consideraremos tres tipos de espectro de turbulencia, 
a saber: el espectro de Kolmogorov con $\gamma=5/3$
(un caso particularmente atractivo, debido a que, de acuerdo con las observaciones \cite{arm81,ruz88}, 
las fluctuaciones en la densidad del medio interestelar siguen este espectro de turbulencia), 
el espectro hidromagn\'etico de Kraichnan con $\gamma=3/2$ \cite{kra65}, y el espectro de Bykov-Toptygin con $\gamma=2$ \cite{byk87}.  

El m\'etodo de c\'alculo consiste, b\'asicamente, en generar una configuraci\'on particular de campo magn\'etico random 
(eligiendo al azar la direcci\'on de propagaci\'on, polarizaci\'on y fase de las $N_m$ ondas planas en la ec.~(\ref{bfourier})), 
y seguir la trayectoria de una part\'{\i}cula que se propaga desde el origen, con una direcci\'on inicial arbitraria, 
en el campo magn\'etico total (es decir, en el campo que resulta de considerar la suma de componentes regular y random). 
Luego, los resultados deben ser promediados sobre un gran n\'umero de diferentes configuraciones de campo para calcular los 
coeficientes de difusi\'on. Por ejemplo, en \cite{gia99} los resultados fueron obtenidos integrando primero las trayectorias
de 2500 part\'{\i}culas para una dada configuraci\'on de campo, y luego promediando sobre 50 realizaciones de campo diferentes. 
Nosotros encontramos, sin embargo, que los resultados muestran una mayor dependencia con la elecci\'on particular
de la configuraci\'on de campo que con la elecci\'on de la velocidad inicial de la part\'{\i}cula. En consecuencia, resulta m\'as 
conveniente promediar sobre un gran n\'umero de configuraciones de campo; en nuestro trabajo, generamos t\'{\i}picamente $\sim 10^5$
configuraciones de campo, siguiendo la trayectoria de una \'unica part\'{\i}cula en cada configuraci\'on, y usamos $N_m=100$ modos
en todas las simulaciones.    

Para la integraci\'on num\'erica de las trayectorias, utilizamos una rutina de Runge-Kutta con un paso temporal $\Delta t=0.1\ r_L/c$.
En \cite{cas02}, los coeficientes de difusi\'on paralelo y perpendicular se calcularon a partir de la tasa de incremento asint\'otica
(a tiempos grandes) de los desplazamientos cuadr\'aticos medios en cada direcci\'on, es decir,
\begin{equation}
D_\parallel=\lim_{\Delta t\to\infty}{{\langle(\Delta z)^2\rangle}\over{2\Delta t}}
\label{parplat}
\end{equation}
y
\begin{equation}
D_\perp=\lim_{\Delta t\to\infty}{{\langle(\Delta x)^2\rangle}\over{2\Delta t}}=
\lim_{\Delta t\to\infty}{{\langle(\Delta y)^2\rangle}\over{2\Delta t}}\ .
\label{perpplat}
\end{equation}  
Sin embargo, este m\'etodo requiere seguir las trayectorias de part\'{\i}cula para tiempos considerablemente largos con el fin de
alcanzar la regi\'on asint\'otica, t\'{\i}picamente m\'as all\'a de $t=1000\ r_L/c$. En cambio, nosotros encontramos m\'as conveniente 
calcular las funciones de descorrelaci\'on de velocidad $R_{ij}(t)$, luego integr\'andolas en la f\'ormula de Kubo (\ref{kubo}),
dado que este procedimiento requiere seguir las trayectorias s\'olo hasta tiempos t\'{\i}picamente no mayores a $t=100\ r_L/c$. 
Por otro lado, este m\'etodo permite calcular el coeficiente de difusi\'on antisim\'etrico por medio de $R_{yx}$, mientras
que promedios tales como $\langle\Delta x\Delta y\rangle$ se anulan a tiempos largos y, en consecuencia, no resultan \'utiles
para calcular $D_A$. 

\begin{figure}[th!]
\centerline{{\epsfxsize=3.5truein\epsfysize=2.7truein\epsffile{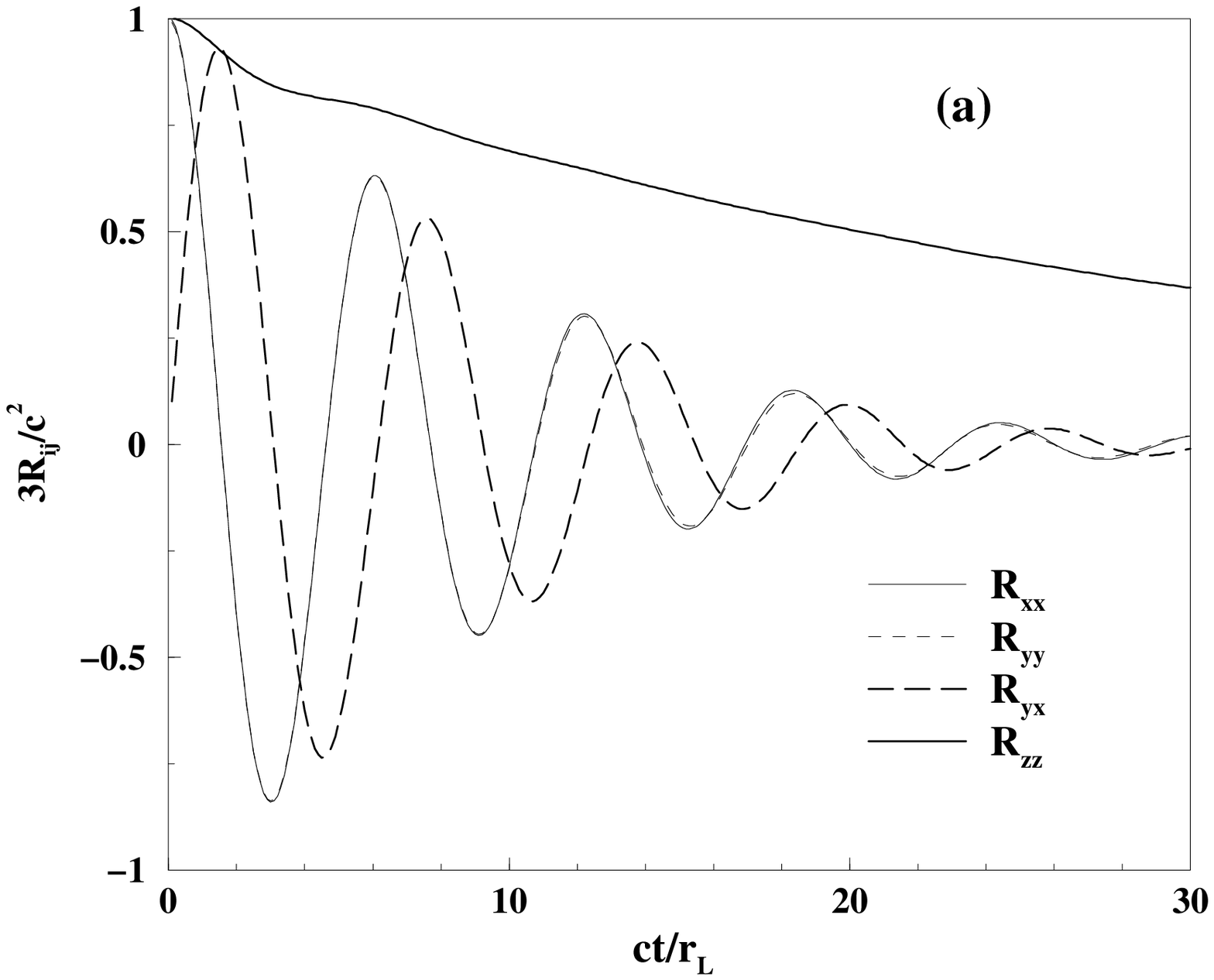}}}
\centerline{{\epsfxsize=3.5truein\epsfysize=2.7truein\epsffile{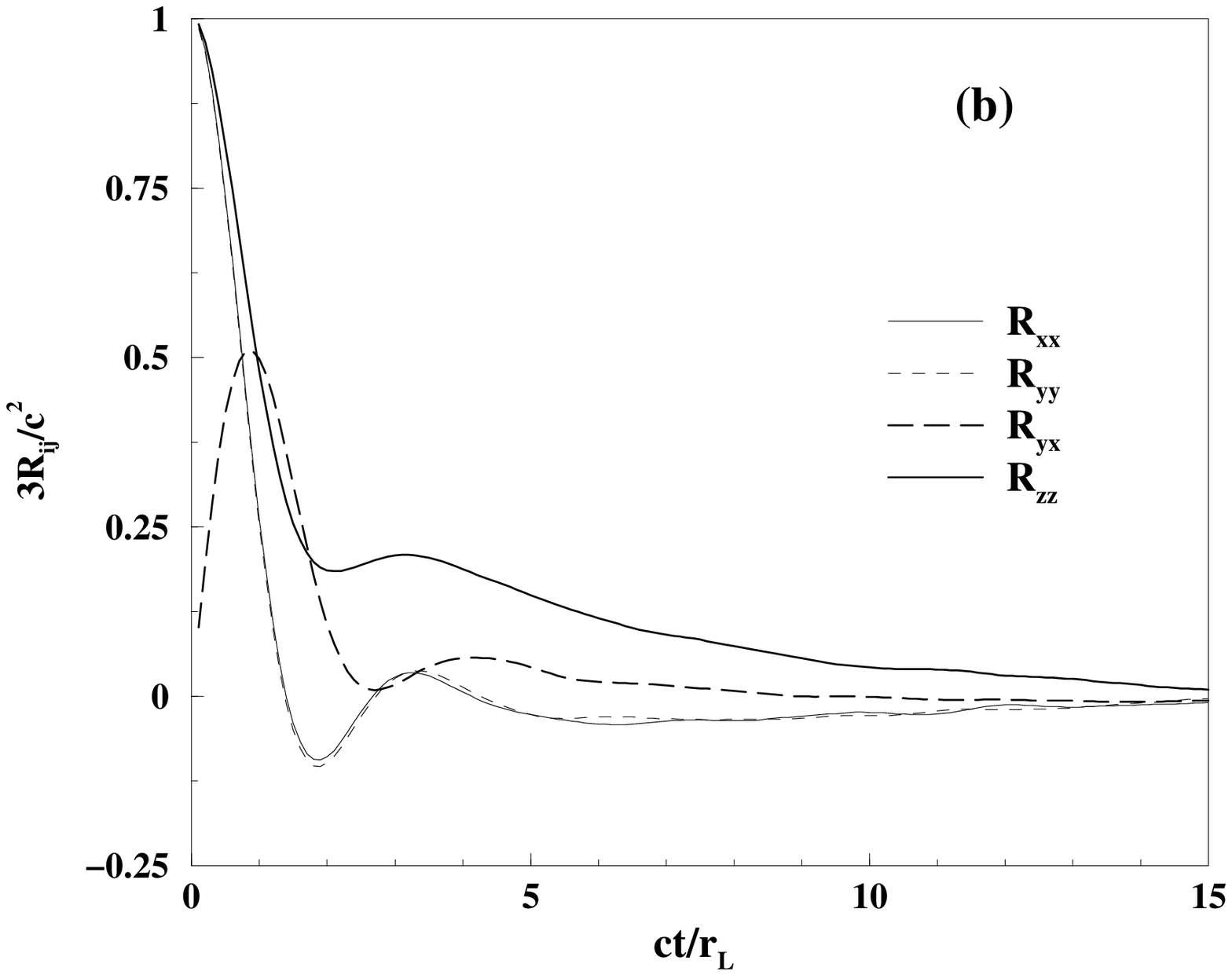}}}
\caption{Dependencia temporal de las funciones de descorrelaci\'on asociadas a diferentes componentes de la velocidad, 
para el espectro de turbulencia de Kolmogorov y dos niveles de turbulencia diferentes: (a) $\sigma^2=0.3$ y 
(b) $\sigma^2=5$, ambos correspondientes a $r_L=0.1\ L_{max}$.}
\label{fig1P6}
\end{figure}

\begin{figure}[t]
\centerline{{\epsfxsize=3.5truein\epsfysize=2.7truein\epsffile{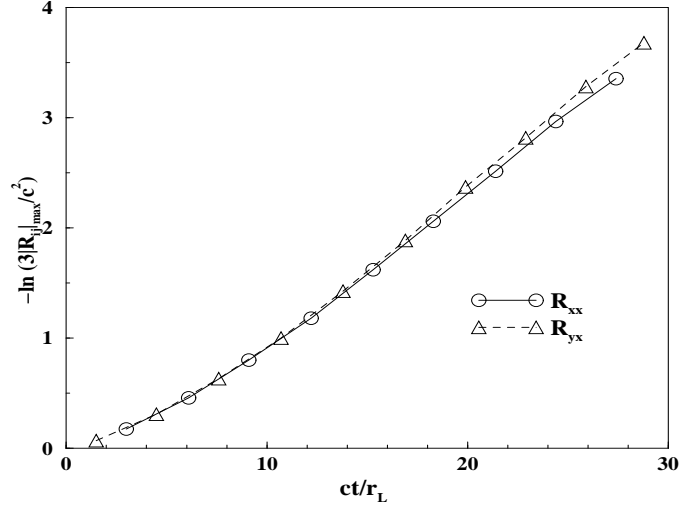}}}
\caption{M\'aximos locales de $|R_{xx}|$ y $|R_{yx}|$ en funci\'on de $t$, graficados en una escala lineal-log,
seg\'un corresponde al caso de la fig.\ref{fig1P6}(a). De aqu\'{\i}, resulta $\tau_\perp=\tau_A$ (ver ecs.~(\ref{rxx})--(\ref{rxy})).}
\label{fig2P6}
\end{figure}

La fig.\ref{fig1P6} muestra la dependencia temporal de las funciones de descorrelaci\'on asociadas a diferentes componentes de la 
velocidad, para el espectro de Kolmogorov y diferentes niveles de turbulencia \footnote{Aqu\'{\i} definimos el nivel de 
turbulencia como $\sigma^2\equiv\langle B_r^2\rangle/B_0^2$ \cite{gia99}. Alternativamente, el nivel de turbulencia puede definirse como 
$\eta\equiv\langle B_r^2\rangle/(\langle B_r^2\rangle+B_0^2)=\sigma^2/(1+\sigma^2)$ \cite{cas02}.}. 
Para baja turbulencia, 
se observa en la fig.\ref{fig1P6}(a) que, luego de un per\'{\i}odo transitorio, las funciones de descorrelaci\'on se comportan 
seg\'un las ecs.~(\ref{rxx})--(\ref{rzz}). \'Esto permite calcular las escalas temporales de descorrelaci\'on; por ejemplo, 
siguiendo los m\'aximos locales de las funciones sinusoidales, se encuentra $\tau_\perp=\tau_A$, tal como muestra la fig.\ref{fig2P6}.
En \cite{bie97}, estas escalas temporales de descorrelaci\'on se consideraban iguales solamente como una hip\'otesis simplificadora,
pero all\'{\i} se suger\'{\i}a que podr\'{\i}an ocurrir nuevos efectos en el caso general en que $\tau_\perp\neq\tau_A$.      
Aqu\'{\i} encontramos, en cambio, que efectivamente estas escalas temporales son iguales, como puede esperarse del hecho de que
el origen f\'{\i}sico de la descorrelaci\'on es el mismo en ambos casos. Para altos niveles de turbulencia, la descorrelaci\'on
es mucho m\'as abrupta, tal como muestra la fig.\ref{fig1P6}(b). Las funciones de descorrelaci\'on se tornan despreciables mientras
a\'un prevalece el per\'{\i}odo transitorio, y las ecs.~(\ref{rxx})--(\ref{rzz}) dejan de ser v\'alidas. 

Definiendo el par\'ametro adimensional $\rho\equiv r_L/L_{max}$, los resultados, expresados en t\'erminos de   
$D/(cL_{max}\rho)$ vs. $\rho$, son universales y satisfacen relaciones de escala que permiten calcular el 
tensor de difusi\'on para diferentes conjuntos de valores para la amplitud del campo regular, las escalas de
longitud caracter\'{\i}sticas asociadas a los modos del campo random, y la energ\'{\i}a de las part\'{\i}culas
que se propagan. Por ejemplo, el rango $0.01\leq\rho\leq 1$ que estudiamos aqu\'{\i} puede considerarse como 
correspondiente a protones de energ\'{\i}a $10^{15}\leq E/\rm{eV}\leq 10^{17}$ propag\'andose en la galaxia 
(con $B_0=1~\mu$G and $L_{max}=100$~pc). Alternativamente, considerando protones propag\'andose en el campo 
interplanetario, de intensidad $B_0=50~\mu$G y $L_{max}=0.01$~AU, el l\'{\i}mite al r\'egimen difusivo para $\rho=1$
corresponde a la energ\'{\i}a cin\'etica $E_k=1.8$~GeV. \footnote{Sin embargo, en \cite{gia99} 
el espectro de ley de potencias de la componente random del campo magn\'etico interplanetario se contin\'ua 
con amplitud constante hasta una escala m\'axima de 1 AU, extendi\'endose considerablemente el r\'egimen difusivo.} Dado que 
la dispersi\'on del \'angulo de pitch ocurre fundamentalmente en condiciones de resonancia, la propagaci\'on es 
esencialmente independiente de la escala m\'{\i}nima de la turbulencia $L_{min}$, en tanto sea $L_{min}\ll r_L$.
Aqu\'{\i} tomamos t\'{\i}picamente $L_{min}=0.1\ r_L$.    

\begin{figure}[t]
\centerline{{\epsfxsize=3.5truein\epsfysize=2.7truein\epsffile{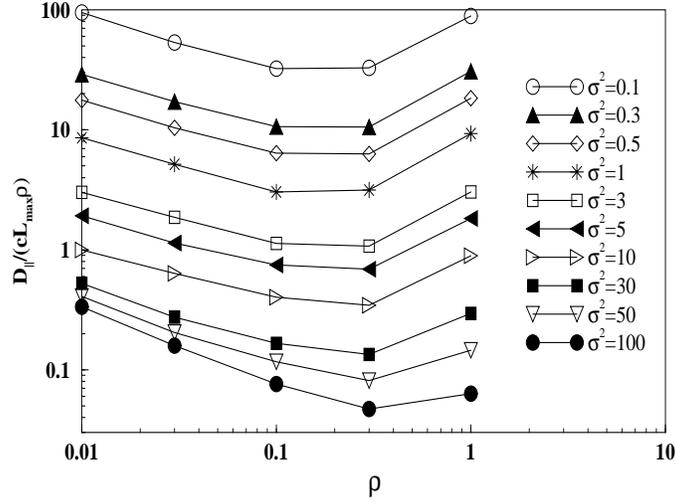}}}
\caption{Resultados correspondientes a la difusi\'on paralela, para el
espectro de Kraichnan y diferentes niveles de turbulencia en el rango $0.1\leq \sigma^2\leq 100$.} 
\label{fig3P6}
\end{figure} 
La fig.\ref{fig3P6} muestra los resultados num\'ericos obtenidos para la difusi\'on paralela, correspondientes al espectro de 
Kraichnan, para diferentes niveles de turbulencia en el rango $0.1\leq \sigma^2\leq 100$.
Como resulta de las ecs.~(\ref{dpar1})--(\ref{dpar2}), en el r\'egimen de dispersi\'on resonante se espera que 
$D_\parallel\propto\rho^{2-\gamma}$, mientras que a mayor rigidez (fuera del r\'egimen de dispersi\'on resonante)
es de esperar un decrecimiento en la efectividad de la dispersi\'on seg\'un $E^{-2}$ \cite{ptu93,ber90}, 
conduciendo as\'{\i} a $D_\parallel\propto\rho^2$, independientemente del espectro de turbulencia considerado.
Los resultados que muestra la fig.\ref{fig3P6} concuerdan muy bien con el comportamiento esperado;
la regi\'on de transici\'on entre el r\'egimen de dispersi\'on resonante y el no resonante ocurre alrededor de
$\rho\simeq 0.2$. En efecto, la longitud de escala que separa ambos reg\'{\i}menes est\'a determinada por la longitud de 
correlaci\'on del espectro de turbulencia $L_c$, definida como 
\begin{equation}
\int_{-\infty}^\infty{\rm d}L\langle{\bf B_r}(0)\cdot{\bf B_r}({\bf r}(L))\rangle\equiv L_c\langle B_r^2\rangle\ ,
\end{equation}
donde el punto ${\bf r}(L)$ est\'a desplazado una distancia $L$ con respecto al origen, a lo largo de una direcci\'on fija.
Entonces, considerando un espectro de fluctuaciones de \'{\i}ndice $\gamma$ que se extiende entre las longitudes de escala 
$L_{min}$ y $L_{max}$, la longitud de correlaci\'on queda dada por \cite{har02}
\begin{equation}
L_c={{1}\over{2}}L_{max}\ {{\gamma-1}\over{\gamma}}\ {{1-\left(L_{min}/L_{max}\right)^\gamma}\over
{1-\left(L_{min}/L_{max}\right)^{\gamma-1}}}\ .
\end{equation}
De aqu\'{\i} puede verse que $L_c/L_{max}\simeq 0.2$ resulta un valor representativo para la 
longitud de correlaci\'on de los espectros de campo random que consideramos en este trabajo, y \'esto 
permite explicar el cambio de r\'egimen observado en $\rho\simeq 0.2$. 

De acuerdo con estas consideraciones, una manera apropiada de ajustar los resultados es interpolando 
$D_\parallel$ entre las leyes de potencia que caracterizan los reg\'{\i}menes de baja y alta rigidez. 
\'Esto puede llevarse a cabo mediante la expresi\'on
\begin{equation}
{{D_\parallel}\over{cL_{max}\rho}}={{N_\parallel}\over{\sigma^2}}
\sqrt{\left({{\rho}\over{\rho_\parallel}}\right)^{2(1-\gamma)}+\left({{\rho}\over{\rho_\parallel}}\right)^2}\ ,
\end{equation}
donde los par\'ametros $N_\parallel$ y $\rho_\parallel$ est\'an dados en la Tabla \ref{tabla2}. 
Esta expresi\'on ajusta bien nuestros resultados num\'ericos hasta $\sigma^2\simeq 10$, que corresponde a un nivel de 
turbulencia sobradamente 
alto para la mayor parte de las aplicaciones astrof\'{\i}sicas de inter\'es. Para niveles de turbulencia
a\'un mayores, los resultados quedan mejor descriptos asint\'oticamente por los par\'ametros $N_\parallel^{asint}$ y 
$\rho_\parallel^{asint}$, tambi\'en dados en la Tabla \ref{tabla2}. 

\begin{figure}[th!]
\centerline{{\epsfxsize=3.5truein\epsfysize=2.7truein\epsffile{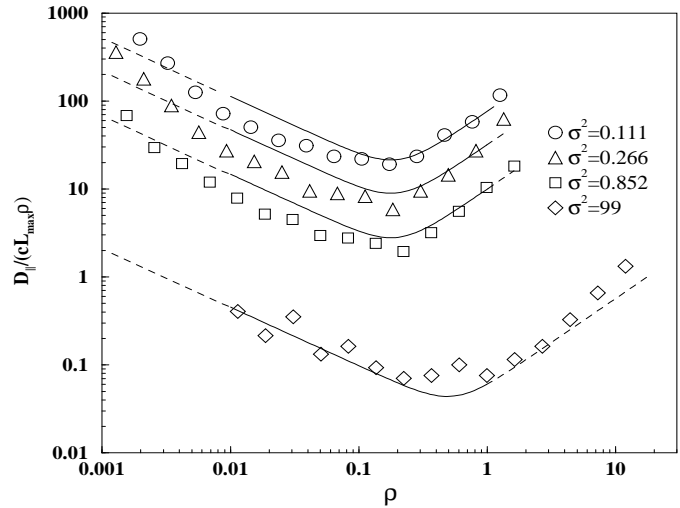}}}
\caption{Comparaci\'on entre el ajuste a $D_\parallel$ dado en este trabajo (v\'alido formalmente s\'olo en el rango
$0.01\leq\rho\leq 1$) y los resultados de \cite{cas02}, ambos correspondientes a un espectro de Kolmogorov, y para
los mismos niveles de turbulencia.} 
\label{fig4P6}
\end{figure} 

\begin{table}
\begin{tabular}{|r|r|r|r|r|r|r|r|r|} \hline
\multicolumn{1}{|c|}{Espectro}&\multicolumn{1}{|c|}{$\gamma$}&\multicolumn{1}{|c|}{$N_\parallel$}&
\multicolumn{1}{|c|}{$\rho_\parallel$}&\multicolumn{1}{|c|}{$N_\parallel^{asint}$}&
\multicolumn{1}{|c|}{$\rho_\parallel^{asint}$}&\multicolumn{1}{|c|}{$N_\perp$}&\multicolumn{1}{|c|}{$a_\perp$}&
\multicolumn{1}{|c|}{$N_A$}\\ \hline
 Kraichnan & 3/2 & 2.0 & 0.22 & 3.5 & 0.65 & 0.019 & 1.37 & 17.6 \\
 Kolmogorov & 5/3 & 1.7 & 0.20 & 3.1 & 0.55 & 0.025 & 1.36 & 14.9 \\
 Bykov-Toptygin & 2 & 1.4 & 0.16 & 2.6 & 0.45 & 0.020 & 1.38 & 14.2 \\ \hline
\end{tabular}
\caption{Par\'ametros de las f\'ormulas de ajuste correspondientes a los coeficientes de difusi\'on paralelo, 
transversal y antisim\'etrico, para diferentes tipos de espectros de turbulencia. Ver m\'as detalles en el texto.}
\label{tabla2}
\end{table}

Con el fin de chequear la consistencia con los resultados num\'ericos previos para la difusi\'on paralela en 
condiciones de alta turbulencia (que fue estudiada en \cite{cas02} asumiendo un espectro de fluctuaciones de Kolmogorov),
la fig.\ref{fig4P6} muestra una comparaci\'on entre nuestra f\'ormula de ajuste y los resultados de \cite{cas02}, 
ambos correspondientes a un espectro de Kolmogorov, y para los mismos niveles de turbulencia. 
Debe notarse que en \cite{cas02} el radio de Larmor\footnote{Para evitar confusiones, adoptamos aqu\'{\i} un
asterisco para referirnos a las cantidades definidas en \cite{cas02}.} $r_L^*$ se define reemplazando 
$B_0\to\sqrt{B_0^2+B_r^2}$ en la ec.~(\ref{rl}), es decir, acoplando en $r_L^*$ la dependencia con la rigidez y con el nivel de
turbulencia, mientras que, por otro lado, el par\'ametro adimensional $\rho^*$ se define como 
$\rho^*=2\pi r_L^*/L_{max}$. En consecuencia, los resultados deben ser reescaleados de acuerdo con las relaciones
$D/(cL_{max}\rho)=D/(r_L^*c)/\sqrt{1+\sigma^2}$ y $\rho=\rho^*\sqrt{1+\sigma^2}/2\pi$. Como puede verse en la fig.\ref{fig4P6},
ambos conjuntos de datos est\'an en buen acuerdo. 

\begin{figure}[t!]
\centerline{{\epsfxsize=3.5truein\epsfysize=2.7truein\epsffile{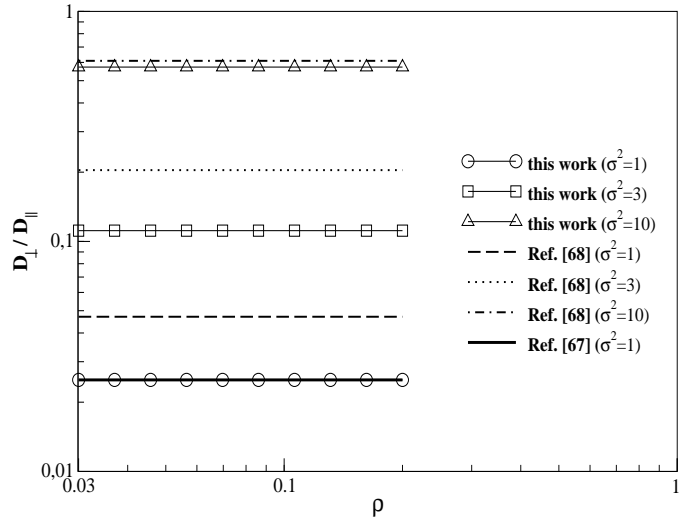}}}
\caption{Ajuste a $D_\perp/D_\parallel$ como funci\'on de la rigidez para el espectro de Kolmogorov en el 
r\'egimen de baja rigidez (es decir, $\rho\leq 0.2$), para diferentes niveles de turbulencia. Para mayor rigidez,  
$D_\perp/D_\parallel\propto\rho^{-2}$ (ver la ec.~(\ref{ratiofit1})).
Para comparar, se muestran tambi\'en los resultados que se obtienen de ajustes a los datos 
dados en \cite{gia99} y \cite{cas02}.}
\label{fig5P6}
\end{figure} 
La manera m\'as simple de parametrizar el coeficiente de difusi\'on perpendicular es a trav\'es de los resultados para el
cociente $D_\perp/D_\parallel$, dado que este cociente exhibe una dependencia sencilla con la rigidez. 
Con el fin de evitar el r\'egimen subdifusivo (que discutimos brevemente m\'as adelante), nos restringimos a la regi\'on 
delimitada por $\rho\geq 0.03$ y $\sigma^2\geq 1$. 
Como en las investigaciones num\'ericas previas \cite{gia99,cas02}, encontramos que el cociente $D_\perp/D_\parallel$ es independiente
de la rigidez en la regi\'on de baja rigidez (es decir, para $\rho\leq 0.2$), mientras que exhibe una dependencia seg\'un 
$\rho^{-2}$ en la regi\'on de alta rigidez. En consecuencia, nuestros resultados pueden ser parametrizados
convenientemente por medio de la expresi\'on 
\begin{equation} 
{{D_\perp}\over{D_\parallel}}=N_\perp\times(\sigma^2)^{a_\perp}\times
\left\{\begin{array}{cl}1\ \ \ \ \ \ \ \ \ \ \ (\rho\leq 0.2)\\
\left(\rho/0.2\right)^{-2}\ (\rho>0.2)\end{array}\right.\ ,
\label{ratiofit1}
\end{equation}     
donde los par\'ametros $N_\perp$ y $a_\perp$ est\'an dados en la Tabla \ref{tabla2}. Hemos visto que los datos de \cite{cas02} (para
el r\'egimen de baja turbulencia) pueden representarse mediante 
\begin{equation}  
{{D_\perp}\over{D_\parallel}}={{1}\over{1+4.5^2/\left(\sigma^2\right)^{1.5}}} 
\label{ratiofit2}
\end{equation}  
(ver la ec.~(\ref{dpdp}) y el p\'arrafo que le sigue); an\'alogamente, los resultados que se presentan en \cite{gia99},
correspondientes al rango de turbulencia $0.03\leq\sigma^2\leq 1$, pueden ajustarse por medio de la expresi\'on 
\begin{equation}  
{{D_\perp}\over{D_\parallel}}=0.025\times(\sigma^2)^{1.835}\ .
\label{ratiofit3}
\end{equation}  

La fig.\ref{fig5P6} muestra el ajuste dado por la ec.~(\ref{ratiofit1}) para el caso de Kolmogorov y diferentes niveles de turbulencia.
Para comparar, tambi\'en se muestran los resultados de \cite{cas02} (seg\'un se obtienen de la ec.~(\ref{ratiofit2})) y de \cite{gia99} 
(dados por la ec.~(\ref{ratiofit3})). Para niveles de turbulencia moderados ($\sigma^2=1$) nuestros resultados muestran un 
acuerdo excelente con los de \cite{gia99}, mientras que el acuerdo con \cite{cas02} tiende a mejorar para turbulencias mayores. 

\begin{figure}[t!]
\centerline{{\epsfxsize=3.5truein\epsfysize=2.7truein\epsffile{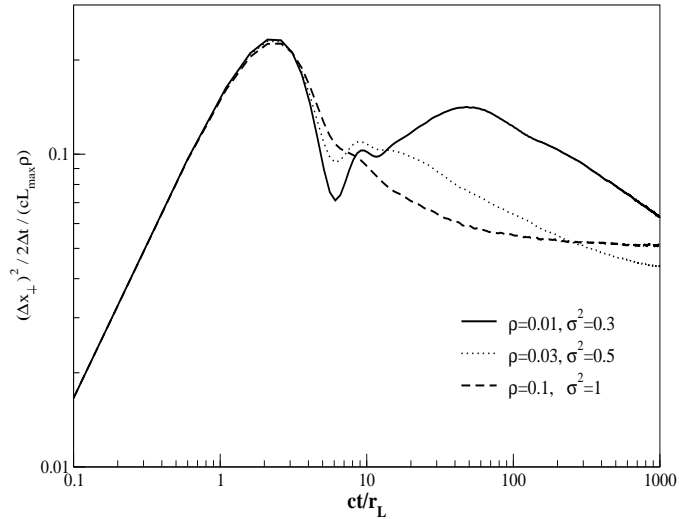}}}
\caption{$\langle\left(\Delta x_\perp\right)^2\rangle/2\Delta t/(cL_{max}\rho)$ en funci\'on del tiempo, para diferentes valores
de rigidez y niveles de turbulencia, en el caso de un espectro de fluctuaciones de Kolmogorov. El r\'egimen subdifusivo aparece
en la regi\'on de rigidez y nivel de turbulencia peque\~nos.} 
\label{fig6P6}
\end{figure} 
Para rigidez y nivel de turbulencia peque\~nos (por ejemplo, $\rho\simeq 0.01$ y $\sigma^2\leq 1$), hemos encontrado evidencia 
del fen\'omeno de subdifusi\'on, en concordancia con lo reportado en \cite{cas02}. Una part\'{\i}cula de rigidez
peque\~na, propag\'andose en condiciones de baja turbulencia, tiende a permanecer siempre sujeta a una misma l\'{\i}nea de campo;
entonces, la difusi\'on transversal ocurre fundamentalmente debido a la caminata aleatoria transversal de las propias l\'{\i}neas
de campo. En consecuencia, se espera que el desplazamiento perpendicular medio evolucione con el tiempo m\'as lentamente, en 
comparaci\'on con el caso de la difusi\'on normal (es decir, $\langle\Delta x_\perp^2\rangle\propto t^m$, con $m<1$). 
La fig.\ref{fig6P6} muestra $\langle\left(\Delta x_\perp\right)^2\rangle/2\Delta t/(cL_{max}\rho)$ como funci\'on del tiempo, 
para diferentes valores de rigidez y nivel de turbulencia, en el caso de un espectro de fluctuaciones de Kolmogorov.
Un {\it plateau} se obtiene a tiempos grandes para $\rho=0.1$ y $\sigma^2=1$ (correspondiendo, por lo tanto, a la relaci\'on
de difusi\'on usual $\langle\Delta x_\perp^2\rangle\propto t$), mientras que, para valores m\'as peque\~nos de rigidez y
nivel de turbulencia, aparece el fen\'omeno de la subdifusi\'on. Como una primera aproximaci\'on a la subdifusi\'on, 
se ha propuesto e investigado el caso l\'{\i}mite conocido como ``difusi\'on compuesta'', en donde 
las part\'{\i}culas est\'an estrictamente restringidas a moverse a lo largo de las l\'{\i}neas de campo  
\cite{get63,for77,urc77,kot00}. Para el caso de la difusi\'on compuesta, se encuentra que $m=1/2$, mientras que, para el 
transporte de part\'{\i}culas en tres dimensiones, se espera que $m$ tenga una dependencia suave con la rigidez y la turbulencia, 
tal que $1/2\leq m(\rho,\sigma^2)<1$. 

Con respecto al coeficiente de difusi\'on antisim\'etrico, los resultados num\'ericos pueden ajustarse adecuadamente mediante 
la expresi\'on 
\begin{equation}
{{D_A}\over{cL_{max}\rho}}={{1}\over{3}}{{1}\over{\sqrt{1+\left(\sigma^2/\sigma^2_0\right)^2}}}\ ,
\label{danum1}
\end{equation}
donde 
\begin{equation}
\sigma_0^2(\rho)=N_A\times\left\{\begin{array}{ll}\rho^{0.3}\ \ \ \ \ \ (\rho\leq 0.2)\\
1.9\ \rho^{0.7}\ (\rho>0.2)\end{array}\right.\ , 
\label{danum2}
\end{equation}
y donde los valores para el par\'ametro $N_A$ est\'an dados en la Tabla \ref{tabla2}. En el l\'{\i}mite de turbulencia muy baja, esta
expresi\'on tiende apropiadamente al valor $D_A\simeq cr_L/3$ \cite{ptu93} 
(ver tambi\'en la ec.~(\ref{daeq}) en el l\'{\i}mite de $\tau_A$ grande), mientras que, como es de esperar, 
se anula en el l\'{\i}mite de turbulencia muy alta. 
\begin{figure}[t!]
\centerline{{\epsfxsize=3.5truein\epsfysize=2.7truein\epsffile{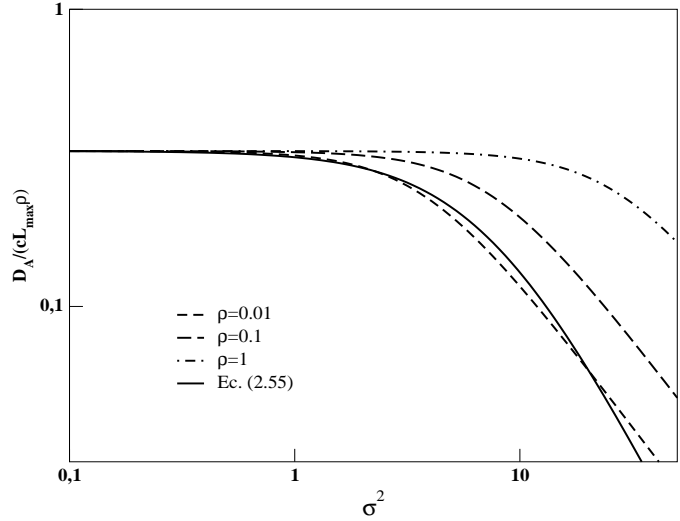}}}
\caption{Comparaci\'on entre el ajuste a $D_A/\rho$ obtenido aqu\'{\i}
(para el espectro de Kolmogorov y los valores de rigidez indicados) y la prescripci\'on independiente de $\rho$ 
adoptada previamente (ecs.~(\ref{da}) y (\ref{presc})).}
\label{fig7P6}
\end{figure} 
Anteriormente, hab\'{\i}amos adoptado una simple prescripci\'on para el coeficiente
antisim\'etrico (ver la ec.~(\ref{da})); en t\'erminos de $\sigma^2$, la expresi\'on que hab\'{\i}amos usado resulta 
\begin{equation}
{{D_A}\over{cL_{max}\rho}}={{1}\over{3}}\ {{1}\over{1+\left(\sigma^2\right)^{1.5}/4.5^2}}\ .
\label{presc}
\end{equation}
Puede verse que, seg\'un esta prescripci\'on, $D_A/cL_{max}\rho$ es independiente de la rigidez. 
En la fig.\ref{fig7P6} se compara esta prescripci\'on con el ajuste dado por las ecs.~(\ref{danum1})--(\ref{danum2}), 
calculado para el caso de Kolmogorov y diferentes valores para la rigidez.
Se observa un excelente acuerdo entre la prescripci\'on (\ref{presc}) y el ajuste correspondiente a 
$\rho=0.01$; la dependencia de $D_A/\rho$ con $\rho$ es, en cualquier caso, solamente significativa para niveles
de turbulencia muy altos ($\sigma^2\gg 1$). Finalmente, resulta importante destacar que
la adopci\'on de las nuevas expresiones que presentamos aqu\'{\i}
para los coeficientes de difusi\'on en condiciones de alta turbulencia no modifica sustancialmente los resultados
obtenidos en las Secciones precedentes.

%% file: capitulo3.tex
\chapter{Neutrinos de muy alta energ\'{\i}a inducidos por rayos c\'osmicos en escenarios dependientes de la rigidez}

\begin{center}
\begin{minipage}{5.6in}
\textsl{
En este Cap\'{\i}tulo, estudiamos el flujo de neutrinos de altas energ\'{\i}as en
escenarios en los que la rodilla del espectro de rayos c\'osmicos es un efecto dependiente de la rigidez magn\'etica; 
en esta clase de escenarios, el fondo de neutrinos atmosf\'ericos 
se reduce significativamente, facilitando la b\'usqueda de fuentes astrof\'{\i}sicas.
As\'{\i}, se pone en evidencia la necesidad de resolver el origen de la rodilla, con el fin de 
interpretar correctamente las observaciones en los nuevos telescopios de neutrinos.
Luego de investigar este fen\'omeno en el canal usual de trazas lept\'onicas, asociado a la
detecci\'on de neutrinos mu\'on, proponemos  
un nuevo m\'etodo para aislar el flujo de neutrinos atmosf\'ericos prompt a trav\'es 
del canal de las cascadas hadr\'onica y electromagn\'etica, asociado a la detecci\'on de neutrinos electr\'on.} 
\end{minipage}
\end{center}  

\section{Introducci\'on}

La existencia de campos magn\'eticos en la regi\'on de propagaci\'on de los rayos c\'osmicos tiene como efecto 
la alteraci\'on de sus trayectorias, dificultando, en consecuencia, la identificaci\'on de fuentes
puntuales. M\'as all\'a del r\'egimen difusivo, que hemos estudiado en detalle en el Cap\'{\i}tulo anterior, 
a mayores energ\'{\i}as se hacen presentes otros fen\'omenos, asociados al {\it lensing} magn\'etico (la magnificaci\'on
o atenuaci\'on de las fuentes observadas, el desplazamiento de sus posiciones aparentes, la aparici\'on
de pares de im\'agenes adicionales, etc. \cite{har99,har00a,har00b,har02}); 
eventualmente, la astronom\'{\i}a de rayos c\'osmicos podr\'{\i}a resultar
d\'{\i}ficil de ser llevada a la pr\'actica. 
Las mismas fuentes astrof\'{\i}sicas de altas energ\'{\i}as podr\'{\i}an producir tambi\'en rayos $\gamma$, 
que conservan la direcci\'on de propagaci\'on, pero que, para altas energ\'{\i}as y grandes distancias, 
ser\'{\i}an absorbidos por la reacci\'on $\gamma\gamma\rightarrow e^+e^-$ con el fondo infrarrojo de radiaci\'on 
(por ejemplo, el camino libre medio es de 100 Mpc para fotones con energ\'{\i}as alrededor 
de $10^4$~GeV \cite{kne02,kne04}). En muchos modelos de fuentes de altas energ\'{\i}as, 
tambi\'en los neutrinos son producidos en abundancia. Los neutrinos tienen la ventaja de no ser deflectados ni 
absorbidos a\'un recorriendo vastas distancias; adem\'as, son capaces de escapar de fuentes de gran densidad.    
Una desventaja evidente es que, precisamente debido a su elusivo car\'acter, resultan tambi\'en muy dif\'{\i}ciles
de detectar, ya que s\'olo interact\'uan d\'ebilmente. 

Sin embargo, por primera vez, los telescopios de neutrinos actualmente en operaci\'on o en construcci\'on 
tendr\'an la sensibilidad requerida para poner a prueba modelos realistas de las fuentes de mayor energ\'{\i}a
del Universo \cite{lea00,alb02,hal02,spi03,tor04}.  
Por ejemplo, provenientes de varios n\'ucleos gal\'acticos activos (AGN) cercanos, 
fueron detectados rayos $\gamma$ de alta energ\'{\i}a hasta alrededor de  
$10^4$~GeV \cite{aha99,kre01,kre02,aha03}. Si estos rayos $\gamma$ est\'an originados en el decaimiento 
de piones neutros, entonces deben esperarse flujos similares de neutrinos originados en el decaimiento
de piones cargados. La producci\'on de piones resulta natural en modelos en los que un flujo de protones
de altas energ\'{\i}as interact\'ua, en la regi\'on de las fuentes, con otros nucleones, o bien con fotones en el campo
de radiaci\'on ambiente. El detector AMANDA est\'a comenzando a poner a prueba estos modelos a un nivel
competitivo con los telescopios de rayos $\gamma$ \cite{and01,ahr02,ahr03a,ahr03b,kow03,ack04,ahr04}.

M\'as all\'a de las fuentes puntuales, los telescopios de neutrinos tambi\'en pueden medir el fondo 
difuso que proviene de fuentes m\'as distantes y de mayor energ\'{\i}a, que no son directamente visibles mediante
rayos $\gamma$ debido a la opacidad del fondo c\'osmico infrarrojo. Sin embargo, ser\'a un gran desaf\'{\i}o poder
distinguir un fondo extragal\'actico difuso del flujo de neutrinos producido por la colisi\'on de los 
rayos c\'osmicos con la atm\'osfera terrestre. El espectro de neutrinos atmosf\'ericos decrece seg\'un  
$E^{-\gamma}$, con el \'{\i}ndice espectral en el rango $\gamma \simeq 3-3.7$, 
mientras que, para la componente extragal\'actica, se asume usualmente un espectro 
$E^{-2}$. En consecuencia, una componente no atmosf\'erica podr\'{\i}a ser descubierta a partir de un
quiebre en el espectro medido. Debajo del quiebre, el espectro estar\'{\i}a dominado por el fondo, y arriba del
quiebre, por la se\~nal. 

Los neutrinos atmosf\'ericos han sido bien identificados a bajas energ\'{\i}as por   
Super - Kamiokande y otros detectores \cite{kaj01,gai02}, y recientemente tambi\'en fueron medidos 
a mayores energ\'{\i}as por AMANDA~\cite{wos04}. En efecto, 
el estudio de neutrinos atmosf\'ericos de energ\'{\i}as sub-GeV y multi-GeV ha sido de enorme importancia; 
en particular, ha provisto la primera evidencia clara en favor de las oscilaciones de neutrinos, demostrando 
que su masa es no nula. Estos neutrinos se producen principalmente en el decaimiento de piones y kaones   
(por ejemplo, $\pi^-\to\mu^-\overline{\nu}_\mu$, con $\mu^-\to e^-\overline{\nu}_e\nu_\mu$) que son producidos por rayos 
c\'osmicos que inciden sobre la atm\'osfera, generando lluvias atmosf\'ericas con una componente hadr\'onica significativa. 

La vida media de los mesones aumenta con la energ\'{\i}a, por  
el efecto relativista de dilataci\'on temporal; 
debido a la interrelaci\'on entre el decaimiento de los mesones y su interacci\'on en el aire, 
a mayores energ\'{\i}as los mesones sufren una atenuaci\'on mayor antes de decaer. De un modo similar, los muones producidos
en la atm\'osfera pueden propagarse sin decaer y alcanzar la superficie terrestre. 
Los efectos relativistas explican que el \'{\i}ndice asociado al espectro de neutrinos atmosf\'ericos sea mayor 
al correspondiente a los rayos c\'osmicos; un espectro de rayos c\'osmicos con \'{\i}ndice 
$\gamma_{RC}\simeq 2.7$ produce un espectro de neutrinos atmosf\'ericos 
caracterizado por un \'{\i}ndice en el rango mencionado, $\gamma \simeq 3-3.7$.
De la longitud de decaimiento de los mesones cargados, $L\equiv\gamma c\tau$, donde, en particular, 
$L_\pi\simeq 5.6$~km($E_\pi/100$~GeV) y $L_K\simeq 7.5$~km($E_K/{\rm TeV}$), resulta que, por encima de 100~GeV, 
los piones atraviesan varias longitudes de atenuaci\'on en la atm\'osfera antes de decaer \footnote{Una longitud de 
atenuaci\'on a esta energ\'{\i}a corresponde aproximadamente a 120~g/cm$^2$.}; en consecuencia, a estas energ\'{\i}as los 
neutrinos son producidos principalmente en el decaimiento de kaones. A la vez, para energ\'{\i}as por encima del TeV, 
los kaones tambi\'en se aten\'uan significativamente antes de decaer, y el decaimiento de part\'{\i}culas charm se 
convierte en el canal de producci\'on de neutrinos m\'as relevante. 
Esta se\~nal, que recibe el nombre de neutrinos atmosf\'ericos {\it prompt charm}, no ha sido a\'un 
identificada experimentalmente. Por otra parte, existen grandes incertezas desde el lado te\'orico. 
La producci\'on de part\'{\i}culas charm por rayos c\'osmicos
requiere tener en cuenta procesos NLO (que incrementan
la producci\'on de part\'{\i}culas charm por un factor mayor a 2) y secciones eficaces que involucran 
el uso de funciones de distribuci\'on part\'onica para valores muy peque\~nos de la variable de Bj{\o}rken $x$,
m\'as all\'a del rango explorado experimentalmente. 

En el momento en que aparezca, eventualmente, un quiebre en el espectro
medido por los telescopios de neutrinos, 
deber\'a analizarse cuidadosamente si la observaci\'on corresponde, efectivamente,   
al descubrimiento de una nueva componente en el flujo de neutrinos de muy alta energ\'{\i}a, 
o bien si se trata s\'olo de una fluctuaci\'on. 
En este sentido, disponer de diferentes canales de detecci\'on ser\'a muy valioso. 
Si la se\~nal es real, podr\'{\i}a constituir un importante indicio de nueva f\'{\i}sica, 
en uno de dos sentidos: (i) como el flujo de neutrinos atmosf\'ericos prompt
(y, por lo tanto, como un nuevo canal que provee informaci\'on sobre rayos c\'osmicos
y QCD), o (ii) como un flujo de neutrinos extragal\'acticos (que dar\'{\i}a nuevos indicios sobre
el Universo de altas energ\'{\i}as). Distinguir entre estas posibilidades tambi\'en requerir\'a
el uso complementario de diferentes canales de detecci\'on.   

Las investigaciones sobre los telescopios de neutrinos se han enfocado, hasta aqu\'{\i},
principalmente hacia el canal de detecci\'on de 
interacciones de corriente cargada de $\nu_\mu$, que involucra la medici\'on de
muones desplaz\'andose hacia arriba. 
Dado que, por encima de algunos cientos de GeV, el rango de los muones en hielo 
excede 1 km (la dimensi\'on lineal caracter\'{\i}stica de IceCube), 
el volumen de detecci\'on efectivo no es igual al volumen del detector, sino, 
en cambio, al \'area del detector multiplicada por el rango de los muones, 
que es una funci\'on que crece con la energ\'{\i}a. Este efecto, combinado con 
el incremento de la secci\'on eficaz de interacci\'on de los neutrinos, 
tiende a compensar el decrecimiento del flujo de los neutrinos incidentes.     

En la Secci\'on 3.2, analizaremos en detalle el canal de trazas lept\'onicas, asociado a la
detecci\'on de neutrinos mu\'on, y enfocaremos nuestra atenci\'on en las componentes de neutrinos atmosf\'ericos. 
Si bien este canal de detecci\'on ha merecido la atenci\'on de numerosas investigaciones previas, 
aqu\'{\i} haremos notar que el flujo de neutrinos de altas energ\'{\i}as (para   
$E_\nu>10^{14}$~eV) depende significativamente del escenario asumido para explicar la rodilla del
espectro de rayos c\'osmicos. En particular, contrastaremos los resultados que se obtienen de suponer 
que los rayos c\'osmicos consisten s\'olo de protones (una hip\'otesis simplificadora usualmente 
asumida en los trabajos previos) con las predicciones que resultan de escenarios m\'as realistas, 
en los que la rodilla es un efecto dependiente de la rigidez magn\'etica. Como veremos, en esta clase de escenarios, 
el flujo difuso de neutrinos de altas energ\'{\i}as producido en la incidencia de rayos c\'osmicos 
sobre la atm\'osfera se reduce significativamente; 
en consecuencia, su detecci\'on en los observatorios de neutrinos se torna m\'as dif\'{\i}cil, pero a la vez facilita la
b\'usqueda de fuentes astrof\'{\i}sicas. As\'{\i}, ponemos en evidencia la necesidad de resolver el origen de la 
rodilla del espectro de rayos c\'osmicos, con el fin
de interpretar correctamente las observaciones de neutrinos de muy
alta energ\'{\i}a en los nuevos telescopios.

Luego, en la Secci\'on 3.3, estudiaremos el canal de las cascadas hadr\'onica y electromagn\'etica, 
asociado a la detecci\'on de neutrinos electr\'on. All\'{\i} proponemos un nuevo m\'etodo para aislar el
flujo de neutrinos atmosf\'ericos prompt, el cual, como ya hemos mencionado, es importante tanto en s\'{\i} mismo,
como en su condici\'on de fondo de la se\~nal extragal\'actica.  
Nuestra propuesta est\'a enfocada en el canal de eventos de corriente cargada de $\nu_e$, enfatizando la
importancia de considerar el espectro de eventos en t\'erminos de la energ\'{\i}a detectable, en lugar
del producto de flujo por secci\'on eficaz como funci\'on de la energ\'{\i}a del neutrino. Aunque este canal 
ha sido, por lo general, ignorado en la literatura, tiene importantes ventajas. La fracci\'on de $\nu_e$
en el flujo atmosf\'erico prompt y en el flujo extragal\'actico es mayor que 
para el flujo atmosf\'erico convencional de altas energ\'{\i}as. 
Veremos que, en este canal, el quiebre espectral ocurre a energ\'{\i}as aproximadamente un orden de magnitud 
menor que para el canal de trazas lept\'onicas; \'esta es una ventaja importante, 
porque, a bajas energ\'{\i}as, los flujos son mayores y los efectos de absorci\'on terrestre son menores. 
Varios autores se han
centrado en la detecci\'on de $\nu_\tau$; sin embargo, a energ\'{\i}as por debajo de 
$5 \times 10^6$~GeV, resulta dif\'{\i}cil separar las interacciones individuales de $\nu_\tau$
de aqu\'ellas de otros sabores. El canal $\nu_e$ deber\'{\i}a ser particularmente efectivo 
para los neutrinos atmosf\'ericos prompt, debido a que su espectro decrece m\'as r\'apidamente 
que el espectro extragal\'actico. De este modo, si el quiebre espectral ocurre a 
menor energ\'{\i}a, el flujo atmosf\'erico prompt puede detectarse m\'as f\'acilmente.  
Por otra parte, la fidelidad espectral entre la energ\'{\i}a del neutrino y la energ\'{\i}a detectada 
es mucho mayor que en el caso de interacciones de $\nu_\mu$ de corriente cargada, 
algo importante para la b\'usqueda de un quiebre espectral.  
Siguiendo el enfoque de la Secci\'on precedente, estudiaremos el efecto que tiene la
composici\'on asumida para los rayos c\'osmicos, comparando nuevamente el caso en el que  
los rayos c\'osmicos consisten s\'olo de protones, 
con escenarios en los que la rodilla es un efecto dependiente de la rigidez magn\'etica.
Finalmente, tambi\'en estudiaremos el flujo difuso de neutrinos gal\'acticos, producidos por la interacci\'on de los rayos
c\'osmicos con el medio interestelar, analizando sus componentes y su importancia relativa.  

\section{El canal de trazas lept\'onicas: neutrinos mu\'on} 

\subsection{El espectro de rayos c\'osmicos}

Una manera sencilla de describir el flujo de rayos c\'osmicos gal\'acticos en un escenario dependiente
de la rigidez es asumir que cada componente de carga $Z$ est\'a dada por
\begin{equation}
{{\rm d}N_Z\over{\rm d}E}\equiv \phi_Z=
{{\phi^<_Z\cdot \phi^>_Z}\over{\phi^<_Z+\phi^>_Z}}\ ,
\label{crfit1}
\end{equation}
donde $\phi^<_Z$ ($\phi^>_Z$) es el flujo de rayos c\'osmicos a bajas (altas) energ\'{\i}as \footnote{En este escenario, 
el espectro asociado a cada componente tiene una rodilla, cuya energ\'{\i}a es proporcional a la carga; 
en relaci\'on con ella, quedan definidas las regiones de bajas y altas energ\'{\i}as que mencionamos aqu\'{\i}.},
dado por
\begin{equation}
\left\{ \begin{array}{cl}\phi^<_Z \\
\phi^>_Z\end{array}\right\}
=f_Z\phi_0\left(\frac{E}{E_0}\right)^{-\alpha_Z}\times
\left\{ \begin{array}{cl}1 \\
\left(E/ZE_k\right)^{-\Delta\alpha}\end{array}\right\}\ .
\label{crfit2}
\end{equation}

\begin{figure}[t]
\centerline{{\epsfxsize=4.5truein\epsfysize=3.1truein\epsffile{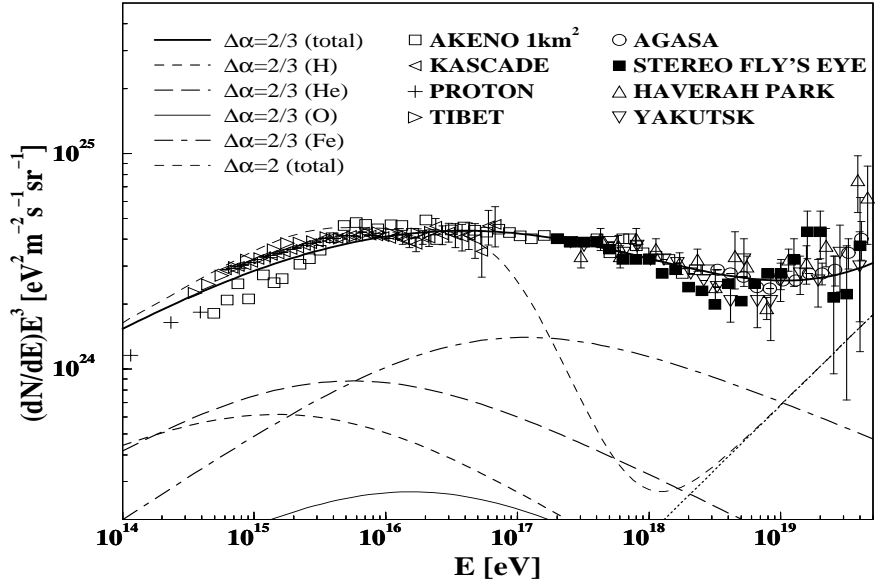}}}
\caption{Espectros de rayos c\'osmicos en escenarios dependientes de la rigidez, seg\'un la parametrizaci\'on 
dada por las ecs.~(\ref{crfit1})--(\ref{crfit2}). Para el cambio de \'{\i}ndice espectral $\Delta\alpha=2/3$, 
tambi\'en se muestran las principales contribuciones al flujo total, correspondientes a n\'ucleos de H, He, O y Fe.
La l\'{\i}nea recta punteada corresponde al flujo de rayos c\'osmicos extragal\'acticos (ec.~(\ref{xgflux})). 
Para comparar, tambi\'en se muestran las observaciones experimentales.}  
\label{f1P5}
\end{figure}

En esta expresi\'on, $\phi_0$ es la normalizaci\'on del flujo total  
y $f_Z$ la abundancia fraccional, ambos correspondientes a la energ\'{\i}a de referencia $E_0$.
Eligiendo $E_0=1$~TeV, la normalizaci\'on y las abundancias son las dadas en la Secci\'on 2.2; 
all\'{\i} tambi\'en est\'an tabulados los \'{\i}ndices espectrales $\alpha_Z$ (ver la Tabla \ref{tabla1}).  
Solamente dos par\'ametros quedan libres: $\Delta\alpha$, el cambio de \'{\i}ndice espectral a trav\'es 
de la rodilla (que es el mismo para todas las componentes), y $E_k$, que fija la posici\'on de la rodilla
(y se elige convenientemente para ajustar el espectro total a las observaciones). 
La ec.~(\ref{crfit1}) provee una interpolaci\'on suave \footnote{Eventualmente, la ec.~(\ref{crfit1}) puede generalizarse 
como $\phi=\phi^<\cdot\phi^>/(\phi^{<n}+\phi^{>n})^{1/n}$, 
donde el par\'ametro $n$ regula el ``ancho'' de la rodilla; tomando $n\gg 1$, resulta un cambio de pendiente muy abrupto. 
Para el caso $\Delta\alpha=2/3$ que consideramos aqu\'{\i}, $n=1$ conduce a un acuerdo excelente con 
los datos experimentales (ver fig.\ref{f1P5}).} entre el flujo $\phi^<_Z$ a bajas energ\'{\i}as $(E<ZE_k)$ y $\phi^>_Z$ a
energ\'{\i}as altas $(E>ZE_k)$. 

Asumiremos, adem\'as, una componente extragal\'actica dada por la ec.~(\ref{xgflux}) (ver la Secci\'on 2.3), y formada s\'olo por
protones \footnote{De todas formas, la composici\'on de la componente extragal\'actica de rayos c\'osmicos s\'olo 
afecta las predicciones para neutrinos de energ\'{\i}a $E_\nu>10^{17}$~eV.}. 
La fig.\ref{f1P5} muestra el ajuste a las observaciones experimentales, seg\'un resulta de tomar 
$\Delta\alpha=2/3$ (correspondiente, por ejemplo, al modelo de la difusi\'on turbulenta y drift desarrollado en el 
Cap\'{\i}tulo 2), tomando el par\'ametro $E_k=2.2\times 10^{15}$eV, y $\Delta\alpha=2$ (correspondiente al modelo asumido en
\cite{hoe03}), con $E_k=3.1\times 10^{15}$eV. 
Mientras el caso $\Delta\alpha=2/3$ exhibe un acuerdo excelente con las observaciones, puede verse que el caso $\Delta\alpha=2$
conduce a la supresi\'on abrupta del flujo total por encima de $\sim 10^{17}$~eV; en lo que sigue, enfocaremos 
nuestra atenci\'on en el caso $\Delta\alpha=2/3$. 

\subsection{El flujo de neutrinos mu\'on atmosf\'ericos}
Los rayos c\'osmicos que inciden sobre la atm\'osfera terrestre producen flujos secundarios de hadrones y leptones, 
que pueden describirse por medio de un sistema acoplado de ecuaciones diferenciales que recibe el nombre de 
``ecuaciones de cascada''. Bajo ciertas hip\'otesis simplificadoras, las ecuaciones de
cascada pueden ser resueltas anal\'{\i}ticamente.
En esta Secci\'on, investigamos las componentes atmosf\'ericas de los flujos de neutrinos y antineutrinos mu\'on. 
Debe notarse que, al no distinguir part\'{\i}culas y antipart\'{\i}culas, aqu\'{\i} nos referimos, en todo momento, 
a la suma de neutrinos y antineutrinos. 

En el rango de inter\'es de este trabajo ($10^3$~GeV$\leq E_\nu\leq 10^8$~GeV), las principales contribuciones al flujo de 
neutrinos provienen del decaimiento de los mesones $\pi^\pm, K^\pm, K_S^0$ y $K_L^0$, que producen los  
neutrinos atmosf\'ericos {\it convencionales} que dominan a bajas 
energ\'{\i}as \cite{vol80,bug87,bug89,bar89,hon90,lip93,hon95,agr96,lip98,bat00,fio01,barr03,hon04}, 
y del decaimiento de los mesones charm $D^0,D^\pm,\Lambda_c$ y $D_s$, 
que dan lugar al flujo de neutrinos atmosf\'ericos {\it prompt charm} que domina a energ\'{\i}as 
altas \cite{zas93,thu96,pas98,pas99,vol01,cos01,cos02,mar03,gel03,hoo03,sta04}.
La contribuci\'on de los muones al flujo de neutrinos atmosf\'ericos es completamente despreciable en este contexto;
debido a efectos relativistas, por encima de $E\sim 10^3$~GeV los muones llegan hasta la superficie terrestre sin decaer.
Finalmente, debe tenerse en cuenta que, para las energ\'{\i}as consideradas aqu\'{\i}, 
las oscilaciones de sabor no tienen ning\'un efecto apreciable en los neutrinos atmosf\'ericos.

La ecuaci\'on de transporte asociada a la componente $j-$\'esima de la cascada puede escribirse como
\begin{equation}
{{{\rm{d}}\phi_j}\over{{\rm{d}}X}}(E,X)=-{{\phi_j}\over{\lambda_j}}(E,X)-
{{\phi_j}\over{\lambda_j^d}}(E,X)+\sum_kS_{kj}(E,X)\ ,
\label{casc}
\end{equation}      
donde $X$ es el slant depth definido por la ec.~(\ref{slant}) (ver la Secci\'on 1.5), $\lambda_j$ la longitud de interacci\'on en el aire
y $\lambda_j^d$ la longitud de decaimiento. Estas dos longitudes caracter\'{\i}sticas son medidas en unidades de  
g/cm$^2$ (es decir, incluyen el factor $\rho(X)$ que corresponde a la densidad local de la atm\'osfera). 
El \'ultimo t\'ermino corresponde al proceso de producci\'on/regeneraci\'on dado por  
\begin{equation}
S_{kj}(E,X)=\int_E^\infty{\rm{d}}E'{{\phi_k(E',X)}\over{\lambda_k(E')}}
{{1}\over{\sigma_{kA}(E)}}{{{\rm{d}}\sigma_{kA\to j}(E,E')}\over{{\rm{d}}E}}\ ,
\label{coupling1}
\end{equation}
que acopla las ecuaciones de transporte asociadas a las diferentes componentes de la cascada. 
Puede notarse que $\sigma_{kA}$ aqu\'{\i} se refiere a la secci\'on eficaz total $k+$aire, mientras que $\sigma_{kA\to j}$ 
corresponde al proceso $k+$aire $\to$ $j+$``...''. Como las energ\'{\i}as de ligadura entre los nucleones de un 
n\'ucleo son mucho menores que las energ\'{\i}as t\'{\i}picamente involucradas en las colisiones nucleares entre rayos 
c\'osmicos y n\'ucleos atmosf\'ericos, en este contexto s\'olo es relevante la interacci\'on nucle\'on-nucle\'on.  
En consecuencia, las diferentes componentes de la cascada que tomaremos en cuenta son la componente de nucleones $N$ 
(que agrupa a protones y neutrones) y las diferentes componentes de mesones $M=\pi^\pm, K^\pm, K_L^0,D^0,D^\pm,\Lambda_c$ y $D_s$. 
Como en \cite{lip93}, la contribuci\'on de $K_S^0$ puede incluirse en el flujo de piones. Por simplicidad, consideraremos
adem\'as que los \'unicos procesos relevantes son aqu\'ellos que corresponden a la regeneraci\'on (es decir, $S_{NN}$ y 
$S_{MM}$) y a la producci\'on de mesones por nucleones (es decir, $S_{NM}$) \cite{lip93,thu96}. 

Consideremos, en particular, flujos de la forma $\phi_k(E,X)= E^{-\beta_k}g(X)$ (es decir, planteamos la separaci\'on de 
variables en los flujos, y escribimos el factor dependiente de la energ\'{\i}a como una ley de potencias). 
Entonces, resulta que
\begin{equation}
S_{kj}(E,X)={{\phi_k(E,X)}\over{\lambda_k(E)}}Z_{kj}(E)\ ,
\label{coupling2}
\end{equation} 
donde $Z_{kj}$ es el momento de producci\'on/regeneraci\'on definido por
\begin{equation}
Z_{kj}(E)=\int_0^1{\rm{d}}x\ x^{\beta_k-1}{{1}\over{\sigma_{kA}(E)}} 
{{{\rm{d}}\sigma_{kA\to j}(E,x)}\over{{\rm{d}}x}}\ ,
\label{zmoment}
\end{equation}
con $x\equiv E/E'$. 

Asumiendo que las longitudes de interacci\'on son independientes de la energ\'{\i}a, y que la distribuci\'on diferencial 
de producci\'on satisface el scaling de Feynman, los propios momentos de producci\'on/regeneraci\'on se tornan independientes 
de la energ\'{\i}a. As\'{\i}, para las longitudes de interacci\'on relevantes y los momentos de 
producci\'on/regeneraci\'on que involucran nucleones y mesones no charm, consideramos los valores independientes de la 
energ\'{\i}a dados en \cite{gai90,lip93}. Adem\'as, tomamos la dependencia de los momentos con $\beta_k$ de ajustes 
a los datos compilados en \cite{gai90}. Por otra parte, las longitudes de 
interacci\'on y los momentos de regeneraci\'on asociados a todas las componentes charm pueden tomarse como iguales a las 
correspondientes a los kaones \cite{pas99}.

Sin embargo, para la producci\'on de part\'{\i}culas charm debe tenerse en cuenta el momento dependiente de la energ\'{\i}a 
dado por la ec.~(\ref{zmoment}), ya que los resultados son muy sensibles a la secci\'on eficaz que se adopta para 
describir el proceso de producci\'on charm. En efecto, las diferentes estimaciones de los flujos atmosf\'ericos
prompt calculados hasta el momento difieren en m\'as de un orden de magnitud \cite{gel03}.

Debido a su masa relativamente grande, se considera usualmente que los quarks charm son producidos en procesos duros que
pueden ser bien descriptos por la cromodin\'amica cu\'antica perturbativa (pQCD).      
A orden dominante (LO) en la constante de acoplamiento, los procesos que contribuyen a la secci\'on
eficaz de producci\'on charm son el proceso de fusi\'on gluon-gluon $gg\to c\overline{c}$ y el proceso de aniquilaci\'on 
quark-antiquark $q\overline{q}\to c\overline{c}$. Sin embargo, en el siguiente orden de aproximaci\'on perturbativa 
(NLO) aparece el proceso de dispersi\'on de gluones $gg\to gg$, aportando una contribuci\'on 
muy significativa a la secci\'on eficaz total, que de esta manera se incrementa en un factor 
$\sim 2-2.5$ \cite{nas88,bee89,mat89}.
Con respecto a correcciones de mayor orden, se espera que su contribuci\'on sea relativamente peque\~na, dado que 
cualitativamente no hay nuevos canales que puedan aparecer. Varias funciones de distribuci\'on part\'onica han sido 
propuestas, demostrando que proveen predicciones te\'oricas de pQCD que reproducen razonablemente bien los datos disponibles 
de los aceleradores. Sin embargo, para calcular el flujo atmosf\'erico charm se necesita extrapolar las funciones de 
distribuci\'on de partones hasta fracciones de momento part\'onico muy peque\~nas, t\'{\i}picamente 
$x\simeq 4$~GeV/$E$ \cite{mar03}. Para las energ\'{\i}as que debemos considerar aqu\'{\i}, 
\'esto queda fuera de la regi\'on accesible experimentalmente ($x>10^{-5}$), y la incerteza en la extrapolaci\'on
afecta los resultados finales significativamente \cite{pas99,gel03}.     
Con el fin de ilustrar este rango de incertidumbre, mostramos los resultados que se obtienen con dos funciones de 
estructura diferentes, a saber la funci\'on de distribuci\'on part\'onica CTEQ3 \cite{lai95} y el modelo Golec-Biernat, 
W\"usthoff (GBW) \cite{gol99,bar02}, que incluye efectos de saturaci\'on de gluones. Las secciones eficaces de producci\'on de
charm correspondientes fueron obtenidas, ya sea directamente del ajuste dado en \cite{mar03} (para los resultados basados 
en el modelo de saturaci\'on GBW), o bien ajustando los resultados dados en \cite{pas99} para las funciones
de estructura CTEQ3 (evaluadas con los par\'ametros $M=2m_c=2\mu$, con la masa del quark charm $m_c=1.3$~GeV), 
e interpolando los resultados para diferentes energ\'{\i}as. Para la secci\'on eficaz nucle\'on-aire, 
usamos la parametrizaci\'on dada por \cite{mie94},
\begin{equation}
\sigma_{NA}(E)=\left[280-8.7\ln\left({\frac{E}{\rm{GeV}}}\right)+
1.14\ln^2\left({\frac{E}{\rm{GeV}}}\right)\right]{\rm{mb}}\ .
\label{sigmaN}
\end{equation}

Una vez que se determinan todas las longitudes de interacci\'on y los momentos de producci\'on/regeneraci\'on relevantes, los
flujos atmosf\'ericos de nucleones y mesones pueden calcularse de las ecuaciones acopladas de cascada dadas por (\ref{casc}) 
y (\ref{coupling2}). Recordando que estas expresiones fueron obtenidas suponiendo que los nucleones y mesones 
tienen espectros de ley de potencias, discutiremos, en primer lugar, los resultados para un flujo inicial de nucleones dado por
$\phi_N(E,X=0)=\phi_{0N}E^{-\gamma}$, y dejaremos la discusi\'on del caso general para m\'as adelante. 

Bajo las suposiciones que hemos mencionado, el flujo de nucleones se desarrolla independientemente de los flujos de los 
mesones, y resulta
\begin{equation}
\phi_N(E,X)=e^{-X/\Lambda_N}\phi_{0N}E^{-\gamma}\ ,
\label{nucleons}
\end{equation}  
donde la longitud de atenuaci\'on de los nucleones se define como
\begin{equation}
\Lambda_N={{\lambda_N}\over{1-Z_{NN}}}\ .
\label{LambdaN}
\end{equation} 

Con respecto a las ecuaciones de cascada de los mesones, \'estas se resuelven usualmente considerando por separado la 
soluci\'on de bajas energ\'{\i}as $\phi_M^L$ (que desprecia los t\'erminos de interacci\'on y de regeneraci\'on, dado que  
$\lambda_M\gg\lambda_M^d$ a bajas energ\'{\i}as) y la soluci\'on de altas energ\'{\i}as $\phi_M^H$ (que, en cambio, 
desprecia el t\'ermino de decaimiento, pues $\lambda_M\ll\lambda_M^d$ a energ\'{\i}as muy altas) \cite{gai90,lip93,thu96}.

La soluci\'on de altas energ\'{\i}as est\'a dada por   
\begin{equation}
\phi_M^H(E,X)={{Z_{NM}}\over{1-Z_{NN}}}{{e^{-X/\Lambda_M}-e^{-X/\Lambda_N}}
\over{1-\Lambda_N/\Lambda_M}}\phi_{0N}E^{-\gamma}\ ,
\label{mesonsH}
\end{equation} 
donde la longitud de atenuaci\'on de los mesones $\Lambda_M$ se define an\'alogamente a la ec.~(\ref{LambdaN}). 

Para el caso de bajas energ\'{\i}as, la soluci\'on se obtiene de reemplazar $\Lambda_M\to\lambda_M^d$ en la \'ultima 
ecuaci\'on. Debido al factor de Lorentz de dilataci\'on temporal, la longitud de decaimiento es lineal con la energ\'{\i}a; 
luego, vemos que la expresi\'on resultante acopla la dependencia con la 
energ\'{\i}a y el slant depth, y no conduce a un simple espectro de ley de potencias.  
Sin embargo, para valores de $X$ no demasiado peque\~nos, la soluci\'on se reduce a 
\begin{equation}
\phi_M^L(E,X)={{Z_{NM}}\over{1-Z_{NN}}}{{\lambda_M^d}\over{\Lambda_N}}
e^{-X/\Lambda_N}\phi_{0N}E^{-\gamma}\ .
\label{mesonsL}
\end{equation}

En las ecs.~(\ref{nucleons})--(\ref{mesonsL}), $Z_{NN},Z_{NM},$ y $Z_{MM}$ deben ser evaluados con el \'{\i}ndice espectral 
$\gamma$, correspondiente al flujo de nucleones. Puede notarse tambi\'en que el flujo de mesones para altas energ\'{\i}as
tiene el mismo \'{\i}ndice espectral que el flujo de nucleones, mientras que la soluci\'on de mesones para bajas 
energ\'{\i}as da lugar a un espectro m\'as duro, debido a un factor adicional (lineal en la energ\'{\i}a) 
impl\'{\i}cito en $\lambda_M^d$. 

El flujo de neutrinos producido por el decaimiento d\'ebil de mesones puede calcularse siguiendo el mismo esquema 
que ya hemos desarrollado para obtener el flujo atmosf\'erico de nucleones y mesones. En efecto, el flujo de neutrinos
est\'a dado por una ecuaci\'on de transporte (an\'aloga a la ec.~(\ref{casc})) que s\'olo contiene los t\'erminos de fuente que 
resultan del decaimiento de mesones, es decir,
\begin{equation}
{{{\rm{d}}\phi_\nu}\over{{\rm{d}}X}}(E,X)=\sum_MS_{M\nu}(E,X)\ ,
\label{nuflux1}
\end{equation}      
donde 
\begin{equation}
S_{M\nu}(E,X)=\int_E^\infty{\rm{d}}E'{{\phi_M(E',X)}\over{\lambda_M^d(E')}}
{{1}\over{\Gamma_M(E)}}{{{\rm{d}}\Gamma_{M\nu}(E,E')}\over{{\rm{d}}E}}\ .
\label{nuflux2}
\end{equation}
La distribuci\'on de decaimiento puede ponerse en t\'erminos de $F_{M\nu}$, el espectro inclusivo de neutrinos en el 
decaimiento del meson $M$, por medio de la relaci\'on  
\begin{equation}  
{{1}\over{\Gamma_M(E)}}{{{\rm{d}}\Gamma_{M\nu}(E,E')}\over{{\rm{d}}E}}=B_{M\nu}F_{M\nu}(E,E')\ ,
\end{equation}  
donde $B_{M\nu}$ es el {\it branching ratio} asociado al decaimiento del meson $M$ en un estado con el dado neutrino $\nu$. 
En el l\'{\i}mite ultrarrelativista, el espectro inclusivo de neutrinos satisface la relaci\'on de escala  
$F_{M\nu}(E,E')=F_{M\nu}(E/E')/E'$ \cite{gai90,lip93}. Por otro lado, ya hemos derivado las soluciones asint\'oticas 
para el flujo de los mesones (ecs.~(\ref{mesonsH}) y (\ref{mesonsL})), que son de la forma $\phi_M(E,X)\propto E^{-\beta}g(X)$, con 
$\beta=\gamma$ ($\beta=\gamma-1$) para los flujos de altas (bajas) energ\'{\i}as. 
De esta manera, los t\'erminos de fuente dados por la ec.~(\ref{nuflux2}) pueden reescribirse como
\begin{equation}
S_{M\nu}(E,X)={{\phi_M(E,X)}\over{\lambda^d_M(E)}}Z_{M\nu}^{\beta+1}(E)\ ,
\label{nuflux3}
\end{equation}  
donde los momentos de decaimiento de los mesones (dependientes de $\beta$) son definidos mediante  
\footnote{Debe aclararse que el branching ratio aqu\'{\i} est\'a inclu\'{\i}do en la definici\'on de los momentos de 
decaimiento, en analog\'{\i}a con los momentos de producci\'on/regeneraci\'on que incluyen la multiplicidad de los estados 
finales. \'Esto coincide con \cite{thu96}, pero difiere de \cite{lip93}.}
\begin{equation}
Z_{M\nu}^\beta(E)=B_{M\nu}\int_0^1{\rm{d}}x\ x^{\beta-1} F_{M\nu}(x)\ .
\label{zdec}
\end{equation}
Los momentos de decaimiento que usamos en este trabajo fueron calculados ajustando la dependencia con $\beta$ de los
momentos relevantes tabulados en \cite{thu96} para $2.7\leq\beta\leq 4$, que, a la vez, fueron determinados mediante los
programas Lund de simulaci\'on Monte Carlo. 

Entonces, los flujos de neutrinos asint\'oticos para bajas y altas energ\'{\i}as pueden ahora ser determinados de las 
ecs.~(\ref{nuflux1}) y (\ref{nuflux3}), que involucran a los momentos de decaimiento de los mesones a y los correspondientes
flujos de mesones para altas y bajas energ\'{\i}as, dados por las ecs.~(\ref{mesonsH}) y (\ref{mesonsL}). La soluci\'on
de bajas energ\'{\i}as resulta
\begin{equation}
\phi_\nu^L(E,X)=Z_{M\nu}^\gamma{{Z_{NM}}\over{1-Z_{NN}}}
\left(1-e^{-X/\Lambda_N}\right)\phi_{0N}E^{-\gamma}\ .
\label{atmosL1}
\end{equation}
Como es de esperar, el flujo de neutrinos se desarrolla r\'apidamente, en la escala de una longitud de interacci\'on de 
nucle\'on, y luego permanece estable. Luego de atravesar toda la atm\'osfera (es decir, en el l\'{\i}mite $X\gg\Lambda_N$) 
el flujo de bajas energ\'{\i}as queda dado por
\begin{equation}
\phi_\nu^L(E)=Z_{M\nu}^\gamma{{Z_{NM}}\over{1-Z_{NN}}}\phi_{0N}E^{-\gamma}\ .
\label{atmosL2}
\end{equation}
De un modo similar, el flujo de altas energ\'{\i}as al nivel de la superficie terrestre es
\begin{equation}
\phi_\nu^H(E)=Z_{M\nu}^{\gamma+1}{{Z_{NM}}\over{1-Z_{NN}}}{{\ln\left(\Lambda_M/
\Lambda_N\right)}\over{1-\Lambda_N/\Lambda_M}}{{\epsilon_M}\over{\cos\theta}}
\phi_{0N}E^{-(\gamma+1)}\ ,
\label{atmosH}
\end{equation}
donde $\theta$ es el \'angulo zenital asociado a la direcci\'on de incidencia del rayo c\'osmico primario, y donde  
\begin{equation}
\epsilon_M={{m_Mch_0}\over{\tau_M}}\ ,
\end{equation}
con $m_M$ y $\tau_M$ la masa y la vida media del meson, respectivamente, y donde $h_0=6.4$~km es una altura caracter\'{\i}stica
de la escala de variaciones de la densidad en la atm\'osfera. 
Dado que los momentos de producci\'on/regeneraci\'on ($Z_{NN},Z_{NM}$ y $Z_{MM}$) son iguales a los que aparecen en los flujos 
de los mesones, deben ser evaluados usando el \'{\i}ndice espectral $\gamma$ asociado al flujo de nucleones. 
La ec.~(\ref{atmosH}) asume un perfil de densidad que corresponde a una atm\'osfera 
isot\'ermica (es decir, $\rho(h)\propto\exp(-h/h_0)$), que es apropiado en la estrat\'osfera ($h\geq 11$~km), la zona en la 
que ocurre la mayor parte de las interacciones \cite{lip93,thu96}. 
Adem\'as, en la ec.~(\ref{atmosH}) se desprecia la curvatura de la tierra, de modo que, en principio, s\'olo ser\'{\i}a 
v\'alida para \'angulos zenitales peque\~nos. Sin embargo, a\'un para \'angulos grandes podemos seguir usando la misma 
expresi\'on, con la simple prescripci\'on 
de reemplazar $\theta\to\theta^*(\theta,h=30$~km), donde, para una dada l\'{\i}nea de visi\'on que corresponde a un \'angulo
$\theta$ en la posici\'on del observador, el \'angulo $\theta^*(\theta,h)$ es el \'angulo zenital que ser\'{\i}a observado 
en una posici\'on en la que esta misma l\'{\i}nea de visi\'on est\'a a una altura $h$ con respecto a la 
superficie \cite{lip93}.
Por ejemplo, para rayos c\'osmicos que inciden en la direcci\'on horizontal ($\theta=90^\circ$), resulta  
$\theta^*\simeq \arcsin(1+h/R_\oplus)^{-1}\simeq 84.5^\circ$.     

Para obtener una expresi\'on general que describa los flujos de neutrinos luego de atravesar toda la atm\'osfera, las 
soluciones asint\'oticas (\ref{atmosL2})--(\ref{atmosH}) pueden ser unificadas por medio de la 
funci\'on de interpolaci\'on 
\begin{equation}
\phi_\nu={{\phi^L_\nu\cdot \phi^H_\nu}\over{\phi^L_\nu+\phi^H_\nu}}\ ,
\label{interpol}
\end{equation}
an\'aloga a la funci\'on de interpolaci\'on dada en la ec.~(\ref{crfit1}) para unificar los flujos de bajas y altas energ\'{\i}as 
asociados a cada componente de rayos c\'osmicos gal\'acticos.

Hasta aqu\'{\i}, hemos descripto un esquema de c\'alculo que provee los flujos de nucleones, mesones y neutrinos producidos 
por un flujo inicial de nucleones de ley de potencias, $\phi_N(E,X=0)=\phi_{0N}E^{-\gamma}$, que incide sobre la parte superior
de la atm\'osfera. Ahora obtendremos los resultados correspondientes al espectro total de rayos c\'osmicos que llega a la
tierra, tomando en cuenta la contribuci\'on de las diferentes especies nucleares que intervienen en su composici\'on.

Sea $\phi_Z$ el flujo de rayos c\'osmicos asociado a la componente nuclear de carga $Z$ y 
n\'umero m\'asico promedio $A$. Esta componente nuclear aporta un flujo de nucleones 
dado por $\phi_{N,Z}(E_N)=A^2\phi_Z(E=AE_N)$. Si el flujo de cada componente de rayos c\'osmicos de carga $Z$ es 
$\phi_Z(E)=\phi_{0Z}(E/E_0)^{-\gamma_Z}$, el flujo de nucleones correspondiente est\'a dado por
\begin{equation}
\phi_N(E)=\sum_Z A^{2-\gamma_Z}\phi_{0Z}\left({E\over E_0}\right)^{-\gamma_Z}\ .
\label{crnucleon}
\end{equation}
Consideremos, en particular, el flujo de rayos c\'osmicos
gal\'acticos parametrizado por las ecs.~(\ref{crfit1}) y (\ref{crfit2}). All\'{\i}, cada
componente de rayos c\'osmicos de carga $Z$ est\'a representada por dos flujos de ley de potencias $\phi_{Z}^<$ 
y $\phi_{Z}^>$, con \'{\i}ndices espectrales $\alpha_Z$ y $\alpha_Z+\Delta\alpha$, respectivamente. Para cada flujo de ley
de potencias, podemos calcular el flujo de nucleones correspondiente (dado por la ec.~(\ref{crnucleon})) y usarlo
para calcular el flujo de neutrinos asociado. Entonces, el flujo final de neutrinos producido por esa dada componente
de rayos c\'osmicos puede obtenerse mediante la interpolaci\'on de las dos soluciones, usando nuevamente una funci\'on de
interpolaci\'on an\'aloga a la ec.~(\ref{crfit1}). Finalmente, el flujo total de neutrinos resulta de sumar la contribuci\'on 
de todas las componentes de rayos c\'osmicos gal\'acticos, m\'as la contribuci\'on adicional de la componente extragal\'actica.

\begin{figure}[t]
\centerline{{\epsfxsize=4.7truein\epsfysize=3.2truein\epsffile{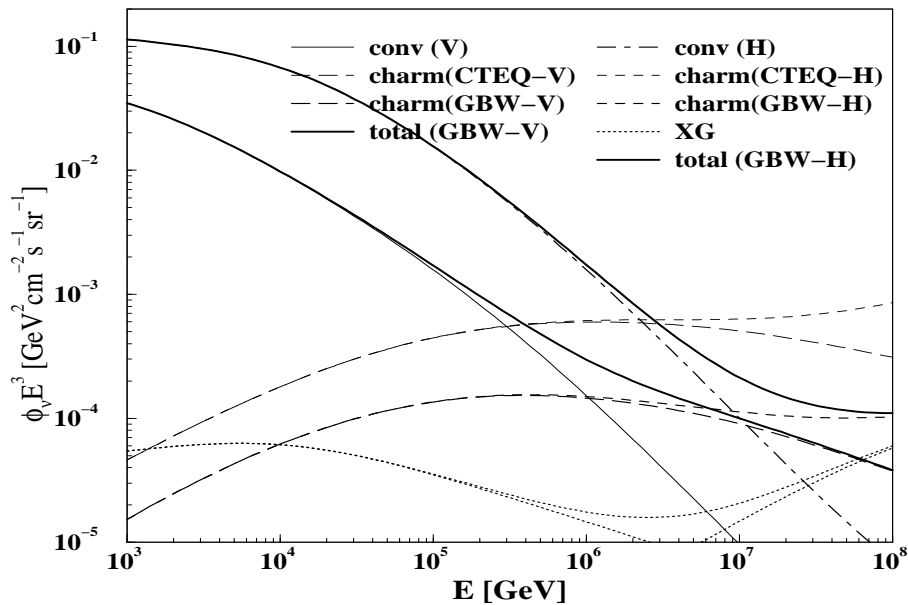}}}
\caption{Flujos atmosf\'ericos ($\nu_\mu+\bar\nu_\mu$) para un escenario dependiente de la rigidez con $\Delta\alpha=2/3$
(que corresponde al espectro de rayos c\'osmicos de la fig.\ref{f1P5}).
Se muestran los flujos convencionales (horizontales y verticales), prompt charm/GBW y prompt charm/CTEQ3, como tambi\'en 
los flujos de neutrino totales en las direcciones horizontal y vertical para el caso prompt charm/GBW.
En este \'ultimo caso, se indican tambi\'en los flujos de neutrinos convencional, charm y total producidos por
la componente extragal\'actica de rayos c\'osmicos.} 
\label{f2P5}
\end{figure}

\begin{figure}[t]
\centerline{{\epsfxsize=4.5truein\epsfysize=3.1truein\epsffile{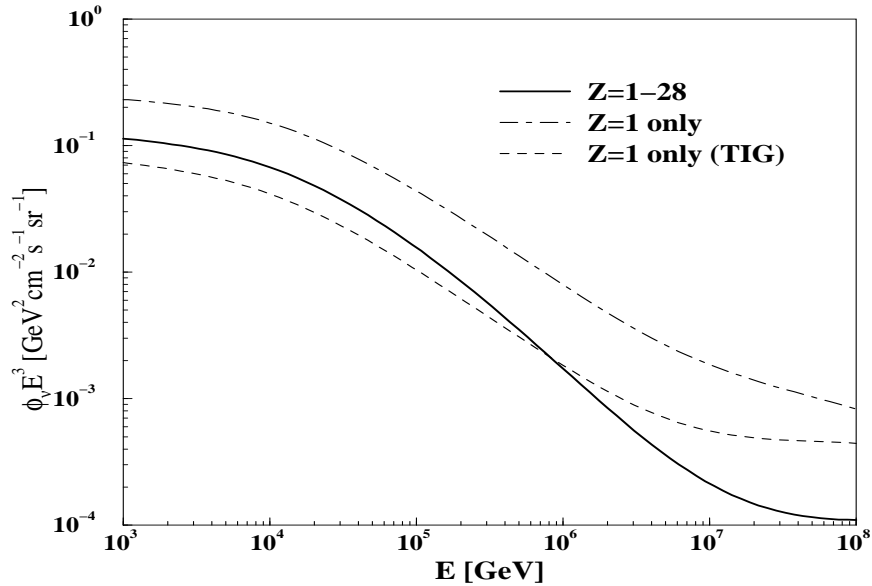}}}
\caption{Comparaci\'on del flujo total horizontal atmosf\'erico ($\nu_\mu+\bar\nu_\mu$) 
para prompt charm/GBW producido por rayos c\'osmicos bajo diferentes suposiciones con respecto al espectro y a la 
composici\'on: el escenario dependiente de la rigidez con $\Delta\alpha=2/3$, que considera la contribuci\'on de las
componentes con carga $1\leq Z\leq 28$; el mismo espectro total de rayos c\'osmicos, pero asumiendo que consiste s\'olo de
protones; y el espectro de rayos c\'osmicos usado en \cite{thu96,ing96} (TIG), que tambi\'en tiene en cuenta s\'olo a los
protones, pero adem\'as con una normalizaci\'on m\'as baja.}
\label{f3P5}
\end{figure}

La fig.\ref{f2P5} muestra los flujos de neutrinos atmosf\'ericos que corresponden al escenario dependiente de la rigidez
con $\Delta\alpha=2/3$, y con las f\'ormulas de parametrizaci\'on y la elecci\'on de par\'ametros dadas en la Secci\'on 3.2.1.
La figura muestra la contribuci\'on de diferentes componentes y para diversos casos, a saber los flujos convencionales 
(horizontales y verticales), prompt charm/GBW y prompt charm/CTEQ3. Para el flujo prompt charm calculado mediante el 
modelo GBW, la figura tambi\'en muestra los flujos de neutrino totales en las direcciones horizontal y vertical. 
Finalmente, en el \'ultimo caso tambi\'en se indican los flujos de neutrinos convencional, charm y total producidos por
la componente extragal\'actica de rayos c\'osmicos. 
De la ec.~(\ref{crnucleon}), vemos que el espectro de nucleones que resulta del espectro de rayos c\'osmicos es sensible
a la composici\'on asumida; en consecuencia, tambi\'en el flujo de neutrinos atmosf\'ericos 
depende de la composici\'on de los rayos c\'osmicos. En \cite{thu96,ing96} se asume, por simplicidad, que la componente de rayos c\'osmicos 
dominante consiste s\'olo de protones, con un cambio de \'{\i}ndice espectral $\Delta\alpha=0.3$ que da cuenta de la 
rodilla del espectro. 
Para observar el efecto de la composici\'on de rayos c\'osmicos en los flujos de neutrinos, la fig.\ref{f3P5} compara el flujo
atmosf\'erico horizontal total (con la componente prompt charm calculada con el modelo GBW) seg\'un se obtiene para diferentes 
suposiciones sobre el espectro de rayos c\'osmicos, a saber, el escenario dependiente de la rigidez con $\Delta\alpha=2/3$
(que corresponde a los resultados dados ya en la fig.\ref{f2P5}), el mismo espectro total pero asumiendo que consiste s\'olo
de protones, y el espectro de rayos c\'osmicos usado en \cite{thu96,ing96} (donde se asume que el espectro est\'a 
constitu\'{\i}do s\'olo por protones, y que tiene un \'unico cambio de \'{\i}ndice $\Delta\alpha=0.3$ para 
$E_r=5\times 10^{15}$eV).     
En este \'ultimo caso, el espectro de rayos c\'osmicos tiene una normalizaci\'on menor, en un factor de $\sim 2.6$, 
comparado con los anteriores, y est\'a algo suprimido en comparaci\'on con las observaciones
actuales. Comparando los resultados para el mismo espectro total de rayos c\'osmicos, el flujo
de neutrinos correspondiente a la composici\'on de diferentes especies nucleares resulta menor que el 
producido por un espectro de rayos c\'osmicos formado s\'olo por protones; adem\'as, puede verse que este efecto es m\'as 
significativo a energ\'{\i}as m\'as altas. La ec.~(\ref{crnucleon}) muestra que la componente nuclear  
de carga $Z$ y masa $A$ da una contribuci\'on al flujo de nucleones que est\'a suprimida por un factor  
$A^{2-\alpha_Z}\sim A^{-0.7}$ a bajas energ\'{\i}as, mientras que aparece suprimido por un factor 
$A^{2-\alpha_Z-\Delta\alpha}\sim A^{-1.4}$ a energ\'{\i}as altas. Este efecto no fue tratado adecuadamente en \cite{mar03}, 
donde la supresi\'on fue considerada como dada por un factor $A^{-2}$; por otra parte, el impacto del cambio en la 
composici\'on en la regi\'on de la rodilla no fue discutido antes en la literatura. La fuerte supresi\'on del flujo
de neutrinos producido por las componentes m\'as pesadas implica que las componentes livianas (H y He) son responsables
de una gran fracci\'on de los neutrinos a\'un por encima de su rodilla, de modo que el cambio en la pendiente de las 
componentes individuales aparece reflejado tambi\'en en el cambio de pendiente de los flujos de neutrinos.    
 
Finalmente, tambi\'en es de inter\'es considerar los flujos de muones atmosf\'ericos producidos por rayos c\'osmicos. 
Por ejemplo, ha sido sugerido recientemente \cite{gel03} que la observaci\'on (mediante telescopios de neutrinos) 
de muones atmosf\'ericos propag\'andose hacia abajo podr\'{\i}a proveer una medida indirecta del flujo de neutrinos 
atmosf\'ericos prompt, de modo que \'esto podr\'{\i}a ser usado para confrontar las predicciones NLO de pQCD. En efecto, 
debido a la cinem\'atica del decaimiento semilept\'onico de part\'{\i}culas charm, el flujo de muones prompt 
coincide con el flujo de neutrinos prompt dentro de un $\sim 10\%$, independientemente de la energ\'{\i}a y del modelo 
usado para tratar la producci\'on de charm atmosf\'erico \cite{gel03}. Por otro lado, el flujo de muones 
atmosf\'ericos convencionales es alrededor de un factor $\sim 5$ mayor que el flujo de neutrinos muon convencionales 
dentro del rango de energ\'{\i}a de inter\'es en este trabajo, y exhibe aproximadamente la misma dependencia con la 
energ\'{\i}a \cite{gel03}. Efectos completamente an\'alogos (a los descriptos aqu\'{\i} para los flujos de neutrinos)
deber\'{\i}an esperarse para los muones atmosf\'ericos producidos por los rayos c\'osmicos que llegan a la Tierra.    

\section{El canal de cascadas hadr\'onicas y electromagn\'eticas: neutrinos electr\'on}

\subsection{El sabor de los neutrinos}

\begin{figure}[t]
\centerline{{\epsfxsize=4.5truein\epsfysize=3.1truein\epsffile{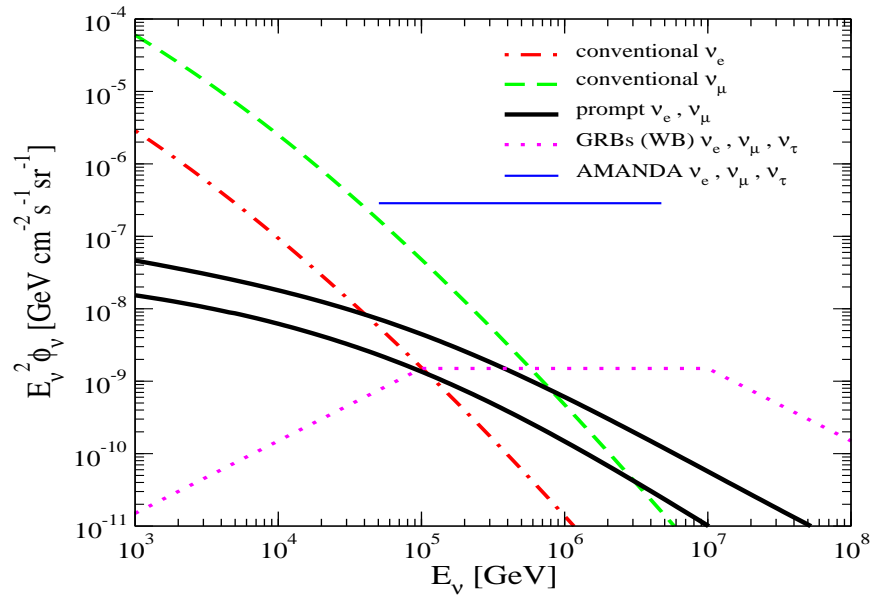}}}
\caption{Principales componentes del espectro de neutrinos de altas energ\'{\i}as, 
discriminadas por sabor. El flujo atmosf\'erico convencional est\'a promediado sobre el \'angulo zenital. 
Se incluye la predicci\'on de Waxman-Bahcall (WB) para el flujo de GRBs \cite{wax99}, 
y el l\'{\i}mite m\'as reciente de AMANDA \cite{ack04}.}
\label{f1P7}
\end{figure}   

Comencemos por reconsiderar las componentes de neutrinos de altas energ\'{\i}as, 
incluyendo ahora las contribuciones relevantes de todos los sabores de neutrino. 
De aqu\'{\i} en adelante, nos referiremos a los flujos de neutrinos y antineutrinos 
discriminados por sabor. La fig.\ref{f1P7} muestra las principales componentes, 
y en la Tabla 3.1 se identifican sus caracter\'{\i}sticas m\'as salientes.  
 
La componente atmosf\'erica convencional de neutrinos electr\'on es menor, en m\'as de un orden 
de magnitud, que la correspondiente a los neutrinos mu\'on; su principal contribuci\'on proviene
del decaimiento de kaones. Por otra parte, una componente de neutrinos tau s\'olo
puede aparecer, en el flujo atmosf\'erico convencional, a trav\'es de oscilaciones de sabor.
Debido a que este mecanismo est\'a muy suprimido a altas energ\'{\i}as, aqu\'{\i} podemos
ignorar esta contribuci\'on (a modo de ilustraci\'on, ver la fig. 7 de \cite{mar03}). 

Con respecto a la componente atmosf\'erica prompt, las tasas de producci\'on de 
neutrinos electr\'on y mu\'on en el decaimiento de los mesones charm son muy similares. 
En cambio, el flujo prompt de neutrinos tau es alrededor de un orden de magnitud menor, y aqu\'{\i} ser\'a ignorado.
Como en la Secci\'on precedente, la incerteza en la extrapolaci\'on de funciones de distribuci\'on
part\'onica para $x$ peque\~no se ilustra mediante el uso de dos funciones de estructura 
diferentes (a saber, CTEQ3 y el modelo GBW).  

El modelo de rayos c\'osmicos empleado en el c\'alculo de las componentes atmosf\'ericas es, como
en la Secci\'on anterior, un escenario dependiente de la rigidez con $\Delta\alpha=2/3$. M\'as
adelante, volveremos a considerar los efectos de variar la composici\'on de los rayos c\'osmicos. 
Debe notarse que, mientras que los neutrinos atmosf\'ericos prompt son esencialmente isotr\'opicos, 
los neutrinos atmosf\'ericos convencionales tienen una marcada dependencia zenital, con su m\'aximo
en la direcci\'on del horizonte; aqu\'{\i} presentamos los flujos convencionales promediados sobre
el hemisferio superior. 

En la fig.\ref{f1P7} tambi\'en mostramos el l\'{\i}mite m\'as reciente de AMANDA sobre el flujo de 
neutrinos de altas energ\'{\i}as, obtenido de sus an\'alisis de cascadas \cite{kow03,ack04}. 
Algunos de los 
c\'alculos previos del flujo de neutrinos atmosf\'ericos prompt son hasta unos 2 \'ordenes de magnitud
mayores que los flujos que consideramos aqu\'{\i}. Aunque probablemente no sean realistas, 
a\'un estos grandes flujos ser\'{\i}an consistentes con el l\'{\i}mite actual de AMANDA.    
Aqu\'{\i}, en cambio, asumimos el caso menos favorable (pero m\'as realista) de un flujo prompt peque\~no.   
Por otra parte, tambi\'en consideramos un flujo extragal\'actico peque\~no, dado por el modelo de 
Waxman y Bahcall de {\it Gamma Ray Bursts} (GRBs) \cite{wax99,ahr04}. 
Para una fuente astrof\'{\i}sica de neutrinos gen\'erica, 
se espera que la raz\'on de los flujos de neutrinos producidos sea $1:2:0$, transformada por las oscilaciones
de neutrinos, durante la propagaci\'on, en $1:1:1$ (aunque no puede descartarse que la existencia de nueva f\'{\i}sica
en el sector del neutrino pueda alterar tanto los flujos como los cocientes de 
sabor \cite{bea03a,bea03b,bare03,ker03,bea04a,bea04b}). 
Si los flujos reales son mayores a los considerados aqu\'{\i}, entonces la t\'ecnica que proponemos es a\'un m\'as
f\'acil de implementar.    

\begin{table}[t]
\begin{center}
\caption{Breve resumen de las caracter\'{\i}sticas esenciales de las principales componentes 
del flujo de neutrinos. Los espectros correspondientes aparecen en la fig.\ref{f1P7}.} 
\medskip
\begin{tabular}{|l|c|l|}
\hline
Flujo de neutrinos & Sabores ($ \nu_e : \nu_\mu : \nu_\tau$)
& Dependencia angular \\
\hline
atmosf\'erico convencional & $\frac{1}{20} : 1 : 0$ & m\'aximo hacia el horizonte \\
atmosf\'erico prompt & $1 : 1 : \frac{1}{10}$ & isotr\'opico \\
gal\'actico & $1 : 1 : 1$ & m\'aximo hacia el centro gal\'actico \\
extragal\'actico & $1 : 1 : 1$ & isotr\'opico; fuentes puntuales/transitorias \\
\hline
\end{tabular}
\end{center}
\label{tabla1P7}
\end{table}

\subsection{Los espectros detectados}

En la fig.~\ref{f1P7}, las componentes prompt y extragal\'actica emergen 
sobre el flujo de neutrinos atmosf\'ericos convencionales, que constituye el fondo 
dominante a bajas energ\'{\i}as, reci\'en a energ\'{\i}as por encima de  
$10^6$ GeV. Para ser precisos, \'esto ocurre para el espectro de $\nu_\mu$ 
y el canal de corriente cargada correspondiente, basado en la detecci\'on de trazas de
muones de largo alcance. Si el espectro de $\nu_e$ pudiera aislarse, 
entonces el cambio del espectro podr\'{\i}a ocurrir a energ\'{\i}as menores en alrededor de un orden de magnitud, 
donde los flujos son mucho mayores y facilitan la detecci\'on. Nuestra estrategia, entonces, es la de 
reducir el fondo de neutrinos atmosf\'ericos convencionales excluyendo los 
eventos de $\nu_\mu$ de corriente cargada (caracterizados por las trazas 
lept\'onicas), y concentr\'andonos en los eventos de $\nu_e$ de corriente cargada, 
que generan lluvias (o cascadas) en el detector. Como muestran la fig.~\ref{f1P7} 
y la Tabla 3.1, las se\~nales tienen flujos de $\nu_e$ y $\nu_\mu$ similares, 
mientras que el fondo debido a los neutrinos atmosf\'ericos convencionales     
tiene un contenido de $\nu_e$ comparativamente menor. 
A\'un cuando tambi\'en existe una contribuci\'on de $\nu_\mu$ atmosf\'ericos convencionales
al canal de las cascadas, debido a interacciones mediadas por la corriente neutra,
veremos que su impacto es reducido.

En la interacci\'on de un neutrino con un nucle\'on, la energ\'{\i}a del neutrino 
$E_\nu$ se comparte entre el quark saliente, al que le corresponde una fracci\'on $y$,
y el lept\'on saliente, que se lleva una fracci\'on $1-y$. La secciones eficaces diferenciales
para interacciones de corriente cargada y neutra tienen su m\'aximo en  
$y=0$. En una interacci\'on $\nu_e$ de corriente cargada, el quark inicia una lluvia hadr\'onica 
de energ\'{\i}a $\simeq y E_\nu$, y el electr\'on una lluvia electromagn\'etica de energ\'{\i}a 
$\simeq (1-y)E_\nu$, de modo que la energ\'{\i}a visible total resulta $E_{vis} \simeq E_\nu$  
(asumiendo que las lluvias hadr\'onica y electromagn\'etica son indistinguibles en el detector). 
En una interacci\'on de corriente neutra, $E_{vis}$ es menor en un factor 
$\langle y\rangle\simeq 0.4-0.3$ (decreciente con el aumento de la energ\'{\i}a) \cite{gan98}.
Adem\'as, las secciones eficaces totales para la corriente neutra son menores 
que las correspondientes a la corriente cargada, 
$\sigma_{NC}/\sigma_{CC}\simeq 0.4$~\cite{gan98}.  
Tomando en cuenta que el espectro de neutrinos atmosf\'ericos convencionales decrece muy r\'apidamente
(con un \'{\i}ndice espectral $\gamma \sim 3-3.7$), el flujo de lluvias originadas en interacciones
de corriente neutra de $\nu_\mu$ queda suprimido por un factor 
$\sim\langle y\rangle^{(\gamma-1)}\times\sigma_{NC}/\sigma_{CC}$, es decir, aproximadamente un orden de 
magnitud en relaci\'on con el flujo de lluvias originadas en interacciones de corriente cargada de $\nu_e$.

\begin{figure}[t]
\begin{center}
\includegraphics[width=9cm,clip]{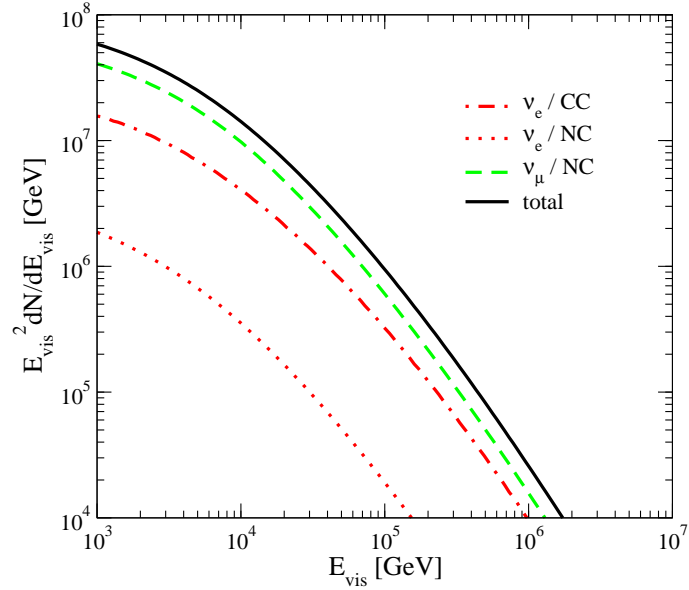}
\caption{\label{f2P7} Flujos diferenciales de cascadas en funci\'on de la energ\'{\i}a visible $E_{vis}$, 
correspondientes a un detector de 1 km$^3$ de tama\~no luego de 10 a\~nos de mediciones, 
utilizando solo neutrinos propag\'andose hacia abajo. Los flujos incidentes sobre el detector son
los que aparecen en la fig.\ref{f1P7}. Aqu\'{\i} s\'olo se muestran 
las componentes del espectro de neutrinos convencionales atmosf\'ericos, 
con la contribuci\'on por separado de las interacciones de corriente cargada (CC)
y corriente neutra (NC). La importancia relativa del canal $\nu_e$ CC crece con respecto al canal $\nu_\mu$ NC
a mayores energ\'{\i}as, debido al decrecimiento de $\langle y\rangle$.} 
\end{center}
\end{figure}

En consecuencia, en el espectro detectable de lluvias producidas por neutrinos
atmosf\'ericos convencionales, las contribuciones de $\nu_e$ y $\nu_\mu$ son
comparables: la diferencia en el flujo (ver la fig.\ref{f1P7}) se compensa con la 
diferencia en la detectabilidad.  Nuestros resultados para los neutrinos atmosf\'ericos 
convencionales se muestran en la fig.\ref{f2P7}. Como hemos mencionado, 
estamos excluyendo eventos de $\nu_\mu$ de corriente cargada, que pueden ser 
reconocidos mediante trazas de muones de largo alcance. Los espectros que
muestra la figura se calcularon convolucionando los
espectros de neutrinos asumidos con la secci\'on eficaz diferencial (promediada 
entre neutrinos y antineutrinos) \cite{gan98}.  La fig.\ref{f2P7} muestra 
que las t\'ecnicas que describimos aqu\'{\i} pueden reducir significativamente 
el fondo de neutrinos atmosf\'ericos convencionales. 

Dado que la contribuci\'on de $\nu_e$ y $\nu_\mu$ a los flujos de
neutrinos atmosf\'ericos prompt y extragal\'acticos es comparable,  
el flujo de las cascadas correspondientes quedar\'a 
dominado por las interacciones de corriente cargada de $\nu_e$. 
Aunque aqu\'{\i} incluimos las interacciones de corriente neutra de todos los sabores
relevantes, estas contribuciones podr\'{\i}an ser ignoradas; por ejemplo, 
\'esto puede verse en la fig.\ref{f2P7}, comparando las contribuciones 
de las corrientes neutras y cargadas de $\nu_e$.  
 
Hasta aqu\'{\i}, no hemos mencionado las interacciones de $\nu_\tau$, en el caso en que 
su contribuci\'on sea relevante (ver la tabla 3.1). A energ\'{\i}as 
por debajo de $E_\nu \simeq 5 \times10^6$~GeV, su interacci\'on de corriente cargada
produce solamente lluvias (con $E_{vis}\simeq E_\nu$), debido a la corta vida media 
del lept\'on tau. Por encima de esa energ\'{\i}a, la longitud de la traza lept\'onica es
suficiente para ser identificada y separada de la lluvia; sin embargo, a energ\'{\i}as 
tan altas, los flujos son muy peque\~nos, de modo que 
la separaci\'on de esos eventos es esencialmente irrelevante. Cuando $\nu_\tau$ 
est\'a presente en el flujo, inclu\'{\i}mos sus contribuciones de corriente cargada y neutra
en el espectro de lluvias detectadas. De todos modos, dado que la fracci\'on
de $\nu_\tau$ en el flujo de neutrinos atmosf\'ericos prompt es muy peque\~na, la identificaci\'on
directa de cualquier evento de $\nu_\tau$ ser\'{\i}a un firme indicador de una se\~nal 
extragal\'actica. Una descripci\'on m\'as detallada de las caracter\'{\i}sticas de los diferentes
tipos de eventos en un detector de neutrinos, y de su detectabilidad relativa, se brinda en   
\cite{bea03a}.

\begin{figure}[t]
\begin{center}
\includegraphics[width=9cm,clip]{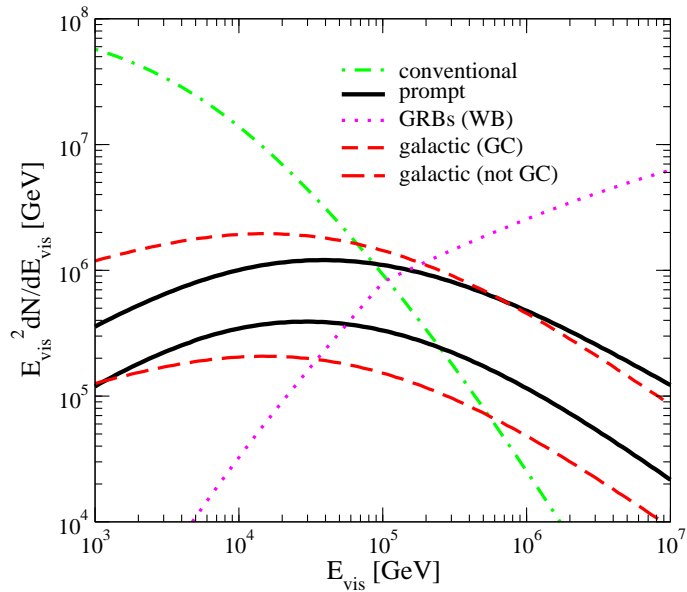}
\caption{\label{f3P7} Espectros diferenciales de cascadas en funci\'on de la energ\'{\i}a visible  
$E_{vis}$, para un detector de tama\~no km$^3$ luego de 10 a\~nos de operaci\'on,
usando s\'olo neutrinos propag\'andose hacia abajo, y los espectros de la fig.\ref{f1P7}. 
La definici\'on de las componentes gal\'acticas, y la normalizaci\'on de los flujos, se explican en el texto.} 
\end{center}
\end{figure}

La fig.\ref{f3P7} muestra los espectros de lluvias detectables correspondientes a los 
flujos de la fig.\ref{f1P7}. La energ\'{\i}a a la que la se\~nal prompt atmosf\'erica, o bien 
la extragal\'actica, podr\'{\i}a emerger del fondo de neutrinos atmosf\'ericos convencionales, 
es ahora un orden de magnitud menor. \'Esto no hubiera sido evidente si, en cambio, 
hubi\'eramos presentado el flujo (o el producto del flujo por la secci\'on eficaz) 
en funci\'on de la energ\'{\i}a del neutrino. 
Hooper et al.~\cite{hoo03} tambi\'en propusieron la medici\'on de los neutrinos atmosf\'ericos prompt y 
extragal\'acticos por medio de la detecci\'on de lluvias. 
Sin embargo, existen ciertas diferencias importantes entre su propuesta y la nuestra. 
Aqu\'{\i} asumimos que todos los eventos de 
$\nu_\mu$ de corriente cargada pueden ser exclu\'{\i}dos, identificando las trazas mu\'onicas de largo alcance;
por el contrario, en \cite{hoo03} se incluyen todos los eventos en que la interacci\'on ocurre en el volumen
del detector. Por otra parte, nosotros calculamos los espectros en t\'erminos de la energ\'{\i}a visible, 
en lugar de la energ\'{\i}a del neutrino; \'esto tiene un gran impacto en la reducci\'on del fondo 
de neutrinos atmosf\'ericos convencionales. As\'{\i}, tomar en cuenta estos efectos nos permite considerar 
un caso m\'as realista, en el que los flujos asumidos para los neutrinos atmosf\'ericos prompt y extragal\'acticos  
son sustancialmente menores a los estudiados en \cite{hoo03}.  

El canal de las lluvias no provee informaci\'on espec\'{\i}fica acerca del sabor de los neutrinos, 
y no puede distinguir entre eventos de corriente cargada y neutra. De todos modos, \'esto no representa una 
gran desventaja, y est\'a compensado por la mayor fidelidad que se consigue entre la energ\'{\i}a del neutrino y la
visible, un aspecto esencial para resolver un posible quiebre en el espectro. La resoluci\'on angular es 
moderada ($\simeq 20^\circ$, comparada con la resoluci\'on $\simeq 1^\circ$ para el canal de $\nu_\mu$
de corriente cargada), pero es perfectamente adecuada para un flujo isotr\'opico. 

Una de las desventajas del canal de detecci\'on de lluvias es que los muones atmosf\'ericos pueden producir 
un fondo significativo cuando, pasando cerca del detector, inician una lluvia a partir de 
un evento de {\it bremsstrahlung}. En efecto, el l\'{\i}mite actual de lluvias de AMANDA est\'a dado, precisamente,
por el fondo producido por este tipo de eventos \cite{kow03,ack04}.  
Sin embargo, dado que \'este es un efecto de superficie, el tama\~no (mucho mayor) de IceCube deber\'{\i}a 
permitir una reducci\'on sustancial de este fondo, conservando a la vez un volumen fiducial suficientemente
grande. Una exclusi\'on similar de la regi\'on externa del detector ser\'a tambi\'en necesaria, para eliminar 
los eventos de $\nu_\mu$ de corriente cargada en los que la traza del mu\'on escapa a la detecci\'on (de modo
que s\'olo queda registrada la lluvia hadr\'onica). Naturalmente, 
estos cortes reducir\'an la exposici\'on que hemos asumido aqu\'{\i}. 
Para evitar los efectos de absorci\'on en la Tierra, hemos considerado solamente neutrinos propag\'andose hacia abajo. 
Los neutrinos que atraviesan un di\'ametro terrestre completo son absorbidos para energ\'{\i}as alrededor 
de $4 \times 10^4$ GeV; para distancias menores, la energ\'{\i}a de absorci\'on es significativamente mayor 
(por ejemplo, ver la fig.2 de \cite{lab04}). Luego, la exposici\'on en el rango relevante de energ\'{\i}as podr\'{\i}a 
incrementarse tomando tambi\'en en cuenta una fracci\'on considerable de los eventos de propagaci\'on hacia arriba. 

Aunque el flujo de neutrinos atmosf\'ericos convencionales tiene su m\'aximo hacia el horizonte, aqu\'{\i} ha sido promediado 
sobre todo el hemisferio superior. Eliminando los eventos cercanos al horizonte, podr\'{\i}amos obtener se\~nales 
(no convencionales) identificables a menores energ\'{\i}as. 
Un estudio m\'as completo de la sensibilidad y la capacidad de separar 
las componentes del flujo de neutrinos, usando, por ejemplo, las t\'ecnicas Monte Carlo desarrolladas
en \cite{kow03}, ser\'{\i}a de gran inter\'es. 

\subsection{El flujo de neutrinos gal\'acticos}

Hasta aqu\'{\i}, hemos discutido los flujos de neutrinos producidos en la atm\'osfera o en fuentes extragal\'acticas, 
omitiendo el flujo gal\'actico difuso que resulta del decaimiento de piones y muones, producidos en la interacci\'on
de los rayos c\'osmicos con el medio interestelar (ISM) \cite{ste79,ber93,dom93,ing96,ath03} (ignoraremos un posible 
flujo de neutrinos originado en una fuente puntual en el centro gal\'actico). 

El flujo gal\'actico difuso puede ser determinado siguiendo el mismo procedimiento que ya hemos descripto, en la
Secci\'on precedente, para la producci\'on de los neutrinos atmosf\'ericos.  
Por simplicidad, asumiremos que los rayos c\'osmicos est\'an distribu\'{\i}dos homog\'eneamente dentro del disco 
gal\'actico. Las part\'{\i}culas ultrarrelativistas interact\'uan con el gas ambiente del ISM, que constituye 
un plasma no relativista de muy baja densidad, formado fundamentalmente por hidr\'ogeno at\'omico y molecular.
Debemos tener en cuenta que, debido a la densidad extremadamente baja del ISM, 
la interacci\'on de las part\'{\i}culas secundarias producidas en las interacciones nucle\'on-nucle\'on son 
completamente despreciables; en consecuencia, todas las part\'{\i}culas 
secundarias decaen mucho antes de interactuar, independientemente de la energ\'{\i}a considerada. As\'{\i}, la 
principal contribuci\'on al flujo de neutrinos difusos producido en la galaxia proviene del decaimiento de piones y muones; 
las contribuciones adicionales que provienen de la producci\'on y el decaimiento de mesones m\'as pesados 
pueden ser despreciadas. 

Consideremos un flujo inicial de nucleones dado por $\phi_N(E,X=0)=\phi_{0N}E^{-\gamma}$ 
(el flujo de neutrinos inducidos por el espectro total de rayos c\'osmicos puede obtenerse 
siguiendo el procedimiento que se us\'o en el c\'alculo de los flujos atmosf\'ericos).
En este contexto, el slant depth atmosf\'erico  
debe ser reemplazado por la columna de densidad atravesada en el ISM a lo largo de la l\'{\i}nea de visi\'on; 
debido a la densidad muy baja del ISM, es 
\footnote{En este caso, $\lambda_N$ corresponde a la longitud de interacci\'on para nucleones propag\'andose a trav\'es 
del ISM. N\'otese que $\lambda_N$, expresado en g/cm$^2$, resulta muy similar al correspondiente a la propagaci\'on de 
nucleones en el aire. En cambio, la longitud de decaimiento de los mesones $\lambda_M^d$, siendo lineal en la densidad 
de masa del medio de propagaci\'on, resulta mucho menor en el ISM que en la atm\'osfera.}    
$X\ll\lambda_N$. Entonces, de la ec.~(\ref{atmosL1}), el flujo de neutrinos mu\'on producido por el decaimiento de
piones en el ISM es
\begin{equation}
\phi_\nu(E,X)=Z_{M\nu}^\gamma Z_{NM}{{X}\over{\lambda_N}}\phi_{0N}E^{-\gamma}\ .
\label{gal}
\end{equation} 

Para calcular la contribuci\'on (al flujo de neutrinos mu\'on) 
que resulta del decaimiento de muones, es suficiente con una sencilla estimaci\'on
a partir de los resultados ya obtenidos para el decaimiento de piones. De la cinem\'atica del decaimiento, 
la fracci\'on media de energ\'{\i}a (relativa a la part\'{\i}cula primaria) en el decaimiento $\pi\to\mu+\nu_\mu$ es 
$K_\pi=0.21$ para el $\nu_\mu$ y 0.79 para el mu\'on \cite{vol80}. An\'alogamente, en el decaimiento 
$\mu\to e+\nu_e+\nu_\mu$ la fracci\'on efectiva de energ\'{\i}a para el $\nu_\mu$ resultante es 0.35 \cite{vol80}, 
es decir, una fracci\'on $K_\mu=0.28$ relativa al pi\'on original. En consecuencia, se espera que la contribuci\'on del 
decaimiento de muones sea aproximadamente un factor $(K_\mu/K_\pi)^\gamma$ mayor que el flujo de neutrinos proveniente 
del decaimiento de piones (por ejemplo, un factor de $2.1$ para un \'{\i}ndice $\gamma=2.7$ en el espectro de nucleones, 
en buen acuerdo con c\'alculos m\'as detallados \cite{ing96}).  

Un c\'alculo similar nos permite estimar el flujo de neutrinos electr\'on.  
En el decaimiento $\mu\to e+\nu_e+\nu_\mu$, la fracci\'on de energ\'{\i}a
efectiva para el $\nu_e$ es 0.3 \cite{vol80}. Entonces, el flujo de $\nu_e$ es aproximadamente un factor $(0.3/0.35)^\gamma$ 
del flujo de $\nu_\mu$ producido en el decaimiento de muones (es decir, un factor $\sim 0.7$ para un \'{\i}ndice $\gamma=2.7$ 
en el espectro de nucleones, nuevamente en buen acuerdo con \cite{ing96}). 
Finalmente, las oscilaciones de sabor de los neutrinos gal\'acticos redistribuyen los flujos  
producidos de $\nu_\mu$ y $\nu_e$ uniformemente entre los tres sabores. 

Dado que el flujo gal\'actico es lineal en la columna de densidad atravesada en el ISM a lo largo de la 
l\'{\i}nea de visi\'on, su m\'aximo est\'a en la direcci\'on del centro gal\'actico, y es, por lo tanto, 
anisotr\'opico en coordenadas gal\'acticas. En \cite{ber93}, la densidad del ISM est\'a dada en funci\'on de 
coordenadas gal\'acticas, y se asume una densidad m\'{\i}nima $n=0.087$~cm$^{-3}$. 
Considerando que el halo gal\'actico contiene esta densidad de materia m\'{\i}nima, las anisotrop\'{\i}as 
en la columna de densidad resultan despreciables, excepto en la direcci\'on del centro gal\'actico. 
Asumiendo un halo de 20 kpc de radio y una escala vertical caracter\'{\i}stica de 2 kpc, la columna de densidad 
en una direcci\'on t\'{\i}pica es $x_{\rm no\ CG} \simeq 10^{21}$ cm$^{-2}$, mientras que, en la direcci\'on del centro 
gal\'actico, es $x_{\rm CG} \simeq 10^{22}$ cm$^{-2}$.   
Consideraremos por separado la contribuci\'on de la regi\'on del centro gal\'actico 
($|b|\leq 10^\circ$ y $|l|\leq 10^\circ$, que corresponde a un \'angulo s\'olido 
$\Delta\Omega_{{\rm CG}}=0.12$~sr) y la contribuci\'on promedio de todas las dem\'as direcciones
en el hemisferio superior (que corresponde a $\Delta\Omega_{{\rm no\ CG}}=6.16$~sr).  
Esta separaci\'on es consistente con la resoluci\'on angular esperada para la detecci\'on de cascadas en IceCube.

\begin{figure}[t]
\begin{center}
\includegraphics[width=9.cm,clip]{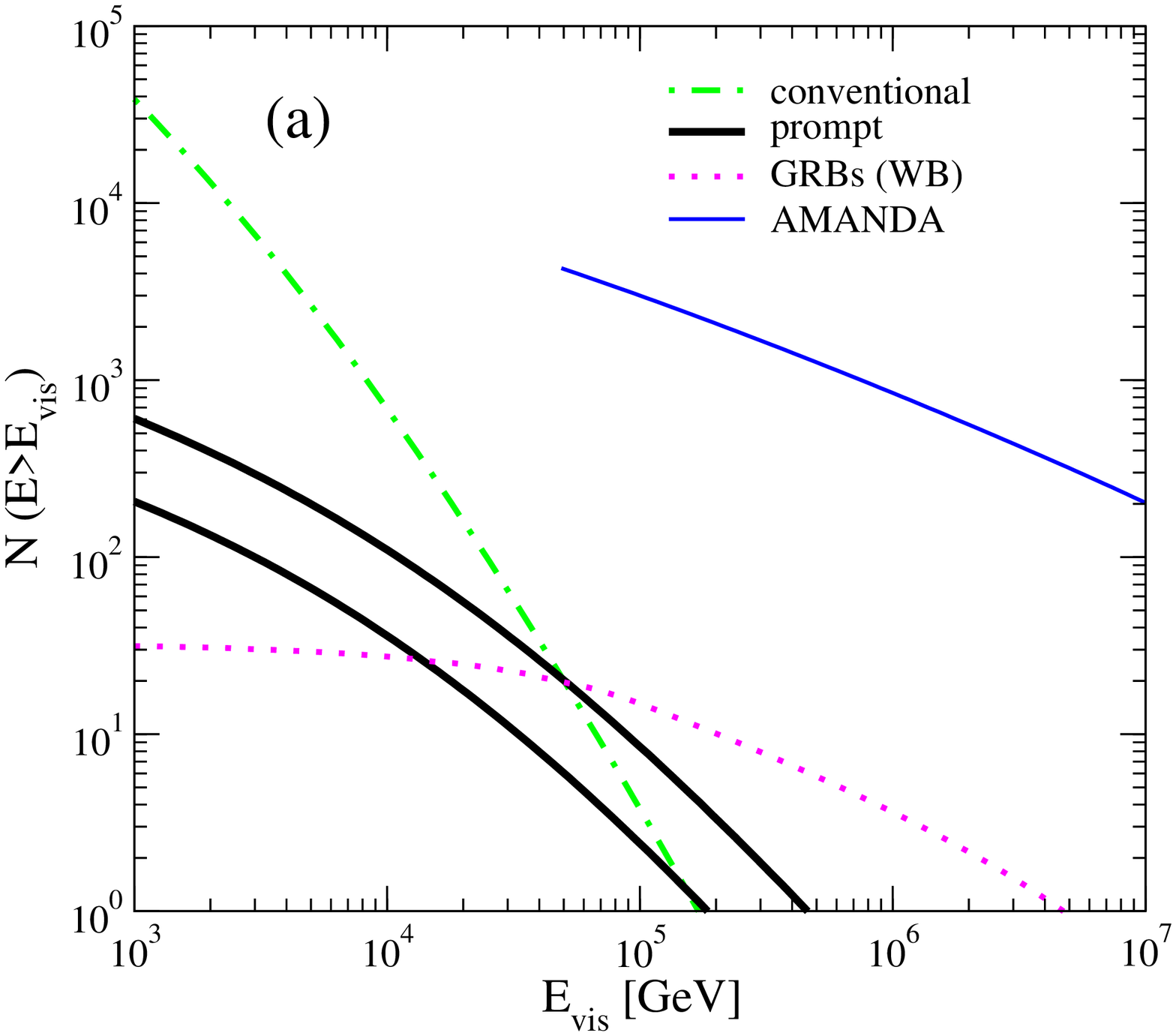}
\includegraphics[width=9.cm,clip]{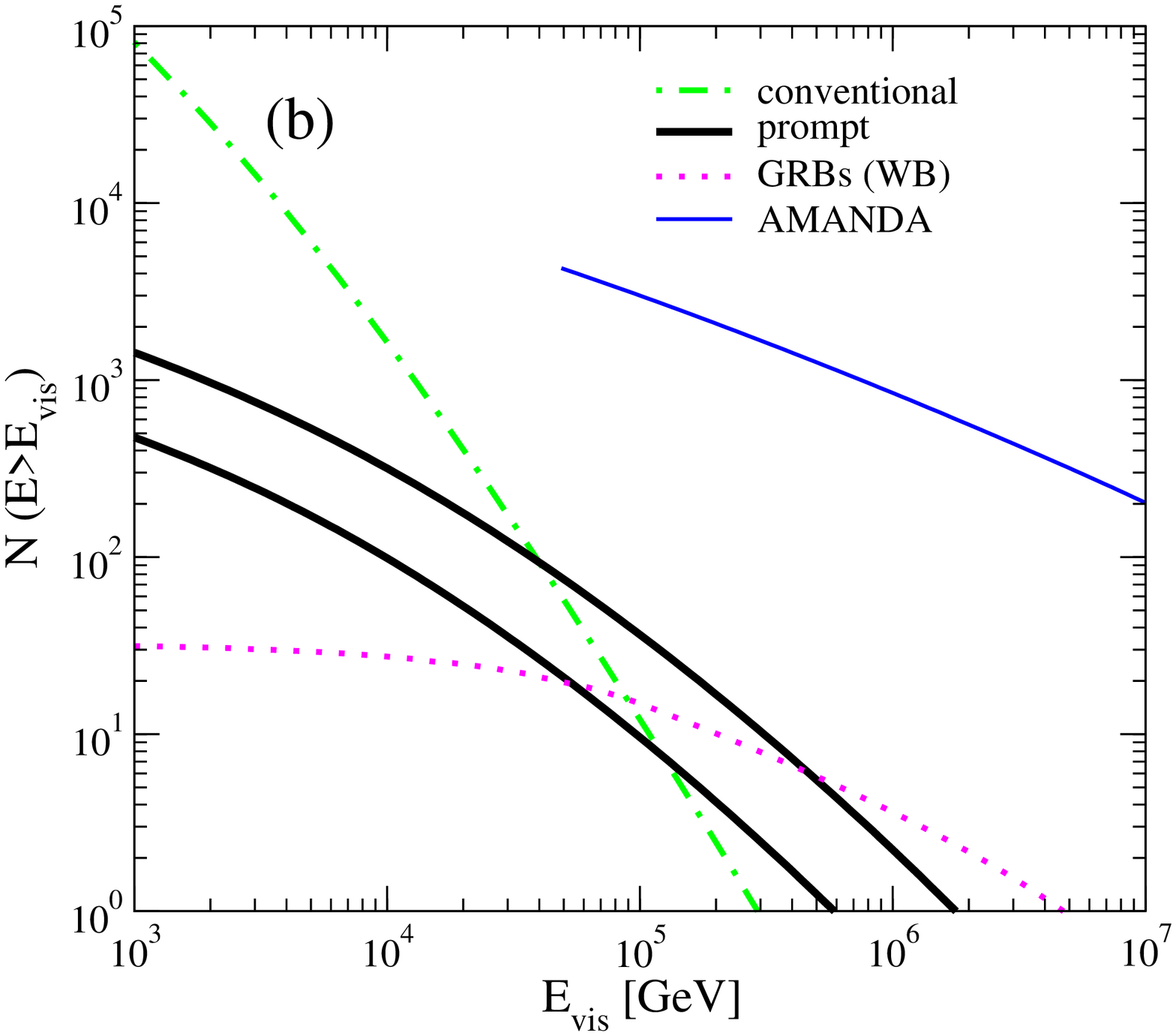}
\caption{\label{f4P7} Tasas integrales de eventos de cascada en funci\'on de la energ\'{\i}a visible 
$E_{vis}$, para un detector de tama\~no km$^3$ luego de 10 a\~nos de operaci\'on,
asumiendo diferentes composiciones de rayos c\'osmicos:  
(a) el escenario dependiente de la rigidez con $\Delta\alpha=2/3$, que considera la contribuci\'on de componentes con 
$1\leq Z\leq 28$; (b) el mismo espectro total, pero asumiendo que consiste s\'olo de protones. 
La l\'{\i}nea marcada ``AMANDA'' indica el espectro integral resultante asumiendo una ley de potencias 
$E^{-2}$, sin una energ\'{\i}a de corte superior, normalizado por el l\'{\i}mite diferencial de AMANDA.} 
\end{center}
\end{figure}

La fig.\ref{f3P7} muestra los flujos diferenciales de las cascadas correspondientes a las componentes gal\'acticas, 
comparados con los flujos de las componentes que ya hemos discutido. Excluyendo la direcci\'on del centro gal\'actico, 
el flujo gal\'actico puede despreciarse, en comparaci\'on con las otras componentes del espectro. 
En la direcci\'on del centro gal\'actico, en cambio, el flujo gal\'actico es comparable con las otras componentes. 
Sin embargo, en la fig.\ref{f3P7}, el flujo en la direcci\'on del centro gal\'actico aparece como si fuera 
isotr\'opico. Para calcular la tasa real de eventos en la direcci\'on del centro gal\'actico, \'este y los otros
flujos deben renormalizarse, a fin de tener en cuenta la reducci\'on en la aceptancia angular; 
al multiplicar todos los flujos por  
$\Delta\Omega_{\rm CG}/\Delta\Omega_{\rm no\ CG}\simeq 0.02$, 
los flujos resultantes son demasiado peque\~nos para ser observados. 

\subsection{Los efectos de la composici\'on de rayos c\'osmicos}

Las figs.\ref{f4P7}(a)-(b) muestran las tasas integrales de eventos de cascada, correspondientes a las
componentes relevantes del espectro total de neutrinos de altas energ\'{\i}as. En la fig.\ref{f4P7}(a), se
asume que todas las especies nucleares en el rango 
$1\leq Z\leq 28$ contribuyen a la composici\'on en un escenario dependiente de la rigidez; en la fig.\ref{f4P7}(b), 
se asume que el mismo espectro de rayos c\'osmicos est\'a compuesto s\'olo por protones. 
Debe notarse que la composici\'on asumida afecta a los flujos de neutrinos atmosf\'ericos, tanto convencionales como
prompt. Como en figuras anteriores, se utilizan dos curvas para las predicciones asociadas a los neutrinos 
atmosf\'ericos prompt, con el fin de indicar el rango de incerteza que resulta de usar distintas prescripciones 
para la regi\'on de $x$ peque\~no.    

Como ya hemos observado en la Secci\'on anterior, 
al considerar una composici\'on de rayos c\'osmicos en la que interviene una mezcla de especies nucleares, 
el flujo de neutrinos atmosf\'ericos tiene un impacto relativamente menor, y resulta 
en un fondo significativamente reducido para la detecci\'on de un flujo extragal\'actico difuso originado en
fuentes no resueltas. 

Si bien las tasas integrales que obtenemos son peque\~nas, 
el n\'umero de eventos obtenido en estas predicciones es realista y razonable.  
Por otra parte, el l\'{\i}mite de AMANDA permite la existencia de flujos mayores en alrededor de 2 \'ordenes de magnitud, 
que incrementar\'{\i}an las tasas de eventos detectados en un factor similar.  A\'un antes de descubrir los flujos de
neutrinos atmosf\'ericos prompt o extragal\'acticos, la t\'ecnica que describimos aqu\'{\i} permitir\'{\i}a una medici\'on 
del flujo atmosf\'erico convencional con gran estad\'{\i}stica, algo esencial para determinar confiablemente
su extrapolaci\'on a mayores energ\'{\i}as.  

El quiebre espectral entre el fondo de neutrinos atmosf\'ericos convencionales, que domina en la regi\'on 
de bajas energ\'{\i}as, y la nueva se\~nal, que domina a energ\'{\i}as altas, ocurre para una energ\'{\i}a menor, en 
alrededor de un orden de magnitud, que al considerar el canal usual de detecci\'on de trazas lept\'onicas (comparar
las figs.\ref{f4P7}(a)-(b) con la fig.\ref{f1P7}). Una vez que el quiebre en el espectro sea observado,   
diversas caracter\'{\i}sticas permitir\'an distinguir el flujo de neutrinos atmosf\'ericos prompt de un flujo
extragal\'actico; por ejemplo, el \'{\i}ndice del espectro observado, la detecci\'on de una componente de neutrinos tau, 
o la identificaci\'on de fuentes puntuales o transitorias. 

%% file: capitulo4.tex
\chapter{Fotodesintegraci\'on de rayos c\'osmicos: una explicaci\'on alternativa para la rodilla del espectro}

\begin{center}
\begin{minipage}{5.6in}
\textsl{
Como una propuesta alternativa para explicar la rodilla del espectro de rayos c\'osmicos,
en este Cap\'{\i}tulo exploramos el escenario en el que la rodilla resulta de la interacci\'on 
de los rayos c\'osmicos con un fondo de fotones
\'opticos y UV blandos que puede estar presente en el entorno de las fuentes. En este escenario, la rodilla del 
espectro refleja la energ\'{\i}a umbral asociada a los procesos de fotodesintegraci\'on nuclear y 
de producci\'on de fotomesones por protones, y la composici\'on exhibe una tendencia hacia componentes m\'as livianas 
por encima de la rodilla.}
\end{minipage}
\end{center}  

\section{El escenario de la fotodesintegraci\'on nuclear}
En el sistema en reposo de un rayo c\'osmico de energ\'{\i}a $E$ y masa $Am_p$, la energ\'{\i}a de un fot\'on de energ\'{\i}a 
$\epsilon$ aparece incrementada por el factor relativista $\gamma=E/Am_pc^2$. Teniendo en cuenta que la fotodesintegraci\'on 
nuclear tiene t\'{\i}picamente una energ\'{\i}a umbral en torno de algunos MeV (en el sistema en reposo del n\'ucleo), 
resulta que un fot\'on en el rango \'optico ($\epsilon=1-10$~eV) puede desintegrar n\'ucleos con energ\'{\i}as 
$E\geq A\times 10^{15}$~eV. En consecuencia, la presencia de un fondo de fotones \'opticos y UV blandos en el entorno de las
fuentes podr\'{\i}a explicar el origen de la rodilla del espectro de rayos c\'osmicos. Este escenario fue propuesto 
por Hillas \cite{hil79} en 1979, y luego desarrollado por Karakula y Tkaczyk \cite{kar93}. 

Aqu\'{\i} presentamos una reconsideraci\'on de este escenario, reevaluando en detalle la propagaci\'on de n\'ucleos
y protones en la regi\'on de las fuentes, e incorporando elementos adicionales que modifican las predicciones de \cite{kar93}.
En particular, en la reevaluaci\'on de las tasas de fotodesintegraci\'on nuclear incorporamos los procesos de emisi\'on 
de m\'as de un nucle\'on (que, en efecto, constituyen 
el canal dominante para n\'ucleos pesados), y mejoramos el tratamiento de la
energ\'{\i}a umbral en la resonancia dipolar gigante asociada a cada n\'ucleo estable (adoptando los recientes resultados de 
Stecker y Salamon \cite{ste99}). Tambi\'en discutimos el impacto del llamado 
``mecanismo del neutr\'on'' \cite{ber77a,ber77b},   
que permite que los neutrones escapen de la fuente sin p\'erdidas de energ\'{\i}a adicionales. Finalmente, introduciendo 
una energ\'{\i}a de corte ({\it cutoff}) inferior en la distribuci\'on de fotones de ley de potencias, 
eliminamos la supresi\'on abrupta que se observa en el espectro calculado en \cite{kar93} por encima de $10^{17}$~eV. 
Con la implementaci\'on de estas consideraciones adicionales, la columna de
densidad de energ\'{\i}a requerida para reproducir la rodilla del espectro de rayos c\'osmicos en este escenario resulta menor, 
en un orden de magnitud, que las estimadas previamente en \cite{kar93}. 
   
\section{La propagaci\'on de rayos c\'osmicos}
\subsection{La fuente y el espectro del fondo de radiaci\'on}
En este trabajo, asumiremos que la fuente emite n\'ucleos en el rango de n\'umero m\'asico $1\leq A\leq 56$.
Siendo que, por lo general, existe un \'unico is\'otopo estable para cada n\'ucleo de masa $Am_p$ en todo el rango que
va desde $^1$H hasta $^{56}$Fe, consideraremos una \'unica carga $Z$ asociada a cada valor de $A$. Asumiremos que los 
flujos diferenciales de rayos c\'osmicos emitidos por la fuente son espectros de ley de potencias, 
\begin{equation}
\phi_i^0(E)=\Phi_i^0\ E^{-\beta_i}\ \ (0\leq i\leq 55)\ ,
\end{equation}   
donde $i\equiv 56-A$. Las intensidades $\Phi_i^0$ e \'{\i}ndices espectrales $\beta_i$ se obtienen a partir de las
observaciones a bajas energ\'{\i}as, compiladas en \cite{wie98,hoe03} (para m\'as detalles, ver la Secci\'on 2.2).

Para el fondo de fotones que rodea a la fuente, estudiaremos el caso de una distribuci\'on de Planck, 
\begin{equation}
n(\epsilon)={\rho\over 2\zeta(3)(k_BT)^3}\frac{\epsilon^2}{\exp(\epsilon/k_BT)-1}\ ,
\label{planck}
\end{equation}
donde $k_B$ es la constante de Boltzmann, $T$ la temperatura absoluta y $\zeta$ la funci\'on zeta de Riemann 
(tal que $\zeta(3)\approx 1.202$), y una distribuci\'on de ley de potencias,
\begin{equation}
n(\epsilon)=\rho(\alpha-1)\left(\frac{1}{\epsilon_m^{\alpha-1}}-\frac{1}
{\epsilon_M^{\alpha-1}}\right)^{-1}\epsilon^{-\alpha}\ ,
\label{powerlaw}
\end{equation} 
donde $\epsilon_m$ y $\epsilon_M$ son los cutoffs inferior y superior, respectivamente,  
y donde $\alpha$ es el \'{\i}ndice espectral. En ambas distribuciones, 
$n(\epsilon)$ representa el espectro diferencial correspondiente a la densidad de fotones de energ\'{\i}a 
$\epsilon$, mientras que $\rho=\int{\rm d}\epsilon\ n(\epsilon)$ es el n\'umero total de fotones por unidad de volumen. 
Llamando $\rho_E$ a la densidad total de energ\'{\i}a, resulta 
\begin{equation}
\rho_E\approx 2.631k_BT\rho 
\end{equation}
para el espectro de Planck, mientras que 
\begin{equation}
\rho_E={\alpha-1\over 2-\alpha}\left(\epsilon_M^{2-\alpha}-\epsilon_m^{2-\alpha}\right)
\left({1\over\epsilon_m^{\alpha-1}}-{1\over\epsilon_M^{\alpha-1}}\right)^{-1}\rho
\end{equation}
corresponde a la distribuci\'on de ley de potencias. 
Siguiendo a \cite{kar93}, asumiremos $\alpha=1.3$. Sin embargo, a diferencia de \cite{kar93}, aqu\'{\i} adoptamos un cutoff 
inferior $\epsilon_m$ que excluye la presencia de un abundante n\'umero de fotones de baja energ\'{\i}a (infrarrojos), 
que de otro modo ser\'{\i}an dominantes en la fotodesintegraci\'on por encima de $10^{17}$~eV.        

Con respecto a la distribuci\'on espacial del fondo de radiaci\'on, consideraremos que es homog\'enea en
toda una regi\'on (de dimensi\'on lineal $L$) circundante a la fuente.          
De cualquier modo, los resultados depender\'an s\'olo de la columna de densidad de fotones total, integrada a lo largo
de la trayectoria de los rayos c\'osmicos desde su producci\'on en la fuente hasta su escape de la regi\'on del fondo de 
fotones. Despreciando efectos de confinamiento en la fuente, la propagaci\'on de los rayos c\'osmicos es esencialmente 
rectil\'{\i}nea; la columna de densidad total atravesada, expresada en t\'erminos de la densidad de energ\'{\i}a, es $\rho_EL$.  

\subsection{Propagaci\'on de n\'ucleos: fotodesintegraci\'on}
El principal mecanismo de p\'erdida de energ\'{\i}a para n\'ucleos con $E\leq 10^{18}$~eV propag\'andose en un fondo de 
radiaci\'on \'optico es el proceso de fotodesintegraci\'on. Un n\'ucleo
(de n\'umero m\'asico $A=56-i$ y factor de Lorentz $\gamma$) que se propaga a trav\'es de una distribuci\'on de fotones $n(\epsilon)$ 
(dada, por ejemplo, por las ecs.~(\ref{planck}) \'o (\ref{powerlaw})) tiene una probabilidad de fisi\'on con 
emisi\'on de $j$ nucleones dada por 
\begin{equation}
R_{ij}(E)={1\over 2\gamma^2}\int_{\epsilon'_{thr,ij}/2\gamma}^\infty{\rm d}\epsilon \ \
{n(\epsilon)\over\epsilon^{2}}\int_{\epsilon'_{thr,ij}}^{2\gamma\epsilon}
{\rm d}\epsilon'\epsilon'\sigma_{ij}(\epsilon')\ ,
\label{rij}
\end{equation} 
donde $\sigma_{ij}$ es la secci\'on eficaz de fotodesintegraci\'on correspondiente al proceso, $\epsilon$ la energ\'{\i}a del
fot\'on en el sistema del observador y $\epsilon'$ su energ\'{\i}a en el sistema en reposo del n\'ucleo. 
Para calcular $R_{ij}$, ajustamos $\sigma_{ij}$ a los par\'ametros dados en las Tablas 1 y 2 de \cite{pug76}, 
mientras que los umbrales $\epsilon'_{thr,ij}$ fueron tomados de la Tabla 1 de \cite{ste99}. Para tener en cuenta todos los 
canales de la reacci\'on, resulta \'util definir la tasa de emisi\'on efectiva,
\begin{equation}
R_{i,ef}=\sum_{j\geq 1}jR_{ij}\ .
\label{reff}
\end{equation} 

\begin{figure}[t!]
\centerline{{\epsfxsize=3.8truein \epsffile{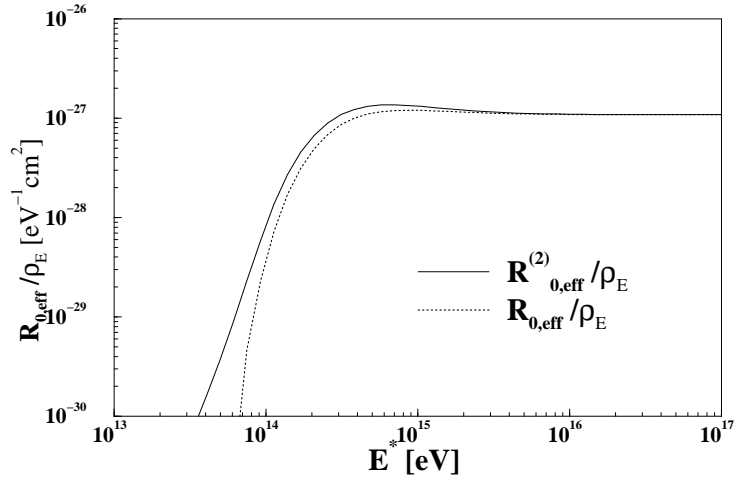}}}
\caption{Tasas de emisi\'on efectiva versus energ\'{\i}a por nucle\'on, para n\'ucleos de $^{56}$Fe propag\'andose en 
un espectro de fotones de Planck (con $k_BT=10$~eV). $R_{0,ef}^{(2)}$ fue calculado usando la energ\'{\i}a umbral fija 
$\epsilon'_{thr,ij}=2$~MeV, mientras que $R_{0,ef}$ corresponde a los nuevos resultados \cite{ste99} que involucran 
energ\'{\i}as umbrales espec\'{\i}ficas.}
\label{f1P1}
\end{figure}

A bajas energ\'{\i}as ($\epsilon'_{thr,ij}\leq\epsilon'\leq 30$~MeV), la secci\'on eficaz $\sigma_{ij}(\epsilon')$ 
est\'a dominada por la resonancia dipolar gigante, y la fotodesintegraci\'on ocurre fundamentalmente mediante la emisi\'on 
de uno o dos nucleones. En \cite{kar93,pug76} se asume una \'unica energ\'{\i}a umbral $\epsilon'_{thr,ij}=2$~MeV para todos 
los canales de la reacci\'on y todos los n\'ucleos. Sin embargo, de acuerdo a recientes consideraciones \cite{ste99}, 
la energ\'{\i}a umbral de las reacciones depende tanto del n\'ucleo como del n\'umero de nucleones emitidos. 
Incorporando los nuevos resultados, la energ\'{\i}a umbral se desplaza a valores mayores;
en la emisi\'on de un \'unico nucle\'on, la energ\'{\i}a umbral t\'{\i}pica resulta $\sim 10$~MeV, 
mientras que la correspondiente a la emisi\'on de dos nucleones es ahora $\sim 20$~MeV. 
Aunque la aparici\'on de la rodilla en este escenario es, en \'ultima instancia, un efecto de umbral, el nuevo c\'alculo,
mejorado con la inclusi\'on del cambio en los umbrales de reacci\'on, no modifica significativamente las
tasas de emisi\'on resultantes en comparaci\'on con aqu\'ellas obtenidas con la energ\'{\i}a umbral fija 
$\epsilon'_{thr,ij}=2$~MeV. \'Esto puede apreciarse en la fig.\ref{f1P1}, donde se muestran gr\'aficos de $R_{i,ef}/\rho_E$ en
funci\'on de la energ\'{\i}a por nucle\'on $E^*$ para un n\'ucleo de $^{56}$Fe ($i=0$) que se propaga a trav\'es de un 
espectro de fotones de Planck (con $k_BT=10$~eV). La tasa de emisi\'on efectiva $R_{0,ef}$  
(calculada con las energ\'{\i}as umbrales espec\'{\i}ficas) exhibe s\'olo un peque\~no desplazamiento hacia mayores energ\'{\i}as
con respecto a aqu\'ella obtenida por medio de la energ\'{\i}a umbral fija 
(denotada $R_{0,ef}^{(2)}$). La raz\'on de un efecto tan poco significativo es que la fotodesintegraci\'on est\'a dominada por
la regi\'on del pico de la resonancia gigante, situado t\'{\i}picamente alrededor de $\sim 20$~MeV para la emisi\'on de un
nucle\'on, y en torno de $\sim 26$~MeV para la emisi\'on de dos nucleones, es decir, bien por encima de las energ\'{\i}as 
umbrales.          

\begin{figure}[t!]
\centerline{{\epsfxsize=3.5truein \epsffile{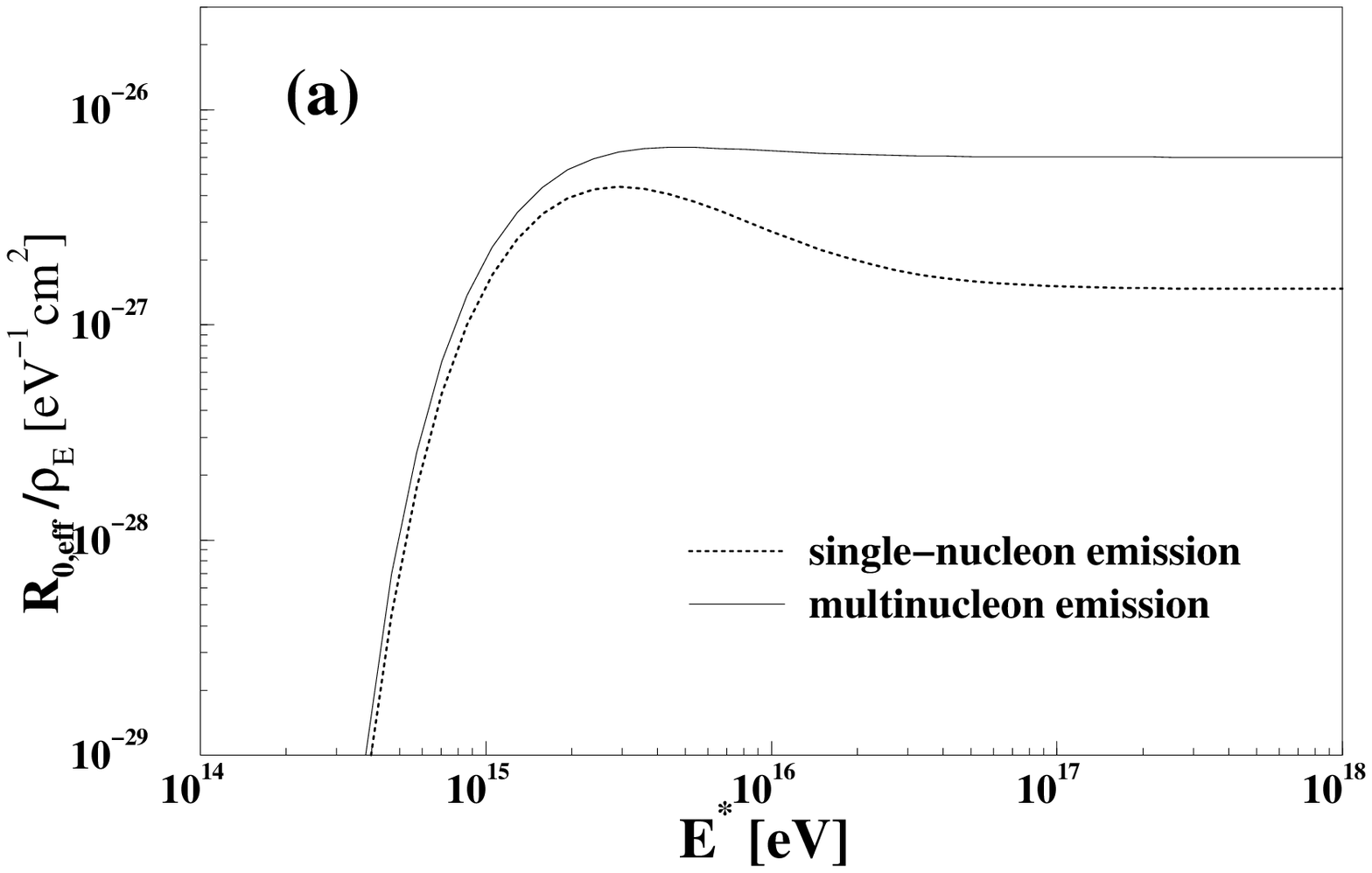}}}
\centerline{{\epsfxsize=3.5truein \epsffile{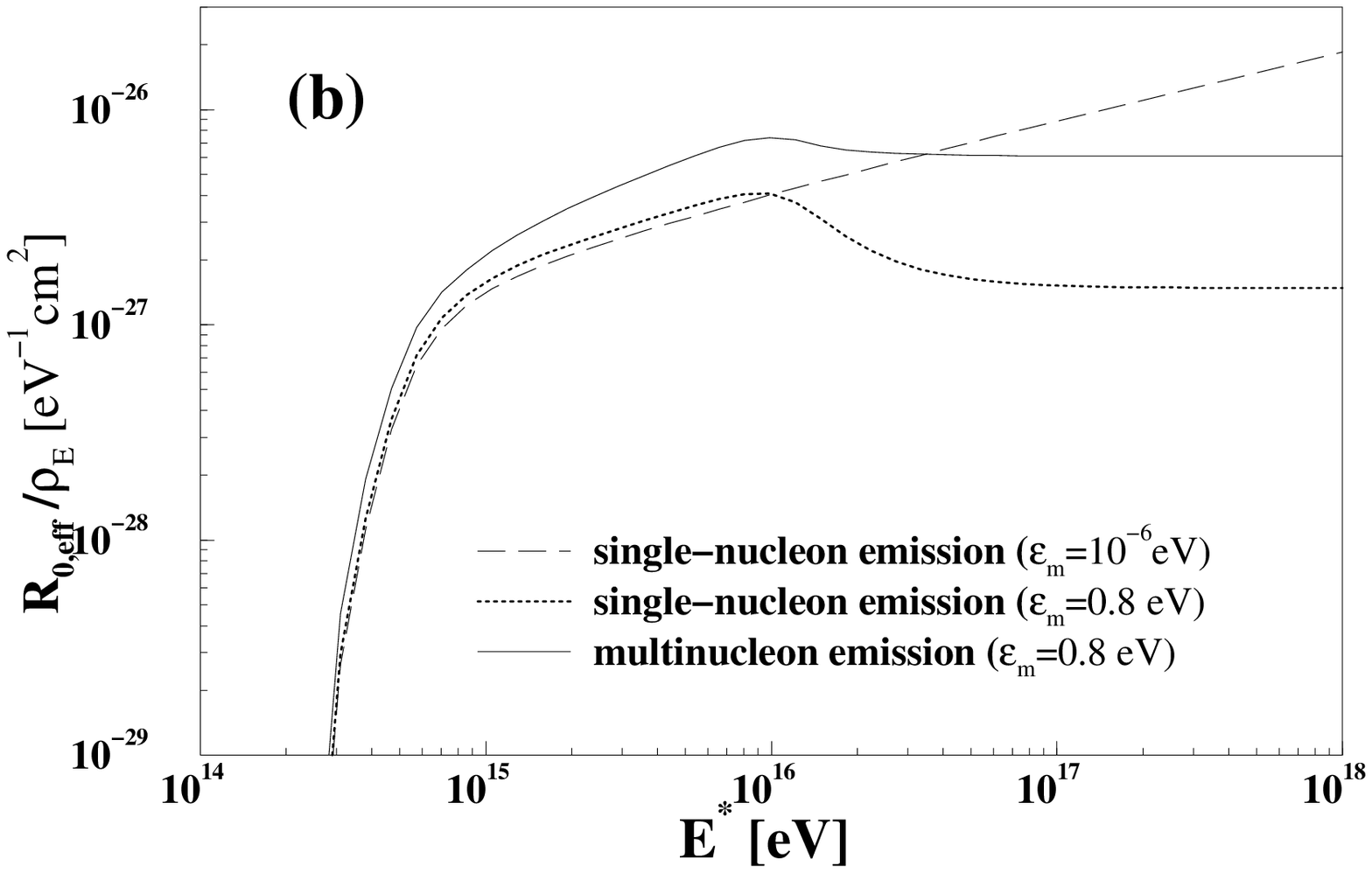}}}
\caption{Tasas de emisi\'on efectiva versus energ\'{\i}a por nucle\'on, para n\'ucleos de $^{56}$Fe,
(a) para un espectro de Planck (con $k_BT=1.8$~eV), y (b) para una distribuci\'on de ley de potencias con \'{\i}ndice 
espectral $\alpha=1.3$ y un cutoff superior $\epsilon_M=20$~eV. La tasa de emisi\'on de varios nucleones corresponde
a un cutoff inferior $\epsilon_m=0.8$~eV, mientras que las tasas de emisi\'on de un \'unico nucle\'on son para
$\epsilon_m=10^{-6}$ y 0.8~eV, como se indica.}
\label{f2P1}
\end{figure}

A mayores energ\'{\i}as, $\sigma_{ij}(\epsilon')$ es aproximadamente plana y la emisi\'on de varios nucleones adquiere una 
mayor probabilidad, torn\'andose dominante para el caso de los n\'ucleos pesados, para los cuales la probabilidad de 
emisi\'on de un \'unico nucle\'on es de s\'olo $10\%$. Para estimar la relevancia de los procesos de emisi\'on de varios 
nucleones, comparamos la tasa efectiva que resulta de las ecs. (\ref{rij}) y (\ref{reff}) con la tasa de emisi\'on que 
desprecia los procesos que involucran m\'as de un nucle\'on (es decir, la que se emplea en \cite{kar93}).      
La fig.\ref{f2P1}(a) muestra $R_{i,ef}/\rho_E$ en funci\'on de la energ\'{\i}a por nucle\'on $E^*$ para 
un n\'ucleo de $^{56}$Fe 
($i=0$) que se propaga a trav\'es de un espectro de fotones de Planck (con $k_BT=1.8$~eV). Resulta evidente que la emisi\'on
de varios nucleones es significativa, incrementando la tasa de emisi\'on a energ\'{\i}as altas por un factor $\sim 4$ con 
respecto a los resultados que involucran la emisi\'on de un \'unico nucle\'on. An\'alogamente, la fig.\ref{f2P1}(b) muestra los
resultados correspondientes a un espectro de fotones de ley de potencias con un cutoff superior $\epsilon_M=20$~eV. 
En esta figura tambi\'en puede verse el efecto de introducir un cutoff inferior no despreciable en la distribuci\'on. 
Se observa que, con un cutoff despreciable (por ejemplo, $\epsilon_m=10^{-6}$~eV) la tasa de emisi\'on crece
mon\'otonamente, debido a que la contribuci\'on dominante a la fotodesintegraci\'on viene dada, en este caso, por los fotones 
de baja energ\'{\i}a (infrarrojos), que son m\'as abundantes para mayores valores de $E^*$.
Estos fotones mantienen el predominio de la resonancia dipolar gigante a\'un a energ\'{\i}as nucleares altas, y la emisi\'on 
de varios nucleones resulta poco significativa. Sin embargo, si consideramos cutoffs inferiores
no despreciables (por ejemplo, $\epsilon_m=0.8$~eV), encontramos que, a partir de un cierto valor de $E^*$,
la resonancia dipolar gigante queda exclu\'{\i}da, y la contribuci\'on dominante a las tasas de emisi\'on es la del
plateau de altas energ\'{\i}as de $\sigma_i$. En consecuencia, a energ\'{\i}as altas,
la tasa de emisi\'on satura y adquiere tambi\'en un plateau. Comparando las dos curvas correspondientes a la tasa de
emisi\'on con cutoffs no despreciables, observamos aqu\'{\i} nuevamente que los procesos de emisi\'on de varios nucleones 
conducen a un incremento significativo (por el mismo factor $\sim 4$ que ya hab\'{\i}amos encontrado antes) en la tasa de 
emisi\'on. De esta manera, vemos que asumir un cutoff inferior en el espectro de fotones de ley de potencias debe 
conducir a resultados similares a los que resultan de considerar un espectro de Planck. En este caso, siendo que las 
tasas de emisi\'on de varios nucleones son considerablemente mayores (que las correspondientes a la emisi\'on de un \'unico
nucle\'on), es de esperar que se requiera una densidad de fotones menor a la estimada previamente en \cite{kar93} para 
producir efectos similares.   

Debe notarse que, en las figs.\ref{f1P1} y \ref{f2P1}, las tasas de emisi\'on aparecen 
graficadas en funci\'on de la energ\'{\i}a 
por nucle\'on $E^*$, en lugar de la energ\'{\i}a por n\'ucleo $E$. \'Esto es debido a que $E^*$ 
permanece constante durante la fotodesintegraci\'on y, por otra parte, porque la forma de las 
secciones eficaces de los diferentes n\'ucleos son similares, con sus respectivos m\'aximos ubicados aproximadamente en 
los mismos valores de $E^*$.            

Dado que el proceso dominante que afecta el transporte de los n\'ucleos a trav\'es del fondo de radiaci\'on es la 
fotodesintegraci\'on nuclear, asumiremos un modelo simplificado en el que los n\'ucleos se propagan en forma
rectil\'{\i}nea, despreciando otros efectos (tales como difusi\'on, convecci\'on, fragmentaci\'on nuclear (spallation),
reaceleraci\'on, etc.). Entonces, las ecuaciones de transporte que describen la propagaci\'on de los n\'ucleos resultan
\begin{equation}
\frac{\partial\phi_i(E^*,x)}{\partial x}=-\phi_i(E^*,x)\sum_{j\geq 1}R_{ij}(E^*)+
(1-\delta_{0i})\sum_{j=1}^iR_{(i-j)j}(E^*)\phi_{i-j}(E^*,x)\ ,
\end{equation}
donde $\phi_i(E^*,x)$ es el flujo diferencial correspondiente a un n\'ucleo de masa $A=56-i$ ($0\leq i\leq 54$) con
una energ\'{\i}a por nucle\'on $E^*$ a la distancia de propagaci\'on $x$ (medida desde la fuente), $R_{ij}(E^*)$ es la 
tasa de emisi\'on, y $\delta_{ij}$ es la delta de Kronecker. 
Con las simplificaciones asumidas, este sistema de 55 ecuaciones diferenciales acopladas admite soluciones 
anal\'{\i}ticas exactas; la soluci\'on (es decir, los flujos diferenciales de todas las especies nucleares a la 
distancia de propagaci\'on $L$) viene dada por
\begin{equation}
\phi_i(E^*,L)=\sum_{j=0}^ib_{ij}(E^*)\exp(-\sum_{k\geq 1}R_{jk}(E^*)L)\ ,
\label{nuc1}
\end{equation}
donde
\begin{equation}
b_{ii}=\phi_i^0(E^*)-(1-\delta_{0i})\sum_{j=0}^{i-1}b_{ij}(E^*)
\label{nuc2}
\end{equation}
y 
\begin{equation}
b_{ij}={\sum_{k=1}^{i-j}b_{(i-k)j}(E^*)R_{(i-k)k}(E^*)\over 
\sum_{k\geq 1}R_{ik}(E^*)-\sum_{k\geq 1}R_{jk}(E^*)}\ ,\ \ {\rm para}\ \ i>j\ .
\label{nuc3}
\end{equation}

Puede notarse que, dado que las tasas de emisi\'on $R_{ij}$ son lineales en la densidad de energ\'{\i}a de fotones $\rho_E$, 
los flujos resultantes dependen s\'olo de la columna de densidad $\rho_EL$, como era de esperar. N\'otese adem\'as que, dado que las
ecuaciones de transporte dependen de la energ\'{\i}a por nucle\'on $E^*$, los espectros finales de las diferentes 
componentes nucleares de rayos c\'osmicos requieren recalcular las soluciones (ecs. (\ref{nuc1})-(\ref{nuc3})) en 
t\'erminos de la energ\'{\i}a por n\'ucleo $E$.  

\subsection{Propagaci\'on de protones: producci\'on de fotomesones}

El principal mecanismo de p\'erdida de energ\'{\i}a que debe tenerse en cuenta en la propagaci\'on de los protones 
es el proceso de producci\'on de fotomesones. Deberemos distinguir entre el flujo de protones primarios 
producidos en la fuente ($\phi^0_{i=55}$) y los nucleones liberados como producto de la fotodesintegraci\'on 
nuclear ($\phi^{ps}$ y $\phi^n$ para los protones secundarios y neutrones, respectivamente). Sea $\phi_i^N(E,x)$
el flujo diferencial de nucleones de energ\'{\i}a $E$ producidos dentro de una distancia $x$ desde la fuente
por la fotodesintegraci\'on de n\'ucleos de masa $A=56-i$ y energ\'{\i}a por nucle\'on $E^*=E/A$. Entonces,
la ecuaci\'on que gobierna la evoluci\'on de los flujos es 
\begin{equation}
\frac{\partial\phi_i^N(E,x)}{\partial x}=\phi_i(E^*,x)\sum_{j\geq 1}jR_{ij}(E^*)\ ,\ \ {\rm para}\  0\leq i\leq 54\ ,
\end{equation}
con la condici\'on inicial $\phi_i^N(E,x=0)=0$. La soluci\'on es
\begin{equation}
\phi_i^N(E,L)=\sum_{j=0}^i\left(\frac{\sum_{k\geq 1}kR_{ik}(E^*)}{\sum_{k\geq 1}R_{jk}(E^*)}\right) 
b_{ij}(E^*)\left[1-\exp\left(-\sum_{k\geq 1}R_{jk}(E^*)L\right)\right]\ .
\label{nucleon}
\end{equation}

Los flujos totales de protones secundarios y neutrones se obtienen mediante las sumas pesadas sobre el \'{\i}ndice $i$:  
\begin{equation}
\phi^{ps}(E,L)=\sum_{i=0}^{54}\frac{Z_i}{56-i}\phi_i^N(E,L)
\end{equation}
y
\begin{equation}
\phi^n(E,L)=\sum_{i=0}^{54}\left(1-\frac{Z_i}{56-i}\right)\phi_i^N(E,L)\ ,
\label{neut1}
\end{equation}
donde $Z_i$ es la carga del n\'ucleo $i-$\'esimo. 

En presencia de un campo magn\'etico en la regi\'on de la fuente, las part\'{\i}culas cargadas quedan retenidas all\'{\i} 
por alg\'un tiempo, mientras que las no cargadas pueden escapar m\'as r\'apidamente. Este 
fen\'omeno, el ``mecanismo del neutr\'on'', ha sido propuesto en el contexto del estudio de mecanismos de producci\'on de
rayos c\'osmicos ultra-energ\'eticos \cite{ber77a,ber77b}; 
aqu\'{\i} discutiremos la relevancia de este mecanismo en el escenario que estamos
estudiando. Bajo estas consideraciones, asumiremos que el flujo de neutrones $\phi^n$ se suma al flujo saliente total de 
rayos c\'osmicos, despreciando cualquier p\'erdida de energ\'{\i}a adicional en su propagaci\'on a trav\'es del fondo de fotones que
rodea a la fuente. En cambio, el flujo de protones secundarios $\phi^{ps}$, combinado con el flujo inicial de protones primarios 
$\phi^0_{i=55}$, experimenta p\'erdidas de energ\'{\i}a debido al proceso de producci\'on de fotopiones, a trav\'es de las 
reacciones
\begin{equation}
p+\gamma\to p+\pi^0
\label{picero}
\end{equation}
y
\begin{equation}
p+\gamma\to n+\pi^+\ .
\label{pimas}
\end{equation}
La probabilidad de interacci\'on viene dada por
\begin{equation}
g(E)={1\over 2\gamma^2}\int_{\epsilon'_{thr}/2\gamma}^\infty{\rm d}\epsilon\ {n(\epsilon)\over\epsilon^2}
\int_{\epsilon'_{thr}}^{2\gamma\epsilon}{\rm d}\epsilon'\epsilon'\sigma_{\gamma p}(\epsilon')K(\epsilon')\ ,
\label{gE}
\end{equation}
donde, an\'alogamente a la ec. (\ref{rij}), $\gamma=E/m_pc^2$ es el factor de Lorentz del prot\'on, $\sigma_{\gamma p}$ es 
la secci\'on eficaz correspondiente a la interacci\'on $\gamma p$ (obtenida de ajustes a los datos experimentales 
compilados en \cite{gro00}), $K$ es el coeficiente de inelasticidad (definido como la p\'erdida de energ\'{\i}a relativa promedio
del prot\'on), $\epsilon$ es la energ\'{\i}a del fot\'on en el sistema del observador, y $\epsilon'$ su energ\'{\i}a en el 
sistema en reposo del prot\'on. La energ\'{\i}a umbral para la producci\'on de fotopiones es $\epsilon'_{thr}=145$~MeV. 

Entonces, llamando $\phi^p(E,x)$ y $\phi^{*n}(E,x)$ a los flujos diferenciales de protones y neutrones de energ\'{\i}a
$E$ (producidos por las reacciones (\ref{picero}) y (\ref{pimas}), respectivamente) luego de 
atravesar una distancia $x$ desde la fuente, las ecuaciones de propagaci\'on correspondientes quedan dadas por
\begin{equation}
\frac{\partial\phi^p(E,x)}{\partial x}=-g(E)\phi^p(E,x)+\frac{1}{2}\frac{1}{(1-K)}g\left(\frac{E}{1-K}\right)
\phi^p\left(\frac{E}{1-K},x\right)
\label{prot}
\end{equation}
y (despreciando p\'erdidas de energ\'{\i}a en el flujo de neutrones)
\begin{equation}
\frac{\partial\phi^{*n}(E,x)}{\partial x}=\frac{1}{2}\frac{1}{(1-K)}g\left(\frac{E}{1-K}\right)
\phi^p\left(\frac{E}{1-K},x\right)\ ,
\label{neut}
\end{equation}
donde los factores $\frac{1}{2}$ resultan de la probabilidad relativa de ocurrencia de las reacciones (\ref{picero}) y 
(\ref{pimas}). Al coeficiente de inelasticidad se le dio el valor t\'{\i}pico $K=0.3$, independiente de la energ\'{\i}a
del prot\'on \cite{kar93}. Las condiciones iniciales correspondientes a las ecuaciones de propagaci\'on, (\ref{prot}) y 
(\ref{neut}), son
\begin{equation}
\phi^p(E,x=0)=\phi_{i=55}^0(E)+\phi^{ps}(E,L)
\end{equation}
y
\begin{equation}
\phi^{*n}(E,x=0)=0\ ,
\end{equation}       
respectivamente. En la primera se asume, por simplicidad, que los protones secundarios producidos por la 
fotodesintegraci\'on nuclear se suman directamente al espectro inicial de protones. Esta suposici\'on est\'a justificada
debido a que, en los casos en que la producci\'on de fotopiones es relevante, la fotodesintegraci\'on es muy
eficiente y los n\'ucleos son desintegrados r\'apidamente. La soluci\'on a la ec. (\ref{prot}) es
\begin{equation}
\phi^p(E,L)=\sum_{l=0}^{\infty}\phi_l^p(E,L)\ , 
\label{prot1}
\end{equation}
donde
\begin{equation}
\phi_0^p(E,L)=\phi^p(E,0)\exp(-g(E)L)\ ,
\label{prot2}
\end{equation}
mientras que, para $l>0$,
\begin{equation}
\phi_l^p(E,L)={\phi^p\left(\frac{E}{(1-K)^l},0\right)\over(2(1-K))^l}\prod_{j=1}^lg\left(\frac{E}{(1-K)^j}\right)
\sum_{n=0}^l{\exp\left(-g\left(\frac{E}{(1-K)^n}\right)L\right)\over\prod_{^{m=0}_{m\neq n}}^l\left[
g\left(\frac{E}{(1-K)^m}\right)-g\left(\frac{E}{(1-K)^n}\right)\right]}\ .
\label{prot3}
\end{equation}

Para la componente de neutrones, la soluci\'on a la ec. (\ref{neut}) resulta
\begin{equation}
\phi^{*n}(E,x)=\sum_{l=0}^{\infty}\phi_l^{*n}(E,x)\ , 
\label{neut2}
\end{equation}
donde
\begin{equation}
\phi_0^{*n}(E,L)=\frac{\phi^p\left(\frac{E}{1-K},0\right)}{2(1-K)}
\left(1-\exp\left(-g\left(\frac{E}{1-K}\right)L\right)\right)\ ,
\label{neut3}
\end{equation}
mientras que, para que $l>0$, 
$$ 
\phi_l^{*n}(E,L)=\frac{\phi^p\left(\frac{E}{(1-K)^{l+1}},0\right)}{(2(1-K))^{l+1}}
\prod_{j=0}^{l} g\left(\frac{E}{(1-K)^{j+1}}\right)\times
$$
\begin{equation}
\sum_{n=0}^{l}{1-\exp\left(-g\left(\frac{E}{(1-K)^{n+1}}\right)L\right)\over g\left(\frac{E}{(1-K)^{n+1}}\right)
\prod_{^{m=0}_{m\neq n}}^l\left[g\left(\frac{E}{(1-K)^{m+1}}\right)-g\left(\frac{E}{(1-K)^{n+1}}\right)\right]}\ .
\label{neut4}
\end{equation}

La linealidad de la tasa de producci\'on de fotopiones $g(E)$ con respecto a la densidad de energ\'{\i}a de fotones $\rho_E$
(ver la ec.~(\ref{gE})) hace que los flujos de protones y neutrones dependan s\'olo de la columna de densidad $\rho_EL$,
tal como ocurre en el caso de los n\'ucleos.    

\section{Resultados y comparaci\'on con las observaciones}
 
Hasta aqu\'{\i}, hemos planteado las ecuaciones que describen la propagaci\'on de los rayos c\'osmicos en 
el entorno de las fuentes, y encontramos su soluci\'on exacta: ecs.~(\ref{nuc1})--(\ref{nuc3}) para los n\'ucleos, 
ecs.~(\ref{nucleon}), (\ref{neut1}) y (\ref{neut2})--(\ref{neut4}) para los neutrones 
(que, eventualmente, decaen en protones fuera de la regi\'on 
de la fuente), y ecs.~(\ref{prot1})--(\ref{prot3}) para los protones 
(que sufren p\'erdidas de energ\'{\i}a debido a la producci\'on de
fotopiones), siendo la suma de todas estas contribuciones el flujo total de rayos c\'osmicos $\phi_{total}$. Ahora podemos
examinar c\'omo depende $\phi_{total}$ de la distribuci\'on de fotones asumida para el entorno de la fuente, 
es decir, c\'omo es su dependencia en t\'erminos de la columna de densidad $\rho_EL$ y de los par\'ametros involucrados en las
distribuciones espectrales de fotones (ecs.~(\ref{planck}) y (\ref{powerlaw})).

\begin{figure}[t!]
\centerline{{\epsfxsize=3truein \epsffile{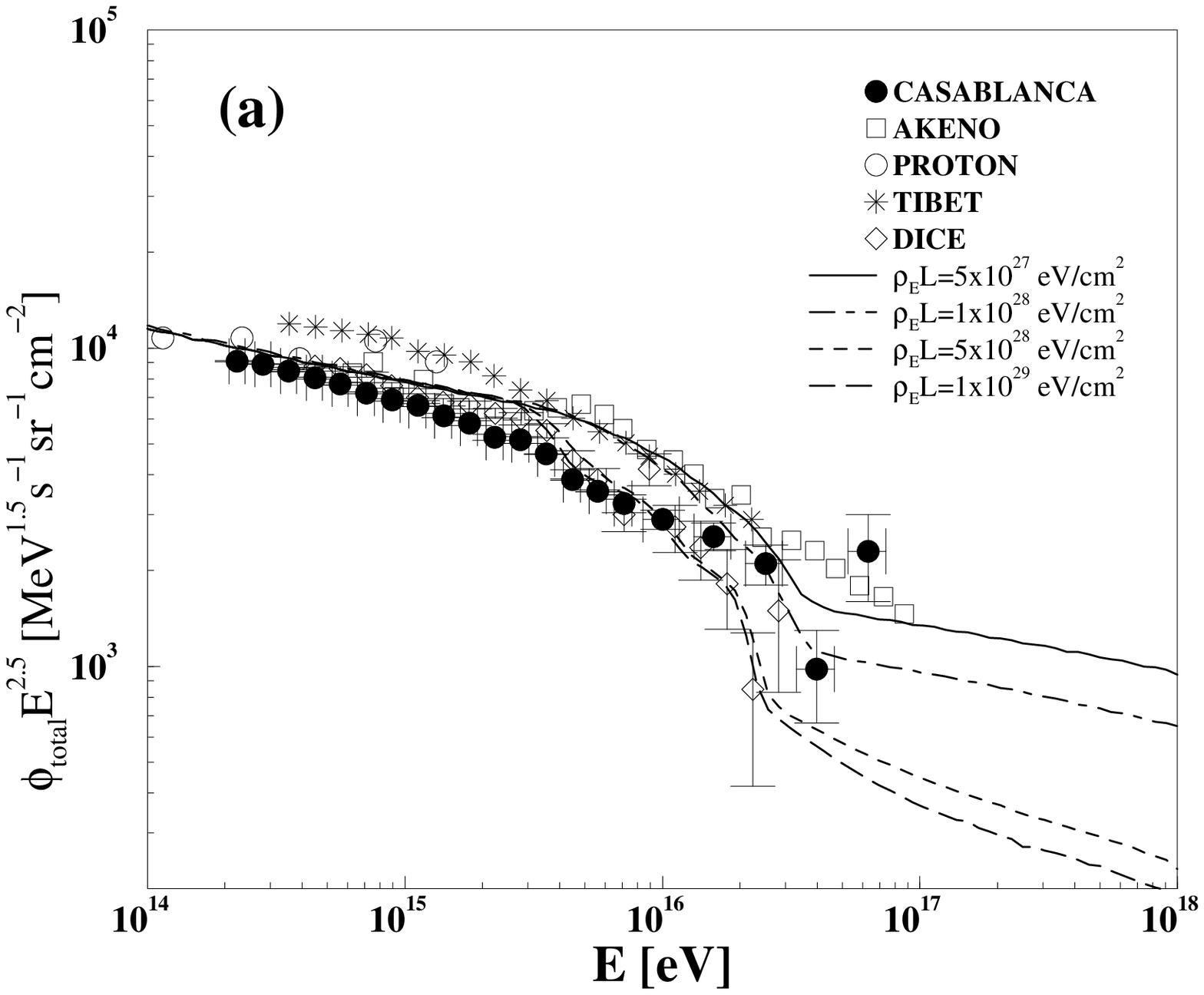}}}
\centerline{{\epsfxsize=3truein \epsffile{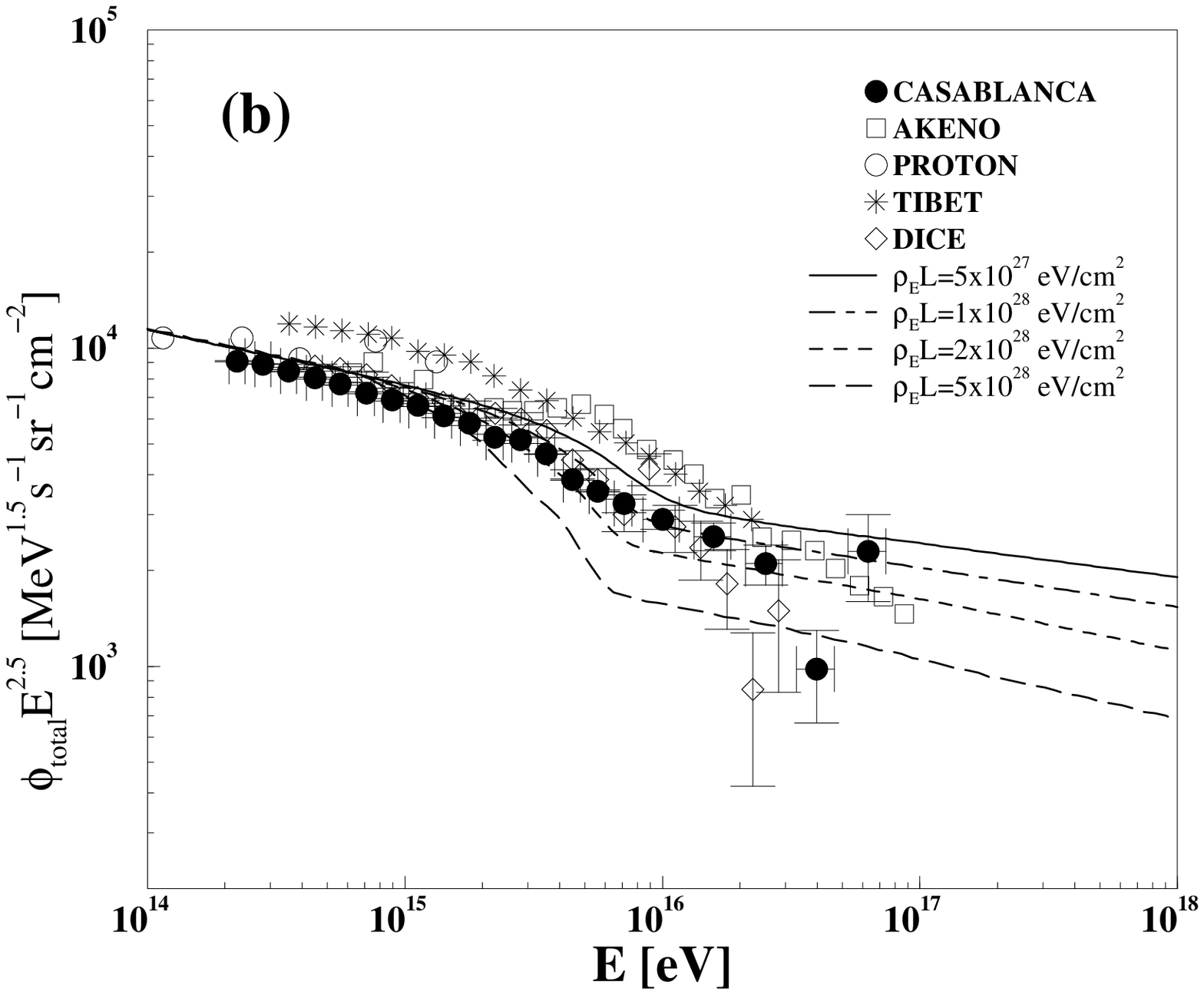}}}
\caption{Flujo total de rayos c\'osmicos para espectros de fotones de Planck con (a) $k_BT=1.8$~eV y (b) $k_BT=10$~eV,
y diferentes valores para la columna de densidad $\rho_EL$, como se indica. Los resultados se comparan con los datos 
observados por diferentes experimentos.}
\label{f3P1}
\end{figure}

La fig.\ref{f3P1}(a) muestra el flujo diferencial total de rayos c\'osmicos (en funci\'on de la energ\'{\i}a por 
part\'{\i}cula) para un espectro de Planck con $k_BT=1.8$~eV y diferentes valores de la columna de densidad $\rho_EL$. 
An\'alogamente, la fig.\ref{f3P1}(b) muestra los resultados correspondientes a un espectro de Planck con $k_BT=10$~eV.
En las figuras tambi\'en aparecen los espectros observados mediante diferentes experimentos (CASABLANCA, DICE, Tibet,
PROTON y AKENO) \cite{swo00,fow01,ame96}. 
Seg\'un se observa claramente en las figuras, al aumentar la temperatura la rodilla se desplaza 
hacia menores energ\'{\i}as, mientras que el efecto de incrementar $\rho_EL$ es el de intensificar
la supresi\'on del flujo por encima de la rodilla. Este comportamiento resulta del hecho de que la fotodesintegraci\'on 
nuclear es el proceso responsable de la formaci\'on de la rodilla (debido a que los procesos de producci\'on de fotopiones aparecen
reci\'en a mayores energ\'{\i}as). Entonces, si la temperatura de la distribuci\'on (y, por lo tanto, la energ\'{\i}a media de
los fotones) es mayor, los n\'ucleos comienzan a desintegrarse a energ\'{\i}as m\'as bajas. Por otra parte, la 
propagaci\'on de rayos c\'osmicos no puede verse afectada por $\rho_EL$ debajo de la rodilla, a energ\'{\i}as en las que 
no hay fotodesintegraci\'on; en cambio, por encima de la rodilla, 
las tasas de desintegraci\'on son lineales en $\rho_E$. En consecuencia, debe esperarse que $\rho_EL$ sea la cantidad 
que regule la supresi\'on del espectro en la regi\'on de la rodilla, tal como se observa en las figuras.   

El mejor ajuste a las observaciones para el espectro de fotones de Planck, reportado en \cite{kar93}, ocurre para
$k_BT=1.8$~eV y $\rho_EL=2.25\times 10^{29}$~eV/cm$^2$; para comparar, 
la fig.\ref{f3P1}(a) muestra nuestros resultados para la misma temperatura.      
Como puede observarse, los nuevos resultados son compatibles con los datos experimentales para valores de columna
de densidad alrededor de $\rho_EL\simeq 5\times 10^{27}$~eV/cm$^2$, es decir, menores en un factor $\sim 50$ que el 
resultado previo. La fig.\ref{f3P1}(b) muestra que los resultados correspondientes a $k_BT=10$~eV est\'an en buen acuerdo
con los datos experimentales para columnas de densidad en el rango 
$\rho_EL=5\times 10^{27}-2\times 10^{28}$~eV/cm$^2$. 

\begin{figure}[t!]
\centerline{{\epsfxsize=3truein \epsffile{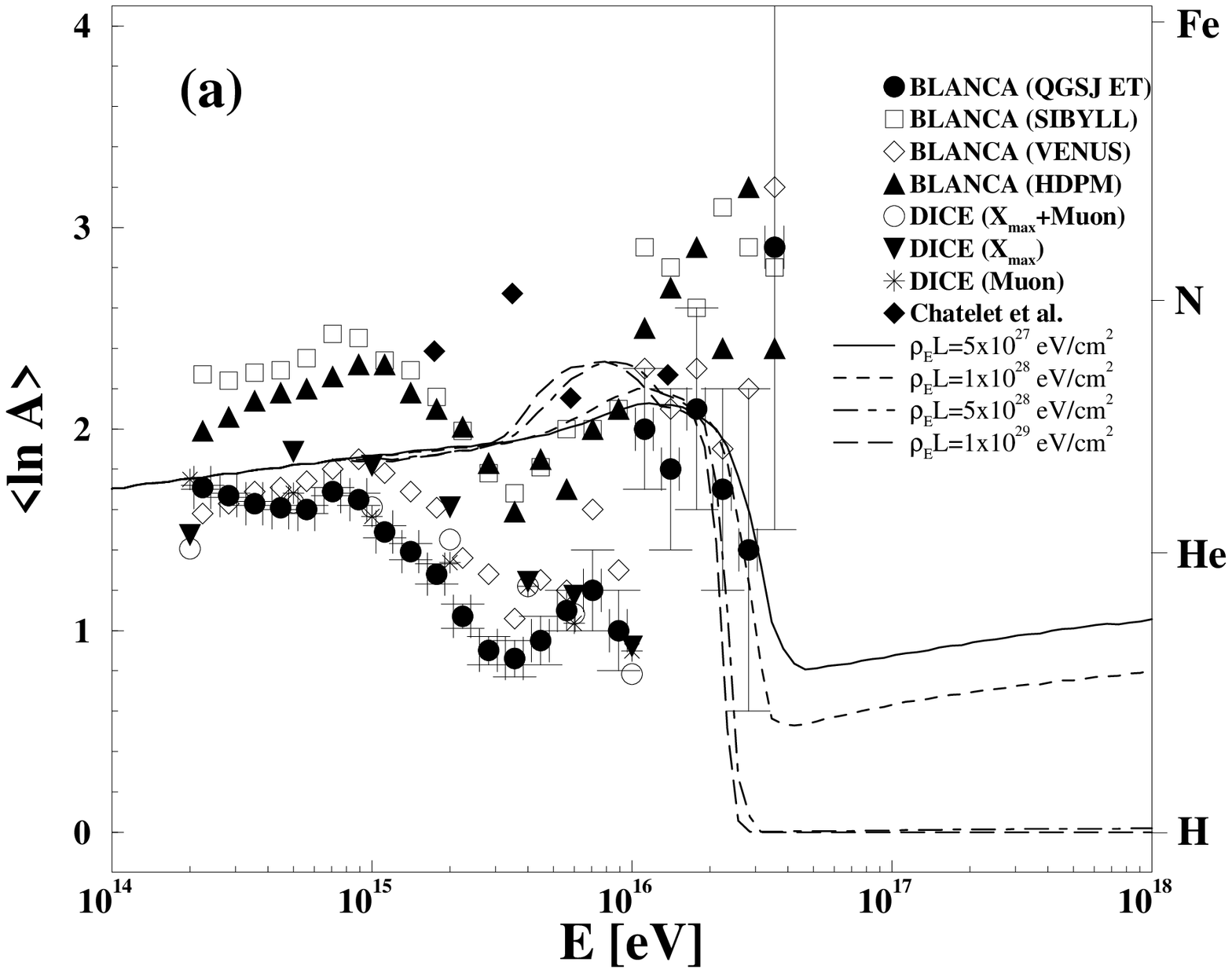}}}
\centerline{{\epsfxsize=3truein \epsffile{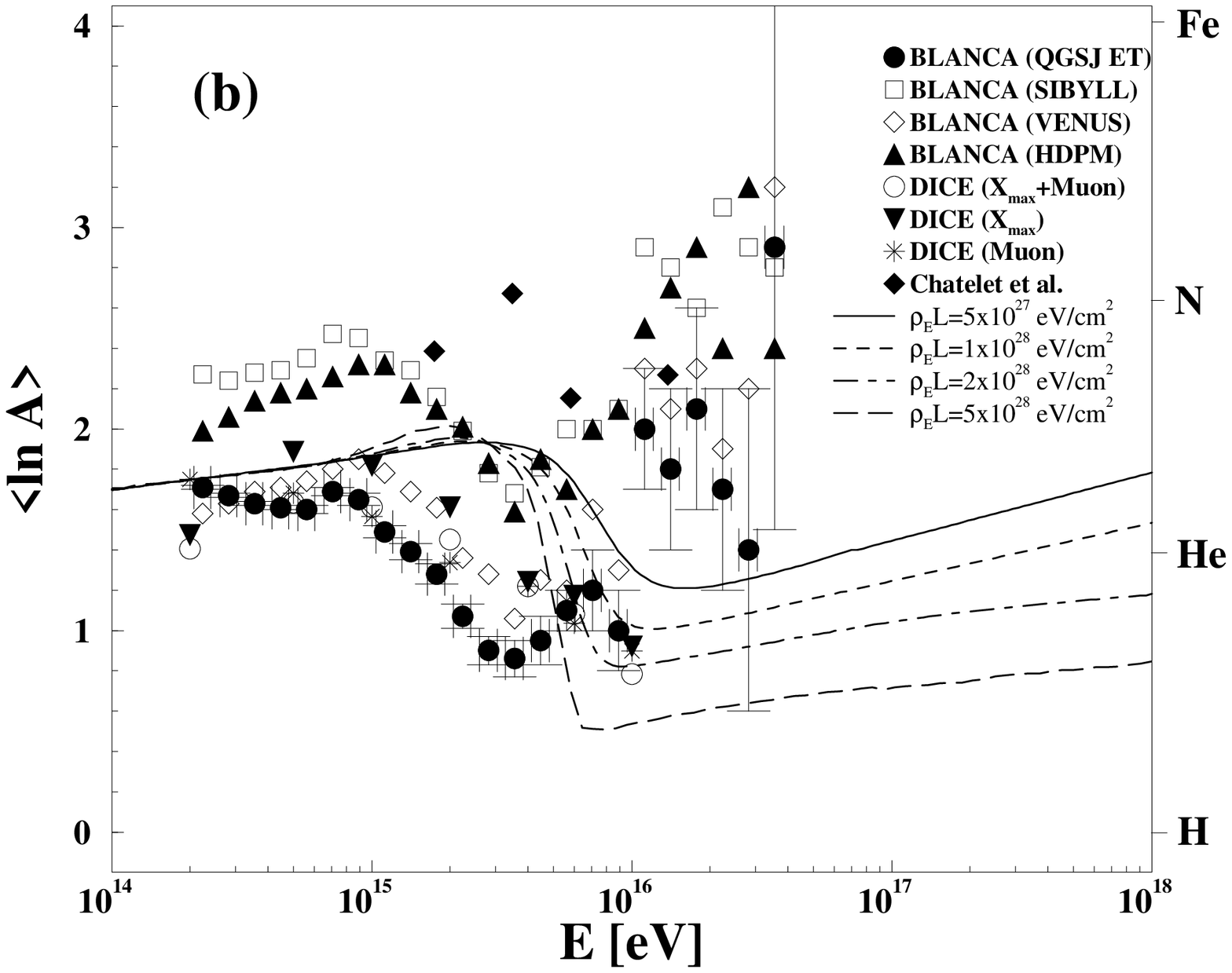}}}
\caption{Composici\'on m\'asica de rayos c\'osmicos para espectros de fotones de Planck con (a) $k_BT=1.8$~eV y 
(b) $k_BT=10$~eV, y diferentes valores para la columna de densidad $\rho_EL$, como se indica. 
Los resultados se comparan con las observaciones de diferentes experimentos.}
\label{f4P1}
\end{figure}

Las figs.\ref{f4P1}(a) y (b) muestran la composici\'on m\'asica media, $\langle{\rm ln}A\rangle$, correspondiente a espectros de
Planck con $k_BT=1.8$~eV y $k_BT=10$~eV, respectivamente, en comparaci\'on con los datos de diferentes 
experimentos \cite{swo00,fow01,cha91}. 
Las soluciones con los mayores valores de la columna de densidad $\rho_EL$ 
predicen que los rayos c\'osmicos arriba de la rodilla est\'an mayormente constitu\'{\i}dos por protones, en desacuerdo 
con los datos experimentales. La contribuci\'on no despreciable de componentes m\'as pesadas a 
energ\'{\i}as por encima de la rodilla impone una cota superior sobre la columna de densidad alrededor de 
$(\rho_EL)_{max}\approx 2\times 10^{28}$~eV/cm$^2$. Este resultado hace ver nuevamente que, en este escenario,
los procesos de emisi\'on de varios nucleones juegan un rol muy significativo en la propagaci\'on de los rayos c\'osmicos 
en la regi\'on de la fuente. 

Podemos observar tambi\'en que, aunque las soluciones con columnas de densidad relativamente bajas 
($\rho_EL\leq (\rho_EL)_{max}$) est\'an en un acuerdo razonable con algunos conjuntos de datos experimentales dentro de
las barras de error, los resultados que se obtienen para la composici\'on arriba de la rodilla parecen estar demasiado
suprimidos. Sin embargo, debe tenerse en cuenta que aqu\'{\i} hemos considerado s\'olo una fuente con una columna de 
densidad fija, mientras que en una situaci\'on realista se tendr\'{\i}an muchas fuentes, cada una, en principio, con un
entorno de radiaci\'on diferente. Evitando c\'alculos detallados, consideremos el efecto que resultar\'{\i}a de agregar 
una fuente adicional. As\'{\i}, 
si esta segunda fuente tuviera una columna de densidad $\rho_EL$ algo menor, dar\'{\i}a lugar a
un espectro menos suprimido; entonces, mientras la primera fuente ser\'{\i}a responsable de la formaci\'on de la rodilla,
la segunda mantendr\'{\i}a una componente nuclear m\'as all\'a de la rodilla. Por otra parte, en nuestros c\'alculos 
hemos asumido que todos los n\'ucleos recorren la misma distancia $L$ en la regi\'on que contiene la radiaci\'on; es
decir, no hemos tenido en cuenta que, para cualquier dada energ\'{\i}a, los campos magn\'eticos presentes en el entorno 
de la fuente retendr\'{\i}an m\'as 
efectivamente a los rayos c\'osmicos m\'as pesados. Tomando \'esto en cuenta (por ejemplo,
asumiendo que un n\'ucleo recorre una distancia proporcional a su carga) encontrar\'{\i}amos que, 
mientras la desintegraci\'on podr\'{\i}a ser muy efectiva para n\'ucleos pesados (y ser\'{\i}a, de hecho, la responsable 
de la formaci\'on de la rodilla), ser\'{\i}a mucho menos efectiva para los n\'ucleos m\'as livianos (ya que la columna de
densidad para ellos ser\'{\i}a mucho menor) y entonces podr\'{\i}an sobrevivir sin desintegrarse hasta 
energ\'{\i}as m\'as all\'a de la rodilla. 

De este modo, vemos que este escenario puede reproducir el comportamiento observado por algunos experimentos,  
en el que la composici\'on cambia hacia el predominio de componentes m\'as livianas por encima de la rodilla.  
Sin embargo, como hemos comentado anteriormente, existe una marcada discrepancia en el conjunto de datos 
experimentales disponibles. Los resultados de los experimentos m\'as recientes (KASCADE \cite{kam01b} y EAS-TOP/MACRO \cite{agl03})
parecen concordar en que la composici\'on se torna m\'as pesada por encima de la rodilla, favoreciendo as\'{\i} a 
los escenarios dependientes de la rigidez (como, por ejemplo, el escenario de la difusi\'on turbulenta y drift estudiado
en el Cap\'{\i}tulo 2), aunque otros experimentos muestran la tendencia contraria (por ejemplo, DICE \cite{swo00}). 

\begin{figure}[t!]
\centerline{{\epsfxsize=3.5truein \epsffile{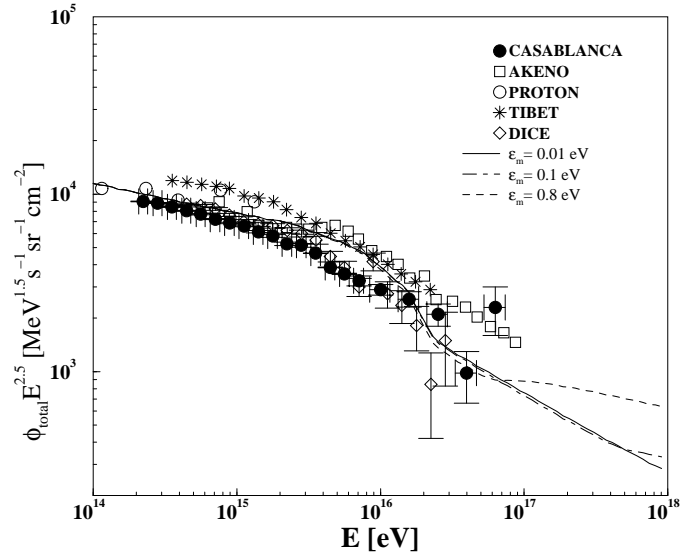}}}
\caption{Flujo total de rayos c\'osmicos para espectros de fotones de ley de potencias con $\alpha=1.3$, 
$\epsilon_M=20$~eV, $\rho_EL=10^{28}$~eV y diferentes valores para el cutoff inferior de la distribuci\'on de
fotones, como se indica. Los resultados se comparan con las 
observaciones de diferentes experimentos.}
\label{f5P1}
\end{figure}

Los resultados para un espectro de fotones de ley de potencias (con un cutoff inferior en el rango infrarrojo)       
no presentan mayores diferencias con los obtenidos para espectros de Planck.
En lo que sigue, fijaremos el cutoff superior, tomando el valor $\epsilon_M=20$~eV. 
La fig.\ref{f5P1} muestra el flujo total de rayos c\'osmicos (en funci\'on de la energ\'{\i}a por 
part\'{\i}cula) para distribuciones de ley de potencias con una columna de densidad $\rho_EL=10^{28}$~eV/cm$^2$ y 
varios valores para el cutoff inferior del espectro; los espectros
obtenidos reproducen adecuadamente la rodilla. Analizando la dependencia de los resultados con el 
cutoff inferior del espectro de fotones, vemos que, m\'as all\'a de peque\~nas diferencias en la normalizaci\'on, 
los valores m\'as bajos de $\epsilon_m$ producen flujos m\'as suprimidos en la regi\'on de energ\'{\i}as altas. 
De la expresi\'on para la tasa de emisi\'on de la fotodesintegraci\'on (ec. (\ref{rij})), puede verse que el rol de 
$\epsilon_m$ es el de determinar un valor para la energ\'{\i}a de los rayos c\'osmicos $E$, tal que, debajo de $E$, domina la 
resonancia dipolar gigante, mientras que, por encima de $E$, prevalece el plateau de $\sigma$ propio del r\'egimen de altas
energ\'{\i}as. A medida que bajamos $\epsilon_m$, disponemos de un n\'umero mayor de fotones de baja energ\'{\i}a, de modo
que la supresi\'on del flujo impuesta por la resonancia gigante se extiende a energ\'{\i}as de part\'{\i}cula mayores.         

Este an\'alisis permite explicar apropiadamente los resultados de la fig.\ref{f5P1}, como as\'{\i} tambi\'en los que 
muestra la fig.\ref{f2P1}(b). Los resultados obtenidos previamente para espectros de fotones de ley de potencias \cite{kar93} 
exhiben una supresi\'on del flujo extremadamente abrupta a altas energ\'{\i}as (por encima de $E=10^{17}$~eV) que parecen
estar en desacuerdo con las observaciones del espectro y, especialmente, de la composici\'on de los rayos c\'osmicos.
Como vemos aqu\'{\i}, \'esto se atribuye al valor despreciable que se adopt\'o all\'{\i} para $\epsilon_m$; el uso de un 
cutoff no despreciable evita la supresi\'on abrupta del flujo de rayos c\'osmicos y provee un mejor acuerdo con las 
observaciones en esta regi\'on de energ\'{\i}as. Por otra parte, debe notarse que una soluci\'on que exhiba un flujo de 
rayos c\'osmicos excesivo a energ\'{\i}as mayores que $10^{17}$~eV no es problem\'atica, ya que es de esperar que a 
energ\'{\i}as tan altas el escape de rayos c\'osmicos de la galaxia comience a ser un mecanismo efectivo de supresi\'on del 
flujo. Adicionalmente, la aceleraci\'on de los rayos c\'osmicos en la fuente podr\'{\i}a ser menos efectiva a altas 
energ\'{\i}as, contribuyendo tambi\'en a suprimir el flujo. En consecuencia, teniendo en cuenta la existencia de estos 
mecanismos de supresi\'on adicionales, no ser\'{\i}a razonable esperar una explicaci\'on del espectro de rayos c\'osmicos s\'olo 
mediante procesos de fotodesintegraci\'on m\'as all\'a de $E\sim 10^{17}$~eV.

\begin{figure}[t!]
\centerline{{\epsfxsize=3.8truein \epsffile{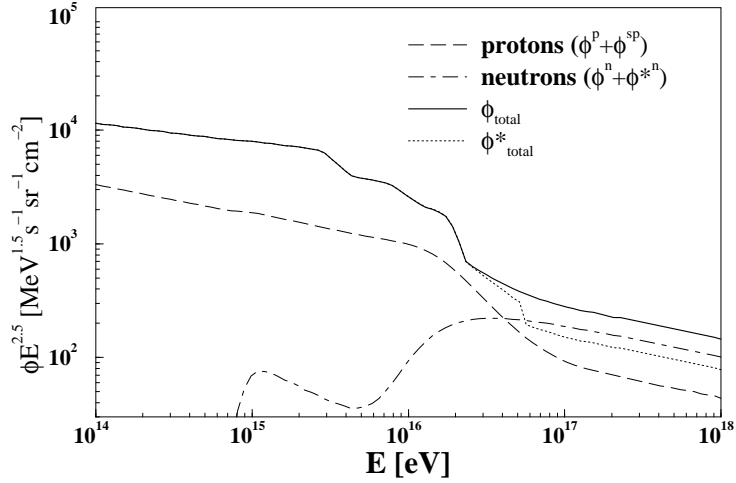}}}
\caption{Flujo de protones ($\phi^p+\phi^{ps}$), neutrones ($\phi^n+\phi^{*n}$) y total ($\phi_{total}$) para un espectro 
de fotones de Planck con $k_BT=1.8$~eV y $\rho_EL=2\times 10^{29}$~eV/cm$^2$. Tambi\'en se muestra el flujo total de 
rayos c\'osmicos ($\phi^*_{total}$) que se obtiene ignorando el escape de neutrones.} 
\label{f6P1}
\end{figure}

La fig.\ref{f6P1} muestra las componentes de protones y neutrones ($\phi^p+\phi^{ps}$ y $\phi^n+\phi^{*n}$, respectivamente)
y el flujo total de rayos c\'osmicos $\phi_{total}$, para el espectro de fotones de Planck con $k_BT=1.8$~eV 
y $\rho_EL=2\times 10^{29}$~eV/cm$^2$. Con el fin de estimar el posible impacto del mecanismo del neutr\'on, tambi\'en 
aparece graficado el flujo total de rayos c\'osmicos, $\phi^*_{total}$, que se obtiene ignorando el escape de neutrones. 
En la figura se observa que el mecanismo del neutr\'on incrementa el flujo total de rayos c\'osmicos
en $\sim 100\%$ para $E\geq 5\times 10^{16}$~eV. Si, en cambio, hubi\'eramos adoptado una columna de densidad  
$\rho_EL=10^{29}$~eV/cm$^2$, el incremento de flujo correspondiente hubiera sido de s\'olo $\sim 10\%$. Sin 
embargo, debe notarse que estos resultados corresponden a columnas de densidad muy grandes (de hecho, incompatibles 
con la composici\'on de rayos c\'osmicos observada); los efectos que resultan del mecanismo de escape del neutr\'on 
resultan, en cambio, despreciables para menores valores de $\rho_EL$, como tambi\'en las p\'erdidas de energ\'{\i}a 
de los protones por producci\'on de fotopiones.      
    
Finalmente, estimaremos las condiciones que deben imponerse sobre las fuentes para que puedan producir las 
densidades de fotones que requiere este escenario. Una fuente isotr\'opica con una densidad de energ\'{\i}a de fotones 
$\rho_{E,0}$ en su superficie (con radio $r_0$) tendr\'a a su alrededor una densidad de energ\'{\i}a de fotones dada por
$\rho_E(r)=\rho_{E,0}(r_0/r)^2$. \'Esto significa que los rayos c\'osmicos que abandonan la fuente radialmente 
atraviesan t\'{\i}picamente una columna de densidad de fotones $\langle\rho_Er\rangle\simeq\rho_{E,0}r_0$.
Por otra parte, si los rayos c\'osmicos est\'an confinados magn\'eticamente cerca de la fuente, el camino que
recorrer\'an a trav\'es de la regi\'on con la radiaci\'on ser\'a mayor, es decir, 
$\langle\rho_EL\rangle=\lambda\langle\rho_Er\rangle$, con $\lambda\gg 1$.
Si, con el fin de estimar las m\'aximas densidades que es posible producir, asumimos que la fuente rad\'{\i}a con una
luminosidad $L=4\pi r_0^2\rho_{E,0}c$ menor que el l\'{\i}mite de Eddington (es decir,  
$L<L_{Edd}=4\pi GMm_pc/\sigma_T\simeq 1.3\times 10^{38}M/M_\odot$~erg/s, donde $M$ es la masa de la fuente
y $\sigma_T$ la secci\'on eficaz de Thomson), entonces el radio de la fuente debe
satisfacer
\begin{equation}
r_0<\frac{GMm_p}{\sigma_T\rho_{E,0}r_0}\simeq \left(\frac{\lambda}{10}\right)\left(\frac{10^{28}eV/cm^2}
{\langle \rho_E L\rangle}\right)\left( \frac{M}{M_\odot}\right)2\times 10^9 {\rm m}\ ,
\end{equation}
de modo que la fuente debe ser compacta (como un p\'ulsar) y, en general, m\'as peque\~na que el tama\~no t\'{\i}pico de una 
enana blanca. Sin embargo, debe tenerse en cuenta que, en situaciones no estacionarias (como, por ejemplo, en explosiones de
supernova), se pueden alcanzar luminosidades mayores que $L_{Edd}$. Otra consideraci\'on de inter\'es es que, 
si la fuente produce un espectro de cuerpo negro con temperatura $T$ y est\'a radiando en el l\'{\i}mite de Eddington, 
la relaci\'on $4\pi r_0^2\sigma T^4=L_{Edd}$ implica que $T\simeq 1.8\times 10^3$ eV$ (M/M_\odot)^{1/4}(10$~km/$r_0)^{1/2}$, 
de modo que, para que la energ\'{\i}a de los fotones est\'e en la regi\'on \'optica/UV, los radios algo mayores que aqu\'ellos 
de las estrellas de neutrones estar\'{\i}an favorecidos (excepto que la luminosidad sea menor que $L_{Edd}$).   

%% file: conclusiones.tex
\chapter{Conclusiones}

En esta Tesis Doctoral estudiamos diversos aspectos relacionados con la propagaci\'on de rayos c\'osmicos
gal\'acticos de muy alta energ\'{\i}a, con particular \'enfasis en la regi\'on que comprende las dos 
rodillas del espectro (es decir, en el rango $E\simeq 10^{15}-10^{18}$~eV). 

En el {\bf Cap\'{\i}tulo 2}, consideramos en detalle la difusi\'on y los efectos de drift en la 
propagaci\'on de rayos c\'osmicos en configuraciones plausibles de campos magn\'eticos gal\'acticos, 
adoptando coeficientes de difusi\'on apropiados para los niveles de alta turbulencia presentes en
la galaxia, y asumiendo un espectro de Kolmogorov de fluctuaciones. En este escenario, 
cada componente nuclear de rayos c\'osmicos adquiere una rodilla 
(con un cambio de \'{\i}ndice espectral $\Delta\alpha\simeq 2/3$) que ocurre a una energ\'{\i}a proporcional 
a la carga correspondiente; de este modo, la rodilla del espectro resulta del escape de las 
componentes livianas (esencialmente,
protones y part\'{\i}culas alfa), mientras que el escape de las componentes m\'as pesadas 
(b\'asicamente, los n\'ucleos de hierro)
forma la segunda rodilla.
Encontramos que el escenario de la difusi\'on turbulenta y drift describe adecuadamente el espectro observado
desde energ\'{\i}as por debajo de la primer rodilla hasta las correspondientes a la regi\'on del tobillo, en donde 
tiene lugar un cambio hacia el predominio de la componente extragal\'actica de rayos c\'osmicos. La composici\'on que predice 
este escenario tiende a ser m\'as pesada a mayores energ\'{\i}as a trav\'es de la rodilla, y est\'a en excelente acuerdo
con los datos de algunos experimentos, como, por ejemplo, las m\'as recientes observaciones de KASCADE y EAS-TOP/MACRO.  

Por otro lado, la amplitud de anisotrop\'{\i}a experimenta un {\it crossover} desde el r\'egimen dominado por la 
difusi\'on transversal al r\'egimen dominado por los drifts, que aparece alrededor de la regi\'on de la rodilla. En efecto, 
\'esta es una caracter\'{\i}stica particular que diferencia claramente las predicciones de este escenario de las de otras 
propuestas. Adem\'as, examinando la fase de anisotrop\'{\i}a del primer arm\'onico en ascensi\'on recta, encontramos que
el m\'aximo de la anisotrop\'{\i}a podr\'{\i}a estar orientado hacia la direcci\'on del centro gal\'actico para energ\'{\i}as cercanas
a $10^{18}$~eV, tal como fue observado por AGASA y SUGAR. Las predicciones de este escenario resultan tambi\'en de
potencial inter\'es para las futuras observaciones en el hemisferio sur con el observatorio AUGER. 

Finalmente, complementamos este estudio con la determinaci\'on num\'erica, mediante simulaciones extensivas Monte Carlo, 
de los coeficientes de difusi\'on paralela, transversal y antisim\'etrica 
que describen la propagaci\'on de rayos c\'osmicos en condiciones de alta turbulencia. 
Extendiendo investigaciones previas, analizamos diferentes tipos de espectro de turbulencia y proponemos parametrizaciones
sencillas que describen los resultados obtenidos en cada caso. 

En el {\bf Cap\'{\i}tulo 3}, 
calculamos el flujo difuso de altas energ\'{\i}as de los neutrinos 
producidos por rayos c\'osmicos que inciden sobre la atm\'osfera, o bien en su
propagaci\'on en el medio interestelar de la galaxia. Los resultados dependen 
significativamente del espectro y de la composici\'on asumidos. En particular,
considerando que la composici\'on se torna m\'as pesada por encima de la rodilla
(es decir, como resulta de considerar escenarios que explican la rodilla como debida a un efecto dependiente de la rigidez)
el fondo de neutrinos atmosf\'ericos se reduce significativamente, facilitando la b\'usqueda de fuentes astrof\'{\i}sicas.
Luego de investigar este fen\'omeno en el canal usual de trazas lept\'onicas, asociado a la
detecci\'on de neutrinos mu\'on, proponemos  
un nuevo m\'etodo para aislar el flujo de neutrinos atmosf\'ericos prompt a trav\'es 
del canal de las cascadas hadr\'onica y electromagn\'etica, asociado a la detecci\'on de neutrinos electr\'on.
Este estudio muestra la importancia de resolver la composici\'on para estimar apropiadamente 
el fondo difuso de neutrinos de altas energ\'{\i}as producido por los rayos c\'osmicos, y es particularmente relevante en
la posible detecci\'on de neutrinos extraterrestres en experimentos tales como AMANDA y ICECUBE.       

En el {\bf Cap\'{\i}tulo 4}, consideramos en detalle la fotodesintegraci\'on nuclear de rayos c\'osmicos, en el 
escenario en que su interacci\'on con fotones \'opticos y UV blandos en el entorno de la fuente da origen a la supresi\'on del
flujo y al cambio en la composici\'on en la regi\'on de la rodilla. El tratamiento m\'as preciso de los procesos nucleares,
particularmente la inclusi\'on de procesos que involucran la emisi\'on de varios nucleones, implica que las columnas de 
densidad de fotones requeridas sean menores (por un orden de magnitud) a las calculadas en investigaciones previas. 
Como una propuesta alternativa para explicar la rodilla del espectro, este modelo predice que la composici\'on 
se torna m\'as liviana a partir de $\sim 10^{16}$~eV; \'esto constituye una diferencia sustancial con los escenarios 
dependientes de la rigidez, que predicen la tendencia opuesta, y podr\'a ponerse a prueba con las mediciones 
m\'as precisas que se obtendr\'an pr\'oximamente en KASCADE-Grande.

En conclusi\'on, enfocando nuestra atenci\'on hacia la propagaci\'on de rayos c\'osmicos gal\'acticos de muy alta energ\'{\i}a,
hemos elaborado propuestas que explican las rodillas del espectro, 
la composici\'on y las anisotrop\'{\i}as; encontramos, adem\'as, 
que una soluci\'on a estos problemas podr\'{\i}a tambi\'en contribuir significativamente al conocimiento de la distribuci\'on 
de las fuentes de rayos c\'osmicos en la galaxia, de la regi\'on en que los rayos c\'osmicos se aceleran, de la estructura del 
campo magn\'etico gal\'actico, y de los flujos de neutrinos de muy alta energ\'{\i}a que podr\'an detectarse en los nuevos 
observatorios.

%% file: apendices.tex
\appendix
\chapter{El tensor de difusi\'on}
El car\'acter difusivo del transporte de rayos c\'osmicos en la 
galaxia resulta de considerar al campo random como un mecanismo efectivo de 
colisi\'on. Las componentes
diagonales del tensor de difusi\'on corresponden a la ``difusi\'on browniana'',
es decir, a la aparici\'on de un flujo de part\'{\i}culas en respuesta a un 
gradiente de concentraci\'on, ambos en la misma direcci\'on 
(${\bf J}=-D\nabla N$). Como el campo regular rompe la isotrop\'{\i}a, debe
distinguirse la direcci\'on transversal al campo regular de la direcci\'on
paralela al mismo; 
considerando que el campo magn\'etico est\'a localmente dirigido en la
direcci\'on $\hat{z}$, 
las componentes diagonales quedan definidas como $D_{xx}=D_{yy}\equiv D_\perp$,
$D_{zz}\equiv D_\parallel$.
 
\begin{figure}[th!]
\centerline{{\epsfxsize=5.5truecm\epsfysize=5truecm\epsffile{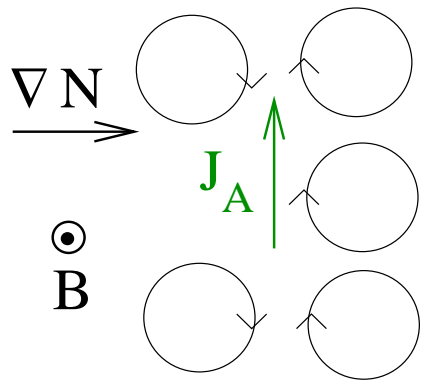}}}
\caption{Esquema que ilustra el origen f\'{\i}sico del drift macrosc\'opico (de \cite{rou03}).}
\label{driftfig}
\end{figure}

El origen f\'{\i}sico del drift macrosc\'opico queda ilustrado por la
fig.\ref{driftfig}, que muestra las \'orbitas helicoidales que 
corresponden a part\'{\i}culas de carga positiva, proyectadas sobre un plano
perpendicular al campo magn\'etico regular (por simplicidad, aqu\'{\i} ignoramos
la distorsi\'on que introduce la componente random del campo magn\'etico). 
El resultado de promediar el flujo sobre una distribuci\'on de part\'{\i}culas
es la aparici\'on de una corriente macrosc\'opica perpendicular al gradiente 
de concentraci\'on; debe notarse que este efecto resulta del comportamiento 
colectivo de la distribuci\'on de part\'{\i}culas, y est\'a ausente a nivel 
microsc\'opico (es decir, considerando trayectorias de part\'{\i}culas individuales).
Por otra parte, es inmediato observar que la componente de drift es 
antisim\'etrica: $D_{xy}=-D_{yx}\equiv D_A$. Finalmente,
siendo que la fuerza de Lorentz es perpendicular al campo magn\'etico que la 
produce, este efecto est\'a ausente en los planos $x-z$, $y-z$. 

En t\'erminos de las componentes de ${\bf b}={\bf B_0}/|{\bf B_0}|$, 
el vector unitario en la direcci\'on del campo magn\'etico
regular, el tensor de difusi\'on queda dado por la ec.~(\ref{dij}),
$$
D_{ij}=\left(D_{\parallel}-D_{\perp}\right)b_ib_j+D_{\perp}\delta_{ij}+D_A\epsilon_{ijk}b_k\ .
$$

\chapter{El formalismo de Kubo}

La difusi\'on de part\'{\i}culas cargadas en campos magn\'eticos random puede 
calcularse a partir de la descorrelaci\'on entre diferentes componentes de la velocidad de una part\'{\i}cula 
(una vez promediada sobre una distribuci\'on isotr\'opica de muchas part\'{\i}culas, y sobre muchas configuraciones diferentes
de campo random), en lo que se conoce como el formalismo de Kubo \cite{kub57,for77,bie97}. 

La funci\'on de correlaci\'on de velocidades puede definirse como  
\begin{equation}
R_{ij}(t)\equiv\langle v_i(t_0)v_j(t_0+t)\rangle\ \ (t\geq 0)\ ,
\end{equation}
donde $\langle ...\rangle$ denota el promedio sobre un ensemble de
configuraciones de campo random. Esta definici\'on asume la ergodicidad 
del sistema (es decir, que $R_{ij}$ es independiente de $t_0$).
Asumiremos que $R_{ij}(t)\to 0$ suficientemente r\'apido en el l\'{\i}mite $t\to\infty$, de modo que 
\begin{equation}
\int_0^\infty{\rm d}t\ R_{ij}(t)<\infty\ ,  
\end{equation}
y definiremos $T_{ef}$ como alg\'un valor de $t$ suficientemente grande,
tal que
\begin{equation}
\int_0^{T_{ef}}{\rm d}t\ R_{ij}(t)\gg\int_{T_{ef}}^\infty{\rm d}t\ R_{ij}(t)\ .
\end{equation}
Calculemos el l\'{\i}mite
\begin{equation}
D_{ij}\equiv lim_{T\to\infty}{{\langle\Delta x_i(T)\Delta x_j(T)\rangle}\over{2T}}\ .
\end{equation}
Siendo que 
$$
{\langle\Delta x_i(T)\Delta x_j(T)\rangle}=
\int_0^T{\rm d}t\int_0^T{\rm d}t^\prime\langle v_i(t)v_j(t^\prime)\rangle=
$$
\begin{equation}
\int_0^T{\rm d}t\left[\int_0^t{\rm d}t^\prime R_{ji}(t^\prime)+
\int_0^{T-t}{\rm d}t^\prime R_{ij}(t^\prime)\right]\ ,
\end{equation}
resulta que, si $i\neq j$ y $R_{ij}(t)=-R_{ji}(t)\equiv R(t)$, entonces
\begin{equation}
{\langle\Delta x_i(T)\Delta x_j(T)\rangle}=
\left[\int_0^{T/2}{\rm d}t\int_t^{T-t}{\rm d}t^\prime R(t^\prime)-
\int_{T/2}^T{\rm d}t\int_{T-t}^t{\rm d}t^\prime R(t^\prime)\right]=0\ .
\end{equation}

Si $R_{ij}(t)=R_{ji}(t)\equiv R(t)$, entonces
$$
{\langle\Delta x_i(T)\Delta x_j(T)\rangle}=
$$
\begin{equation}
\left(\int_0^{T_{ef}}{\rm d}t+\int_{T_{ef}}^{T-T_{ef}}{\rm d}t
+\int_{T-T_{ef}}^T{\rm d}t\right)\left[\int_0^t{\rm d}t^\prime R(t^\prime)+
\int_0^{T-t}{\rm d}t^\prime R(t^\prime)\right]\ .
\end{equation}
Asumiendo $T\gg T_{ef}$, el primer t\'ermino queda 
$$
\int_0^{T_{ef}}{\rm d}t\left[\int_0^t{\rm d}t^\prime R(t^\prime)+
\int_0^{T-t}{\rm d}t^\prime R(t^\prime)\right]\leq T_{ef}\left[\int_0^{T_{ef}}{\rm d}t R(t)+
\int_0^T{\rm d}t R(t)\right]
$$
\begin{equation}
\simeq 2\ T_{ef}\int_0^{T_{ef}}{\rm d}t\ R(t)\ ;
\end{equation}
an\'alogamente, se llega al mismo resultado para el tercer t\'ermino. La contribuci\'on del
segundo t\'ermino es 
\begin{equation}
\int_{T_{ef}}^{T-T_{ef}}{\rm d}t\left[\int_0^t{\rm d}t^\prime R(t^\prime)+
\int_0^{T-t}{\rm d}t^\prime R(t^\prime)\right]\simeq 2\ T\int_0^{T_{ef}}{\rm d}t\ R(t)\ ;
\end{equation}
en el l\'{\i}mite $T\to\infty$, \'esta resulta la \'unica contribuci\'on no nula al coeficiente de difusi\'on. 
En este caso, llegamos finalmente a la f\'ormula de Kubo (\ref{kubo}), 
\begin{equation}
D_{ij}=\int_0^\infty{\rm d}t\ R_{ij}(t)\ .
\end{equation}

\chapter{Soluciones exactas de la ecuaci\'on de difusi\'on}
Adoptando un modelo simplificado en el que los campos, los coeficientes de difusi\'on y las fuentes tienen una dependencia espacial 
sencilla, la ecuaci\'on de difusi\'on puede resolverse anal\'{\i}ticamente \cite{ptu93}. Las soluciones exactas que se obtienen en este
modelo simple muestran el origen de la rodilla en el escenario de la difusi\'on turbulenta y drift.   

Teniendo en cuenta una \'unica componente de campo magn\'etico regular, 
asumiremos que su amplitud es $|{\bf B_0}(r,z)|\sim 1/r$. De las ecs.~(\ref{dperp}) y (\ref{da}), 
resulta esencialmente que $D_\perp\propto r_L^{1/3}$ y $D_A\propto r_L$; entonces, podemos despreciar la dependencia espacial
de la difusi\'on transversal ($D_\perp (r,z)=$~cte.) y asumir $D_A(r,z)\propto r$. Para las fuentes, consideramos 
$Q(r,z)=2h_sq(r)\delta(z)$.
 
Si el campo magn\'etico regular no tiene inversiones espaciales (es decir, $b_\phi(r,z)=-1$ en toda la galaxia),
las velocidades de drift (\ref{ureq})-(\ref{uzeq}) quedan reducidas a 
\begin{equation}
u_r=0\ ,\ \ \ u_z=-2D_A/r={\rm cte.}\ ,
\end{equation}
y la ecuaci\'on de difusi\'on (\ref{difeqcil}) resulta
\begin{equation}
\left[{{\partial^2}\over{\partial r^2}}+{{\partial^2}\over{\partial z^2}}+{{1}\over{r}}{{\partial}\over{\partial r}}-
{{u_z}\over{D_\perp}}{{\partial}\over{\partial z}}\right]N(r,z)=-{{2h_s}\over{D_\perp}}q(r)\delta(z)\ .
\label{flathalo}
\end{equation} 
Como una simplificaci\'on adicional, asumiremos que la altura del halo es mucho menor que el radio de la galaxia ($H\ll R$),
de modo que los gradientes radiales puedan despreciarse frente a los gradientes en $z$ \footnote{Otra manera de hacer 
nulos los gradientes radiales es restringirse al caso en el que las fuentes est\'an uniformemente distribuidas en todo el 
disco gal\'actico (es decir, tomando $q(r)={\rm cte.}$).}.
Para $z\neq 0$, la ecuaci\'on se reduce a 
\begin{equation}
{{\partial^2 N}\over{\partial z^2}}={{u_z}\over{D_\perp}}{{\partial N}\over{\partial z}}\ ;
\end{equation}
la soluci\'on es
\begin{equation}
N(r,z)={\mathcal{F}}(r)\times
\left\{\begin{array}{cl}\left(e^{w\left(1-z/H\right)}-1\right)\left(1-e^{-w}\right)\ (z>0)\\
\left(e^w-1\right)\left(1-e^{-w\left(1+z/H\right)}\right)\ (z<0)\end{array}\right.\ ,
\label{fhsol}
\end{equation}
donde $w\equiv 2HD_A/rD_\perp$.
De integrar la ec.~(\ref{flathalo}) en el entorno del disco gal\'actico 
(es decir, calculando $lim_{\epsilon\to 0}\int_{-\epsilon}^\epsilon dz\ ...$) y reemplazar la soluci\'on~(\ref{fhsol}), resulta 
\begin{equation}
{\mathcal{F}}(r)={{h_s\ H/D_\perp}\over{w\ senh(w)}}\ q(r)\ .
\end{equation}

\begin{figure}[t]
\centerline{{\epsfxsize=9truecm\epsfysize=7truecm\epsffile{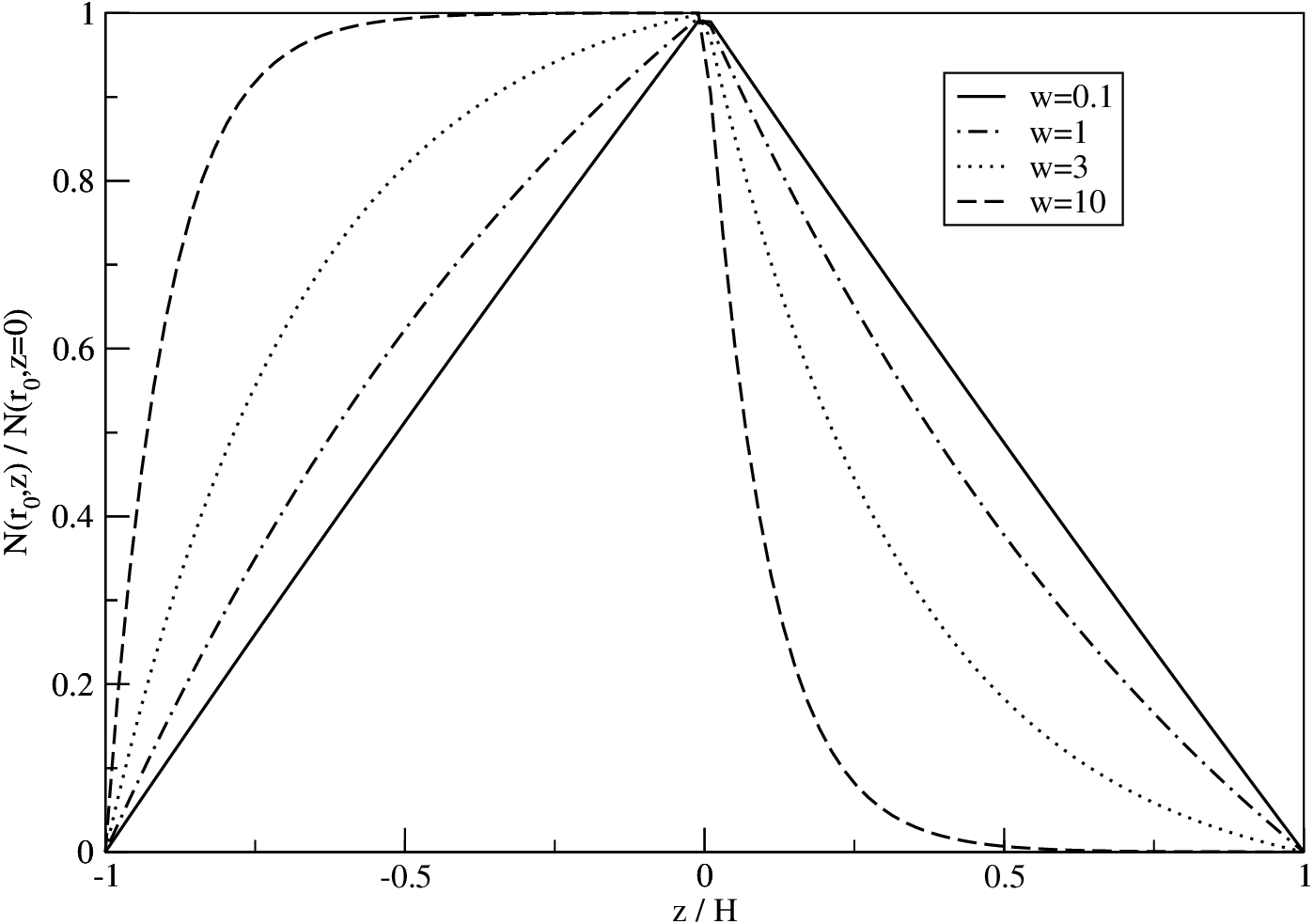}}}
\caption{Perfil vertical de la densidad de rayos c\'osmicos, en el caso de un campo magn\'etico regular sim\'etrico 
con respecto al plano gal\'actico, para diferentes valores de $w\equiv 2HD_A/rD_\perp$.}
\label{fhsim}
\end{figure}

Debido a la ausencia de drifts radiales, la dependencia radial de la densidad de rayos c\'osmicos 
es id\'entica a la distribuci\'on radial de fuentes; en particular, si $q(r_0)=0$ en un dado $r_0$, all\'{\i} es tambi\'en $N(r_0,z)=0$. 
Para $r_0$ tal que $q(r_0)\neq 0$, el perfil vertical de la densidad de rayos c\'osmicos refleja la influencia de la
corriente macrosc\'opica $u_z<0$, tal como muestra la fig.\ref{fhsim}.   

En el plano gal\'actico, la soluci\'on es
\begin{equation}
N(r,z=0)={{h_s}\over{D_A/r}}\ q(r)\ tanh\left({{HD_A/r}\over{D_\perp}}\right)\ .
\end{equation}
Teniendo en cuenta que $D_\perp\propto E^{1/3}$ y $D_A\propto E$, y tomando $q\propto E^{-\beta}$ para el espectro diferencial de 
producci\'on en las fuentes, vemos que, a bajas energ\'{\i}as (para $HD_A/rD_\perp\ll 1$) es $N\propto E^{-\beta-1/3}$, mientras que
a energ\'{\i}as altas (para $HD_A/rD_\perp\gg 1$) resulta $N\propto E^{-\beta-1}$. En consecuencia, la soluci\'on exhibe una
``rodilla'' (correspondiente a un cambio de \'{\i}ndice espectral igual a $2/3$) que ocurre en torno de $HD_A/rD_\perp\approx 1$.   

\begin{figure}[t]
\centerline{{\epsfxsize=9truecm\epsfysize=7truecm\epsffile{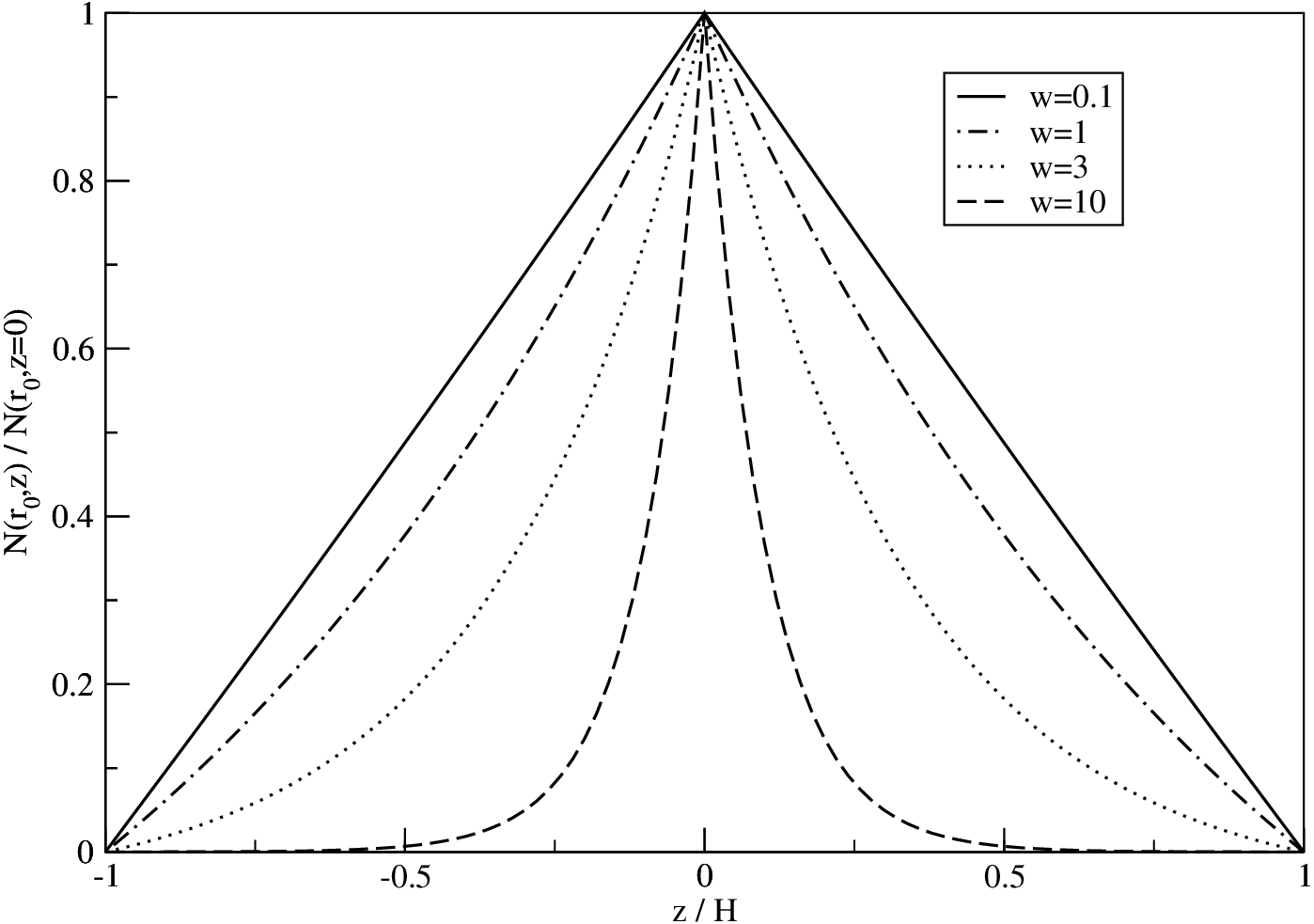}}}
\caption{Perfil vertical de la densidad de rayos c\'osmicos, en el caso de un campo magn\'etico regular antisim\'etrico 
con respecto al plano gal\'actico, para diferentes valores de $w\equiv 2HD_A/rD_\perp$.}
\label{fhasim}
\end{figure} 

En el caso de un campo regular antisim\'etrico respecto del plano gal\'actico (es decir, tal que $b_\phi(r,z)=-sgn(z)$),
los drifts resultan 
\begin{equation}
u_r=2D_A\delta(z)\ ,\ \ \ u_z=-{{2D_A}\over{r}}sgn(z)\ ;
\end{equation} 
es decir, los drifts verticales est\'an ahora dirigidos hacia el plano gal\'actico, y sobre el plano aparece un drift radial
singular, dirigido hacia el exterior de la galaxia. La soluci\'on (en la aproximaci\'on $H\ll R$) es
\begin{equation}
N(r,z)={{h_s}\over{D_A/r}}\left({{1-e^{w(1-|z|/H)}}\over{1-e^w}}\right)\int_0^1{\rm d}y\ q(yr)y^{-1+2/\left(1-e^{-w}\right)}\ .  
\end{equation}
Debido al drift radial, la densidad en un dado $r_0$ recibe la contribuci\'on de todas las fuentes ubicadas en $r\leq r_0$,
mientras que los drifts verticales tienden a concentrar los rayos c\'osmicos en la regi\'on del plano gal\'actico;   
la fig.\ref{fhasim} muestra un perfil vertical caracter\'{\i}stico.